\documentclass[a4paper,11pt]{article}
\pdfoutput=1

\usepackage{jheppub}

\usepackage{axodraw2}
\usepackage{amssymb}
\usepackage{amsmath}
\usepackage{bbold}
\usepackage{graphicx}
\usepackage{hyperref}
\usepackage{cancel}

\allowdisplaybreaks

\newcommand{\revision}[1]{\textcolor{black}{#1}}

\makeatletter

\divide\@tempdima\baselineskip
\@tempcnta=\@tempdima
\def\paragraph{\@startsection{paragraph}{4}{\z@}{+2.00ex plus
 +1ex minus +.2ex}{1.5ex plus .2ex}{\it\normalsize}}

\def\section{\@startsection {section}{1}{\z@}{+3.0ex plus +1ex minus
  +.2ex}{2.3ex plus .2ex}{\normalsize\bf}}
\def\subsection{\@startsection{subsection}{2}{\z@}{+2.5ex plus +1ex
minus +.2ex}{1.5ex plus .2ex}{\normalsize\bf}}
\def\subsubsection{\@startsection{subsubsection}{3}{\z@}{+3.25ex plus
 +1ex minus +.2ex}{1.5ex plus .2ex}{\normalsize\bf}}

\@addtoreset{equation}{section}

\expandafter\ifx\csname mathrm\endcsname\relax\def\mathrm#1{{\rm #1}}\fi

\makeatletter

\def\asymp#1%
{\mathrel{\raisebox{-.4em}{$\widetilde{\scriptstyle #1}$}}}

\def\Nequal#1%
{\mathrel{\raisebox{-.5em}{$\stackrel{=}{\scriptstyle\rm#1}$}}}
\def\dsl{\mathpalette\make@slash}
\def\make@slash#1#2{\setbox\z@\hbox{$#1#2$}%
  \hbox to 0pt{\hss$#1/$\hss\kern-\wd0}\box0}

\def\beq#1\eeq{\begin{equation}#1\end{equation}}
\def\beqar{\begin{eqnarray}}
\def\eeqar{\end{eqnarray}}
\def\barr#1{\begin{array}{#1}}
\def\earr{\end{array}}
\def\bfi{\begin{figure}}
\def\efi{\end{figure}}
\def\btab{\begin{table}}
\def\etab{\end{table}}
\def\bce{\begin{center}}
\def\ece{\end{center}}
\def\nn{\nonumber}

\def\de{\delta}
\def\De{\Delta}
\def\eps{\epsilon}

\def\la{\lambda}

\def\refeq#1{\mbox{(\ref{#1})}}

\def\reffi#1{\mbox{Fig.~\ref{#1}}}

\def\refse#1{\mbox{Sec.~\ref{#1}}}

\def\refapp#1{\mbox{App.~\ref{#1}}}
\def\citere#1{\mbox{Ref.~\cite{#1}}}
\def\citeres#1{\mbox{Refs.~\cite{#1}}}

\newcommand{\GeV}{\unskip\,\mathrm{GeV}}
\newcommand{\MeV}{\unskip\,\mathrm{MeV}}

\newcommand{\ri}{{\mathrm{i}}}

\newcommand{\rd}{{\mathrm{d}}}

\newcommand{\rT}{{\mathrm{T}}}
\newcommand{\rw}{{\mathrm{w}}}

\newcommand{\M}{\mathcal{M}}

\renewcommand{\L}{\mathcal{L}}

\def\mathswitchr#1{\relax\ifmmode{\mathrm{#1}}\else$\mathrm{#1}$\fi}

\newcommand{\PW}{\mathswitchr W}
\newcommand{\Pw}{\mathswitchr w}
\newcommand{\PZ}{\mathswitchr Z}

\newcommand{\Ph}{\mathswitchr h}
\newcommand{\PH}{\mathswitchr H}

\newcommand{\Pe}{\mathswitchr e}

\newcommand{\Pep}{\mathswitchr {e^+}}

\def\mathswitch#1{\relax\ifmmode#1\else$#1$\fi}

\newcommand{\MW}{\mathswitch {M_\PW}}

\newcommand{\MZ}{\mathswitch {M_\PZ}}
\newcommand{\MH}{\mathswitch {M_\PH}}
\newcommand{\Mh}{\mathswitch {M_\Ph}}

\newcommand{\sw}{\mathswitch {s_\Pw}}
\newcommand{\cw}{\mathswitch {c_\Pw}}

\newcommand{\GF}{\mathswitch {G_\mu}}

\newcommand{\sa}{s_\alpha}
\newcommand{\ca}{c_\alpha}

\def\Re{\mathop{\mathrm{Re}}\nolimits}

\newcommand{\LO}{\mathrm{LO}}
\newcommand{\NLO}{\mathrm{NLO}}
\newcommand{\MSbar}{{\overline{\mathrm{MS}}}}

\newcommand{\PRTS}{{\mathrm{PRTS}}}
\newcommand{\FJTS}{{\mathrm{FJTS}}}
\newcommand{\GIVS}{{\mathrm{GIVS}}}
\newcommand{\ct}{\mathrm{ct}}
\newcommand{\dhe}{\delta_h}

\newcommand{\deft}{\delta^\mathrm{EFT}}

\newcommand{\eff}{\mathrm{eff}}
\newcommand{\tree}{\mathrm{tree}}
\newcommand{\ren}{\mathrm{ren}}
\newcommand{\SM}{\mathrm{SM}}
\newcommand{\OS}{\mathrm{OS}}
\newcommand{\SESM}{\mathrm{SESM}}
\newcommand{\BSM}{{\mathrm{BSM}}}
\newcommand{\EFT}{{\mathrm{EFT}}}
\newcommand{\SMEFT}{{\mathrm{SMEFT}}}
\newcommand{\nonSMEFT}{{\,\cancel{\text{SMEFT}}}}

\newcommand{\HEFT}{{\mathrm{HEFT}}}
\newcommand{\EOM}{{\text{EOM}}}

\newcommand{\virt}{\mathrm{virt}}
\newcommand{\real}{\mathrm{real}}

\newcommand{\lin}{{\text{lin}}}

\renewcommand{\max}{\mathrm{max}}
\newcommand{\z}{\setbox0\hbox{+}\hbox to \wd0{\hss0\hss}}
\def\limfunc#1{\mathop{\rm #1}}

\def\tr{\limfunc{tr}}
\def\Re{\limfunc{Re}}

\def\tfrac#1#2{{\textstyle {#1 \over #2}}}

\def\slash#1{{\setbox0=\hbox{$#1$}
  \rlap{\ifdim\wd0>.7em\kern.22\wd0\else\kern.1\wd0\fi /}#1}}

\def\braket#1#2{\left\langle #1\vphantom{#2}
  \right. \kern-2.5pt\left| #2\vphantom{#1}\right\rangle }
\def\L{{\cal L}}

\def\M{{\cal M}}

\def\oneloop{\text{1-loop}}

\def\rT{{\mathrm{T}}}

\def\barW{\overline{W}}

\newcommand{\eq}[1]{Eq.~\eqref{eq:#1}}
\newcommand{\eqs}[2]{Eqs.~\eqref{eq:#1} and \eqref{eq:#2}}
\newcommand{\eqss}[3]{Eqs.~\eqref{eq:#1}, \eqref{eq:#2}, and \eqref{eq:#3}}

\newcommand{\eqsm}[2]{Eqs.~\eqref{eq:#1}\,--\,\eqref{eq:#2}}
\renewcommand{\sec}[1]{Sec.~\ref{sec:#1}}
\newcommand{\secs}[2]{Secs.~\ref{sec:#1} and \ref{sec:#2}}
\newcommand{\secsm}[2]{Secs.~\ref{sec:#1}\,--\,\ref{sec:#2}}

\newcommand{\subsec}[1]{Sec.~\ref{subsec:#1}}
\newcommand{\subsecs}[2]{Secs.~\ref{subsec:#1} and \ref{subsec:#2}}
\newcommand{\subsecsm}[2]{Secs.~\ref{subsec:#1}\,--\,\ref{subsec:#2}}
\newcommand{\fig}[1]{Fig.~\ref{fig:#1}}

\newcommand{\app}[1]{App.~\ref{app:#1}}

\newcommand{\rcites}[1]{Refs.~\cite{#1}}
\newcommand{\rcite}[1]{Ref.~\cite{#1}}

\newcommand{\ord}{\mathcal{O}}

\DeclareMathAlphabet{\mathbbold}{U}{bbold}{m}{n}
\newcommand{\bbid}{\mathbbold{1}}

\DeclareFontFamily{OT1}{pzc}{}
\DeclareFontShape{OT1}{pzc}{m}{it}{<-> s * [1.20] pzcmi7t}{}
\DeclareMathAlphabet{\mathpzc}{OT1}{pzc}{m}{it}

\begin{document}

\thispagestyle{empty}
\def\thefootnote{\fnsymbol{footnote}}
\setcounter{footnote}{1}
\null
\strut\hfill FR-PHENO-2026-001
\vskip 0cm
\vfill
\begin{center}
{\Large \bf
\boldmath{Integrating out a heavy Higgs singlet: \\
on the edge between SMEFT and HEFT
}
\par} \vskip 2.5em
{\large
{\sc Stefan Dittmaier, Sebastian Schuhmacher, \\ and Maximilian Stahlhofen}\\[1ex]
{\normalsize
\it
Albert-Ludwigs-Universit\"at Freiburg,
Physikalisches Institut, \\
Hermann-Herder-Stra\ss{}e 3,
79104 Freiburg, Germany
}
}

\par \vskip 1em
\end{center} \par
\vskip 2cm
\noindent
{\bf Abstract:} \par
We use a functional approach based on the background-field formalism and
the expansion by regions to integrate out the
heavy Higgs field (associated with the mass eigenstate~H)
\revision{at the one-loop level
in a singlet extension of the Standard Model (SM),
which features an additional real scalar singlet and a spontaneously
broken $\mathbb{Z}_2$ symmetry.
We obtain} an effective Lagrangian to $\mathcal{O}(1/\MH^2)$ in the limit of large Higgs mass ($\MH \gg \Mh \approx 125$\,GeV) providing a consistent treatment of effects from Higgs mixing and the renormalization of the underlying model.
\revision{The scaling behaviour}
of the model parameters in the large-$\MH$ limit determines whether the effective Lagrangian can be accommodated in the
SM Effective Field Theory (SMEFT)
or involves non-SMEFT operators within the more general Higgs
Effective Field Theory (HEFT) framework.
\revision{We choose a limit that ensures decoupling of beyond-SM (BSM) effects}
at $\mathcal{O}(1/\MH^0)$
by demanding that the Higgs mixing angle~$\alpha$ is of $\mathcal{O}(\Mh/\MH)$ and putting minimal constraints \revision{on the other
input parameters.}
The considered model is restricted to massless fermions.
\revision{
Any attempt of a bottom-up (diagrammatic) matching
with only bosonic SMEFT operators at $\mathcal{O}(1/\MH^2)$ necessarily fails,
although the BSM sector does not directly couple to massless fermions.
However, it is possible to match SMEFT with bosonic and fermionic operators
to the SESM in the considered large-$\MH$ limit.
For the emerging Effective Field Theory (EFT) we give two alternative Lagrangians:
one that includes only bosonic BSM EFT operators, but necessarily involves
operators of non-SMEFT type, and another one that is of SMEFT form, but involves
also fermionic EFT operators.}
We validate our results for the effective Lagrangians at next-to-leading order
in the coupling expansion by verifying that
the difference between EFT and full-theory predictions
vanishes faster than $1/\MH^2$
\revision{for several electroweak observables in the large-$\MH$ limit.}

\par
\vfill
\vskip .5cm
\null
\setcounter{page}{0}
\clearpage
\def\thefootnote{\arabic{footnote}}
\setcounter{footnote}{0}

\tableofcontents
\clearpage

\section{Introduction}

The Standard Model (SM) of particle physics has so far been very successful in describing
data from all kinds of collider experiments.
However, observations beyond colliders, such as the evidence for Dark Matter, neutrino masses, the
matter--antimatter asymmetry in the observable universe,
as well as the fact that the SM does not include gravity
suggest that
there is physics beyond the SM (BSM).
While the
characteristic energy scale of potential BSM physics is unknown,
the lack of clear signals in the data indicates that
it might well lie beyond the reach of high-energy colliders available now and in the near future,
rendering direct discoveries uncertain if not impossible.
Under these conditions, model-independent approaches that
probe the current and future (low-energy) precision data for
hints towards high-mass BSM effects have become essential tools in particle theory.

The most popular techniques are formulated in the Effective Field Theory (EFT)
framework which parametrizes potential
effects of heavy new particles
by adding effective operators to the SM Lagrangian.
The operators are accompanied by Wilson coefficients suppressed by factors
$\sim \Lambda^{4-n}$ with $n>4$,
where $\Lambda$ denotes the characteristic mass scale of BSM physics.
The BSM scale $\Lambda$ is assumed to be much larger than the
electroweak (EW) scale represented by the SM vacuum expectation value (vev),
$v=246\GeV \ll \Lambda$, which at the same time shall serve as a proxy for the typical (low-)energy scale, where the new physics is being probed.
This provides a suppression of the effects of BSM operators by factors
$(v/\Lambda)^{n-4}\ll1$ and suggests an EFT expansion for $\Lambda\gg v$
with the SM at leading mass order (or ``leading power'').
In EFTs for BSM theories,
the power~$n$ is often identical to the mass dimension of
the BSM operators, but this is not necessarily the case.
Although existing precision data suggest that the BSM operators should respect the symmetries of the SM,
this does not uniquely fix the set of BSM operators appearing in the EFT without further assumptions.
On the other hand, a model-independent confrontation of the EFT with experimental data to
constrain or determine the Wilson coefficients necessarily requires an appropriate BSM operator basis.
The question which BSM operators appear in the large-mass limits of BSM theories is, thus, of
great importance, both theoretically and phenomenologically.

Adjusting the Wilson coefficients of generic EFTs such as to reproduce the low-energy behaviour of explicit BSM models
has become a focal point of interest and an important component of precision BSM phenomenology
in recent years.
This {\it matching} between full theory and EFT can be accomplished
basically in two different ways: Either in a \textit{bottom-up} approach,
i.e.,
by making an ansatz for the EFT operator basis to some given power in
$\Lambda^{4-n}$ and fixing
the Wilson coefficients by equating EFT and
(large-mass expanded)
full-theory results for a representative set of physical S-matrix elements or
Green functions,
a method known as {\it diagrammatic matching}
(see, e.g. \citeres{Haisch:2020ahr,Carmona:2021xtq,DeAngelis:2023bmd,Li:2023edf,Chala:2024llp,LopezMiras:2025gar}),%
\footnote{We will frequently use the terms \textit{bottom-up matching} and  \textit{diagrammatic matching} interchangeably for any kind of matching procedure (not necessarily involving Feynman diagrams) that is based on making an ansatz for the effective Lagrangian, as opposed to functional (``top-down'') matching that does not require an ansatz.}
or by directly integrating out the fields of the heavy degrees of freedom in the path integral,
which is known as {\it functional matching}~\cite{Dittmaier:1995cr,Dittmaier:1995ee,Drozd:2015rsp,
Boggia:2016asg,Henning:2016lyp,Fuentes-Martin:2016uol,Buchalla:2016bse,Zhang:2016pja,Ellis:2017jns,
DasBakshi:2018vni,Jiang:2018pbd,Cohen:2020fcu,%
Dittmaier:2021fls,Fuentes-Martin:2022jrf,Fuentes-Martin:2024agf}.
While diagrammatic matching requires an ansatz for the EFT operator basis, functional matching
is agnostic in this respect and produces the operator basis as part of its result,
a fact that we consider a great advantage.
Nevertheless, an EFT Lagrangian obtained via functional matching
is not unambiguously determined by the physically unique S-matrix of the theory.
The form of Lorentz-invariant Lagrangians can always be changed
by integration by parts (IBP) applied to field monomials or field
redefinitions, which leave the S-matrix invariant.
For complex field theories this can make the question quite non-trivial whether two different Lagrangians
are physically equivalent or not.

Moreover, we want to stress already at this point that the EFT operator basis emerging in a given large-mass
limit of an underlying theory might significantly depend on the precise way in which the large-mass limit
is realized, i.e., it is not sufficient to name the heavy degrees of freedom. It is rather necessary
to precisely state how all independent BSM parameters of the full theory behave in the large-mass limit.
The question about the influence of the precise form of a large-mass limit
on the emerging EFT operator basis and the approximative quality of EFT predictions
in reproducing results from UV-complete models has already been discussed in the literature
from various different angles (see, e.g.,
\citeres{Boggia:2016asg,Buchalla:2016bse,Cohen:2020xca,Dittmaier:2021fls,
Dawson:2023oce,Ge:2026qfa,Asiain:2026sio}).

Identifying an appropriate EFT operator basis is of particular relevance in
BSM models with modified Higgs sectors due to the delicate nature of spontaneous symmetry breaking.
In the proper low-energy EFTs for such models, the SM-like Higgs field might not be part of the full SM Higgs doublet $\phi_{\mathrm{lin}}$,
but rather be a Higgs singlet field~$h$, while the Goldstone fields $\varphi_a$ $(a=1,2,3)$ separately furnish a non-linear realization of SU(2).
Both variants are fully compatible with Lorentz invariance
as well as the gauge and discrete symmetries of the SM.
In fact, the SM itself admits either formulation, as the two field parametrizations are related by a non-linear field redefinition~(see, e.g., \citeres{Lee:1972yfa,Grosse-Knetter:1992tbp,Dittmaier:2022maf}).
While the SM Lagrangian is of the usual polynomial type if
the physical Higgs field is embedded into a complex SU(2) doublet
$\phi_{\mathrm{lin}}$,
the Lagrangian contains arbitrarily
high powers of Goldstone fields $\varphi_a$ if
the physical Higgs field is represented by a gauge singlet.
Adding BSM operators in the two different field parametrizations of the SM while retaining the
underlying SM gauge symmetry, opens more possibilities in the non-linear realization where the Higgs
field is a gauge singlet:
On the one hand, every Lagrangian with a linearly realized Higgs sector can be transformed to the non-linear realization by
an appropriate non-linear field
transformation. On the other hand, the opposite direction, i.e., applying
the inverse field
transformation to a Lagrangian containing arbitrary combinations of the Higgs singlet $h$ and the Goldstone fields $\varphi_a$, does not necessarily lead to a Lagrangian that is expressible as a polynomial of a linearly realized Higgs doublet $\phi_{\mathrm{lin}}$.

Assuming the EFT Lagrangian is expressible in terms of $\phi_{\mathrm{lin}}$,
the most general EFT Lagrangian with SM symmetries and particle content defines the
{\it Standard Model Effective Field Theory (SMEFT)}~\cite{%
Buchmuller:1985jz,Grzadkowski:2010es,%
  LHCHiggsCrossSectionWorkingGroup:2013rie,LHCHiggsCrossSectionWorkingGroup:2016ypw,%
  Brivio:2017vri},
where the power~$n$ in the large-mass suppression of a Wilson coefficient
($\sim \Lambda^{4-n}$)
consequently corresponds to the mass dimension of its  operator.
Lifting this assumption defines the {\it Higgs Effective Field Theory (HEFT)}~\cite{%
Grinstein:2007iv,Alonso:2012px,Contino:2013kra,Buchalla:2013rka,%
Brivio:2013pma,Ghezzi:2015vva,Brivio:2016fzo,%
LHCHiggsCrossSectionWorkingGroup:2016ypw},
where $n$ is not necessarily related to the operator dimension.
Since the difference between SMEFT and HEFT concerns the existence of the distinct field parametrization
in terms of $\phi_{\mathrm{lin}}$,
the question whether an EFT is of either type might be answered by
geometrical methods applied to field space.
Steps in this direction have been taken in
\rcites{Alonso:2015fsp,Alonso:2016oah,Helset:2020yio,Cohen:2020xca},
but a general straightforward and algorithmic way to answer the
question is currently unkown.

In this work we apply functional matching to derive
the low-energy EFT of a BSM model which entails a simple extension of the
SM Higgs sector, in order to showcase subtleties related to mixing of
BSM degrees of freedom with the
neutral CP-even field of the SM-type Higgs doublet.
Concretely, we consider a Higgs Singlet Extension of the SM (SESM)~\cite{Schabinger:2005ei,Patt:2006fw,Bowen:2007ia,Pruna:2013bma,%
Kanemura:2015fra,Bojarski:2015kra,Altenkamp:2018bcs,Denner:2018opp} which
extends the SM by adding a single real scalar singlet field with non-zero vev, which in general mixes with the Higgs field
contained in the SM-type scalar SU(2) doublet.
In the present work we only consider the SESM with massless fermions.
Switching to mass eigenstates, the SESM thus contains two
CP-even Higgs bosons, h and H. We will assume that h is the lighter one and corresponds to the known
Higgs boson with mass $\approx 125\GeV$, while the mass $\MH$ of H is taken to be much larger than the
EW scale~$v$.
In \rcite{Dittmaier:2021fls} we have introduced our calculational approach and derived the EFT of this model (for a well-motivated large-$\MH$ limit) to mass order $\MH^0$, thereby investigating under which circumstances ``decoupling'' holds, i.e., the SM is reproduced at leading power.

To a great deal, our method described in \rcite{Dittmaier:2021fls}
is a further development of the method introduced in \rcites{Dittmaier:1995cr,Dittmaier:1995ee},
where a heavy Higgs field was integrated out directly in the path integral
of an SU(2) gauge theory and the SM, respectively. The use of the background-field method (BFM)~\cite{DeWitt:1967ub,DeWitt:1980jv,Abbott:1980hw,Denner:1994xt}
had already been advocated there, but the method described in \rcite{Dittmaier:2021fls}
incorporates several generalizations and improvements. Some of these generalizations
had already been proposed in
\rcite{Fuentes-Martin:2016uol} (see also \rcites{Zhang:2016pja,Cohen:2020fcu}),
such as the consistent use of the method of regions~\cite{Beneke:1997zp,Smirnov:2002pj},
which separates soft (small-momentum) and hard (large-momentum) modes in loop integrals, or the systematic heavy-mass expansion
of inverse operators via the Neumann series.
Another feature of \rcite{Dittmaier:2021fls} taken over from \rcites{Dittmaier:1995cr,Dittmaier:1995ee}, but
not considered in \rcites{Fuentes-Martin:2016uol,Zhang:2016pja,Cohen:2020fcu},
concerns the combination of background-field gauge invariance with
a non-linear realization of the SM-type Higgs doublet.
This leads to further technical
simplifications, because intermediate manipulations can be
carried out in the unitary background gauge, while full gauge invariance is restored
at the end of the calculation.
The method~\cite{Dittmaier:2021fls} we apply is fully algorithmic and, being based on functional matching,
flexible in the sense that the underlying low-energy operators need not be specified in advance.
Whether the emerging EFT is of SMEFT or HEFT type is part of the result without initial
assumptions; the approach is even applicable to BSM theories with non-decoupling large-mass effects
in which the SM is not reproduced at low energies.
\smallskip

\noindent
\textbf{Structure of the article:}
\smallskip

The main body of this paper consists of three parts. The first introductory part,
comprising \secs{recap}{prelim},
introduces the SESM and the considered EFT limit as well as
the conventions necessary to understand the final result.
The middle, rather technical part, covers the derivation of the effective Lagrangian by explicitly integrating out the heavy mass eigenstate in the SESM along the lines of~\rcite{Dittmaier:2021fls} and encompasses \secsm{calculation}{fieldredef}.
Readers less interested in the technical aspects of the derivation
may directly proceed, after reading the first part,
to the third part in \secsm{final}{LeffSMEFT}.
\revision{In this part, we present two ready-to-use forms of the EFT Lagrangian:
one that involves only bosonic BSM EFT operators, albeit not all of SMEFT type,
and another one that is of SMEFT form, but involves both bosonic and fermionic SMEFT operators.
The EFT is validated against next-to-leading order (NLO) predictions for several EW precision observables
within the full SESM in the large-$\MH$ limit.
The third part also contains detailed
discussions of our findings and a comparison to existing literature.}

\smallskip
\noindent
\textbf{Detailed outline of the sections:}
\smallskip

After recapitulating the salient features and individual conceptual steps of our approach
in \sec{recap}, we set the starting point of the EFT derivation in \sec{prelim}
by further discussing the SMEFT and HEFT frameworks, describing details of the SESM, and
discussing details of the chosen large-mass limit.
Section~\ref{sec:calculation} contains the actual solution of the functional integral for the
heavy Higgs field up to the one-loop level, with the unrenormalized effective Lagrangian as a result.
The contributions to the effective Lagrangian related to the renormalization of the underlying theory
are derived in \refse{sec:renormalization}.
Section~\ref{sec:fieldredef} subsequently describes the elimination of redundant operators via
field redefinitions of the light fields and the transition from the BFM formulation back to
the conventional formalism of Faddeev--Popov quantization without the splitting into background and quantum fields.
\revision{In \sec{final} we present and briefly discuss the resulting effective Lagrangian, which includes bosonic BSM EFT operators only, but is not of SMEFT form.}
In \sec{pheno} we show phenomenological applications
which also serve to validate our effective Lagrangian.
Concretely, we compare NLO predictions
obtained in the full SESM with corresponding EFT predictions for several precision observables
in the large-$\MH$ limit:
the W-boson mass derived from muon decay, some W/Z-boson decay widths, the effective weak mixing angle,
and the leptonic four-body Higgs decay $\Ph\to\PW\PW\to\nu_\Pe\Pe^+\mu^-\overline{\nu}_\mu$.
Moreover,
we use these results to show that assuming massless fermions no EFT with bosonic SMEFT operators only
can reproduce the large-$\MH$ limit of all considered observables simultaneously.
\revision{Section~\ref{sec:LeffSMEFT} extends the construction of the EFT Lagrangian by allowing
for fermionic BSM EFT operators, revealing that the emerging EFT can eventually be brought
into SMEFT form.
After discussing our final result in detail and comparing it to the literature} in \subsecs{discussion}{Comparison}, respectively, we present our conclusions in \sec{conclusions}.
The appendices provide further details and (intermediate) results that seem too lengthy for the main text, but allow for following our derivation at a deeper level and offer possibilities for cross-checks with other calculations.

\section{Recap of the general method}
\label{sec:recap}

Let us briefly summarize the salient features and
particular strengths of our method~\cite{Dittmaier:2021fls} to derive the effective Lagrangian
at the one-loop level, some of them common to the approaches
of \rcites{Fuentes-Martin:2016uol,Zhang:2016pja,Cohen:2020fcu}, others
imported from \rcites{Dittmaier:1995cr,Dittmaier:1995ee} or
specific to \rcite{Dittmaier:2021fls}:
\begin{enumerate}
	\renewcommand{\labelenumi}{\theenumi}
	\renewcommand{\theenumi}{(\roman{enumi})}
	\item
	a clear separation of tree-level and loop effects of the heavy field modes by employing the BFM;
	\item
	the possibility to fix the (background) gauge in intermediate steps of the
	calculation and to restore gauge invariance of the effective Lagrangian at the end;
	\item
	transparency in the sense that at each stage of the calculation it is possible to
	identify the origin of all contributions to the effective Lagrangian in terms of (classes of)
	Feynman diagrams;
	\item
	flexibility due to the fact that no ansatz is made for the effective Lagrangian.
	Whether the emerging EFT is of SMEFT or HEFT type is part of the result.
	\item
	An automation of the method is possible, since it is fully algorithmic. In principle,
	given a BSM Lagrangian, a proper definition of the large-mass limit with a corresponding power-counting scheme,
	and some details on the renormalization
	of the BSM sector, the actual determination of the effective Lagrangian
	at the one-loop level can be carried out by computer algebra.
\end{enumerate}

These features of the matching approach~\cite{Dittmaier:2021fls} are partly due to the following
 generalizations and
optimizations w.r.t.\ \rcite{Fuentes-Martin:2016uol} and the original approach
of \rcites{Dittmaier:1995cr,Dittmaier:1995ee}:
\begin{enumerate}
\item
\textit{Background-field formalism and
	non-linear Higgs realization.}
\\
Formulating the theory within the BFM
splits all fields into
background (i.e.\ in some sense semi-classical) and quantum parts,
which correspond to tree-like lines and lines within loops in Feynman
graphs, respectively.
Since vertices that are part of a one-loop subdiagram, thus, involve exactly
two quantum fields, the BFM nicely separates field modes according to
their role in higher-order corrections.

Employing a non-linear realization of the SU(2) gauge symmetry in the scalar
sector~\cite{Dittmaier:1995cr,Dittmaier:1995ee},
it is possible to absorb
all background Goldstone-boson fields into the background
gauge fields by a straightforward St\"uckelberg transformation
which reduces the amount of algebraic work
in the subsequent steps considerably.

\item
\textit{Separation of hard and soft field modes.}
\\
Following the idea of the method of regions~\cite{Beneke:1997zp,Smirnov:2002pj}
for large-mass expansions of Feynman diagrams, we split both quantum and
background fields into soft and hard field modes.%
\footnote{In \rcite{Dittmaier:2021fls} we also referred to these modes as ``light'' and ``heavy'' respectively, although this terminology is less standard and indeed somewhat misleading, because strictly speaking ``soft'' and ``hard'' refers to the momenta of the modes and not their masses.
Accordingly, we use here the subscript labels
``$s$'' and ``$h$'' (instead of ``$l$'' and  ``$h$'' as in \rcite{Dittmaier:2021fls}) to indicate soft and hard field modes ($\phi_s$ and $\phi_h$), respectively.}
This separation disentangles EFT contributions that correspond to
higher-order effects from hard interactions
ending up in Wilson coefficients of effective operators on one side and effective operators
entering loops built from soft fields on the other.

The splitting of one-loop diagrams into two integration
domains of small and large momenta can be interpreted as a splitting
of the path integral into two functional integrals extending over
soft and hard field modes.
The consistent mode separation according to the method of regions
is a conceptual generalization of the procedure of
\rcites{Dittmaier:1995cr,Dittmaier:1995ee},
where only hard
field modes appeared in the calculation of the
non-decoupling effects in the
effective Lagrangian to leading mass order at one loop.
This mode separation has already been introduced in
\rcite{Fuentes-Martin:2016uol} (and applied in  \rcites{Zhang:2016pja,Cohen:2020fcu}) within
 a functional approach to calculate the effective Lagrangian.

\item
\textit{Integrating out the hard modes of the heavy quantum fields in the path integral.}
\\
Since the part of the Lagrangian that is relevant at one-loop order
is only quadratic in the quantum fields, the path integral over the
hard field modes of the
heavy quantum fields is of Gaussian type and can be done
analytically. The major complication in this step is the fact that there are
also terms that are linear in the heavy quantum fields.
These terms can be removed by a field redefinition of the heavy quantum field modes
in a fully algorithmic manner.
This means that the resulting part of the Lagrangian quadratic in the heavy quantum fields can be directly identified based again on a simple power-counting argument.
This algorithmic handling, which has also been realized in \rcites{Fuentes-Martin:2016uol} (see also \rcite{Zhang:2016pja} and, for a slightly different approach, \rcite{Cohen:2020fcu}),
establishes an important technical improvement
over the procedure described in \rcites{Dittmaier:1995cr,Dittmaier:1995ee}.

\item
\textit{Equations of motion for the soft modes of the
	heavy fields and renormalization.}
\\
After the hard modes
of the heavy field have been integrated out, the effective
Lagrangian still involves the soft modes
of the quantum and background fields of the heavy particle.
As their momenta are much smaller than their mass,
they do not represent
dynamical degrees of freedom in the EFT.
In fact, they can conveniently be removed from the effective Lagrangian by
solving the equations of motion for the heavy fields
up to the relevant order in the large-mass expansion and inserting the
solution into the effective Lagrangian.

This step requires a proper power-counting of all parameters and fields in the heavy-mass limit.
As emphasized in \rcite{Dittmaier:2021fls},
in order to obtain a consistent effective Lagrangian the
large-mass expansion must be carefully performed taking into account that the full-theory renormalization constants
may have a different scaling behaviour
than the corresponding renormalized quantities.
``Good schemes'' should respect the decoupling behaviour of the theory at
leading mass order in the EFT expansion.
In this context, the choice of a proper
renormalization of the Higgs vevs, i.e.\ a proper tadpole scheme, is crucial in $\MSbar$
renormalization~\cite{Dittmaier:2021fls} of full-theory parameters.

\item
\textit{Final form of the effective Lagrangian.}
\\
The effective Lagrangian resulting from the previous steps
only involves soft background and quantum fields,
but none of the modes of the heavy fields.
The effects of the latter are absorbed into the Wilson coefficients
of the effective operators.

Finally, redefinitions of the light quantum fields and IBP in the effective Lagrangian
might be used to eliminate redundant operators that only influence
off-shell Green functions, but no physical scattering amplitudes. This step
is, in particular, necessary to bring the effective Lagrangian into a standard form,
i.e.\ to SMEFT or (more generally) HEFT form.
\revision{However, finding an appropriate ansatz for the necessary field transformations
can be rather non-trivial in the general case,
where the question of SMEFT versus HEFT is unclear, is not yet available.}

\end{enumerate}

\section{Preliminaries}
\label{sec:prelim}

\subsection{SMEFT vs. HEFT}
\label{subsec:SMEFTvsHEFT}

In the bottom-up approach,
SMEFT is constructed by supplementing the SM Lagrangian with $ k_n  $ operators
$ \mathcal{O}^{(n)}_{i} $ of mass dimension~$n$ $(4<n\leq N)$
suppressed by $n-4$ powers of some high-energy scale $ \Lambda $:
\begin{equation}
  \label{eq:EFTexp}
	\L_\eff \,=\, \L_\SM
	+\sum_{n=5}^{N}\sum_{i=1}^{k_n}\frac{C^{(n)}_{i}}{\Lambda^{n-4}}
    \;\mathcal{O}^{(n)}_{i}
	+\mathcal{O} \bigl(\Lambda^{3-N} \bigr) \,,
\end{equation}
where the effective operators $\mathcal{O}_i^{(n)}$ are composed of SM fields and
are invariant under the SM gauge group.
Here, the Wilson coefficients $C^{(n)}_{i} \sim \Lambda^0$ are normalized to be dimensionless.
The key difference between SMEFT and the more general HEFT lies in the embedding of the field corresponding to the physical Higgs boson into a scalar SU(2) multiplet.
SMEFT assumes the Higgs field and the Goldstone fields to form a complex doublet
under SU(2), just like in the SM.
By contrast, the Goldstone fields in HEFT furnish a real non-linear SU(2) realization, while the Higgs field is a singlet under the full SM gauge group.
The constraints imposed on HEFT are more lenient, allowing for effective operators that encompass all SMEFT operators, so that to any given order in $n$ we have
$\L_\HEFT\supset\L_\SMEFT$~\cite{%
Grinstein:2007iv,Alonso:2012px,Contino:2013kra,Buchalla:2013rka,%
Brivio:2013pma,Brivio:2016fzo,%
LHCHiggsCrossSectionWorkingGroup:2016ypw}.
This additional freedom in HEFT can be used to accommodate models with
alternative symmetry breaking mechanisms, replacing the Higgs mechanism
as employed in the SM.

In many popular extensions of the scalar sector of the SM, like the SESM or
Two-Higgs-Doublet Models,
the Higgs component of the complex SM-type SU(2) doublet does not directly correspond to a mass eigenstate.
Instead, these models typically include mixing of scalar fields after the spontaneous breaking of the SU(2) symmetry.
Integrating out the field of a heavy Higgs boson can, thus, lead to an EFT where the SM-like Higgs field does not entirely originate from the complete doublet with the Goldstone bosons.
The EFT is therefore not necessarily of SMEFT type but may rather require non-SMEFT operators within the HEFT framework.
As the Higgs mixing happens after the spontaneous symmetry breaking, it is intrinsically a low-energy ($E \ll \Lambda$) effect.
On the other hand,
for energy scales well above the SM vev,
but still far below the new physics scale~$\Lambda$,
one is tempted to treat the difference between gauge and mass eigenstates as negligible.
In that case, the derivation of an effective Lagrangian simplifies significantly,
and the result can be cast in SMEFT form, see e.g.\ \rcites{Cohen:2020xca,Ellis:2017jns,Jiang:2018pbd,Haisch:2020ahr}.
We stress, however, that such mixing effects are not \revision{necessarily} suppressed by any large mass scale (as compared to other new-physics effects)
and therefore need to be taken into account to derive a consistent EFT that is valid at energies of the order of or below the SM vev.
The SM-like Higgs field corresponding to the mass eigenstate (with mass $\Mh \approx 125$ GeV) might then not transform as genuine
part of a doublet with the Goldstone bosons under SU(2),
and the corresponding low-energy EFT
\revision{might}
not necessarily be of SMEFT type.
In the present work, we will analyze this situation
for a simple SESM in the phenomenologically most interesting variant of the
large-Higgs-mass ($\MH$) limit.

\subsection{The Higgs singlet extension of the Standard Model}
\label{subsec:SESM}

The SESM is arguably the simplest BSM model with an extended Higgs
sector~\cite{Schabinger:2005ei,Patt:2006fw,Bowen:2007ia,Pruna:2013bma,%
Kanemura:2015fra,Bojarski:2015kra,Altenkamp:2018bcs,Denner:2018opp}.
Here we give the details of the SESM
version~\cite{Altenkamp:2018bcs,Denner:2018opp,Dittmaier:2021fls,Dittmaier:2022ivi}
that we use for our study.
We follow the conventions of \rcites{Bohm:1986rj,Denner:1991kt,Denner:2019vbn} for the SM-type part of the Lagrangian,
and employ a non-linear parametrization
of the SU(2) Higgs doublet, as e.g.\ in \rcites{Grosse-Knetter:1992tbp,Dittmaier:1995cr,Dittmaier:1995ee,Dittmaier:2021fls,Dittmaier:2022ivi,Dittmaier:2022maf}:
\begin{equation}
	\Phi=\frac{1}{\sqrt{2}}\left(v_{2}+h_{2}\right)U \,, \qquad
	U=\exp\biggl(2\ri\frac{\varphi}{v_{2}}\biggr) \,, \qquad
	\varphi=\frac{1}{2}\varphi_a\tau_a \,,
	\label{eq:doublet}
\end{equation}
where the $ \tau_a $ are the Pauli matrices, and  the $ \varphi_a $ denote real Goldstone fields. These are related to the usual Goldstone fields $ \phi^\pm,\, \chi $ of the Higgs doublet in the linear
realization by~\cite{Dittmaier:2021fls,Dittmaier:1995ee}
$ \phi^\pm= (1+h_2/v_2)(\varphi_2\pm\ri\varphi_1)/\sqrt{2} + \ord(\varphi_a^3) $, $\chi=-(1+h_2/v_2)\varphi_3 + \ord(\varphi_a^3)$.
 The non-linearly realized Higgs field $ h_2 $ and its vacuum expectation value $ v_2 $ are singlets under the SM gauge group by construction. We emphasize that although $ h_2 $ transforms as a singlet,
in the SESM (as in the SM) it only appears
within the combination $ \Phi $ representing an SU(2) doublet.
This can be made explicit by switching (back) to the linear Higgs realization via a field redefinition without changing physical predictions.

The covariant derivative of the matrix-valued fields $\Phi$ and $U$ is
\begin{equation}
	D_{\mu}X=\partial_{\mu}X-\ri g_{2}W_{\mu}X-\ri g_{1} X \frac{\tau_{3}}{2}B_{\mu}\,, \qquad X=\Phi, U ,
  \label{eq:CovDivMatrixfield}
\end{equation}
where $ W_\mu=W_\mu^a \tau_a/2$ and $g_2$ are the SU(2) gauge field and coupling, and $B^\mu$, $g_1$ are the U(1) gauge field and coupling,  respectively.
In terms of these gauge couplings the electromagnetic coupling $e$ is
expressed as
\begin{equation}
e = g_2 \,\sw = g_1 \,\cw \,,
\qquad
\sw\equiv\sin \theta_\mathrm{w}\,,  \quad \cw\equiv\cos \theta_\mathrm{w} \,,
\end{equation}
with the weak mixing angle $\theta_\mathrm{w}$.
Throughout this work we will use the acronyms ``LO'' and ``NLO'' for leading and next-to-leading order exclusively in the context of the perturbation series in the coupling constant $\alpha_\mathrm{em} \equiv e^2/(4\pi) \sim g_1^2 \sim g_2^2$, which for virtual corrections coincides with the loop expansion (as opposed to the large-mass expansion).
We use the symbol ``$\sim$'' to indicate the scaling of
quantities in the loop or the large-mass expansion introduced below (rather than the full asymptotics).

In the SESM the SM Higgs Lagrangian
is extended to
\begin{align}
	\L_\mathrm{Higgs}^\SESM={}&
	\frac{1}{2}\tr\left[\left(D_{\mu}\Phi\right)^\dagger\left(D^{\mu}\Phi\right)\right]
	+\frac{1}{2}\mu_{2}^2\tr\left[\Phi^\dagger\Phi\right]
	-\frac{1}{16}\lambda_{2}  \Big(\tr\left[\Phi^\dagger\Phi\right]\Big)^2
	\nn\\
	&+\frac{1}{2}\left(\partial_{\mu}\sigma\right)\left(\partial^{\mu}\sigma\right)+\mu_{1}^2\sigma^2-\lambda_{1}\sigma^4-\frac{1}{2}\lambda_{12}\sigma^2\tr\left[\Phi^\dagger\Phi\right],
	\label{eq:SESM}
\end{align}
with a new real singlet
scalar field $\sigma$, while the gauge sector equals the one of the SM.
For simplicity,
we consider the fermionic sector of the SESM in the approximation of massless fermions
and leave the (straight-forward) inclusion of
massive fermions in the effective Lagrangian to future work.
The Higgs Lagrangian \eqref{eq:SESM}
represents the most general renormalizable choice with a $ \mathbb{Z}_2 $ symmetry under $ \sigma\rightarrow-\sigma $.
Vacuum stability requires
\begin{equation}
	\lambda_1>0 \,,
	\qquad
	\lambda_2>0 \,,
	\qquad
	\lambda_1\lambda_2>\lambda_{12}^2 \,.
\end{equation}
The mass parameters $ \mu_{1,2}^2>0 $ then induce non-vanishing vevs $ v_{1,2} $ for the SM-type Higgs doublet as well as the new singlet,
motivating the decomposition
\begin{align}
	\sigma=v_1+h_1 \,,
	\label{eq:singlet}
\end{align}
with field excitation $h_1$.
The vevs $ v_{1,2} $ are fixed at tree level by the tadpole conditions
$t_1 = t_2 = 0$ with
\begin{equation}
	t_1 =v_1(2\mu_1^2-v_2^2\lambda_{12}-4v_1^2\lambda_1) \,,
	\qquad
	t_2=\frac{v_2}{4}(4\mu_2^2-4v_1^2\lambda_{12}-v_2^2\lambda_2) \,,
\label{eq:tadpoleeq}
\end{equation}
where $t_1h_1$ and $ t_2h_2 $ are the terms linear in $ h_1$ and $h_2 $ after inserting
\eqs{doublet}{singlet}
into the Higgs Lagrangian \eqref{eq:SESM}.
The non-zero vevs together with the ``mixing coupling''
$ \lambda_{12}\neq0 $ lead to a non-diagonal mass matrix for the scalar fields $ h_1,h_2 $.
We diagonalize this matrix at tree level by a rotation about the mixing angle $ \alpha $ $ (s_\alpha\equiv\sin\alpha,\,c_\alpha\equiv\cos\alpha) $,
\begin{equation}\label{eq:HfieldRotation}
	\begin{pmatrix}H\\h\end{pmatrix} =
	\begin{pmatrix}c_\alpha&s_\alpha\\-s_\alpha&c_\alpha\end{pmatrix}
	\begin{pmatrix}h_1\\h_2\end{pmatrix},
\end{equation}
defining the Higgs fields $ H,h $ corresponding to mass eigenstates with masses $ \MH $ and $ \Mh $, respectively.
We choose $ h $ to be the ``SM-like'' Higgs field of mass $ \Mh\approx125\GeV $, while $ H $ represents a BSM field with a large mass possibly outside the direct reach of current and previous experiments.
The mass hierarchy $ \MH>\Mh $ is enforced by choosing the range for the mixing angle according to
\begin{equation}
	0\leq\alpha<\frac{\pi}{2}\quad\text{for }\lambda_{12}\geq0
	\quad\text{and}\quad-\frac{\pi}{2}<\alpha<0\quad\text{for } \lambda_{12} < 0 \,,
\end{equation}
see \rcites{Altenkamp:2018bcs,Dittmaier:2021fls,Dittmaier:2022ivi}.
This determines the masses of the physical Higgs bosons at tree level,
\begin{equation}
	M_\mathrm{h}^2=\frac{1}{2}v_2^2\lambda_2-2v_1v_2\lambda_{12} \frac{s_\alpha}{c_\alpha} \,, \qquad M_\mathrm{H}^2=\frac{1}{2}v_2^2\lambda_{2}
	+2v_1v_2\lambda_{12}\frac{c_\alpha}{s_\alpha} \,,
	\label{eq:HMasses}
\end{equation}
together with the corresponding tree-level tadpole terms
\begin{equation}
	\begin{pmatrix} t_\PH \\ t_\Ph \end{pmatrix} =
	\begin{pmatrix} c_\alpha & s_\alpha \\ -s_\alpha & c_\alpha \end{pmatrix}
	\begin{pmatrix} t_1 \\ t_2 \end{pmatrix},
\end{equation}

The mass $ \MH $ of $ H $ represents the high energy scale $ \Lambda \sim \MH $ which indicates the onset of new physics in the EFT, cf.~\eq{EFTexp}.
We introduce a dimensionless power-counting parameter
\begin{equation}
  \label{eq:zeta}
   \zeta\sim \frac{\MH}{\Mh} \sim \frac{\MH}{v_2} \; \gg 1 \,,
\end{equation}
to carefully define the EFT (large-mass) expansion in our derivation of the effective Lagrangian in \subsec{EFTlimit}.
It will allow us to assign a certain scaling in the large-mass limit to parameters of different mass dimension.
To facilitate the contact to measurements, we will work with phenomenologically inspired input parameters in the following. For the BSM part we trade the new
fundamental BSM parameters $\{\mu_1^2,\lambda_1,\lambda_{12}\}$
for the experimentally more accessible set $ \{\MH,s_\alpha,\lambda_{12}\} $. Relations between the fundamental and input parameters are given
in \rcites{Altenkamp:2018bcs,Dittmaier:2021fls,Dittmaier:2022ivi} and read
\begin{align}
v_1 &= \frac{c_\alpha s_\alpha}{2\lambda_{12} v_2}\, (\MH^2 - \Mh^2),
\nn\\
\mu_{1}^2 &=
\frac{1}{4v_1}\!\left[
s_\alpha \Mh^2 (s_\alpha v_1 - c_\alpha v_2)
+ c_\alpha \MH^2 (c_\alpha v_1 + s_\alpha v_2)
+ 3c_\alpha\, t_\PH - 3s_\alpha\, t_\Ph
\right],
\nn\\
\mu_{2}^2 &=
\frac{1}{2v_2}\!\left[
c_\alpha \Mh^2 (c_\alpha v_2 - s_\alpha v_1)
+ s_\alpha \MH^2 (c_\alpha v_1 + s_\alpha v_2)
+ 3c_\alpha\, t_h + 3s_\alpha\, t_\PH
\right],
\nn\\
\lambda_1 &=
\frac{1}{8v_1^3}\!\left[
v_1(c_\alpha^2 \MH^2 + s_\alpha^2\Mh^2)
+ c_\alpha\, t_\PH - s_\alpha\, t_\Ph
\right],
\nn\\
\lambda_2 &=
\frac{2}{v_2^3}\!\left[
v_2(c_\alpha^2 \Mh^2 + s_\alpha^2 \MH^2)
+ c_\alpha\, t_\Ph + s_\alpha\, t_\PH
\right],
\label{eq:originalparams}
\end{align}
where we explicitly keep tadpole terms, because they are needed at NLO.

\subsection{Choosing an EFT limit}
\label{subsec:EFTlimit}

In order to define a consistent and phenomenologically viable EFT limit, we first note that the naive choice $ \MH\rightarrow\infty $ with fixed mixing angle $\alpha$ and coupling $\lambda_{12}$, does not reproduce the SM at
leading order in the large-mass
expansion (i.e.\ in inverse powers of $\MH$)  and is therefore experimentally disfavoured.
In that case, the couplings of the SM-like Higgs field $ h $ to the other SM fields only coincide with their SM values for vanishing $\alpha$.
Moreover, there are SESM vertices that are proportional to (positive powers of) the heavy mass scale $\MH$, thus giving rise to enhanced higher-order loop corrections for fixed $\alpha$.
This invalidates the applicability of the decoupling theorem of
Appelquist and Carazzone~\rcite{Appelquist:1974tg}.

A proper EFT limit, which leads to a consistent decoupling of the SM, therefore consists of simultaneous limits of multiple BSM parameters, e.g.\ $\MH\rightarrow \infty$ together with $s_\alpha\rightarrow0$ or $s_\alpha, \lambda_{12}\rightarrow0$.
With this in mind, we assign the generic scaling
\begin{equation}
	\MH\sim\zeta^1\,, \qquad
	s_\alpha\sim \zeta^{-a} \,,
	\qquad\lambda_{12}\sim \zeta^{-\ell}
	\label{eq:scaling}
\end{equation}
to the BSM input parameters, where $\zeta$ is the dimensionless power-counting parameter defined in \eq{zeta}.
The powers $ a,\ell$ parametrize
the behaviour of the mixing angle and the mixing coupling in the large-mass limit.
Different choices for  $a,l$ define different EFT limits with different phenomenological applicability \cite{Buchalla:2016bse,Dittmaier:2021fls,Dawson:2023oce}.
The scaling of the fundamental parameters of the SESM Lagrangian \cite{Dittmaier:2021fls},
\begin{align}
	v_1&\sim \zeta^{2-a+\ell},&&
	\mu_1^2\sim \zeta^{\max\{2,-2a,-\ell\}},&&
	\lambda_1\sim \zeta^{\max\{2a-2\ell-2,-2\ell-4\}},
	\nn\\
	v_2&\sim \zeta^0,&&
	\mu_2^2\sim \zeta^{\max\{4-2a+\ell,2-2a,0\}},&&
	\lambda_2\sim \zeta^{\max\{2-2a,0\}},
	\label{eq:fundiscaling}
\end{align}
can be deduced from \eq{scaling} and the parameter relations given in \eq{originalparams}, or \rcite{Altenkamp:2018bcs}.
To ensure that we stay in the phenomenologically relevant parameter space for arbitrary values of $\MH >\Mh$, we demand the SM-like masses and couplings to scale like $\zeta^0$ in the limit $ \zeta\rightarrow\infty $.
In particular, since $\MW =g_2 v_2/2$  at LO, the vev $ v_2 $ is highly constrained by precision measurements of the $ W $-boson mass. It should thus not be tied to the heavy scale, as this would constrain $ \MH $ to a rather limited interval.
We note that our choice of BSM input parameters immediately yields the desired scaling for the vev $ v_2$, while $\lambda_2 \overset{!}{\sim} \zeta^0$ requires $a\ge 1$.
As in \rcite{Dittmaier:2021fls}, we will work throughout this paper with $ a=1,\,\ell=0, $ i.e.
\begin{equation}
	\MH\sim\zeta\,,
	\qquad
	s_\alpha\sim\zeta^{-1} \,,
	\qquad
	\lambda_{12}\sim\zeta^0 \,,
  \label{eq:EFTlimit}
\end{equation}
as 	$\zeta\rightarrow\infty$.
This limit is minimally restrictive, in the sense that it assumes the weakest suppression of $ s_\alpha $ needed to reproduce the SM in the limit $ \zeta\rightarrow\infty $, and larger values of $a$ and $\ell$ would further suppress the BSM interactions.
The parameters of the Lagrangian \eqref{eq:SESM}
and the emerging vevs then scale as
\begin{equation}
  v_2, \lambda_{1}, \lambda_{2}, \lambda_{12} \sim \zeta^0, \qquad
  v_1^2, \mu_1^2, \mu_2^2 \sim \zeta^2 \,,
  \label{eq:EFTlimit2}
\end{equation}
so that none of the SM-like couplings and masses is tied to the heavy scale $ \MH$.

In \rcite{Buchalla:2016bse} this limit is called the ``weakly-coupled limit (II)", while \rcite{Dawson:2023oce} calls it ``$ \mathrm{PC}^\mathrm{R}_1 $".
By contrast, the ``strongly-coupled limit (I)" in \rcite{Buchalla:2016bse}, which corresponds to the limit ``$ \mathrm{PC}^\mathrm{R}_2 $" in \rcite{Dawson:2023oce}, is characterized by $ a=0,\,l=-2 $, yielding
\begin{equation}
	\MH^2\sim\lambda_1\sim\lambda_2\sim\lambda_{12}\sim\mu_1^2\sim\mu_2^2\sim\zeta^2,\qquad
	v_1\sim v_2\sim\Mh\sim s_\alpha\sim\zeta^0 \,,
\end{equation}
and does not provide decoupling of the heavy Higgs field~$H$.

Although tying the mass $\MH$ and the mixing angle $ \alpha $ together seems to constrain the parameter space that can be probed, we stress that the product
$s_\alpha\MH\sim\zeta^0 $ is essentially unconstrained as long as $ \vert s_\alpha\vert,v_2/\MH\ll1$.
This means we do not lose a full dimension in the parameter space, but rather
parametrize
it in a way that makes it easy to focus on the experimentally allowed region. Measurements of the Higgs signal strength exclude $ s_\alpha>0.26 $ at the $ 95\% $ confidence level \cite{FERBER2024104105},
and LHC data from Run~2 start to challenge $\sa$ values of
$0.2$ (see, e.g., \citeres{cms_collaboration_2026_rn1g0-73405,cms_collaboration_2026_90xht-tpc60}).

\section{Computing the effective Lagrangian}
\label{sec:calculation}

Our functional (top-down) matching calculation follows the same procedure as described in \rcite{Dittmaier:2021fls}.  We will only briefly outline the key elements here, and refer to \rcite{Dittmaier:2021fls} for details.

\subsection{Background-field method, St\"uckelberg transformation,\\and separation of modes}

Employing the
BFM~\cite{DeWitt:1967ub,DeWitt:1980jv,Abbott:1980hw,Denner:1994xt}
we split every gauge and Higgs field,
generically denoted by $\phi$,
into a classical background field $ \hat{\phi} $ and a quantum field $ \phi $,
\begin{equation}\label{eq:BFM}
  \phi\rightarrow\tilde{\phi}=\hat{\phi}+\phi \,.
\end{equation}
The unitary Goldstone matrix associated with the non-linear Higgs realization
is split in a multiplicative way%
\footnote{This can be achieved by a decomposition of the Goldstone fields analogous to \eq{BFM},
i.e.\ $\varphi_a \to \hat{\varphi}_a + \varphi_a$, and a subsequent field redefinition of the $\varphi_a$ according to the Baker--Campbell--Hausdorff formula.}
following \rcites{Dittmaier:1995cr,Dittmaier:1995ee}:
\begin{equation}\label{eq:BFMGoldstones}
  U\rightarrow\tilde{U}=\hat{U}U
  =\exp\left(2\ri\frac{\hat{\varphi}}{v_{2}}\right)
  \exp\left(2\ri\frac{\varphi}{v_{2}}\right).
\end{equation}
We can now write (the bosonic part of)
the SESM Lagrangian as
\begin{align}
  \L_\mathrm{gauge+Higgs}
  =&-\frac{1}{2}\tr[\tilde{W}_{\mu\nu}\tilde{W}^{\mu\nu}]
  -\frac{1}{4}\tilde{B}_{\mu\nu}\tilde{B}^{\mu\nu}
  +\frac{1}{4}(v_{2}+\tilde{h}_{2})^2
  \tr\left[(\tilde{D}_{\mu}\tilde{U})^\dagger(\tilde{D}^{\mu}\tilde{U})\right] \nn\\
  &+\frac{1}{2}(\partial_{\mu}\tilde{h}_{2})^2
  +\frac{1}{2}\mu_{2}^2(v_{2}+\tilde{h}_{2})^2
  -\frac{1}{16}\lambda_{2}(v_{2}+\tilde{h}_{2})^4 \nn\\
  &+\frac{1}{2}(\partial_{\mu}\tilde{h}_{1})^2
  +\mu_{1}^2(v_{1}+\tilde{h}_{1})^2
  -\lambda_{1}(v_{1}+\tilde{h}_{1})^4
  -\frac{1}{2}\lambda_{12}(v_{2}+\tilde{h}_{2})^2(v_{1}+\tilde{h}_{1})^2 \,,
  \label{eq:LSESMgaugeHiggs}
\end{align}
where all deviations from the SM are contained in the last line.
Here $W_{\mu \nu}$ and $B_{\mu \nu}$ denote the usual $W$ and $B$ field-strength tensors.
One of the benefits of working in the background-field formalism is
the possibility to choose separate gauges
for background and quantum fields. For the latter we choose the gauge-fixing Lagrangian \cite{Dittmaier:1995cr,Dittmaier:1995ee}
\begin{equation}
\label{eq:GaugeFixing}
\L_{\mathrm{fix}}=-\frac{1}{\xi_W}\tr\biggl[\biggl(\hat{D}_{W}^{\mu}W_{\mu}+\frac{1}{2}\xi_W g_{2}v_{2}\hat{U}\varphi\hat{U}^\dagger\biggr)^{\!2}\,\biggr]-
\frac{1}{2\xi_B}\left(\partial^{\mu}B_{\mu}+\frac{1}{2}\xi_B g_{1}v_{2}\varphi_{3}\right)^{\!2} \,,
\end{equation}
where
\begin{equation}
  \label{eq:covDW}
  \hat{D}_{W}^{\mu}\phi=\partial^\mu\phi-\ri g_{2}\bigl[\hat{W}^\mu,\phi\bigr]
\end{equation}
for any field $\phi$ in the adjoint representation of SU(2).
The two gauge parameters for the fields $ W^\mu $ and $ B^\mu $ are set equal, $ \xi_{W}=\xi_{B}=\xi, $ to avoid mixing in tree-level propagators.
Note that neither $\L_{\mathrm{fix}}$ nor the corresponding Faddeev--Popov Lagrangian contain
couplings to Higgs fields.
For the background fields we employ the unitary gauge in order to eliminate the background Goldstone fields. This can be achieved by a generalized St\"uckelberg transformation~\cite{Stueckelberg:1938zz,Stueckelberg:1957zz,Kunimasa:1967zza,Lee:1972yfa}%
\footnote{
The accompanying transformation of the quantum
fields is just a (unitary) change of integration variables in the functional integral with unit Jacobian.}
\begin{equation}
  \hat{W}_{\mu}\rightarrow\hat{U}\hat{W}_{\mu}\hat{U}^\dagger
  +\frac{\ri}{g_{2}}\hat{U}\partial_{\mu}\hat{U}^\dagger\,, \qquad
  W_{\mu}\rightarrow\hat{U}W_{\mu}\hat{U}^\dagger\,, \qquad
  \hat{B}_{\mu}\rightarrow\hat{B}_{\mu}\,, \qquad
  B_\mu\rightarrow B_{\mu}\, .
  \label{eq:Stueckelberg}
\end{equation}
The background field-strength tensors and  the covariant derivative of the Goldstone matrix transform as
\begin{equation}
  \tilde{D}_{\mu}\tilde{U}\rightarrow\hat{U}(\tilde{D}_{\mu}U) \,,\qquad
  \tilde{W}_{\mu\nu}\rightarrow\hat{U}\tilde{W}_{\mu\nu}\hat{U}^\dagger \,, \qquad
  \tilde{B}_{\mu\nu} \to \tilde{B}_{\mu\nu} \,.
  %\tag{\stepcounter{equation}\theequation}
\end{equation}
We will invert this St\"uckelberg transformation (see e.g.~\rcite{Dittmaier:1995ee}) after calculating the effective Lagrangian in  \subsec{InvStueckelberg} to restore gauge invariance and explicit Goldstone fields.%
\footnote{
The inverse St\"uckelberg transformation can be used to enforce explicit gauge invariance of a
\revision{Lagrangian}
in the following sense:
Start from a gauge-invariant SU(2)$\times$U(1) gauge theory with massless gauge
bosons (e.g.\ the SM ignoring masses), express the Lagrangian non-linearly (i.e.\ in HEFT form)
in unitary gauge (i.e.\ without Goldstone fields), and add explicit mass terms (and/or similarly non-invariant interaction terms) for the SU(2) gauge bosons.
Although the additional terms seem to break the gauge symmetry, one can now  \mbox{(re-)}in\-tro\-duce Goldstone fields by an inverse St\"uckelberg transformation,
producing a manifestly SU(2)$\times$U(1) gauge-invariant Lagrangian.}

A central ingredient to our
derivation~\cite{Dittmaier:2021fls}
of the effective Lagrangian
is the method of regions~\cite{Beneke:1997zp,Smirnov:2002pj}, which
allows for expanding Feynman integrals at the level of the integrand.
In order to implement this in our functional approach, we formally split all fields $ \phi $ into soft and hard modes $ \phi_s$, $\phi_h$, carrying soft ($p^\mu \sim \Mh$) and hard momenta ($p^\mu \sim \MH$), respectively.
Together with the BFM decomposition into background and quantum fields the separation of modes results in a fourfold split of the fields, $ \phi\rightarrow\hat{\phi}_h+\hat{\phi}_s+\phi_h+\phi_s $.
This allows us to integrate out the hard loop momentum region of radiative corrections in the SESM by solving the path integral over the hard quantum modes $ \phi_h $.
Contributions related to hard loops without a heavy Higgs propagator are scaleless in dimensional regularization and vanish accordingly.
Note also that we aim to construct an EFT valid at low energies/momenta ($\sim \Mh$), where all external momenta are soft. We can therefore drop the hard modes of background fields, which only occur at tree level, immediately, i.e.\ $\hat{\phi}_h \to 0$.
The soft modes of the background and quantum heavy Higgs will be eliminated by solving their equations of motion iteratively order by order in $ \zeta^{-1}$, as described in the following.

\subsection{Integrating out the heavy Higgs field at tree level}
\label{subsec:Lefftree}

The ``tree-level'' Lagrangian is obtained by collecting all terms of the SESM Lagrangian that exclusively depend on soft field modes and expanding them in the EFT parameter
$\zeta^{-1}$ to $\ord{(\zeta^{-2})}$.
For this expansion we count $\tilde{H}_s \sim \zeta^{-1}$ as can be inferred from the heavy Higgs propagator \cite{Dittmaier:2021fls}, and otherwise adopt the power-counting scheme for the (BSM) parameters defined in \subsec{EFTlimit}.
We find%
\footnote{Here $\L_\SM^\tree$ is understood to consist of soft fields and to include the corresponding soft $\L_{\mathrm{fix}}$ from \eq{GaugeFixing}.}
\begin{align}
\L^\tree={}&\L_{\mathrm{SM}}^\tree
-\frac{1}{2}\MH^2\tilde{H}_{s}^2
-\frac{s_{\alpha}\MH^2}{2v_2}\tilde{h}_s^2\tilde{H}_{s}
-\frac{s_{\alpha}^2\MH^2}{8v_2^2}\tilde{h}_s^4
\nn\\ & {}
-\frac{1}{2}\tilde{H}_s\Box\tilde{H}_s
+\frac{3s_{\alpha}^2\Mh^2}{4v_2}\tilde{h}_s^3
-\frac{s_{\alpha}(2\Mh^2-\MH^2s_{\alpha}^2+2\lambda_{12}v_2^2)}{2v_2}\tilde{h}_s^2\tilde{H}_s
-\frac{\MH^2s_{\alpha}^2-2\lambda_{12}v_2^2}{v_2}\tilde{h}_s\tilde{H}_s^2
\nn\\ & {}
-\frac{\lambda_{12}v_2}{s_{\alpha}}\tilde{H}_s^3
+\frac{s_{\alpha}^2(3\Mh^2+2\MH^2s_{\alpha}^2-4\lambda_{12}v_2^2)}{8v_2^2}\tilde{h}_s^4
-\frac{s_{\alpha}(\Mh^2+\MH^2s_{\alpha}^2-2\lambda_{12}v_2^2)}{2v_2^2}\tilde{h}_s^3\tilde{H}_s
\nn\\ & {}
-\frac{\lambda_{12}}{2}\tilde{h}_s^2\tilde{H}_s^2
+\frac{e^2s_{\alpha}}{8\sw^2}(\tilde{h}_s+v_2)(-\tilde{h}_s s_{\alpha}+2\tilde{H}_s)(\tilde{C}_{\mu,s}^a)^2
+\ord\left(\zeta^{-4}\right),
\label{eq:Ltree}
\end{align}
where we have
introduced the field $C_\mu$ for a certain combination of the massive gauge fields,
\begin{equation}
	C_\mu=
	\frac{\tau_a}{2}C^a_\mu
	=W_\mu+\frac{\sw}{\cw}B_\mu\frac{\tau_3}{2}
	=\frac{1}{2}\left(W^1_\mu\tau_1+W^2_\mu\tau_2+\frac{1}{\cw}Z_\mu\tau_3\right).
\end{equation}
The soft modes of the heavy Higgs field
cannot become on-shell and therefore do not occur as external particles in processes described by the low-energy EFT.
In \eq{Ltree} the heavy Higgs field only appears in the original combination $ \tilde{H}_s = \hat{H}_s +  H_s $ and can thus be simply eliminated by iteratively solving the equation of motion (EOM) derived from $\L^\tree$ order by order in $ \zeta^{-1}$ to obtain
\begin{align}
  \label{eq:EOMtree}
	\tilde{H}_{s}={} &
	-\frac{s_{\alpha}}{2v_2}\tilde{h}_s^2
	+\frac{s_{\alpha}}{2v_2\MH^2}\left(\Box-2\Mh^2
	-2\lambda_{12}v_2^2+s_{\alpha}^2\MH^2\right)\tilde{h}_s^2
	-\frac{s_{\alpha}}{2v_2^2\MH^2} \left(\Mh^2-s_{\alpha}^2\MH^2+2\lambda_{12}v_2^2 \right)\tilde{h}_s^3
	\nn\\
	&{}
	-\frac{s_{\alpha}\lambda_{12}}{4v_2\MH^2}\tilde{h}_s^4
	+\frac{e^2 s_{\alpha}}{4\sw^2\MH^2}(\tilde{C}_{\mu,s}^a)^2(v_2+\tilde{h}_s)
	+\mathcal{O}\bigl(\zeta^{-5}\bigr)\,.
\end{align}
For the calculation of the one-loop part of the effective Lagrangian we need this
result even including the $\ord(\zeta^{-5})$ contribution. There, however, effectively only the background part
$\hat{H}_s$ of the full field $\tilde{H}_{s}$ is relevant, since the quantum field
$H_s$ in the one-loop effective Lagrangian would only be relevant at the two-loop level.
The explicit solution for $\hat{H}_s$ to $\ord(\zeta^{-5})$ is given in \app{EOM}.

Inserting \eq{EOMtree} in \eq{Ltree}, the resulting effective Lagrangian reads
\begin{equation}
	\L_{\eff}^\tree=\L_{\SM}^\tree
	-\frac{e^2s_{\alpha}^2}{8\sw^2v_2}\tilde{h}(\tilde{h}+v_2)^2 \bigl(\tilde{C}_\mu^a\bigr)^2
	-\frac{s_{\alpha}^2}{8v_2^2}\tilde{h}^2\Box\tilde{h}^2
	+\frac{\Mh^2s_{\alpha}^2}{8v_2^3}\tilde{h}^3(2\tilde{h}^2+7\tilde{h}v_2+6v_2^2)
    \,+\,\mathcal{O}\bigl(\zeta^{-4}\bigr)
    \,,
	\label{eq:LeffTreelong}
\end{equation}
where the (bosonic part of the) SM tree-level Lagrangian in unitary gauge is given by
\begin{equation}
\L_{\SM}^\tree =
  -\frac{1}{4} \tilde{W}^a_{\mu\nu}\tilde{W}^{a,\mu\nu}
  -\frac{1}{4}\tilde{B}_{\mu\nu}\tilde{B}^{\mu\nu}
  +\frac{g_2^2}{8}(v_{2}+\tilde{h})^2 \bigl(\tilde{C}_\mu^a\bigr)^2
  +\frac{1}{2}(\partial_{\mu}\tilde{h})^2
  -\frac{1}{2} \Mh^2 \tilde{h}^2 \biggl(1+\frac{\tilde{h}}{2v_2}\biggr)^2 \,.
\label{eq:LSM}
\end{equation}
Here and in the following we have suppressed the subscript $s$ of the fields for brevity,
because in the EFT all fields are always understood to consist of soft modes only.

\subsection{Integrating out the heavy Higgs field at the one-loop level}
\label{subsec:IntoutOneloop}

To integrate out the hard field modes at the one-loop level, we only need the part of the Lagrangian that is bilinear in the hard quantum fields, which we generically write  as
\begin{equation}
  \label{eq:L1loop}
	\L^\oneloop %(H_h,\phi_{i,h})
  =
	-\frac{1}{2}H_h\Delta_HH_h
	+H_h\mathcal{X}_{Hi} \phi_{i,h}
	-\frac{1}{2}\phi_{i,h}\mathcal{A}_{ij}\phi_{j,h} \,,
\end{equation}%
where the $ \phi_{i,h} $ denote hard modes of the light fields~$\phi_i$
and summation over $i,j$ is implicit.
The differential operators $ \Delta_H(x,\partial_x)$, $\mathcal{X}_{Hi} (x,\partial_x)$, and $ \mathcal{A}_{ij}(x,\partial_x)$ in general contain (soft) background fields, but no quantum fields by construction.
As we exclusively work with real fields, the building blocks behave under Hermitian conjugation as
\begin{equation}
	\phi_{i,h}=\phi_{i,h}^\dagger \,,\qquad
	\Delta_H=\Delta_H^\dagger \,,\qquad
	\mathcal{X}_{Hi}=\mathcal{X}_{iH}^\dagger \,,\qquad
	\mathcal{A}_{ij}=\mathcal{A}_{ji}^\dagger \,.
\end{equation}
The Lagrangian \eqref{eq:L1loop} can now be diagonalized by the linear field redefinitions
\begin{equation}
	\phi_{i,h} \,\rightarrow\, \phi_{i,h}+(\mathcal{A}^{-1})_{ij}\mathcal{X}_{jH}H_h \,,
\end{equation}
which have unit Jacobian determinant in the functional integral.
The resulting diagonal Lagrangian reads
\begin{equation}
	\L^\oneloop = -\frac{1}{2}H_h \tilde{\Delta}_H H_h -\frac{1}{2}\phi_{i,h}\,\mathcal{A}_{ij}\, \phi_{j,h} \,,
\end{equation}
with the modified operator
\begin{equation}
  \label{eq:DeltaHtilde}
	\tilde{\Delta}_H=\Delta_H-\mathcal{X}_{Hi}(\mathcal{A}^{-1})_{ij}\mathcal{X}_{jH} \,.
\end{equation}
At this point we can perform the integration over $ H_h $ explicitly. The functional integral over $ H_h $ is of Gaussian form, which yields the well-known functional determinant~\cite{Dittmaier:2021fls}
\begin{align}
& \int\mathcal{D}H_h\,\exp\left\{-\frac{\ri}{2}\mu^{D-4} \!\! \int\rd^D x\,
H_h\tilde{\Delta}_H(x,\partial_x) H_h\right\}
{} \,\propto\,
\left\{ \mathpzc{Det}_h\Bigl[ \delta(x-y) \tilde{\Delta}_H(x,\partial_x)
\Bigr]\right\}^{-\frac{1}{2}}
\nn\\
&=\exp \left\{ -\frac12
\mathpzc{Tr}_h \Bigl[ \ln\Bigl(
\delta(x-y) \tilde{\Delta}_H(x,\partial_x) \Bigr) \Bigr]
\right\}
\,\equiv\,
\exp\left\{\ri\,\mu^{D-4} \!\! \int\rd^D x\, \delta\L^{\text{1-loop}}_\eff \right\}.
\label{eq:funcdet}
\end{align}
We drop an
irrelevant infinite constant contribution to $\delta\L^{\text{1-loop}}_\eff$
and evaluate the functional trace in momentum space to arrive at the one-loop effective Lagrangian%
\footnote{The second argument of $\tilde{\Delta}_H$ in this formula implies that all partial derivatives in \eq{DeltaHtilde} are to be replaced as $\partial_x \to \partial_x +\ri p$.}
\begin{equation}
	\delta\L^{\text{1-loop}}_\eff = \frac{\ri}{2}\,\mu^{4-D}\int\frac{\rd^D p}{(2\pi)^D}\,
	\mathcal{T}_{h}(p) \,
	\ln\left( \tilde{\Delta}_H(x,\partial_x+\ri p) \right).
	\label{eq:L1loopeff}
\end{equation}
The operator $\mathcal{T}_{h}(p)$, as defined in \rcite{Dittmaier:2021fls}, formally implements the
Taylor expansion of the integrand assuming the loop momentum to be hard, i.e.~$p\sim\zeta$,
according to the method of regions.
The operator $ \tilde{\Delta}_H $ is expanded in $ \zeta^{-1} $ as follows.
We first compute the inverse of the matrix-valued operator $ \mathcal{A} $ order by order in $ \zeta^{-1} $.
To this end, we split off the dominant, invertible part $ \mathcal{D} $ such that
$\mathcal{A}_{ij}=\mathcal{D}_{ij}-\mathcal{X}_{ij}$
with $\mathcal{X}_{ii}=0$, and $\mathcal{X}_{ij}$ suppressed w.r.t.\ $ \mathcal{D}_{ii} $ by at least one power of $ \zeta^{-1} $.
The inverse of $ \mathcal{A} $ can then be computed as a Neumann series~\cite{Fuentes-Martin:2016uol,Dittmaier:2021fls},
\begin{align}
  \left( \mathcal{A}^{-1} \right)_{ij}  = \left(\mathcal{D}^{-1}\right)_{ij}
  + \left(\mathcal{D}^{-1}\right)_{ik} \mathcal{X}_{kl} \left(\mathcal{D}^{-1}\right)_{lj}
  + \left(\mathcal{D}^{-1}\right)_{ik} \mathcal{X}_{kl} \left(\mathcal{D}^{-1}\right)_{lm}
  \mathcal{X}_{mn} \left(\mathcal{D}^{-1}\right)_{nj}
  + \ldots\,,
  \label{eq:NeumannExp}
\end{align}
where the intermediate field indices such as $k,l,\dots$
(but not the  external indices $i,j$) are summed over.
Similarly, $ \mathcal{D}^{-1} $ can be computed via a Neumann series, where now the leading term in the large-mass expansion is the momentum-space propagator of the corresponding light field carrying the (loop) momentum $p$.

In order to calculate the effective Lagrangian of the SESM to $ \ord(\zeta^{-2}) $, we will need terms to order $ \ord(\zeta^{-4}) $ in $ \tilde{\Delta}_H(x,\partial_x+\ri p) $ as given in \eq{DeltaHtilde}.
For ease of exposition we introduce the shorthand notation
\begin{equation}
	\Omega_{\phi_1\phi_2\dots\phi_n} \equiv
	\mathcal{X}_{H{\phi_1}_a}(\mathcal{D}^{-1})_{{\phi_1}_a {\phi_1}_b}
	\mathcal{X}_{{\phi_1}_b{\phi_2}_c}
	(\mathcal{D}^{-1})_{{\phi_2}_c {\phi_2}_d}\dots
	(\mathcal{D}^{-1})_{{\phi_n}_i {\phi_n}_j}
	\mathcal{X}_{{\phi_n}_j H} \,,
\end{equation}
where $a$, $b$, $\ldots$ denote possible SU(2) indices of the corresponding fields and the $\mathcal{X}$ and $\mathcal{D}^{-1}$ operators may in addition carry appropriately contracted Lorentz indices associated with vector fields, which we, however suppress here.
To the relevant order in the large-mass expansion
we can now write
\begin{align}
\mathcal{X}_{Hi}(\mathcal{A}^{-1})_{ij}\mathcal{X}_{jH}={}&
 \Omega_{h}
+\Omega_{\varphi}
+\Omega_{\barW}
+\Omega_{\varphi h}
+\Omega_{h \varphi}
+\Omega_{\barW h}
+\Omega_{h \barW}
+\Omega_{\barW \varphi}
+\Omega_{\varphi \barW}
\nn\\ & {}
+\Omega_{h \varphi h}
+\Omega_{\varphi h \varphi}
+\Omega_{h \barW h}
+\Omega_{h \varphi \barW}
+\Omega_{\barW \varphi h}
+\Omega_{h \barW \varphi}
+\Omega_{\varphi \barW h}
+\Omega_{\varphi \barW \varphi}
\nn\\ & {}
+\Omega_{h \varphi h \varphi}
+\Omega_{\varphi h \varphi h}
+\Omega_{h \barW \varphi h}
+\Omega_{h \varphi \barW h}
+\Omega_{h \varphi \barW \varphi}
+\Omega_{\varphi \barW \varphi h}
\nn\\ & {}
+\Omega_{h \varphi h \varphi h}
+\Omega_{h \varphi \barW \varphi h}
+\mathcal{O}(\zeta^{-5}) \,,
\label{eq:Chains}
\end{align}
were we defined the field $ \barW_\mu $
following \rcites{Dittmaier:1995ee,Dittmaier:2021fls} as
\begin{equation}
  \barW^\mu=\frac{1}{2} \bigl(W^{1,\mu} \tau_1+W^{2,\mu}\tau_2+Z^\mu\tau_3 \bigr) \,.
  \label{eq:Wbar}
\end{equation}
The explicit expressions for the $\mathcal{X}$ and $\mathcal{D}^{-1}$ operators required in this work are given in \app{XD}.
For details on their derivation we refer to \rcite{Dittmaier:2021fls}.

In addition, we need
\begin{align}
\lefteqn{\Delta_H(x,\partial_x + \ri p)} \qquad&
\nn\\*
={}&
{-}p^2 + \MH^2
+2\ri p\!\cdot\!\partial_x + \Box_x
+ 2\bigg[ -2\lambda_{12} +\frac{\MH^2s_{\alpha}^2}{v_2^2} \bigg]v_2\hat{h}
+6\lambda_{12}v_2\frac{\hat{H}}{s_{\alpha}}
+\lambda_{12}\hat{h}^2
\nn\\& {}
+\bigg[2\lambda_{12}s_{\alpha}^2
-\frac{6\lambda_{12}\Mh^2}{\MH^2}
+\frac{\Mh^2s_{\alpha}^2}{v_2^2}
-\frac{\MH^2s_{\alpha}^4}{v_2^2}\bigg]v_2\hat{h}
+3\bigg[ -2\lambda_{12}s_{\alpha}^2 +\frac{2\lambda_{12}\Mh^2}{\MH^2}
+\frac{\MH^2s_{\alpha}^4}{v_2^2}
\bigg]\frac{v_2\hat{H}}{\sa}
\nn\\& {}
+3\bigg[
-2\lambda_{12}s_{\alpha}^2
+\frac{2\lambda_{12}^2v_2^2}{\MH^2}
+\frac{\Mh^2s_{\alpha}^2}{2v_2^2}
+\frac{\MH^2s_{\alpha}^4}{2v_2^2}\bigg]\hat{h}^2
+6\lambda_{12}\bigg[s_{\alpha}^2
-\frac{2\lambda_{12}v_2^2}{\MH^2}\bigg]\frac{\hat{h}\hat{H}}{\sa}
+\frac{6\lambda_{12}^2v_2^2}{\MH^2}\frac{\hat{H}^2}{s_{\alpha}^2}
\nn\\& {}
-\frac{e^2s_{\alpha}^2(\hat{C}_{\mu}^a)^2}{4\sw^2}
+\bigg[
\frac{\lambda_{12}s_{\alpha}^4}{2}
+\frac{3\lambda_{12}\Mh^2s_{\alpha}^2}{\MH^2}
-\frac{6\lambda_{12}\Mh^4}{\MH^4}
-\frac{\Mh^2s_{\alpha}^4}{2v_2^2}
-\frac{\MH^2s_{\alpha}^6}{4v_2^2}\bigg]v_2\hat{h}
\nn\\& {}
+\frac{6\lambda_{12}\Mh^2}{\MH^2}\bigg[
-s_{\alpha}^2
+\frac{\Mh^2}{\MH^2}
\bigg]\frac{v_2\hat{H}}{\sa}
+3\bigg[
2\lambda_{12}s_{\alpha}^4
+\frac{4\lambda_{12}^2\Mh^2v_2^2}{\MH^4}
-\frac{2\lambda_{12}^2s_{\alpha}^2v_2^2}{\MH^2}
-\frac{\Mh^2s_{\alpha}^4}{v_2^2}
-\frac{\MH^2s_{\alpha}^6}{2v_2^2}\bigg]\hat{h}^2
\nn\\& {}
+3\bigg[
-5\lambda_{12}s_{\alpha}^4
-\frac{8\lambda_{12}^2\Mh^2v_2^2}{\MH^4}
+\frac{6\lambda_{12}^2s_{\alpha}^2v_2^2}{\MH^2}
+\frac{\Mh^2s_{\alpha}^4}{v_2^2}
+\frac{\MH^2s_{\alpha}^6}{v_2^2}\bigg]\frac{\hat{h}\hat{H}}{\sa}
\nn\\& {}
+6\lambda_{12}\bigg[
s_{\alpha}^4
-\frac{2\lambda_{12}s_{\alpha}^2v_2^2}{\MH^2}
+\frac{2\lambda_{12}\Mh^2v_2^2}{\MH^4}
\bigg]\frac{\hat{H}^2}{\sa^2}
+\ord\bigl(\zeta^{-5}\bigr) \,.
\end{align}
Recall that $\hat H$ counts as $\ord(\zeta^{-1})$ here.
For bookkeeping we define operators $\Pi^{(n)}(x,p,\partial_x) $ of $\mathcal{O}(\zeta^{-n})$ by
\begin{equation}
	\tilde{\Delta}_H(x,\partial_x+\ri p)=-(p^2-\MH^2)+\sum_{n=-1}^{\infty}\Pi^{(n)}(x,p,\partial_x) \,,
\end{equation}
with $\Pi^{(-1)} = 2 \ri p_\mu \partial_x^\mu$.
Based on the fact that all $ \Pi^{(n)} $ are suppressed compared to the leading contribution
$-(p^2-\MH^2) $ in the large-mass limit, the logarithm in \eq{L1loopeff} can now be expanded and we have
\begin{align}
  \delta\L_\eff^\oneloop ={}&
  \frac{\ri}{2}\,\mu^{4-D}\int\frac{\rd^D p}{(2\pi)^D}\,\Bigg[
  -\frac{\Pi^{(0)}+\Pi^{(2)}+\Pi^{(4)}}{p^2-\MH^2}
  -\frac{\bigl(\Pi^{(0)}\bigr)^2 + \bigl(\Pi^{(1)}\bigr)^2
  + \Pi^{(0)} \Pi^{(2)} + \Pi^{(2)} \Pi^{(0)}}{2(p^2-\MH^2)^2}
  \nn\\&
  -\frac{\bigl(\Pi^{(0)}\bigr)^3
  + \Pi^{(0)}\Pi^{(-1)} \Pi^{(1)}
  + \Pi^{(1)}\Pi^{(-1)} \Pi^{(0)}}{3(p^2-\MH^2)^3}
  -\frac{\Pi^{(0)}\Pi^{(-1)}\Pi^{(-1)}\Pi^{(0)}}{4(p^2-\MH^2)^4}
  \Bigg]
  + \,\ord(\zeta^{-4}) \,.
  \label{eq:LeffPi}
\end{align}
Here we have dropped another irrelevant constant in the Lagrangian as well as
contributions that integrate to zero by virtue of
antisymmetry under $ p^\mu\rightarrow-p^\mu$ at the integrand level or being a total derivative.
We emphasize that  the $\Pi^{(n)}$ in general contain derivatives
acting to the right.
 The resulting vacuum-type one-loop integrals over $ p $ can be reduced by the covariant tensor decomposition
\begin{equation}
	p^\mu p^\nu \rightarrow \frac{p^2}{D}g^{\mu\nu},\qquad
	p^\mu p^\nu p^\rho p^\sigma \rightarrow \frac{p^4}{D(D+2)}(g^{\mu\nu}g^{\rho\sigma}+g^{\mu\rho}g^{\nu\sigma}+g^{\mu\sigma}g^{\nu\rho}) \,,
\end{equation}
to the scalar integrals
\begin{align}
	I_{ab} &= \frac{(2\pi\mu)^{4-D}}{\ri \pi^2} \!\int \! \rd^D p\; \frac{1}{(p^2 - \MH^2 +\ri 0)^a (p^2 +\ri 0)^b}
	\nn\\
	&=
	(4\pi\mu^2)^{(4-D)/2} \, (-1)^{a+b} \,
	\frac{\Gamma \bigl(\frac{D}{2}-b \bigr)
		\Gamma \bigl(a+b-\frac{D}{2}\bigr)}{\Gamma (a) \Gamma \bigl(\frac{D}{2}\bigr)}\,
	\MH^{D-2a-2 b} \,.
	\label{eq:MI}
\end{align}
These integrals can be expressed in terms of a single master integral, for which we choose
the two-point integral
\begin{equation}
\label{eq:I20}
    I_{20} \equiv B_0(0,\MH^2,\MH^2)
	=\Delta+\ln\left(\frac{\mu^2}{\MH^2}\right)+\mathcal{O}(\eps) \,,
\end{equation}
with $B_0$ denoting the
scalar
one-loop two-point integral
in the notation of \rcites{Denner:2019vbn,Bohm:1986rj,Denner:1991kt},
\begin{equation}
  B_0(p^2,m_1^2,m_2^2) = \frac{(2\pi\mu)^{4-D}}{\ri \pi^2}
  \int \! \frac{\rd^D q}{[q^2-m_1^2 +\ri 0][(q+p)^2-m_2^2 +\ri 0]} \,,
  \label{eq:B0}
\end{equation}
and
\begin{equation}
	\Delta=\frac{2}{4-D}-\gamma_\mathrm{E}+\ln(4\pi)=\frac{1}{\epsilon}-\gamma_\mathrm{E}+\ln(4\pi) \,,
	\qquad
	\epsilon=\frac{4-D}{2}.
\end{equation}
The (intermediate) expression for $\delta\L_\eff^\oneloop$ directly resulting from the evaluation of \eq{LeffPi} is provided
in \eq{Leff1loopBeforeEOM}.
Finally, the (soft) background field $\hat{H}$
occurring in that expression is eliminated  via the (tree-level) EOM in \eq{HEOMFull},
and we arrive at the result for $\delta\L_\eff^\oneloop$  presented in \eq{Leff1loopAfterEOM}.

\section{Renormalization}
\label{sec:renormalization}

The original, {\it bare} parameters of any quantum field theory have no clear physical meaning in the presence of radiative corrections.
In order to make contact to phenomenology, it is necessary to
reparametrize
the Lagrangian in terms of quantities that do have a well-defined physical meaning.
This procedure is called renormalization.
The first step in this procedure is the {\it renormalization transformation} which
splits each bare parameter into a finite {\it renormalized}
part, which will be fixed by experiment, and a {\it renormalization constant},
which may contain ultraviolet (UV) divergences.
The contributions from renormalization constants to the Lagrangian
define the {\it counterterm Lagrangian}, which in turn produces counterterm graphs
that combine with loop graphs in predictions.
The second step in the renormalization procedure fixes the {\it renormalization
scheme} which ties the renormalized parameters to measurable quantities and thereby
determines the renormalization constants.
Allowing for field redefinitions that can reshuffle higher-order contributions,
not only predictions for S-matrix elements but also for Green functions
can be made UV-finite.
In {\it on-shell (OS)} renormalization schemes, it can even be achieved that
the renormalized fields are canonically normalized, so that wave-function
corrections \`a la Lehmann--Symanzik--Zimmermann (LSZ) are not necessary.
Different formulations of the renormalization and different
renormalization schemes
for the SESM have been proposed in the
literature~\cite{Kanemura:2015fra,Bojarski:2015kra,Denner:2017vms,Altenkamp:2018bcs,%
Denner:2018opp,Dittmaier:2021fls,Dittmaier:2022ivi}.
We will closely follow \citere{Dittmaier:2021fls} in spirit and notation, which in turn
is closely related to \citeres{Altenkamp:2018bcs,Denner:2018opp,Dittmaier:2022ivi}.

In this section we derive the contributions to the low-energy effective Lagrangian originating
from the renormalization of the full SESM and clarify the role of renormalization within the
emerging EFT.
Following the bottom-up or the top-down approach in the construction of the EFT entails
profound differences in the renormalization procedure. For this reason we start this section
with some broader discussion to put the two different approaches into perspective, before
going into all the technical details. For readers less familiar with renormalization it
might be useful to read this part again after the technical dust
of \secs{renormalization}{pheno} has settled.
We stress that the following discussion is limited to the one-loop level (NLO), unless explicitly stated
otherwise, and assumes the decoupling of the EFT as
established in~\rcite{Dittmaier:2021fls} for our large-$\MH$ limit.

In the bottom-up approach, all Wilson coefficients in the ansatz for the BSM part of the
EFT are considered as free parameters, so that each of these coefficients has to be fixed
by some renormalization condition.
Both the full-theory Lagrangian and the ansatz for the effective Lagrangian
with undetermined operator coefficients are renormalized independently.
After that, a set
of physical observables is computed in both theories.
The respective results are then matched to a given order in the EFT (large-mass) expansion by adjusting the \textit{renormalized} parameters and Wilson coefficients of the EFT.
In this sense renormalization is carried out twice, once for the full-theory and once for the EFT.
If the necessary care is taken,
the respective renormalization schemes may be even chosen independently.

As for the alternative top-down construction of the EFT, we consider two  variants differing in the implementation of the relevant renormalization conditions.
The ``strict'' top-down approach starts from the fully renormalized
UV-complete theory
(including all soft and hard contributions to the renormalization constants).
For our case this means that all (SM-like and BSM) parameters and fields of the SESM are
renormalized,
as described in \subsecsm{renormalization-trafo}{LeffRen} with suitable renormalization conditions formulated in the full theory.
Integrating out the heavy degrees of freedom,
for instance using the functional method of \sec{calculation},
the NLO contributions to $\L_\eff$ from the SESM counterterm Lagrangian are derived in complete analogy to $\L_\eff^\tree$ in \subsec{Lefftree}.
This top-down derivation yields an effective Lagrangian whose operator coefficients are explicitly expressed in terms of the renormalized full-theory parameters and generally contain explicit $1/\eps$ divergences as well as the associated $\ln(\mu)$ terms.
From the EFT perspective these expressions can be associated with
the (UV-divergent) \textit{bare} SM-like parameters and Wilson coefficients of the Lagrangian.
Still, this bare effective Lagrangian can already be used to predict physical observables to
NLO in the EFT without any further renormalization step.
This requires to employ dimensional regularization with the same $\mu$ scale as in the calculation
of $\L_\eff^\oneloop$ in \subsec{IntoutOneloop}, which will then finally
drop out in the NLO EFT predictions for physical observables.
Having renormalized all parameters in the full theory,
another renormalization of parameters and/or
Wilson coefficients in the EFT is not necessary as long as the EFT is used only for fixed-order calculations.
However, if one aims at resumming large logarithms ($\propto \ln(\MH/\Mh)$) via
renormalization group (RG) equations,
the evolution of the EFT parameters/coefficients, which is governed by
their
anomalous dimensions, is needed.
This requires a dedicated calculation of renormalization constants
(as in the bottom-up approach)
within the EFT in a suitable scheme with a
flexible
renormalization scale, like the $\MSbar$ scheme.
In that case the bare expressions for the parameters/coefficients resulting from the
top-down derivation of the EFT Lagrangian serve as the initial ``matching'' conditions
of the running EFT parameters/coefficients, and
$\mu$ plays the role of the matching scale $\mu_\mathrm{M}$.
In this work we, however, do not perform any resummation.

In the derivation
of our EFT we use a variant of the top-down approach described above,
which is more pragmatic and treats SM-like and BSM parameters of the SESM differently concerning their renormalization.
As explained in detail in \subsec{LeffRen}, the BSM parameters are fully renormalized
in the derivation of the effective Lagrangian, whereas for the SM-like parameters (and fields) only the hard contributions to the renormalization constants are taken into account,
but the soft parts are (mostly) dropped in this step.
In doing so, we aim to simplify the expression for $\L_\eff$ rather than to derive meaningful bare expressions for its SM-like operator coefficients.
We consider this procedure as a kind of ``pre-renormalization'' of the SM-like parameters which has no input from measurable quantities and is strictly speaking not necessary.
However, it is convenient to perform the (hard) pre-renormalization at this point in order to consistently account for hard contributions
to SM-like renormalization constants
(like $\delta \Mh^2$) that are $\zeta$-enhanced compared to the corresponding
tree-level parameters and
would obscure the decoupling form of $\L_\eff$, cf.\ \subsec{tadpoleren} and \rcite{Dittmaier:2021fls}.%
\footnote{The soft part of the renormalization constants cannot be $\zeta$-enhanced due to decoupling.}
The SM-like parameters still have to be properly renormalized according to renormalization conditions relating them to experimental input.
We therefore still regard the SM-like parameters as \textit{bare} after the pre-renormalization.
Technically, we use the renormalization constants in the pre-renormalization of the SM-like parameters and fields to absorb contributions of hard field modes
in combination with
some (field) redefinitions as described in \sec{fieldredef}.
The proper renormalization of the SM-like parameters then has to be performed in the EFT. We will do so explicitly in \sec{pheno}, when we
apply our effective Lagrangian to phenomenological predictions at NLO. In this way, we keep
the flexibility of choosing the renormalization scheme for the SM-like EFT parameters and fields,%
\footnote{The expressions for some of the SM renormalization constants (in particular those of the fields)
also change if the corrections to the observables are calculated
within the conventional quantization instead of the BFM, in which the effective
Lagrangian was derived.}
while the bare BSM (dimension-six) Wilson coefficients
are expressed explicitly in terms of renormalized BSM and bare SM-like parameters, cf.~\sec{final}.
In contrast to the strict top-down approach outlined above, the necessary renormalization of the SM-like parameters is thus postponed and left to the user of the EFT.
We will actually exploit this freedom in the field renormalization of the gauge-boson
fields, where we pre-renormalize in the gauge-symmetric field basis in the full theory, but finally renormalize in the basis corresponding to mass eigenstates in the EFT.
As explained for the strict top-down approach, an optional renormalization of the higher-dimensional Wilson coefficients ($C_i$) may be carried out for the purpose of resummation.

As our ``pragmatic'' top-down approach closely follows the ``strict'' variant described before, which in turn is based
on the complete renormalization of all SESM parameters, we will in the following first discuss the latter in some detail.
We will therefore not distinguish between SM-like and BSM parameters (nor between renormalization and pre-renormalization)
in \subsecsm{renormalization-trafo}{LeffRen}.
In \subsec{BSMRenCond} we focus on the renormalization of the BSM parameters,
while the pre-renormalization of the SM-like quantities is fixed in \sec{fieldredef}.

\subsection{Renormalization transformations}
\label{subsec:renormalization-trafo}

A renormalization transformation splits the parameters of a Lagrangian according to
\begin{equation}
	c_{i,0}=c_i+\delta c_i \,, %=c_i+\dhe c_i+\dle c_i,
\end{equation}
with the subscript ``0'' denoting the original bare quantities.
The $\delta c_i $ are renormalization constants which are fixed by imposing
\textit{renormalization conditions} in the
subsequent sections.
These conditions determine the physical meaning of the renormalized parameters $c_i$.
Likewise, we renormalize the fields schematically according to
\begin{equation}
{\phi}_{i,0}=\Bigl(\delta_{ij}+\frac{1}{2}\delta Z_{ij} \Bigr){\phi}_{j} \,.
\end{equation}
However, the field renormalization constants $\delta Z_{ij}$ do not impact results
for observables, and therefore can be chosen for convenience.
In the BFM, this field renormalization is typically carried out only for the
background fields, which are the sources of the BFM effective action.
A renormalization of quantum fields would not affect Green functions with
external background fields at all and is, therefore, not considered.

Specifically, our renormalization transformation of the input and tadpole
parameters of the SESM is given by
\begin{align}
	t_{\mathrm{h},0}&=t_\mathrm{h}+\delta t_\mathrm{h} \,,\quad
	&M_{\mathrm{h},0}^2&=\Mh^2+\delta\Mh^2 \,,\quad
\nn\\
	M_{\PW,0}^2&=\MW^2+\delta\MW^2 \,,\quad
	&M_{\PZ,0}^2&=\MZ^2+\delta\MZ^2 \,,\quad
	&e_0&=(1+\delta Z_e)e \,,
\label{eq:SMparameter-ren}
\end{align}
for the SM-like parameters and by
\begin{align}
	t_{\mathrm{H},0}&=t_\mathrm{H}+\delta t_\mathrm{H} \,,\quad
	&M_{\mathrm{H},0}^2&=\MH^2+\delta\MH^2 \,,\quad
	&s_{\alpha,0}&=s_\alpha+\delta s_\alpha \,, \quad
	&\lambda_{12,0}&=\lambda_{12}+\delta\lambda_{12} \,,
\label{eq:BSMparameter-ren}
\end{align}
for the BSM parameters.
Since the bare and renormalized versions of the
weak mixing angle $\theta_\mathrm{w}$ and the
vev parameter $ v_2 $ are fixed to all orders by
\begin{equation}
	\cos\theta_\mathrm{w} \equiv \cw=\frac{\MW}{\MZ} \,,
	\qquad
	\sw^2 = 1-\cw^2\,,
	\qquad
	\MW=\frac{g_2v_2}{2}=\frac{ev_2}{2\sw}\,,
\label{eq:SMparameter-dep}
\end{equation}
the corresponding renormalization constants, introduced via
\begin{align}
	v_{2,0}=v_2+\delta v_2 \,,\qquad
	s_{\mathrm{w},0}=\sw+\delta\sw \,,
\end{align}
are not independent, but related to
$\delta\MW^2$, $\delta\MZ^2$, and $\delta Z_e$
according to
\begin{equation}
	\frac{\delta\cw^2}{\cw^2}=\frac{\delta\MW^2}{\MW^2}-\frac{\delta\MZ^2}{\MZ^2}\,,
	\qquad
	\frac{\delta\sw}{\sw}=\frac{\delta\sw^2}{2\sw^2}=-\frac{\delta\cw^2}{2\sw^2}\,,
	\qquad
	\frac{\delta v_2}{v_2}=\frac{\delta\MW^2}{2\MW^2}+\frac{\delta\sw}{\sw}-\delta Z_e\,.
  \label{eq:renconstrel}
\end{equation}
Note that the bare tadpole parameters $t_{\Ph,0}$, $t_{\PH,0}$ correspond to the tadpole parameters introduced (at tree level)  in \subsec{SESM} without
subscript~``0'', while
$t_{\Ph}$, $t_{\PH}$ are the renormalized tadpole parameters emerging after the renormalization transformation.

The parameter renormalization transformation is
supplemented by the following renormalization transformation of the fields,
\begin{align}
	\begin{pmatrix}
		\hat{H}_0\\\hat{h}_0
	\end{pmatrix}
	&=
	\begin{pmatrix}
		1+\frac{1}{2}\delta Z_{\hat H\hat H}
		&\frac{1}{2}\delta Z_{\hat H\hat h}\\
		\frac{1}{2}\delta Z_{\hat h\hat H}
		&1+\frac{1}{2}\delta Z_{\hat h\hat h}
	\end{pmatrix}
	\begin{pmatrix}
		\hat{H}\\\hat{h}
	\end{pmatrix},
\nn\\[1 ex]
	\hat{W}^{a,\mu}_0&=\bigl(1+\tfrac{1}{2}\delta Z_{\hat W}\bigr) \hat{W}^{a,\mu},\qquad
	\hat{B}^\mu_0= \bigl(1+ \tfrac{1}{2}\delta Z_{\hat B}\bigr) \hat{B}^\mu.
\label{eq:field-ren}
\end{align}%
The transformation of the gauge fields $\hat{W}^a_\mu$ and $\hat{B}_\mu$ is
carried out in the (gauge) symmetric field basis, as opposed to the field basis $\hat{W}^\pm_\mu$, $\hat{Z}_\mu$, $\hat{A}_\mu$ corresponding to mass eigenstates,
because we aim at a (background) gauge-invariant renormalized effective Lagrangian.
This means that some non-trivial external LSZ factors
would be encountered with external (physical) gauge bosons in the calculation of S-matrix elements if the
emerging effective Lagrangian were used directly.
Throughout this paper all parameter and field renormalization constants are of one-loop order,
i.e.\ of ${\cal O}(e^2)$ relative to the lowest-order contribution, and all
relations involving renormalization constants are
valid at one-loop level, but in general not beyond.

\subsection{Tadpole renormalization}
\label{subsec:tadpoleren}

Before moving on to the formulation of renormalization conditions,
some considerations on the renormalization of the tadpole parameters are in order.
For the wider context concerning tadpole renormalization, we refer to
\citeres{Denner:2016etu,Denner:2018opp,Denner:2019vbn,Dittmaier:2022maf,Dittmaier:2022ivi}.
The tadpole renormalization constants $\de t_{\Ph/\PH}$, which are usually fixed in order to
cancel all tadpole loop graphs $T^{\hat{h}/\hat{H}}$, so that the renormalized
one-point functions $\Gamma^{\hat h/\hat H}_{\ren}$ of the Higgs fields vanish,
\begin{align}
  \Gamma^{\hat h/\hat H}_{\ren} = T^{\hat{h}/\hat{H}} + \de t_{\Ph/\PH} \overset{!}{=} 0 \,,
  \label{eq:TadpoleCondition}
\end{align}
can be introduced as part of the parameter or field renormalization transformation,
or via both.
Introducing the $\de t_{\Ph/\PH}$,
as in \eqs{SMparameter-ren}{BSMparameter-ren},
together with the other parameter renormalization constants follows the procedure
proposed in \citeres{Sirlin:1985ux,Bohm:1986rj,Denner:1991kt}
for the SM and is dubbed
\textit{Parameter Renormalized Tadpole Scheme (PRTS)} in \citere{Denner:2019vbn}.
\revision{To be concrete, in the PRTS the renormalized tadpole parameters are
set to zero by definition, so that the tadpole renormalization constants
coincide with the respective bare tadpole parameters,
\begin{align}
t_{\Ph/\PH}^\PRTS = 0, \qquad
\de t_{\Ph/\PH}^\PRTS = t_{\Ph/\PH,0}^\PRTS.
\end{align}
The relations \refeq{eq:originalparams} between original parameters and
Higgs-boson masses and mixing angle are valid for bare and renormalized parameters,
but with the tadpole terms $t_{\Ph/\PH}$ set to $\de t_{\Ph/\PH}$ and zero, respectively.
Alternatively, following the {\it Fleischer-Jegerlehner Tadpole Scheme (FJTS)} introduced in
\citere{Fleischer:1980ub}, tadpole renormalization constants can}
be introduced via field shifts $v_i + h_i\to v_i+h_i+\Delta v_i$ for any Higgs
fields $h_i$ developing vevs $v_i$ and adjusting $\Delta v_i$ to produce the
appropriate constants $\delta t_i$,
as required by \eq{TadpoleCondition}.
\revision{%
\begin{align}
\De v_{h} = -\frac{\de t_{\Ph}^\FJTS}{\Mh^2}, \qquad
\De v_{H} = -\frac{\de t_{\PH}^\FJTS}{\MH^2}.
\end{align}
The bare tadpole parameters are set to zero, $t_{\Ph/\PH,0}^\FJTS=0$,
and tadpole terms never enter relations between parameters, i.e.\
the relations \refeq{eq:originalparams} are valid for both bare and renormalized
parameters with the tadpole terms $t_{\Ph/\PH}$ being zero.
}

As long as the renormalization conditions of mass parameters and corresponding
mixing angles are tied to S-matrix elements, as in OS renormalization schemes,
predictions are independent of the tadpole renormalization scheme.
However, in other schemes such as $\MSbar$ renormalization schemes the renormalized
mass or mixing parameters, and thus any higher-order predictions for observables,
in general depend on the tadpole scheme, as for instance discussed in
\citeres{Denner:2016etu,Denner:2018opp,Denner:2019vbn,Dittmaier:2022maf,Dittmaier:2022ivi}
(see also references therein).
In that case both the PRTS and the FJTS are known to have up- and downsides:
While perturbative $\MSbar$-based corrections in the PRTS are moderate in size, they develop a gauge dependence,
so that either all calculations have to be done in one
and the same gauge, or converted properly from one gauge to the other.
The gauge-invariance violation of predictions in the PRTS can be traced back
to the occurrence of the gauge-dependent tadpole contributions $\de t_i$ in the
relations between bare input parameters of the original Lagrangian and the bare
particle masses. The moderate size of corrections results from the fact that technically
the Higgs field excitation is defined as deviation from the corrected (``true'')
minimum of the Higgs potential.
On the other hand, $\MSbar$-based predictions in the FJTS are gauge independent, but
prone to huge corrections that are hard (often impossible) to tame at NLO.
The FJTS avoids gauge dependences in predictions, because the field shift is a mere
change of integration variables in the functional path integral, which has no impact
on predictions.
The potentially large corrections of the FJTS are due to the fact that Higgs field
excitations are parametrized around the uncorrected bare minimum of the Higgs potential
which receives large (tadpole) corrections that enter higher-order predictions.

In order to resolve the problems of the PRTS and FJTS in combination with $\MSbar$ renormalization,
the {\it Gauge-Invariant Vacuum expection value Scheme (GIVS)} was suggested
in \citeres{Dittmaier:2022maf,Dittmaier:2022ivi}. It
avoids gauge dependences and leads to moderate corrections at the same time.
Technically, the GIVS splits the tadpole renormalization constants $\de t_i$
into two additive parts:
\revision{$\de t_i^\GIVS = \de t_{i,1}^\GIVS + \de t_{i,2}^\GIVS$:}
The first part is introduced like in the PRTS,
but gauge independent because it is calculated in the non-linear realization of
the scalar fields, where Higgs fields are singlets,
\revision{%
\begin{align}
\de t_{\Ph/\PH,1}^\GIVS = \de t_{\Ph/\PH,\mathrm{nl}}^\PRTS = -T^{\hat h/\hat H}_{\mathrm{nl}},
\label{eq:dtGIVS1}
\end{align}
where the subscript ``nl'' refers to the non-linear Higgs realization.}
The second part of $\de t_i$ is needed since the gauge-independent first part
does not fully cancel the tadpole loops in the linear realization of the scalar
fields, in which radiative corrections are usually calculated.
This second, gauge-dependent part of the tadpole loops is compensated by shifts in the Higgs fields as in the FJTS, so that its
gauge dependence does not enter predictions,
\revision{%
\begin{align}
\de t_{\Ph,2}^\GIVS = -\Mh^2 \De v_{\Ph} = T^{\hat H}_{\mathrm{nl}} - T^{\hat H}.
\end{align}}
The potentially large corrections connected with the perturbative shift in the
minimum of the Higgs potential are contained in the first, gauge-independent contribution
to $\de t_i$, so that the GIVS leads to moderate perturbative corrections to
observables, very similar to $\MSbar$ predictions based on the PRTS.
In summary, predictions in the GIVS (obtained in linear Higgs realizations)
are by design fully equivalent to predictions obtained in non-linear Higgs realizations
with the PRTS.

Since our construction of the effective theory for the SESM with a heavy
Higgs boson~H consistently proceeds in the non-linear Higgs realization,
and we introduce the tadpole constants $\de t_{\Ph/\PH}$ as in the PRTS,
we, thus, effectively employ the GIVS when we renormalize the mixing angle~$\alpha$.
The second FJTS-like part of $\de t_{\Ph/\PH}$,
employed by the GIVS with linearly realized scalar fields, is zero by construction
if the scalar fields are realized non-linearly.
We recall that in \citere{Dittmaier:2021fls} we have considered the
$\MSbar$ renormalization of the (sine of the) mixing angle $\alpha$
also in the FJTS and found that this breaks the
power-counting in $\zeta\sim\MH/\Mh$ badly.
As a consequence the FJTS had to be classified as ineligible for the construction of a proper EFT
of the SESM with $\MSbar$ mixing angle.
Concretely, in contrast to the FJTS, the PRTS
tadpole contributions do not enter other renormalization constants for
masses and mixings,
which is advantageous as tadpole contributions can lead to renormalized parameters
being enhanced in the EFT power-counting with respect to their bare counterparts,
i.e.\ $c_{i,0}\sim\zeta^{z_0}$ and $c\sim\zeta^z$ for some parameter $c_i$ with $z>z_0$.
This is undesirable, because in combination with $\MSbar$ renormalization conditions
for such parameters $c_i$, this breaks the power-counting derived from
LO relations between input parameters, so that the basic principle for the
EFT construction used at tree level is overthrown.%
\footnote{Note, however, that in any scheme $\delta \Mh^2 \sim \MH^2$ is $\zeta$-enhanced
(which can be regarded as a manifestation of the EW hierarchy problem), cf.~\app{CThard}.}
In \citere{Dittmaier:2021fls}, this effect was explicitly observed for
an $\MSbar$-renormalized mixing angle $\alpha$ in the FJTS.
In this paper, we consistently employ an OS scheme or the $\MSbar$ scheme
within the GIVS (meaning the PRTS within non-linear Higgs realizations), in order to avoid such inconsistencies.

\subsection{Counterterm contributions to the effective Lagrangian}
\label{subsec:LeffRen}

The renormalization transformations \eqss{SMparameter-ren}{BSMparameter-ren}{field-ren}
split the full SESM Lagrangian,
\begin{align}
  \L_\SESM = \L_\SESM^\ren + \delta \L_\SESM^\ct \,,
\end{align}
into a ``renormalized'' Lagrangian $\L_\SESM^\ren$ and a counterterm Lagrangian $\delta \L_\SESM^\ct$, which contains all contributions proportional to the $\delta c_i$ or the (optional) $\delta Z_{ij}$.
By definition, the renormalized Lagrangian $\L_\SESM^\ren$
is explicitly UV finite, when expressed in terms of renormalized parameters and fields.
In our derivation of the effective Lagrangian we have not yet addressed the contribution from $\delta \L_\SESM^\ct$, which we denote $\delta \L_\eff^{\SESM,\ct}$ in the following.
It can be obtained in the same way as $\L_\eff^\tree$
in \subsec{Lefftree}, because at NLO the $\delta c_i$ and $\delta Z_{ij}$, both of $\ord(e^2)$, only occur in tree-level graphs.%
\footnote{%
In fact, if we had $\delta c_i/c_i =\ord(\zeta^{z_i})$ with $z_i \le 0$
(and similarly for field renormalization),
$\delta \L_\eff^{\SESM,\ct}$ could be directly determined from $\L_\eff^\tree$ by using the bare parameters
$\delta c_{i,0} = c_i + \delta c_i$ and expanding to $\ord(e^2)$ and $\ord(\zeta^{-2})$.
This is, however, not the case as, e.g., $\delta \Mh^2 \sim \MH^2$.}
The expression for $\delta \L_\eff^{\SESM,\ct}$ in terms of the relevant
$\delta c_i$ and $\delta Z_{ij}$ to $\ord(\zeta^{-2})$ is given in \app{CTLagrangian}.
The complete NLO result for the effective Lagrangian is then
\begin{align}
  \L_\eff = \L_\eff^\tree+\delta\L_\eff^\oneloop+\delta\L_\eff^{\SESM,\ct} \,,
  \label{eq:Leff}
\end{align}
where $\L_\eff^\tree$ and $\delta\L_\eff^\oneloop$ are given in \eqs{LeffTreelong}{Leff1loopAfterEOM}, respectively.

\subsection{Renormalization conditions for the BSM parameters}
\label{subsec:BSMRenCond}

In contrast to the BSM parameters and fields, all renormalization constants of SM-like
quantities can and will eventually be renormalized within the constructed EFT.
We may, however, make use of the freedom to fix
the undetermined SM-like renormalization constants
contained in $\delta\L_\eff^{\SESM,\ct}$
at any point in the derivation of the effective Lagrangian.
For now we leave their determination still open.
Later we will fix their values (i.e.\ pre-renormalize the associated SM-like quantities) in the course of simplifying the effective
Lagrangian via field redefinitions of the SM-like fields in the next section.

From the EFT perspective, the
BSM parameters $\MH$, $\lambda_{12}$, and $\sa$
are no longer explicit input parameters and only appear implicitly in the
expressions for EFT matching coefficients.
Thus, the BSM renormalization constants have to be imported from the full underlying
model, in order to guarantee that the EFT reproduces predictions for observables
in the large-mass limit based on the same input parameters
as in the full theory. In detail, we have to import those BSM renormalization constants
that contribute to
$\delta\L_\eff^{\SESM,\ct}$ to $\ord(\zeta^{-2})$.

The counterterm Lagrangian $\delta\L_\eff^{\SESM,\ct}$ involves the one-loop
BSM renormalization constants
of the full SESM obtained in the non-linear Higgs realization within the BFM.
The calculation of these renormalization constants in this setup proceeds
along the same lines as in the linear Higgs realization and/or the conventional quantization formalism,
but the results cannot be simply taken over.
As we anyway only need a few terms in the large-mass expansion, we employ once again the method of regions.

As already shown in \citere{Dittmaier:2021fls},
renormalizing Higgs masses and fields in the OS scheme
(see Sec.~5.2 of \citere{Dittmaier:2021fls} and
\citeres{Boggia:2016asg,Denner:2018opp} for further details) and
excluding the FJTS, the BSM renormalization constants scale in the large-$\MH$ limit as
\begin{align}
\hspace{3em}
\de\MH^2 \sim{}& \MH^2 \sim \zeta^2, \qquad
&\de\lambda_{12} \sim{}& \zeta^0,  \qquad
&\de\sa \sim{}& \zeta^{-1},  \qquad
&\de t_\PH \sim{}& \zeta^3.
\hspace{3em}
\\
\de Z_{\hat H\hat H} \sim{}& \zeta^0, \qquad
&\de Z_{\hat H\hat h} \sim{}& \zeta^{-1}, \qquad
&\de Z_{\hat h\hat H} \sim{}& \zeta^{1}.
\end{align}%
With this scaling, the mass renormalization constant $\de\MH^2$ does not appear
in $\delta\L_\eff^{\SESM,\ct}$,
to $\ord(\zeta^{-2})$,
so that its explicit expression is not required in our derivation of the effective Lagrangian,
cf.~\app{CTLagrangian}.

Similarly, $\de\lambda_{12}$ does not appear in $\delta \L_\SESM^\ct$
and is not needed either, so that we need not choose a specific renormalization scheme for $\lambda_{12}$,
as the final renormalized effective Lagrangian will not depend on it.
The only requirement is that the renormalized parameter is not enhanced in the EFT power-counting,
i.e.\ $\lambda_{12}\sim\la_{12,0}\sim\zeta^0 $.
A good choice that respects this is the $\MSbar$ scheme for the renormalization of
$\lambda_{12}$.
In this context, it is worthwhile to mention that the free parameter $\lambda_{12}$
could be traded, e.g., for $\lambda_{1}$ without any problems, as also discussed in
\citere{Denner:2018opp} for the full SESM.

Concerning the tadpole renormalization constant $\de t_\PH$,
the explicit expression of its hard part $\de_h t_\PH$
and the scaling behaviour of its soft part,
$\de_s t_\PH\sim\zeta^1$, are required for the effective Lagrangian to $\ord(\zeta^{-2})$.
The hard part $\dhe t_{\PH}$ can be directly
read off as the coefficient of the field monomial $\hat H$
in $\delta\L_\eff^\oneloop$ as given in \eq{Leff1loopBeforeEOM}.
The soft part, which is given for completeness below,
is obtained from a simple calculation in the full SESM, where the
leading-order term
in the large-$\MH$ expansion comes from a tadpole with a light Higgs loop.
In total we have
\begin{align}
\delta t_\PH ={} & \delta_h t_\PH + \delta_s t_\PH \,,
\\
\delta_h t_\PH ={} & -\frac{3\lambda_{12} \MH^2 v_2 I_{20}}{8(D-2)\pi^2\sa}
	- \frac{3(\MH^4\sa^4+2\lambda_{12}\Mh^2 v_2^2 -2\lambda_{12}\MH^2 \sa^2 v_2^2)I_{20}}{16(D-2)\pi^2\sa v_2}
	+ \ord(\zeta^{-1}) \,,
\\
\delta_s t_\PH ={} & -\frac{e\,\sa\Mh^2\MH^2}{64\MW\pi^2\sw} B_0(0,0,\Mh^2)
	+ \ord(\zeta^{-1}) \,,
  \label{eq:dtHsoft}
\end{align}
with the two-point integral as defined in \eq{B0},
\begin{equation}
    B_0(0,0,\Mh^2)
	=\Delta+\ln\left(\frac{\mu^2}{\Mh^2}\right)+1+\mathcal{O}(\eps) \,.
\end{equation}%

The field renormalization constants $\de Z_{\hat H\hat H}$, $\de Z_{\hat H\hat h}$ systematically disappear from the counterterm Lagrangian in the step of eliminating the
heavy field~$\tilde H$ from the Lagrangian, so that they are not needed.
The renormalization constant $\de Z_{\hat h\hat H}$ is still present in $\delta \L_\SESM^\ct$
at this point, but disappears from the final effective Lagrangian in the course of the
field redefinition of the light field~$\tilde h$ described in the next section.
In principle, $\de Z_{\hat h\hat H}$ could, thus, be set to zero, but we keep it in the
actual calculation in order to check that it drops out as expected.

The renormalization constant $\delta \sa$ of the mixing angle~$\alpha$ is the only
renormalization constant of a BSM input parameter the explicit expression of which really matters in our final effective Lagrangian.
Its hard part cannot simply be determined from $\delta \L_\eff^\oneloop$ of
\eq{Leff1loopBeforeEOM} as (depending on the renormalization scheme) some
ingredients need to be computed with external momenta $\sim \MH\sim\zeta$.
Moreover, unlike for the renormalization constants of the SM-like parameters,
its soft part cannot be conveniently obtained by an EFT calculation, where the heavy
Higgs field
is no more a dynamical degree of freedom.
We, thus, have to compute $\delta \sa$ using the full SESM Lagrangian with the non-linear Higgs realization and the BFM.
In the OS scheme of \citere{Denner:2018opp}, $\delta \sa$ reads
in terms of Higgs self-energies%
\footnote{The OS renormalization condition imposed on $\alpha$~\cite{Denner:2018opp} ensures
that the ratio of transition matrix elements for the couplings of h/H to a fictitious singlet ``fake fermion''
with infinitesimal coupling to the singlet field~$\sigma$ is exactly given by $-\tan\alpha$.}
\begin{align}\label{eq:dsaOS}
	\delta\sa^\OS={}&\frac{c_\alpha}{2}
	\mathrm{Re}\bigg[
	c_\alpha\sa\left(\Sigma'^{\hat{h}\hat{h}}(\Mh^2)
	-\Sigma'^{\hat{H}\hat{H}}(\MH^2)\right)
	+\frac{2}{\MH^2-\Mh^2}\left(c_\alpha^2\Sigma^{\hat{H}\hat{h}}(\Mh^2)
	+\sa^2\Sigma^{\hat{h}\hat{H}}(\MH^2)\right)
	\bigg],
\end{align}
where we use the shorthand
\begin{equation}
	\Sigma'^{\hat{\phi}\hat{\phi}}(M^2) \equiv
	\frac{\partial\Sigma^{\hat{\phi}\hat{\phi}}(p^2)}{\partial p^2}\bigg\vert_{p^2=M^2}.
\end{equation}
\revision{In the PRTS and 't~Hooft--Feynman gauge ($\xi=1$), \eq{dsaOS} evaluates to}
\begin{align}
\delta s_\alpha^\OS={}&
\frac{1}{8\pi^2 \sa}
\biggl(
-\frac{6(D-1)\lambda_{12}^2v_2^2}{\MH^2}+(3D-7)\lambda_{12}s_{\alpha}^2
+\frac{2\MH^2s_{\alpha}^4}{v_2^2}
\biggr) \frac{B_0(0,\MH^2,\MH^2)}{ (D-2)}
\nn\\ & {}
+\frac{3\Mh^2s_{\alpha}}{32\pi^2v_2^2}B_0(\Mh^2,\Mh^2,\Mh^2)
+\frac{s_{\alpha}}{32\pi^2}\biggl(\frac{2\Mh^2}{v_2^2}-\frac{e^2}{\sw^2}\biggr)B_0(\Mh^2,\MW^2,\MW^2)
\nn\\ & {}
+\frac{s_{\alpha}}{64\pi^2}\biggl(\frac{2\Mh^2}{v_2^2}-\frac{e^2}{\cw^2\sw^2}\biggr)B_0(\Mh^2,\MZ^2,\MZ^2)
-\frac{e^2s_{\alpha}B_0(0,\MW^2,\MW^2)}{32\pi^2(D-2)\sw^2}
\nn\\ & {}
-\frac{e^2s_{\alpha}B_0(0,\MZ^2,\MZ^2)}{64\pi^2(D-2)\cw^2\sw^2}
+\frac{\MH^4s_{\alpha}^3}{32\pi^2v_2^2}B_0'(0,0,\MH^2)
-\frac{\MH^4s_{\alpha}^3}{16\pi^2v_2^2}B_0'(\MH^2,0,0)
\nn\\ & {}
-\frac{9\lambda_{12}^2v_2^2}{16\pi^2s_{\alpha}}B_0'(\MH^2,\MH^2,\MH^2)
+\frac{9\Mh^4s_{\alpha}}{64\pi^2v_2^2}B_0'(\Mh^2,\Mh^2,\Mh^2)
\nn\\ & {}
+\frac{s_{\alpha}}{32\pi^2}\biggl( \frac{\Mh^4}{v_2^2} -\frac{e^2\Mh^2}{\sw^2} +\frac{(D-1)e^4v_2^2}{4\sw^4} \biggr)
B_0'(\Mh^2,\MW^2,\MW^2)
\nn\\ & {}
+\frac{s_{\alpha}}{64\pi^2\cw^4} \biggl( \frac{\cw^4\Mh^4}{v_2^2} -\frac{\cw^2e^2\Mh^2}{\sw^2}
+\frac{(D-1)e^4v_2^2}{4\sw^4} \biggr) B_0'(\Mh^2,\MZ^2,\MZ^2)
+\ord(\zeta^{-2})
\nn\\
={}&
\bigg[\Delta+\ln\bigg(\frac{\mu^2}{\MH^2}\bigg)\bigg]
\frac{(2\MH^2s_{\alpha}^2+9\lambda_{12}v_2^2)
(\MH^2s_{\alpha}^2-2\lambda_{12}v_2^2)}{16\pi^2\MH^2s_{\alpha}v_2^2}
\nn\\ & {}
+ \bigg[\Delta+\ln\bigg(\frac{\mu^2}{\Mh^2}\bigg)\bigg]
\frac{3s_{\alpha}}{16\pi^2}
\biggl( \frac{\Mh^2}{v_2^2} -\frac{e^2}{4\sw^2} -\frac{e^2}{8\cw^2\sw^2} \biggr)
+ \;\ldots\; +\ord(\zeta^{-2})\,,
\label{eq:dsa}
\end{align}
with
\begin{equation}
B'_0(M^2,m_1^2,m_2^2)\equiv\frac{\partial}{\partial p^2}B_0(p^2,m_1^2,m_2^2)\bigg\vert_{p^2=M^2} \,.
\end{equation}
The ellipses in \eq{dsa} stand for UV-finite, $\MH$-independent terms,
which may include logarithms depending only on the small masses of SM particles.
The renormalization constant $\delta\sa^\MSbar$ in the $\MSbar$ scheme with PRTS tadpole
treatment can easily be obtained by extracting the UV divergent part
\begin{equation}
	\delta\sa^\MSbar=\delta\sa^\OS\big|_{\text{UV-div}}
\end{equation}
from Eq.~\eqref{eq:dsa}, where the subscript ``UV-div'' means that only the UV-divergent terms
proportional to $\Delta$ are kept.
Finally, we emphasize again that the calculational prescription~\eq{dsaOS}
for $\delta\sa^\OS$~\cite{Denner:2018opp}
applies for both the linear and non-linear
Higgs realization, but the explicit result~\eq{dsa} applies only to the non-linear
realization, because the individual Higgs self-energies in the two realizations
are not the same.
The result of \eq{dsa} is valid both in the BFM and in the conventional quantization formalism
to the given order in $\zeta^{-1}$.

At this point, after having fixed the renormalization in the BSM sector,
we emphasize that the resulting
effective Lagrangian is already suitable for phenomenological applications,
assuming the renormalization constants of the SM-like parameters
are fixed by renormalization conditions within the EFT.
This means that predictions for physical observables could be computed now directly following the
procedure outlined at the beginning of \sec{pheno}, without the step of simplifying the
effective Lagrangian via field redefinitions of the light (SM-like) fields
described in the next section.
Results for observables computed from the (fully renormalized) $\L_\eff$~\eqref{eq:Leff}
are finite to (relative) $\ord(e^2)$ and agree with those of the full SESM to $\ord(\zeta^{-2})$.
The full-theory UV divergences in $\delta\L_\eff^\oneloop$ are cancelled by $\delta\L_\eff^{\SESM,\ct}$.
However, $\delta\L_\eff^\oneloop$ also contains divergences arising from IR singularities in the hard region of one-loop integrals with a heavy-Higgs propagator, while $\L_\eff^\tree$ and $\delta\L_\eff^{\SESM,\ct}$ are IR finite.
From the perspective of the EFT these divergences
may be interpreted as UV counterterms and cancel in the computation of physical observables with explicit UV divergences from one-loop EFT diagrams that include single operator insertions from $\L_\eff^\tree$.
Consistency, however, demands to carry out the EFT calculations using dimensional regularization with the same scale $\mu$ that appears in $\L_\eff$.

\section{Field redefinitions}
\label{sec:fieldredef}

As the quantum fields $\phi_i$ represent integration variables in the path integral, we have the freedom to make field transformations of the form
\begin{equation}
  \label{eq:redef}
	\phi_i=\phi'_i+f_i(\hat{\phi}'_j,\phi'_j,\partial\hat{\phi}'_j,\partial\phi'_j,\partial^2\hat{\phi}'_j,\partial^2\phi'_j,\dots) \,,
\end{equation}
where the $ f_i $ are functions
of all the possible quantum and background fields $ \phi'_j,\,\hat{\phi}'_j, $ and their derivatives.
Physical observables like $ S $-matrix elements will be invariant under these field redefinitions, as long as
the functions $ f_i $ are suppressed in either the coupling constant or the EFT power
counting parameter $ \zeta^{-1} $.
The corresponding functional Jacobian of the field redefinition is unity
in dimensional regularization~\cite{Chisholm:1961tha,Kamefuchi:1961sb,Passarino:2016saj,Cohen:2023ekv,Manohar:2018aog,Criado:2018sdb}, i.e.\ the measure of the redefined path integral will have the same form as the original.
We can use this freedom of field redefinitions to bring our EFT
Lagrangian~\eqref{eq:Leff} into more compact
form, which will in particular facilitate its use in
phenomenological applications, see \sec{pheno}.
Our aim is to keep the number of non-SMEFT operators relevant at $\ord(\zeta^{-2})$ as small as possible.
Note that at all stages during processing the effective Lagrangian we
use IBP
to eliminate redundant operators, i.e.\ we bring field monomials to
a unique form using IBP relations and drop all surface terms.

Furthermore, we will exploit the freedom to adjust the corresponding SM-like
parameter and field renormalization constants of our pre-renormalization step
in $\delta\L_\eff^{\SESM,\ct}$
to absorb (hard) one-loop corrections to the effective Lagrangian that have SM form.

\subsection{Field redefinitions for the LO effective Lagrangian}
\label{subsec:FieldRedefLO}

By redefining the Higgs field
(actually the quantum field as integration variable of the path integral)
according to
\begin{equation}
   \tilde h =\tilde{h}'
	+\frac{s_\alpha^2}{2v_2}(v_2 + \tilde{h}')\tilde{h}'
  \label{eq:fieldredeftree}
\end{equation}
and defining
\begin{align}
	\Mh^{\prime\,2}=\Mh^2+s_\alpha^2\Mh^2,
  \label{eq:Mhredef}
\end{align}
the Lagrangian $\L_{\eff}^\tree$ of
\eq{LeffTreelong} could be rendered more compact.
Notably, $\Mh^{\prime\,2}$ plays the same role as $\Mh^2$ in $\L_{\eff}^\tree$ before the substitution.
After the field and parameter redefinitions in \eqs{fieldredeftree}{Mhredef},
$\L_\eff^\tree$ takes the form
\begin{equation}
	\L_{\eff}^{\tree\,\prime}=\L'_{\SM}
	-\frac{s_\alpha^2}{8v_2^2}(v_2+\tilde{h}')^2\Box(v_2+\tilde{h}')^2
	+\mathcal{O}(\zeta^{-4}) \,.
  \label{eq:LeffTree}
\end{equation}%
where the prime on $\L'_{\SM}$ indicates that this is the SM Lagrangian with
$\Mh'$ and $\tilde{h}'$ in place of $\Mh$ and $\tilde{h}$, respectively.
Note that the kinetic term of the Higgs field in $\L_{\eff}^{\tree\,\prime}$
receives corrections of $ \ord(\zeta^{-2}) $ from the second term in \eq{LeffTree},
and is thus not canonically normalized.
This comes from absorbing all BSM contributions in \eq{LeffTreelong} into a single EFT operator
and is at the cost of a non-standard LSZ wave function renormalization of the external Higgs field in calculations for (low-energy) observables.
We note in passing that the EFT operator in \eq{LeffTree} is of SMEFT form;
this will be made more explicit upon restoring the Goldstone fields below.

In spite of its nice compactness, we actually do not make use of $\L_{\eff}^{\tree\,\prime}$, neither in the subsequent processing of $\L_{\eff}$,
nor in practical applications.
Instead, we will work with $\L_{\eff}^{\tree}$ as given in \eq{LeffTreelong},
because it is much easier to perform perturbative calculations with Lagrangians
supporting canonical field normalization at tree level.

\subsection{Field redefinitions for the NLO effective Lagrangian
and pre-renormalization of the SM sector}
\label{subsec:field-redef-NLO}

Let us focus on the simplification of the $\ord(\zeta^{-2})$ part of the effective
Lagrangian \eqref{eq:Leff}. As we wish to retain gauge invariance
under the unbroken electromagnetic $ U(1) $ gauge group, we only consider functions $ f_i $ that
have the same behaviour under the electromagnetic gauge transformation as the field $\phi_i$. Besides that, we make
\revision{a quite general {\it polynomial} ansatz,}
allowing for $f_i$ to have an operator dimension (i.e.\ counting the mass dimension of fields and derivatives) of at most
two units
greater than the mass dimension of $\phi_i$.
Terms of higher operator dimension would give rise to terms of dimension seven or higher in the effective Lagrangian, which we want to avoid.%
\footnote{As an exception we allow for specific field redefinitions of one dimension higher and proportional to $\delta Z_{\Ph\PH}$ in order to eliminate dimension-seven operators proportional to $\delta Z_{\Ph\PH}$ in the counterterm Lagrangian~\eqref{eq:Lct} which survive pre-renormalization, see \eq{RestFieldRedef}. As mentioned earlier, we could have just as well set $\delta Z_{\Ph\PH}$ to zero, such that no terms of higher dimension are needed in the field redefinitions.
}
The coefficients of this ansatz are then systematically chosen to bring the result into a form that is as close to SMEFT as possible.
It turns out that there is no set of coefficients
\revision{of our polynomial ansatz}
that brings our EFT
Lagrangian \eqref{eq:Leff} into SMEFT form completely.
We therefore choose a form where the set of non-SMEFT operators in the final effective Lagrangian is as compact as possible.
The concrete transformations we make are
\begin{equation}\label{eq:RestFieldRedef}
	W^a_\mu=W'^a_\mu,\qquad
	B_\mu=B_\mu',\qquad
	\varphi^a=\varphi'^a,
\end{equation}
and
\begin{align}
\label{eq:HiggsFieldRedef}
h ={}& h'
+\frac{\sa\delta\sa}{v_2}\,\hat h^{\prime\,2}
- \delta Z_{\Ph\PH}\,\frac{\sa}{2\MH^2 v_2}\,\partial^\mu(\hat h'\partial_\mu \hat h')
- \delta Z_{\Ph\PH}\,\frac{e^2\sa}{8\sw^2\MH^2}\,(v_2 + \hat h')\,(C^a_\mu)^2
\nn\\ & {}
+ \delta Z_{\Ph\PH}\,\frac{\sa}{4v_2}
\left[ 1 - \frac{\sa^2}{v_2}(v_2 + \hat h')
+ \frac{\Mh^2}{\MH^2 v_2}(2v_2 + \hat h')
+ \frac{\lambda_{12}}{2\MH^2}(2v_2 + \hat h')^2 \right]
\hat h^{\prime\,2}
\nn\\ & {}
+ \frac{(D-6)(D-4)s_{\alpha}^2I_{20}}{16\pi^2D(D^2-4)v_2^2}\,(\Box\hat h')
- \frac{(D^2-6D+32) e^2 s_{\alpha}^2I_{20}}{64\pi^2D(D^2-4)\sw^2v_2^2}
\, (\hat{C}_{\mu}^a)^2 (v_2+\hat{h}')
\nn\\ & {}
- \frac{(\MH^2\sa^2 - 2\lambda_{12} v_2^2) I_{20}}{16\pi^2 v_2^3}
\left[ \frac{2\sa^2}{(D-2)} + \frac{3(D-1)\lambda_{12}v_2^2}{(D-2)\MH^2} \right]
\hat h^{\prime\,2}
\nn\\ & {}
+ \frac{\sa^2 I_{20}}{16\pi^2 v_2^2}
\biggl[ \frac{(5D^2-6D-80)\Mh^2}{2D(D^2-4) v_2^2}
+ \frac{(D-1)e^2}{D(D-2)\sw^2}
\nn\\ & \qquad {}
+ \frac{(D-4)^2(\MH^2\sa^2-2\lambda_{12}v_2^2)}{2D(D-2)v_2^2} \biggr]
\hat h^{\prime\,2} (3v_2+\hat{h}').
\end{align}
Although we allow for redefinitions of the gauge and Goldstone-boson fields
in our ansatz,
our simplification procedure returns these fields unchanged, i.e.\ redefining them
would render the Lagrangian more complicated rather than simplified.
\revision{In particular, this field transformation does not change any fermionic operator
in the Lagrangian, since the gauge-boson fields remain unchanged and the scalar fields do not couple to massless fermions.}
In \eq{Leff} the background Goldstone fields $\hat{\varphi}^a$ are still absent because of the unitary gauge we have chosen for the background fields via the St\"uckelberg transformation \eqref{eq:Stueckelberg}.
In \subsec{InvStueckelberg} the St\"uckelberg transformation will be inverted and the Goldstone fields will be restored.
This means that requiring manifest SU(2) gauge invariance of the final expression for $\L_\eff$, all terms involving
Goldstone fields are determined by the terms with SU(2) gauge fields.

After applying the
redefinitions \eqref{eq:RestFieldRedef}--\eqref{eq:HiggsFieldRedef},
we obtain a new, more compact expression,
$\delta\L'^{\oneloop}_\eff(\hat{\phi}_i') $, for the NLO part of the effective Lagrangian which is  physically equivalent
to $\delta\L^\oneloop_\eff$ given in \eq{Leff1loopAfterEOM}.
We rename the fields and the Lagrangian,
\begin{equation}
	\hat{\phi}'_i\rightarrow\hat{\phi}_i,\qquad
	\delta\L'^{\oneloop}_\eff(\hat{\phi}_i') \,\rightarrow\, \delta\L^{\oneloop}_\eff(\hat{\phi}_i) ,\,
\end{equation}
effectively dropping the primes for better readability.

In the course of these field redefinitions, we adjust the SM parameter and field renormalization constants
$\de t_{\Ph}$, $\de\Mh^2$, $\de Z_{\hat h\hat h}$,  $\de v_2$, $\de Z_e$, $\de \sw$  $\de Z_{\hat W}$,
and $\de Z_{\hat B}$ of \subsec{renormalization-trafo}
such that the SM operators in \eq{LSM} do not get any corrections from $\delta\L'^{\oneloop}_\eff(\hat{\phi}_i')$, according to \app{CThard}.
Effectively this choice of the renormalization constants plays the role of fixing a renormalization scheme,
where the conditions are chosen according to simplicity (viz.\ to eliminate hard corrections to SM operators)
rather than
requiring certain relations or properties of amplitudes or Green functions.
In order to eventually produce Green functions or amplitudes obeying proper renormalization conditions,
like in OS or $\MSbar$ renormalization, a further renormalization transformation is, thus, needed. For this reason
we call the renormalization of the SM-like quantities in the course of deriving the EFT Lagrangian ``pre-renormalization''.

\subsection{Restoring gauge invariance}
\label{subsec:InvStueckelberg}

In order to restore gauge invariance
of the effective Lagrangian (and thus of the background-field effective action)
w.r.t.\ gauge transformation of the background fields,
we invert the St\"uckel\-berg transformation
\eqref{eq:Stueckelberg} by substituting
\begin{align}
	\hat{W}_{\mu}\rightarrow\hat{U}^\dagger\hat{W}_{\mu}\hat{U}
	+\frac{\ri}{g_{2}}\hat{U}^\dagger\partial_{\mu}\hat{U} \,, \qquad
	&W_{\mu}\rightarrow\hat{U}^\dagger W_{\mu}\hat{U} \,, \qquad
	\hat{B}_{\mu}\rightarrow\hat{B}_{\mu} \,, \qquad
	B_\mu\rightarrow B_{\mu} \,,
	\label{eq:invStueckelberg}
\end{align}
implying
\begin{align}
  \hat{C}_\mu \rightarrow \frac{\ri}{g_2} \hat{U}^\dagger \hat{D}_\mu \hat{U} \,.
\end{align}
The transformation of the quantum fields is again a change of path integration variables, and the change in background fields represents a transformation of the sources of the
background-field effective action.
We stress that the Goldstone fields contained in $\hat U$
are the very same fields that got absorbed into the background gauge fields using \eq{Stueckelberg}.
However, as a part of the SM-like Higgs field $ h_2 $ has been integrated out due to mixing, we can a priori not expect the Goldstone fields to form a doublet with the remaining Higgs field $ \hat{h} $ for non-zero values of the mixing angle $ \alpha\neq0 $.
In the non-linear realization used in
this work this means
\revision{that $ \hat{h} $ cannot be expected to appear
exclusively in the combination $(v_2+\hat{h})\hat{U},$ which
directly translates into the SU(2) doublet of the linear Higgs realization.
In fact, after performing the field transformation of the previous section, we observe
terms involving $ \hat{h} $ that cannot be embedded into a doublet structure,
i.e.\ we get some non-SMEFT operators in the EFT.
As further discussed in \subsec{SMEFTmatch} below, however, these non-SMEFT operators
can still be transformed away by non-polynomial field transformations that lie
beyond the ansatz for the transformation made in the previous section,
include redefinitions of the gauge-boson fields and, thus,
generate fermionic SMEFT operators.}

To restore gauge invariance w.r.t.\ gauge transformations of the quantum fields,
we finally have to explicitly remove
the gauge-fixing Lagrangian
that fixes the gauge of the quantum fields,
which is just inherited from the full-theory
Lagrangian $\L_\mathrm{fix}$ given in \eq{GaugeFixing}:
\begin{equation}
  \L_\eff'(\hat{\phi},\phi)  =  \L_\eff(\hat{\phi},\phi) - \L_\mathrm{fix}(\hat{\phi},\phi) \,,
\end{equation}
where on the r.h.s.\ the Goldstone background fields are understood to be restored
according to \eq{invStueckelberg}.
Once again we will drop the prime in the following and imply the gauge-invariant version of $\L_\eff$ in the following.
Now, all
fields in $\L_\eff^\tree(\hat{\phi},\phi) = \L_\eff^\tree(\tilde{\phi})$ manifestly appear in the original combination $\tilde{\phi} = \hat{\phi} + \phi$ according to \eqs{BFM}{BFMGoldstones}, whereas $\delta\L_\eff^\oneloop$ formally depends on the background fields only.
This is an artefact of our one-loop BFM derivation of the effective Lagrangian.
EFT calculations based on this Lagrangian are by definition only accurate to one-loop level and $\ord(\zeta^{-2})$.
The operators in $\delta\L_\eff^\oneloop$ will thus only contribute to
Green functions via tree-level graphs.
Hence, we can simply promote the
background fields $\hat\phi$
in $\delta\L_\eff^\oneloop$ to the original fields including the quantum part $\phi$,
which does not lead to changes at NLO.

For the remainder of this article we therefore can completely undo the BFM
split of the fields into background and quantum parts and write
\begin{equation}
   \L_\eff(\hat{\phi},\phi)  \to   \L_\eff(\tilde{\phi}) \to \L_\eff(\phi) \,,
\end{equation}
finally dropping also the tilde for simplicity.
Taking the EFT operators in $\L_\eff$ as a basis we could repeat now the matching in a diagrammatic fashion and arrive at the exact same result for the Lagrangian.
Let us stress once again that our effective Lagrangian is fully invariant under the SM gauge group, but the SM-like Higgs field $ h $ and the Goldstone fields $ \varphi_a $ do not transform into each other under the SU(2) gauge group.
\revision{Although the question whether the emerging EFT is of SMEFT type is still open
at this point (unless $\alpha = 0$), see \subsec{SMEFTvsHEFT}, the constructed EFT
certainly fits into the HEFT framework.}

\section{\revision{Effective Lagrangian in terms of bosonic non-SM operators}}
\label{sec:final}

\subsection{\revision{Result for the bosonic effective Lagrangian}}
\label{subsec:nonSMEFTLeff}

\revision{The (bare) effective Lagrangian}
can be split into three parts:
\begin{equation}
\label{eq:FinalResult}
	\L_\eff \,=\,
	\L_\SM
	\,+\, \delta\L_\eff^{\tree}
	\,+\, \delta\L_\eff^{\oneloop,\BSM} \,.
	%\,+\, \delta \L_\eff^\ct \,,
\end{equation}
Recall that $\L_\eff$ corresponds to \eq{Leff} upon field redefinitions, IBP, pre-renormalization, and restoration of gauge invariance as described in \sec{fieldredef}.
On the r.h.s.\ of \eq{FinalResult} we have collected
operators of SM form in $\L_\SM$.
As a result of the processing in \sec{fieldredef}, the operator coefficients in $\L_\SM$ are now, following our approach described in \sec{renormalization}, to be interpreted as those of the bare SM (consisting of bare SM parameters and fields).
In contrast, the BSM parameters of the full theory that now appear in the coefficients of the non-SM operators are understood to be renormalized in the scheme that has been used for the full theory.
For convenience we will give the result for $	\L_\eff$ below in a form where the dependence on the relevant full-theory renormalization constants (in our case only $\delta \sa$) is made explicit, such that the renormalization scheme of the BSM parameters may still be fixed a posteriori.
Note that $\L_\SM$ contains \textit{all} $\ord(\zeta^0)$ terms of $\L_\eff$
as required by decoupling and demonstrated in detail in \citere{Dittmaier:2021fls}.
The non-SM (i.e.\ BSM) operators of $\L_\eff$ are of $\ord(\zeta^{-2})$ and
reside in  $\delta\L_\eff^{\tree}$ and $\delta\L_\eff^{\oneloop,\BSM}$.
LO EFT predictions to $\ord(\zeta^{-2})$ only require $\delta\L_\eff^{\tree}$ for insertions in tree-level diagrams.
The $\ord(e^2)$, i.e.\ NLO, corrections to these predictions involve tree-level graphs with one operator insertion from $\delta\L_\eff^{\oneloop,\BSM}$
and one-loop diagrams with one insertion from $\delta\L_\eff^{\tree}$.

The $\ord(\zeta^{-2})$ part of the effective Lagrangian that emerges from integrating out the heavy Higgs field at tree level has been derived in \subsec{Lefftree}. We have%
\footnote{Here and in the following all expressions for the effective Lagrangian and all phenomenological predictions of the EFT are understood to be valid to $\ord(\zeta^{-2})$ in the large-$\MH$ expansion, i.e.\ we will tacitly drop $\ord(\zeta^{-3})$ contributions, if not specified otherwise.}
	\begin{equation}
		\delta\L_{\eff}^{\tree}= -\frac{\sa^2}{2 v_2^2} \biggl[
		\frac{e^2v_2}{4\sw^2} h(h+v_2)^2 \bigl(C_\mu^a\bigr)^2
		+\frac{1}{4} h^2\Box h^2
		-\frac{\Mh^2}{4v_2} h^3 \bigl(2 h^2+7 h v_2+6v_2^2 \bigr)
        \biggr] .
		\label{eq:LeffTreeFinal}
	\end{equation}
We may now perform the field and parameter redefinitions according to \subsec{FieldRedefLO} without modifying any other term than $\delta\L_{\eff}^{\tree}$ in $\L_{\eff}$ to $\ord(\zeta^{-2})$ (modulo the redefined bare Higgs mass $\Mh$).
As a result $\delta\L_{\eff}^{\tree}$ transforms into
\begin{equation}
  \label{eq:LeffTreeSMEFT}
	\delta\L_{\SMEFT}^{\tree}=
	-\frac{s_\alpha^2}{8v_2^2}(h+v_2)^2\Box(h+v_2)^2
	=C^\tree_{\Phi\Box}\mathcal{O}_{\Phi\Box}\,,
\end{equation}
where $\mathcal{O}_{\Phi\Box}$ is a dimension-six SMEFT operator in the standard Warsaw basis~\cite{Grzadkowski:2010es}
expressed in the non-linear Higgs realization
and
\begin{equation}
	C^\tree_{\Phi\Box}=-\frac{\sa^2}{2 v_2^2}
  \label{eq:Ctree}
\end{equation}
is its Wilson coefficient.
The two Lagrangians in \eqs{LeffTreeFinal}{LeffTreeSMEFT} are thus physically equivalent to the order of interest in the large-$\MH$ expansion.
As argued already in \subsec{FieldRedefLO}, we will, however, not use \eq{LeffTreeSMEFT} for any practical purpose, but rather work with \eq{LeffTreeFinal}, see \sec{pheno}.
We stress once again that in \eq{FinalResult} (and thus everywhere in this section) all SM parameters ($v_2$, $e$, $\Mh$, $\sw$) are bare, while the BSM parameters ($\sa$, $\lambda_{12}$, $\MH$) are renormalized, see \secs{renormalization}{fieldredef}.

The one-loop effective Lagrangian $\delta\L_\eff^{\oneloop,\BSM}$ contains all
$\ord(\zeta^{-2})$ non-SM operators that are generated by integrating out the heavy
Higgs modes at the one-loop level.
\revision{As already mentioned, it is not of SMEFT form after the field transformation
described in \subsec{field-redef-NLO}, but includes only bosonic operators.
We therefore split it in two parts:}
\begin{equation}
\label{eq:SMEFTHEFT}
	\delta\L_\eff^{\oneloop,\BSM} \,=\, \delta\L^\oneloop_\mathrm{SMEFT} \,+\,
    \delta\L^\oneloop_{\nonSMEFT}
\end{equation}
and express $\delta\L^\oneloop_\mathrm{SMEFT}$ in the Warsaw basis~\cite{Grzadkowski:2010es}.
Note, however, that the SMEFT part is fixed uniquely only if the basis
of the non-SMEFT (HEFT) operators is specified as well.

Using standard notation for the SMEFT operators
and defining the matrix-valued Higgs field $\Phi$ as in \eq{doublet}, but with
the physical Higgs field~$h$ instead of $h_2$, $\delta\L^\oneloop_\mathrm{SMEFT}$
is given by
\begin{align}
  \label{eq:LeffSMEFT}
	\delta\L^\oneloop_{\mathrm{SMEFT}}&=
	C_\Phi\mathcal{O}_\Phi
	+C_{\Phi\Box}\mathcal{O}_{\Phi\Box}
	+C_{\Phi D}\mathcal{O}_{\Phi D}
	+C_{\Phi W}\mathcal{O}_{\Phi W}
	+C_{\Phi B}\mathcal{O}_{\Phi B}
	+C_{\Phi WB}\mathcal{O}_{\Phi WB}
	+C_{W}\mathcal{O}_{W} \,,
\end{align}
with the SMEFT operators
\begin{align}
\mathcal{O}_\Phi ={}& \frac{1}{2}\tr\left[(\Phi^\dagger\Phi)^3\right]
= (\phi_{\mathrm{lin}}^\dagger\phi_{\mathrm{lin}})^3 \,,
\nn\\
\mathcal{O}_{\Phi\Box} ={}& \frac{1}{2}\tr\left[(\Phi^\dagger\Phi)\Box(\Phi^\dagger\Phi)\right]
= (\phi_{\mathrm{lin}}^\dagger\phi_{\mathrm{lin}})
	\Box (\phi_{\mathrm{lin}}^\dagger\phi_{\mathrm{lin}}) \,,
\nn\\
\mathcal{O}_{\Phi D} ={}&
\tr\left[\Omega_+\Phi^\dagger D^\mu\Phi\right]
\tr\left[\Omega_-\Phi^\dagger D_\mu\Phi\right]
= (\phi_{\mathrm{lin}}^\dagger D^\mu \phi_{\mathrm{lin}})^*
(\phi_{\mathrm{lin}}^\dagger D_\mu \phi_{\mathrm{lin}}) \,,
\nn\\
\mathcal{O}_{\Phi W} ={}& \frac{1}{2}\tr\left[\Phi^\dagger\Phi\right]W^a_{\mu\nu}W^{a\mu\nu}
= (\phi_{\mathrm{lin}}^\dagger\phi_{\mathrm{lin}}) W^a_{\mu\nu}W^{a\mu\nu} \,,
\nn\\
\mathcal{O}_{\Phi B} ={}& \frac{1}{2}\tr\left[\Phi^\dagger\Phi\right]B_{\mu\nu}B^{\mu\nu}
= (\phi_{\mathrm{lin}}^\dagger\phi_{\mathrm{lin}}) B_{\mu\nu}B^{\mu\nu} \,,
\nn\\
\mathcal{O}_{\Phi WB} ={}&
	\frac{1}{2}\tr\left[\Phi^\dagger\tau_a\Phi\tau_3\right]W^a_{\mu\nu}B^{\mu\nu}
= - (\phi_{\mathrm{lin}}^\dagger \tau_a \phi_{\mathrm{lin}}) W^a_{\mu\nu}B^{\mu\nu} \,,
\nn\\
\mathcal{O}_{W} ={}& -\eps^{abc}W^{a\nu}_{\mu}W^{b\rho}_{\nu}W^{c\mu}_{\rho} \,,
\label{eq:SMEFTops}
\end{align}
where $\Omega_\pm=(1\pm\tau_3)/2$.
Here, we also have made the relation to the standard form of the SMEFT operators $\mathcal{O}_i$
in the linear Higgs realization explicit.
The non-linearly realized matrix-valued Higgs field $\Phi$ is related to the matrix
$(\phi_{\mathrm{lin}}^{\mathrm{c}},\phi_{\mathrm{lin}})$ by a non-trivial field redefinition~\cite{Grosse-Knetter:1992tbp,Dittmaier:2022maf},
with $\phi_{\mathrm{lin}}$ and
$\phi_{\mathrm{lin}}^{\mathrm{c}} \equiv \ri \tau_2 \phi_{\mathrm{lin}}^*$ denoting the  linearly realized Higgs doublet and its charge conjugate, respectively.
In the unitary gauge, both
$\Phi$ and $(\phi_{\mathrm{lin}}^{\mathrm{c}},\phi_{\mathrm{lin}})$
reduce to $(v_2+h)/\sqrt{2}\cdot\bbid$, where $\bbid$ is the $2\times2$ unit matrix.
The covariant derivative acting on $\phi_{\mathrm{lin}}$ or $\phi_{\mathrm{lin}}^{\mathrm{c}}$
(from the left) is defined as
\begin{equation}
	D_{\mu} =\partial_{\mu} - \ri g_{2}W_{\mu} + \ri g_{1} \frac{Y_W}{2} B_{\mu}\,,
\end{equation}
where $Y_W$ denotes the weak hypercharge generator
($Y_W \phi_{\mathrm{lin}} = \phi_{\mathrm{lin}}$,
$Y_W \phi_{\mathrm{lin}}^{\mathrm{c}} = -\phi_{\mathrm{lin}}^{\mathrm{c}}$);
cf.\ \eq{CovDivMatrixfield} for $D_\mu$ acting on $\Phi$.
Concerning the sign factors in \eq{SMEFTops} we emphasize that we adopt the conventions of \rcite{Denner:2019vbn} (cf.~Table~1 there), whereas in the original formulation of the Warsaw basis~\cite{Grzadkowski:2010es} different field conventions are used.

In \eq{LeffSMEFT} we have exploited the fact that not all purely bosonic SMEFT operators
are generated in our large-mass limit of the SESM.
For instance, the P-violating operators naturally do not appear.
For the unrenormalized Wilson coefficients of $\delta\L^\oneloop_{\mathrm{SMEFT}}$
expressed in terms of renormalized BSM parameters we find
\begin{align}
	C_{\Phi}={}&-\frac{(D-1)e^2s_\alpha^2\Mh^2I_{20}}{4\pi^2D(D-2)\sw^2 v_2^4}
	+\frac{(D^2+6D+20)s_\alpha^2\Mh^4I_{20}}{2\pi^2D(D^2-4)v_2^6}
\nn\\
	&{} + \frac{(D-4)(D+2)s_\alpha^2\Mh^2(s_\alpha^2\MH^2-2\la_{12}v_2^2)I_{20}}{4\pi^2D(D-2)v_2^6}
	+\frac{(D-4)(s_\alpha^2\MH^2-2\la_{12}v_2^2)^3I_{20}}{48\pi^2\MH^2v_2^6} \,,
\nn\\
	C_{\Phi\Box}={}&
	-\frac{s_\alpha\delta s_\alpha}{v_2^2}
	+ \frac{(D-1)e^2s_\alpha^2(2\sw^2-3)I_{20}}{16\pi^2D(D-2)\sw^2\cw^2v_2^2}
	+\frac{(D^3-18D^2+128D-96)s_\alpha^4\MH^2I_{20}}{192\pi^2D(D-2)v_2^4}
	\nn\\
	&{} -\frac{(D^3-30D^2+86D-48)\la_{12}s_\alpha^2I_{20}}{48\pi^2D(D-2)v_2^2}
	+\frac{(D^2-42D+44)\la_{12}^2I_{20}}{48\pi^2(D-2)\MH^2} \,,
\nn\\
	C_{\Phi D}={}&-\frac{(D-1)e^2s_\alpha^2I_{20}}{4\pi^2D(D-2)\cw^2v_2^2}\,,
\nn\\
	C_{\Phi W}={}&-\frac{e^2s_\alpha^2I_{20}}{32\pi^2 D(D-2)\sw^2v_2^2}\,,
\nn\\
	C_{\Phi B}={}&-\frac{e^2s_\alpha^2I_{20}}{32\pi^2 D(D-2)\cw^2 v_2^2}\,,
\nn\\
	C_{\Phi WB}={}&
-\frac{e^2s_\alpha^2I_{20}}{16\pi^2 D(D-2)\sw\cw v_2^2}\,,
\nn\\
	C_{W}={}&0\,,
  \label{eq:CoeffsSMEFT}
\end{align}
where the one-loop integral $I_{20}$ is given by \eq{I20}.
Expanding in $\eps = (4-D)/2 $ and introducing $ L_\epsilon=\Delta+\ln(\mu^2/\MH^2), $ we have
\begin{align}
\label{eq:CiSMEFTeps}
	C_{\Phi}={}&-\frac{3e^2\Mh^2s_{\alpha}^2}{32\pi^2\sw^2v_2^4}\left(L_{\epsilon}+\frac{5}{6}\right)
	+\frac{5\Mh^4s_{\alpha}^2}{8\pi^2v_2^6}\left(L_{\epsilon}+\frac{41}{30}\right)
	+\frac{3s_{\alpha}^2\Mh^2\left(2\lambda_{12}v_2^2-\MH^2s_{\alpha}^2\right)}{8\pi^2v_2^6}
\nn\\
	&{}
        -\frac{\left(\MH^2s_{\alpha}^2-2\lambda_{12}v_2^2\right)^3}{24\pi^2\MH^2v_2^6}
	+\mathcal{O(\epsilon)} \,,
\nn\\
	C_{\Phi\Box}={}&-\frac{\delta s_{\alpha}s_{\alpha}}{v_2^2}
  +\frac{3e^2s_{\alpha}^2\left(2\sw^2-3\right)}{128\pi^2\sw^2\cw^2v_2^2} \left(L_{\epsilon}+\frac{5}{6}\right)
	+\frac{9\lambda_{12}\left(\MH^2s_{\alpha}^2-2\lambda_{12}v_2^2\right)}{16\pi^2\MH^2v_2^2} \left(L_{\epsilon}+\frac{10}{27}\right)
\nn\\
	&
  +\frac{s_{\alpha}^2\left(\MH^2s_{\alpha}^2-2\lambda_{12}v_2^2\right)}{8\pi^2v_2^4} \left(L_{\epsilon}+\frac{7}{6}\right)
	+\mathcal{O(\epsilon)} \,,
\nn\\
	C_{\Phi D}={}&-\frac{3e^2s_{\alpha}^2}{32\pi^2\cw^2v_2^2} \left(L_{\epsilon}+\frac{5}{6}\right)
	+\mathcal{O(\epsilon)} \,,
\nn\\
	C_{\Phi W}={}&-\frac{e^2s_{\alpha}^2}{256\pi^2\sw^2v_2^2} \left(L_{\epsilon}+\frac{3}{2}\right)
	+\mathcal{O(\epsilon)} \,,
\nn\\
	C_{\Phi B}={}&-\frac{e^2s_{\alpha}^2}{256\pi^2\cw^2v_2^2} \left(L_{\epsilon}+\frac{3}{2}\right)
	+\mathcal{O(\epsilon)} \,,
\nn\\
	C_{\Phi WB}={}&
-\frac{e^2s_{\alpha}^2}{128\pi^2\sw\cw v_2^2} \left(L_{\epsilon}+\frac{3}{2}\right)
	+\mathcal{O(\epsilon)} \,,
\nn\\
	C_{W}={}&0 \,.
\end{align}
Note that the full-theory renormalization constant
$\delta \sa$ appears in the coefficient $C_{\Phi\Box}$.
Depending on the chosen renormalization condition it may contain soft contributions and in particular (potentially) large logarithms
$\ln (\MH/\Mh)$, cf.~\eq{dsa}.
If desired, this can be avoided by choosing a renormalization scheme for the mixing angle $\sa$ where $\delta \sa$ only includes hard (UV-divergent)
contributions, \revision{see \subsec{discussion} for more details}.

The second part in \eq{SMEFTHEFT}, by our convention, contains eight non-SMEFT operators,
\begin{equation}
	\L_{\nonSMEFT}=\sum_{n=1}^{8}C^{\nonSMEFT}_n\mathcal{O}^{\nonSMEFT}_n,
\end{equation}
which, however, still belong to HEFT as they are invariant under the full SM gauge group.
In the unitary gauge,
they can be compactly written as
\begin{align}
  \mathcal{O}^{\nonSMEFT}_1
  &\,\xrightarrow[]{\text{unitary gauge}}\, (v_2+h)^2(D_B^\mu C_\mu)^a(D_B^\nu C_\nu)^a \,,\nn\\\nn
  \mathcal{O}^{\nonSMEFT}_2
  &\,\xrightarrow[]{\text{unitary gauge}}\, (v_2+h)^2(D_B^\mu C_\nu)^a(D_{B,\mu} C^{\nu})^a \,,\\\nn
  \mathcal{O}^{\nonSMEFT}_3
  &\,\xrightarrow[]{\text{unitary gauge}}\, (v_2+h)^2(D_B^\mu C_\nu)^a(D_B^\nu C_\mu)^a \,,\\\nn
  \mathcal{O}^{\nonSMEFT}_4
  &\,\xrightarrow[]{\text{unitary gauge}}\,
  (v_2+h)^2\epsilon^{ab3}C_\mu^bC_\nu^a B^{\mu\nu}\,, \\\nn
  \mathcal{O}^{\nonSMEFT}_5
  &\,\xrightarrow[]{\text{unitary gauge}}\, (v_2+h)^2\partial^\mu\partial^\nu (C_\mu^aC_\nu^a) \,,\\\nn
  \mathcal{O}^{\nonSMEFT}_6
  &\,\xrightarrow[]{\text{unitary gauge}}\, (v_2+h)^2\Box (C_\mu^a)^2 \,,\\\nn
  \mathcal{O}^{\nonSMEFT}_7
  &\,\xrightarrow[]{\text{unitary gauge}}\, (\partial^\mu h)(\partial^\nu h)C_\mu^aC_\nu^a \,,\\
  \mathcal{O}^{\nonSMEFT}_8
  &\,\xrightarrow[]{\text{unitary gauge}}\, (v_2+h)^2\left[(C_\mu^a)^2(C_\nu^b)^2-C_\mu^aC^{\mu,b}C_\nu^aC^{\nu,b}\right].
  \label{eq:nonSMEFTopsUnitary}
\end{align}
The derivative $ D_B^\mu $ acting on the components of a field
$ \phi=\phi^a\tau_a/2 $
transforming under the adjoint representation of SU(2), is given by
  \begin{equation}
    (D_B^\mu\phi)^a=\partial^\mu\phi^a+\frac{e}{\cw}\eps^{3ab}B^\mu\phi^b.
    \label{eq:Bderiv}
\end{equation}
As explained in \subsec{InvStueckelberg},
the operators in \eq{nonSMEFTopsUnitary} are brought in manifestly gauge-invariant form by an inverse St\"uckelberg transformation.
For the gauge-invariant form we find
\begin{align}
	\mathcal{O}^{\nonSMEFT}_1&=-\frac{2}{g_2^2}(v_2+h)^2\tr\left\{[D_{B}^\mu(U^\dagger (D_\mu U))][D_{B}^\nu(U^\dagger (D_\nu U))]\right\},\nn\\\nn
	&=-\frac{2}{g_2^2}(v_2+h)^2\tr\left\{[D_B^\mu(U^\dagger V_\mu U)][D_B^\nu(U^\dagger V_\nu U)]\right\}, \\\nn
	\mathcal{O}^{\nonSMEFT}_2&=-\frac{2}{g_2^2}(v_2+h)^2\tr\left\{[D_{B}^\mu(U^\dagger (D_\nu U))][D_{B,\mu}(U^\dagger (D^\nu U))]\right\},\\\nn
	&=-\frac{2}{g_2^2}(v_2+h)^2\tr\left\{[D_B^\mu (U^\dagger V_\nu U)][D_{B,\mu}(U^\dagger V^\nu U)]\right\}, \\\nn
	\mathcal{O}^{\nonSMEFT}_3&=-\frac{2}{g_2^2}(v_2+h)^2\tr\left\{[D_{B}^\mu(U^\dagger (D_\nu U))][D_{B}^\nu(U^\dagger (D_\mu U))]\right\},\\\nn
	&=-\frac{2}{g_2^2}(v_2+h)^2\tr\left\{[D_B^\mu (U^\dagger V_\nu U)][D_B^\nu (U^\dagger V_\mu U)]\right\},\\\nn
	\mathcal{O}^{\nonSMEFT}_4&
  =\frac{\ri}{g_2^2}(v_2+h)^2B^{\mu\nu}
		\tr\left\{\tau_3\big[(D_\mu U)^\dagger U ,U^\dagger(D_\nu U)\big]\right\},
\\\nn
	&=-\frac{\ri}{g_2^2}(v_2+h)^2B^{\mu\nu}\tr\left\{T\big[V_\mu,V_\nu\big]\right\} \\\nn
  \mathcal{O}^{\nonSMEFT}_5&=\frac{2}{g_2^2}(v_2+h)^2\partial^\mu\partial^\nu\tr\left\{(D_\mu U)^\dagger(D_\nu U)\right\},\\\nn
	&=-\frac{2}{g_2^2}(v_2+h)^2\partial^\mu\partial^\nu\tr\left\{V_\mu V_\nu\right\} \\\nn
	\mathcal{O}^{\nonSMEFT}_6&=\frac{2}{g_2^2}(v_2+h)^2\Box\tr\left\{(D_\mu U)^\dagger(D^\mu U)\right\},\\\nn
	&=-\frac{2}{g_2^2}(v_2+h)^2\Box\tr\left\{V_\mu V^\mu\right\} \\\nn
	\mathcal{O}^{\nonSMEFT}_7&=\frac{2}{g_2^2}(\partial^\mu h)(\partial^\nu h)\tr\left\{(D_\mu U)^\dagger(D_\nu U)\right\},\\\nn
	&=-\frac{2}{g_2^2}(\partial^\mu h)(\partial^\nu h)\tr\left\{V_\mu V_\nu\right\}, \\\nn
	\mathcal{O}^{\nonSMEFT}_8&
=-\frac{2}{g_2^4}(v_2+h)^2\tr\left\{\big[(D^\mu U)^\dagger U,U^\dagger (D^\nu U)\big]\,
   \big[(D_\mu U)^\dagger U,U^\dagger (D_\nu U)\big]\right\}, \\
	& =-\frac{2}{g_2^4}(v_2+h)^2\tr\left\{\big[V^\mu,V^\nu\big]\,\big[V_\mu,V_\nu\big]\right\},
  \label{eq:opsNonSMEFT}
\end{align}
where
\begin{equation}
	D^\mu_{B}\phi=\partial^\mu\phi+\ri\frac{e}{2\cw}B^\mu[\tau_3,\phi]\,,
\end{equation}
for any field $ \phi $ that transforms in the adjoint representation of SU(2)
in accordance with \eq{Bderiv}.
We also made use of the common shorthand notation introduced in \cite{LONGHITANO1981118}:
\begin{equation}
	V^\mu=(D_\mu U)U^\dagger,\qquad\qquad T=U \tau_3 U^\dagger.
\end{equation}
The corresponding bare Wilson coefficients read
\begin{align}
	&C^{\nonSMEFT}_{1}=-\frac{e^2s_\alpha^2I_{20}}{16\pi^2D(D+2)\sw^2v_2^2}\,,
	&&C^{\nonSMEFT}_{2}=-\frac{(D-6)e^2s_\alpha^2I_{20}}{32\pi^2D(D^2-4)\sw^2v_2^2}\,,\nn\\\nn
	&C^{\nonSMEFT}_{3}=\frac{(2D^2-3D-6)e^2s_\alpha^2I_{20}}{32\pi^2D(D^2-4)\sw^2v_2^2}\,,
	&&C^{\nonSMEFT}_{4}=\frac{(D-3)e^3s_\alpha^2I_{20}}{16\pi^2D(D-2)\sw^2\cw v_2^2}\,,\\\nn
	&C^{\nonSMEFT}_{5}=-\frac{(D^2-3D+6)e^2s_\alpha^2I_{20}}{16\pi^2D(D^2-4)\sw^2v_2^2}\,,
	&&C^{\nonSMEFT}_{6}=-\frac{(D-6)e^2s_\alpha^2I_{20}}{32\pi^2D(D^2-4)\sw^2v_2^2}\,,\\
	&C^{\nonSMEFT}_{7}=\frac{(D^2-2D+4)e^2s_\alpha^2I_{20}}{4\pi^2D(D^2-4)\sw^2v_2^2}\,,
	&&C^{\nonSMEFT}_{8}=\frac{(D-6)e^4s_\alpha^2I_{20}}{64\pi^2D(D^2-4)\sw^4v_2^2}\,,
    \label{eq:HEFTWC}
\end{align}
or expanded for $ D\rightarrow4 $
\begin{align}
	&C^{\nonSMEFT}_{1}=-\frac{e^2s_{\alpha}^2}{384\pi^2\sw^2v_2^2}
	\left(L_{\epsilon}+\frac{5}{6}\right)
	+\mathcal{O(\epsilon)}\,,
	&&C^{\nonSMEFT}_{2}=\frac{e^2s_{\alpha}^2}{768\pi^2\sw^2v_2^2}
	\left(L_{\epsilon}+\frac{17}{6}\right)
	+\mathcal{O(\epsilon)}\,,\nn\\\nn
	&C^{\nonSMEFT}_{3}=\frac{7e^2s_{\alpha}^2}{768\pi^2\sw^2v_2^2}
	\left(L_{\epsilon}-\frac{1}{42}\right)
	+\mathcal{O(\epsilon)}\,,
	&&C^{\nonSMEFT}_{4}=\frac{e^3s_{\alpha}^2}{128\pi^2\sw^2\cw v_2^2}
	\left(L_{\epsilon}-\frac{1}{2}\right)
	+\mathcal{O(\epsilon)}\,,\\\nn
	&C^{\nonSMEFT}_{5}=-\frac{5e^2s_{\alpha}^2}{384\pi^2\sw^2v_2^2}
	\left(L_{\epsilon}+\frac{5}{6}\right)
	+\mathcal{O(\epsilon)}\,,\quad
	&&C^{\nonSMEFT}_{6}=\frac{e^2s_{\alpha}^2}{768\pi^2\sw^2v_2^2}
	\left(L_{\epsilon}+\frac{17}{6}\right)
	+\mathcal{O(\epsilon)}\,,\\
	&C^{\nonSMEFT}_{7}=\frac{e^2s_{\alpha}^2}{16\pi^2\sw^2v_2^2}
	\left(L_{\epsilon}+\frac{5}{6}\right)
	+\mathcal{O(\epsilon)}\,,
	&&C^{\nonSMEFT}_{8}=-\frac{e^4s_{\alpha}^2}{1536\pi^2\sw^4v_2^2}
	\left(L_{\epsilon}+\frac{17}{6}\right)
	+\mathcal{O(\epsilon)} \,.
    \label{eq:CiHEFTeps}
\end{align}
Note that all non-SMEFT coefficients $ C^{\nonSMEFT}_i$
vanish in the no-mixing limit $ \alpha\rightarrow0 $, confirming that the deviation from
\revision{bosonic} SMEFT is a genuine mixing effect.

The $1/\eps$ divergences in the coefficients $ C^\SMEFT_i $ and
$C^\nonSMEFT_i$ emerge in our derivation of the effective Lagrangian as IR singularities of hard loop-momentum regions within the full theory.
These divergences are cancelled by UV divergences of soft EFT
diagrams in physical predictions
and may be absorbed
into EFT counterterms for the purpose of resummation as explained in the beginning of \sec{renormalization}.
The most common choice for the corresponding renormalization of the Wilson coefficients is the $\MSbar$ scheme.
Once an explicit expression for $\delta \sa$
in terms of $L_\eps$ (in a suitable scheme without large logarithms) is inserted,
the matching conditions for the $\MSbar$ renormalized Wilson coefficients at the matching scale $\mu_\mathrm{M} \sim \MH$  are directly obtained from \eqs{CiSMEFTeps}{CiHEFTeps} by the replacement
\begin{equation}
  L_\eps \;\xrightarrow[\;\;\MSbar\;\;]{}\; \ln(\mu_\mathrm{M}^2/\MH^2) \,.
\end{equation}
The leading-logarithmic RG evolution of the $\MSbar$ Wilson coefficients (as well as SM-like $\MSbar$ parameters) to lower scales $\mu \sim \Mh$ is determined by the one-loop UV divergences of the EFT.
In general, it cannot be inferred from the $\mu$-dependence
of the coefficients in
\eqs{CiSMEFTeps}{CiHEFTeps} because of operator mixing.%
\footnote{The one-loop anomalous dimension matrix governing the $\MSbar$ running of all dimension-6 SMEFT operators was computed in \rcites{Grojean:2013kd,Elias-Miro:2013gya,Elias-Miro:2013mua,Jenkins:2013zja,Jenkins:2013wua,Alonso:2013hga,Alonso:2014zka}.
\revision{In the presence of non-SMEFT operators}
the RG running of the SMEFT-type operators may be modified due to mixing with the non-SMEFT operators.
}

Finally, we note that adding the (SM-like) fermion sector to the SESM Lagrangian
leaves the BSM part of the emerging EFT
to the considered level of $\ord(\zeta^{-2})$,
i.e.\ $\delta\L_\eff^{\tree} +\, \delta\L_\eff^{\oneloop,\BSM}$, unaffected as long as the fermions are massless.
This can be seen following our functional derivation of the effective Lagrangian in \subsecs{Lefftree}{IntoutOneloop}:
On the one hand, \eq{Ltree} does not receive fermionic contributions at any order in the large-$\MH$ expansion, because the heavy Higgs field does not couple to massless fermions.
This also excludes fermionic contributions to the EOM for the heavy Higgs field $H$
in \eqs{EOMtree}{HEOMFull}.
On the other hand, since massless fermions ($\psi$) directly couple only to vector bosons, they first enter \eq{Chains} via operator chains arising from a substitution of the form
\begin{equation}
 (\mathcal{D}^{-1})_{\overline{W}_a\overline{W}_b} \quad \longrightarrow \quad
 (\mathcal{D}^{-1})_{\overline{W}_a\overline{W}_c}
\mathcal{X}_{\overline{W}_c  \psi} (\mathcal{D}^{-1})_{\psi \overline{\psi}} \, \mathcal{X}_{\overline{\psi} \, \overline{W}_d}
 (\mathcal{D}^{-1})_{\overline{W}_d\overline{W}_b} \,,
 \label{eq:fermsub}
\end{equation}
in the terms that involve a $\overline{W}$ propagator.
Note that $\mathcal{X}_{\overline{\psi} \,\overline{W}_a}$  ($\mathcal{X}_{\overline{W}_a \psi}$) contains a background (anti)fermion field $\hat{\psi}$ $(\, \hat{\overline{\psi}} \, )$.
For the case of vanishing fermion masses, these operator building blocks are indeed the exclusive source of background fermion fields in the effective Lagrangian.
Owing to the additional $\overline{W}$ and fermion propagators the replacement in \eq{fermsub}, however, comes with a suppression by $\zeta^{-3}$ relative to the operator chains  in \eq{Chains} with a $(\mathcal{D}^{-1})_{\overline{W}_a\overline{W}_b}$ link, which in turn contribute to  $\L_\eff^\oneloop$  only at  $\ord(\zeta^{-2})$.
Longer operator chains encoding the effects of (multiple) virtual fermion (or gluon) lines in hard one-loop diagrams, exist, but are even stronger suppressed than those generated by applying the substitution in \eq{fermsub} to \eq{Chains}.
Hence, massless fermions will first affect the effective Lagrangian beyond the order of interest.
Our field redefinitions in \sec{fieldredef} do not generate additional terms either, because only the light Higgs field,
which likewise does not couple to massless fermions, is transformed.
We can, thus, use the results presented in this section for physical predictions involving, besides photons and massive Higgs and vector bosons, also massless leptons.

\subsection{\revision{Discussion}}

\revision{Before we proceed by working out phenomenological applications of the constructed EFT,
we briefly discuss some interesting features of the \textit{bosonic non-SMEFT}
Lagrangian presented in the previous section as well as general aspects of HEFT-type EFTs:}

\begin{itemize}

\item As explained in \subsec{InvStueckelberg}, we have restored manifest gauge invariance
of the effective Lagrangian after fixing the gauge during its derivation.
In detail, the gauge-fixing of the quantum fields was required for
the quantization and carried out in the BFM as usual to maintain background-field  gauge invariance.
On the other hand, (St\"uckelberg) transforming the background fields to their unitary gauge was merely a matter of convenience.
We stress that this has been possible only because the mass eigenstate we have integrated out was a linear combination of two gauge singlets and thus itself a gauge singlet.
One of these singlets was the Higgs singlet component in the non-linear realization of the SM-type Higgs sector of the SESM.
It should therefore be clear that the non-linear realization  of the
EW gauge symmetry of the SESM is most appropriate
for the construction of a gauge-invariant effective Lagrangian.

\item
The SM-like singlet Higgs field $h$ appears in the effective Lagrangian exclusively in the combination $(v_2 + h)$.
Counting this combination as a field with mass dimension one
and derivatives and all other fields as usual,
all BSM operators of $\ord(\zeta^{-2})$ in unitary gauge ($U=\bbid$) are of dimension six.
While this holds for the SMEFT operators by construction,
the non-SMEFT operators of $\ord(\zeta^{-2})$ could
in principle contain any power of $h/v_2$,
as generally allowed in EFTs of HEFT type.
In our dimension-six Lagrangian, however, the maximal power of $h$ is indeed six (in $\tr\bigl[(\Phi^\dagger\Phi)^3\bigr]$).

\item There are simple linear relations among some of the one-loop results for the Wilson coefficients of the non-SMEFT operators in \eq{HEFTWC}.
For example, we have $C^{\nonSMEFT}_{2} = C^{\nonSMEFT}_{6}$.
This might indicate that the corresponding operators may be linearly combined to form new operators, thereby reducing the number of relevant non-SMEFT operators to correctly describe
the SESM in the large-$\MH$ limit.
In the concrete example one could consider to replace $C^{\nonSMEFT}_{2} \mathcal{O}^{\nonSMEFT}_2 + C^{\nonSMEFT}_{6} \mathcal{O}^{\nonSMEFT}_6 \to C^{\nonSMEFT}_{2} \mathcal{O}^{\nonSMEFT}_{2'}$ with $\mathcal{O}^{\nonSMEFT}_{2'} = \mathcal{O}^{\nonSMEFT}_2+ \mathcal{O}^{\nonSMEFT}_6$.
We are, however, (currently) not aware of any conceptual reason
why the equality $C^{\nonSMEFT}_{2} = C^{\nonSMEFT}_{6}$ or similar relations among the Wilson coefficients should hold to higher loop orders.
We therefore refrain from such merely pragmatic redefinitions.

\item We stress (once again) that the set of operators defining our EFT is
(like for any quantum field theory) not unique.
In particular, the operators may be transformed by field redefinitions or by applying IBP relations
without changing the EFT predictions for physical observables.
In our convention, the SMEFT operators are those of the Warsaw
basis~\cite{Grzadkowski:2010es}, while any other basis related to that by field
redefinitions, IBP relations, and/or linear operator redefinitions
would work just as well.
Our choice of non-SMEFT operators is motivated by simplicity, i.e., (heuristically) minimizing
the set of independent operators, and by their behaviour under
electromagnetic gauge transformations.
At every step of our derivation,
each of the chosen non-SMEFT operators is
electromagnetically gauge invariant by itself, and none of these operators
leads to couplings of photons to neutral particles only.

\item
Generally, the decomposition of the $\ord(\zeta^{-2})$ effective
Lagrangian~\eqref{eq:FinalResult}
into SMEFT and non-SMEFT parts is not unique.
To see this,
consider replacing the non-SMEFT operators in \eq{opsNonSMEFT} by a physically
equivalent selection of operators within the HEFT framework.
A one-loop bottom-up matching of
\revision{bosonic}
SMEFT augmented with the new non-SMEFT operators
onto SESM will in general lead to SMEFT Wilson coefficients that differ from the ones
given in \eq{CoeffsSMEFT}.
For example, a viable alternative set of non-SMEFT operators for our EFT is obtained
by replacing the covariant derivative $D_B^\mu$ in $\mathcal{O}^{\nonSMEFT}_{1,2,3}$
by $D_W^\mu$, which is defined in analogy to \eq{covDW}.
This will lead to modified Wilson coefficients $C_{\Phi W}$, $C_{\Phi B}$, $C_{\Phi WB}$,
$C^{\nonSMEFT}_{2}$, $C^{\nonSMEFT}_{3}$, and $C^{\nonSMEFT}_{8}$,
but leave physical predictions unchanged.
Quantitatively, the SMEFT and non-SMEFT parts of the BSM contributions to
the prediction of observables both change due to this transformation of HEFT operators,
but the sum of SMEFT and non-SMEFT parts in predictions stays the same.

\item
Following up on the previous point:
One can in principle systematically define a basis of dimension-six operators such
that the SMEFT and non-SMEFT parts of the effective Lagrangian decouple in the sense
that the SMEFT Wilson coefficients
are fixed in the bottom-up matching by tying the
SMEFT contribution to predictions for a suitable selection of observables.
The non-SMEFT operators then have to be chosen such that they do not influence
this selection of observables.
While this is always possible by ``brute force''
(i.e., by adding appropriate combinations of SMEFT operators to selected non-SMEFT operators,
similar to Schmidt's orthogonalization procedure in linear algebra),
it is often also achievable by a judicious selection of interaction structures that do not influence the chosen observables used to fix the SMEFT part.

Adopting a similar strategy as in
our proof in \subsec{SMEFTbreakdown}, one could for example construct such a basis of
SMEFT and non-SMEFT operators for our EFT by means of the following procedure:
\begin{enumerate}

    \item Define an ordered set of $N$ physical observables which are independent from each other in the sense that their EFT predictions receive contributions from independent linear combinations of $N$ dimension-six SMEFT operators in the Warsaw basis, where $N$ denotes the number of dimension-six SMEFT operators that are sensitive to the BSM physics of a given new physics model.
    For the SESM in the decoupling limit we consider in this paper, we have $N=7$, see \sec{final}.

    \item
    To start the recursive construction of the operator basis, define the first
    SMEFT operator $O_1$
    as some linear combination of Warsaw basis operators, such that the (prediction for the) first observable on the list $X_1$ receives $\ord(e^2/ \zeta^{2})$ contributions exclusively from this single SMEFT operator $O_1$.
    The $\ord(e^2/ \zeta^{2})$ EFT correction to $X_1$ will thus be proportional to the Wilson coefficient $C_1$ associated with the operator~$O_1$.

    \item Subsequently define operators $O_{n}$ with $n=2,\ldots,N$
in a recursive way such
    that $X_n$ receives $\ord(e^2/ \zeta^{2})$ contributions only from the operators $\{O_1, \ldots, O_n\}$.
    The $\ord(e^2/ \zeta^{2})$ EFT correction to $X_n$ will thus be a linear combination of the Wilson coefficients $C_1, \ldots, C_n$ associated with the operators $\{O_1, \ldots, O_n\}$.
    For $n=N$, the SMEFT part of the operator basis, $\{O_1, \ldots, O_N\}$,
    as well as the corresponding SMEFT Wilson coefficients of the EFT are fixed.

    \item Further ``independent'' observables $X_{N+1}, \ldots, X_{M}$ will receive contributions from non-SMEFT operators, such as the ones in \eq{opsNonSMEFT}.
    One can now recursively define new non-SMEFT operators $O_{N+1},\ldots,O_{N+M}$  in analogy to step 3, such that $X_{N+m}$ (with $0 < m \le M-N$)
    receives $\ord(e^2/ \zeta^{2})$ contributions only from the operators $\{O_1, \ldots, O_{N+m}\}$.
    The total number $M$ of basis operators is determined empirically by the outlined procedure.
    From our findings in \sec{final} we know that $M\le 15$ in our EFT.

  \end{enumerate}

The outlined algorithm provides an operator basis $\{O_1, \ldots, O_M\}$  for our EFT, where the SMEFT part of the basis is defined by $\{O_1, \ldots, O_N\}$.
Matching to the SESM will yield unique Wilson coefficients $C_1, \ldots, C_N$ for any allowed choice of non-SMEFT operators, because the values for $C_1, \ldots, C_N$ are tied to physical observables, which are invariant under
operator redefinitions in the SMEFT sector.
These statements also hold for other EFTs whose dimension-six SMEFT part can be accommodated in the same operator basis.
For EFTs that require more or other SMEFT-type operators one can extend the basis $\{O_1, \ldots, O_N\}$ accordingly, until fixing a complete basis for all SMEFT operators.
The Wilson coefficients of such a SMEFT version will be independent of the choice of
non-SMEFT operators, but will depend on the set of observables chosen to determine the SMEFT operators.
How useful such a SMEFT basis is in practice, given the probably quite complex structure of some of its operators, and how difficult it is to find a suitable defining set of independent observables remains to be investigated.

\item From the above considerations it should be clear that, when working in the
standard (Warsaw) basis, a SMEFT fit to (infinitely) precise (hypothetical) experimental
data will fail only if a sufficiently large number of observables is
included in the fit, even if the underlying theory cannot be accommodated in SMEFT.
If the number of independent observables involved in the fit is too small,
the resulting SMEFT Wilson coefficients may still perfectly
parametrize the data.
However, the connection of the fitted Wilson coefficients to those computed
by matching to some underlying model may be obscured by the ambiguity due to
the choice of non-SMEFT operators.
Fitting and matching the coefficients of a SMEFT operator basis constructed from
observables as described in the previous item would avoid this ambiguity.

\end{itemize}

\section{Phenomenological applications and validation of the EFT}
\label{sec:pheno}

In order to validate the correctness of our EFT and to demonstrate its proper use,
we compute a few EW precision
observables at NLO
both within the full SESM and using the EFT which should reproduce
the full NLO result up to terms of $\ord(\zeta^{-3})$ in the
large-$\MH$ (and \mbox{small-$\sa$)} parameter space.
The EFT predictions are based on the
\revision{bare effective Lagrangian $\L_\eff$ \eqref{eq:FinalResult}.}

\subsection{EFT renormalization}
\label{sec:EFTren}

In order to implement OS renormalization conditions in the EFT
(in the PRTS) for the SM-like input parameters and fields,
at this point we consider again all EFT parameters and fields
(which form the same set as in the SM)
as \textit{bare} and
perform an appropriate renormalization transformation.
In detail, the renormalization transformation as well as the determination of the
SM-like renormalization constants from OS renormalization conditions exactly proceeds
as described in \citere{Denner:2019vbn}.
The transformation of the EFT parameters is formally identical to
\eq{SMparameter-ren}, and the dependent parameters $\sw$ and $v_2$
are treated as given in \eqsm{SMparameter-dep}{renconstrel}.
To distinguish the renormalization constants of this final renormalization
from the pre-renormalization carried out before, we change the $\de$ in this
final renormalization step to $\de^\EFT$.
For the mass parameters, OS renormalization conditions imply
\begin{align}
\delta^\EFT\Mh^2 =\Re\Sigma^{hh}(\Mh^2)\,, \qquad
\delta^\EFT\MW^2 =\Re\Sigma^{WW}_\mathrm{T}(\MW^2)\,, \qquad
\delta^\EFT\MZ^2 =\Re\Sigma^{ZZ}_\mathrm{T}(\MZ^2)\,.
\label{eq:MassRenCond}
\end{align}%
Following the conventions of \citere{Denner:2019vbn},
$\Sigma^{\phi^\dagger\phi}_{\mathrm{(T)}}(p^2)$ denotes the  self-energy of the field $\phi$,
with the subscript~$\mathrm{T}$ indicating the transversal
part for the case that $\phi$ is a vector field.%
\footnote{As mostly done in the literature, we write
$\Sigma^{WW}$
instead of
$\Sigma^{W^+W^-}$
.}
The tadpole renormalization constant is fixed according to \eq{TadpoleCondition}
in the PRTS (which is equivalent to the GIVS in the employed non-linear Higgs realization),
i.e.\ $\de^\EFT t_{\Ph} = - T^{h}$.
\revision{This choice is consistent with using \eq{dsa} for $\delta \sa$ in the Wilson coefficient $C_{\Phi\Box} $, see also \subsec{discussion}.}
The charge renormalization constant
\begin{equation}
\delta^\EFT Z_e = \left.\frac{1}{2}\frac{\partial\Sigma^{AA}(k^{2})}{\partial k^{2}}
\right\vert_{k^{2}=0}-\frac{\sw}{\cw}\frac{\Sigma^{AZ}(0)}{\MZ^{2}}
\end{equation}
is defined in the Thomson limit as usual~\cite{Bohm:1986rj,Denner:1991kt,Denner:2019vbn}.%
\footnote{
  While the explicit forms of the self-energies are not the same in the BFM and the
  conventional quantization formalism, the resulting renormalization constants coincide.
  As a consequence of background-field gauge invariance of the effective
  action, $\Sigma_\rT^{\hat A\hat Z}(0)=0$
  in the BFM~\cite{Denner:1994xt}.}
We note that the BSM sector does not contribute to $\delta^\EFT Z_e$ at the one-loop level, because the Higgs fields are electrically neutral, i.e.\ there are no tree-level
photon--Higgs couplings.
It should also be kept in mind
that the mass renormalization constants depend on the tadpole scheme.

In order to implement the {\it complete OS scheme} for the fields, where LSZ factors are unity
and mixing between OS fields is eliminated, we have to deviate from the renormalization transformation
of the gauge-boson fields given in \eq{field-ren} in the gauge-symmetric field basis.
Instead, we have to switch to some matrix-valued field transformation in the
$\gamma/Z$ sector, as described in \citeres{Denner:1991kt,Denner:2019vbn},
\begin{align}
	h_0 &= \textstyle (1+\frac{1}{2}\delta^\EFT Z_h) h
\nn\\
	W^{\pm,\mu}_0&\textstyle =(1+\frac{1}{2}\delta^\EFT Z_W)W^{\pm,\mu},\qquad
\begin{pmatrix} Z^\mu_{0} \\ A^\mu_{0} \end{pmatrix} =
\begin{pmatrix} 1+\frac{1}{2}\de^\EFT Z_{ZZ} & \frac{1}{2}\de^\EFT Z_{ZA}  \\[1ex]
     \frac{1}{2}\de^\EFT Z_{AZ} &  1+\frac{1}{2}\de^\EFT Z_{AA}\end{pmatrix}
\begin{pmatrix} Z^\mu \\ A^\mu \end{pmatrix},
\label{eq:field-ren2}
\end{align}%
and the OS conditions imply
\begin{align}
\delta^\EFT Z_h ={}& -\Re\Sigma^{\prime\,hh}(\Mh^2),
\nn\\
\delta^\EFT Z_W ={}& -\Re\Sigma_{\rT}^{\prime\,WW}(\MW^2), \quad
&\delta^\EFT Z_{ZZ} ={}& -\Re\Sigma_{\rT}^{\prime\,ZZ}(\MZ^2), \quad
&\delta^\EFT Z_{AA} ={}& -\Sigma_{\rT}^{\prime\,AA}(0), \quad
\nn\\
\delta^\EFT Z_{AZ} ={}& -2\Re\frac{\Sigma_{\rT}^{AZ}(\MZ^2)}{\MZ^2}, \quad
&\delta^\EFT Z_{ZA} ={}& 2\frac{\Sigma_{\rT}^{AZ}(0)}{\MZ^2} .
\end{align}%
Similar to $\delta^\EFT Z_e$, there are no BSM contributions to
$\delta^\EFT Z_{AA}$, $\delta^\EFT Z_{AZ}$, and $\delta^\EFT Z_{ZA}$,
cf.~App.~\ref{app:CTEFT},
because there are no tree-level photon--Higgs couplings.

The EFT counterterm Lagrangian $\delta \L_\eff^{\ct}$ emerging after the described
renormalization transformation
has the same form as the usual
SM counterterm Lagrangian plus the corresponding contribution from
$\delta\L_\eff^{\tree}$,
\begin{equation}
  \L_\eff = \L_\SM^\ren + \delta\L_\eff^{\tree}
  + \delta\L_\eff^{\oneloop,\BSM} + \delta\L_\eff^{\ct} \,,
\label{eq:Leff-final-ren}
\end{equation}
where $\delta\L_\eff^{\tree}$ and $\delta\L_\eff^{\oneloop,\BSM}$
are given in \eqs{LeffTreeFinal}{SMEFTHEFT}, respectively.
The renormalization constants, generically denoted $\delta^{\mathrm{EFT}}Z_i$ in the following,
(just like any NLO EFT result for a given quantity) now receive three types of contributions:
\begin{enumerate}
\renewcommand{\labelenumi}{\theenumi}
\renewcommand{\theenumi}{(\roman{enumi})}
\item
the NLO SM contribution $ \delta^{\mathrm{EFT}}_{\SM} Z_i = \ord(\zeta^0)$, obtained from using $\L_\SM$
alone, including SM counterterms (i.e.\ the $\ord(\zeta^0)$ contribution to the EFT counterterms) at tree level,
\item
the BSM contribution from hard loops, $ \delta^{\mathrm{EFT}}_h Z_i = \ord(\zeta^{-2}) $, obtained from inserting single operators from $\delta\L_\eff^{\oneloop,\BSM}$
in tree-level diagrams,%
\footnote{Tree-level contributions from $\delta\L_\eff^{\oneloop,\BSM}$ and $\delta\L^{\tree}_\eff$ to the $ \delta^{\mathrm{EFT}} Z_i$ exist, because the SM and non-SM operators contain terms of the same form, such that non-SM operators contribute to SM renormalization conditions in the EFT.
\revision{Strictly speaking, (ii) may also contain soft contributions induced by BSM renormalization constants (like our $\delta \sa$) in the Wilson coefficients, see \subsec{discussion} for further discussion.}
}

\item
the BSM contribution from  soft modes, $ \delta^{\mathrm{EFT}}_s Z_i = \ord(\zeta^{-2})$, obtained from single
insertions of $\delta\L^{\tree}_\eff $
in tree-level and one-loop diagrams.
\end{enumerate}
We recall that this procedure completes the rearrangements in the Lagrangian
that have been made before in the course of
the pre-renormalization of the SM parameters
and fields during the calculation of the effective Lagrangian,
where we have actually
only redistributed some hard contributions across the SM counterterms.
The explicit results of the renormalization constants
$\delta^{\mathrm{EFT}}Z_i $ are given in \app{CTEFT}.
This renormalization of all SM parameters and fields of course does not touch the
renormalization of the mixing angle~$\alpha$ and the corresponding renormalization
constant $\delta\sa$,
which does not represent an input parameter of the EFT and only appears
implicitly in some matching conditions for the Wilson coefficients.

\subsection{Calculation of EFT predictions}

Employing the renormalized Lagrangian \eqref{eq:Leff-final-ren} in an NLO
calculation of some quantity $X$
(Green function, S-matrix elements, or observable) leads to four different types of contributions:
$ X^{\mathrm{EFT}}_{\SM} $,
$ X^{\mathrm{EFT}}_h $, and
$ X^{\mathrm{EFT}}_s $ of types
(i)--(iii) analogous to the calculation of
$\delta^{\mathrm{EFT}}Z_i$ described above (where $ X^{\mathrm{EFT}}_{\SM} $ now includes the complete SM tree-level part) plus
(iv) contributions $X^{\mathrm{EFT}}_{\mathrm{ct}}$ from $\delta\L_\eff^{\ct}$
containing
$ \delta^{\mathrm{EFT}}_{\SM} Z_i$, $ \delta^{\mathrm{EFT}}_h Z_i$,
and $ \delta^{\mathrm{EFT}}_s Z_i$.
The full EFT prediction for $X$ then is
\begin{align}
X^{\mathrm{EFT}} = X^{\mathrm{EFT}}_{\SM} + X^{\mathrm{EFT}}_h + X^{\mathrm{EFT}}_s
+ X^{\mathrm{EFT}}_{\mathrm{ct}} .
\end{align}
The counterterm contribution naturally decomposes into subcontributions,
\begin{align}
X^{\mathrm{EFT}}_{\mathrm{ct}} =
X^{\mathrm{EFT}}_{\mathrm{ct},\SM} +
X^{\mathrm{EFT}}_{\mathrm{ct},h} +
X^{\mathrm{EFT}}_{\mathrm{ct},s},
\end{align}
furnished by the respective renormalization constants
$ \delta^{\mathrm{EFT}}_{\SM} Z_i$, $ \delta^{\mathrm{EFT}}_h Z_i$,
and $ \delta^{\mathrm{EFT}}_s Z_i$.
Attributing these counterterm contributions to the respective contributions (i)--(iii)
to $X^{\mathrm{EFT}}$ we define renormalized subcontributions,
\begin{align}
X^{\mathrm{EFT}}_{a,\mathrm{ren}} = X^{\mathrm{EFT}}_{a}
+ X^{\mathrm{EFT}}_{\mathrm{ct},a}, \qquad
a = \SM, h, s.
\end{align}
The EFT prediction for $X$ can now be written as
\begin{align}
X^{\mathrm{EFT}} = X^{\mathrm{EFT}}_{\SM,\mathrm{ren}}
+ X^{\mathrm{EFT}}_{h,\mathrm{ren}}
+ X^{\mathrm{EFT}}_{s,\mathrm{ren}}\,,
\end{align}
where $X^{\SM}\equiv X^{\mathrm{EFT}}_{\SM,\mathrm{ren}}$ is the renormalized
SM prediction.
The genuine BSM contribution is given by
\begin{align}
X^{\mathrm{BSM,EFT}} =
X^{\mathrm{EFT}}_{h,\mathrm{ren}} + X^{\mathrm{EFT}}_{s,\mathrm{ren}} \,.
\end{align}
By construction we, thus, have
$X^{\SM} = \ord(\zeta^0)$ and
$X^{\mathrm{BSM,EFT}} = \ord(\zeta^{-2})$, where the suppression of the BSM part
(i.e.\ decoupling) is
ensured by our choice of parameter scaling in the large-$\MH$ limit,
as discussed in \subsec{EFTlimit}.
If $X$ is an observable, $X^{\SM}$ and $X^{\mathrm{BSM,EFT}}$
are each finite, but
$X^{\mathrm{EFT}}_{h,\mathrm{ren}}$ and $X^{\mathrm{EFT}}_{s,\mathrm{ren}}$
each involve divergences (of IR and UV origin, respectively) that compensate each other.

In order to check the validity of our EFT, we have to verify that the EFT prediction
$X^{\mathrm{EFT}}$ asymptotically approaches the full SESM prediction
$X^{\SESM}$ faster than $\ord(\zeta^{-2})$ in the large-$\MH$ limit.
Defining the genuine BSM contribution $X^{\BSM}$ of $X^{\SESM}$ by subtracting the
SM part, the condition on the asymptotics reads
\begin{align}
X^{\SESM} - X^{\mathrm{EFT}} = X^{\BSM} - X^{\mathrm{BSM,EFT}} = \ord(\zeta^{-3}) \,.
\end{align}

For the numerical results discussed below, we have used the following SM input
parameters,
\begin{align}
\alpha_{\mathrm{em}}(0) ={}& 1/137.0359997 \,, &
\GF ={}& 1.166379\cdot10^{-5}\GeV^{-2},
\nn\\
\MW ={}& 80.385\GeV, & \MZ ={}&  91.1876\GeV,
& \Mh ={}&  125.1\GeV,
\end{align}
with more details on their use given in the respective cases.
Here $\alpha_{\mathrm{em}}(0)$ is the usual fine-structure constant.
If not stated otherwise, we calculate in the $\alpha_{\mathrm{em}}(0)$ scheme for the EW coupling, see e.g.~\rcite{Denner:2019vbn}.
Fermions are treated as massless, except for the top quark.
\revision{In order to be able to consistently apply our EFT to phenomenological predictions
of the SESM at NLO, we have selected processes (and if needed specific observables)
that are insensitive to Yukawa couplings of the top-quark, since all fermions are taken
massless in our EFT.%
\footnote{\revision{Processes that do not involve Higgs bosons or top-quarks at LO are eligible
in this sense, because top quarks then only appear in closed fermion loops which are not connected
to Higgs bosons at NLO. For such processes, the contributions from closed fermion loops
drop out in the difference SESM${}-{}$SM. The only process we consider that is not of this kind is
$\mathrm{h\to WW\to\nu_ee^+\mu^-\bar\nu_\mu}$, where we
guarantee the cancellation of top-quark loops by considering the
difference SESM${}-{}c_\alpha^2\cdot{}$SM, see \subsec{hdecay}.}}
In the numerical studies in the following sections,
we have designed the differences between SESM and SM predictions in such a way that corrections
induced by closed fermion loops drop out, so that
the numerical results shown for these differences
are independent of the top-quark mass.}
The BSM input used for $\sa$ and $\MH$ is varied in phenomenologically
reasonable ranges, and $\lambda_{12}$ is fixed
to the benchmark value $\lambda_{12}=0.17$
(as in scenario BHM400 of \rcite{Altenkamp:2018bcs}),
since its value impacts the results only very weakly.

\subsection{W-boson mass from muon decay}
\label{subsec:MW}

The measurement of the muon lifetime $\tau_\mu$ is usually translated into
the experimental determination of the Fermi constant~$\GF$.
Thus, the theory prediction for $\tau_\mu$ in the SM or
an SM extension
delivers an important constraint on the model parameters.
Neglecting the masses of the external leptons,
this constraint can be formulated as
\begin{equation}
\label{eq:MW-Deltar}
\GF=\frac{\alpha_\mathrm{em}\pi}{\sqrt{2}\MW^2\sw^2}\,(1+\Delta r) \,,
\end{equation}
where $\alpha_\mathrm{em}\equiv e^2/(4\pi)$
is the electromagnetic coupling constant.
The quantity~$\Delta r$ represents the radiative correction to the
muon decay $\mu^-\to\nu_\mu\Pe^-\bar\nu_\Pe$~\cite{Sirlin:1980nh,Denner:2019vbn},
with the photonic corrections of the Fermi model subtracted
(which are thereby absorbed into the definition of $\GF$).
The constraint~\eqref{eq:MW-Deltar}
can be implicitly solved for the W-boson mass $\MW$ in terms of the remaining
input parameters including $\GF$. This predicted value for $\MW$ can then be confronted with the direct
measurement of $\MW$. Employing OS renormalization in the SM and the SESM, the
weak mixing angle is tied to the mass ratio $\MW/\MZ$ according to
$\sw^2=1-\MW^2/\MZ^2$ (to all orders), so that a LO prediction $\MW^{(0)}$ for
$\MW$ can be derived from $\GF$, $\alpha_\mathrm{em}$, and $\MZ$.
NLO corrections to $\MW$ are linear in $\Delta r$, so that the NLO prediction for $\MW$
can be written as,
\begin{equation}
\MW = \MW^{(0)}\left(1 + x_\PW\, \Delta r^{(1)}\right)
\end{equation}
with
\begin{equation}
\MW^{(0)} = \frac{\MZ}{\sqrt{2}}\,\sqrt{1+\sqrt{1-A}} \,, \qquad
x_\PW = \frac{\sqrt{1-A}-1}{4\sqrt{1-A}} \,, \qquad
\qquad
A \equiv \frac{2\sqrt{2}\alpha_\mathrm{em}\pi}{\GF\MZ^2} \,,
\end{equation}
where the coefficient $x_\PW$ is obtained by linearizing the solution of \eq{MW-Deltar}
for $\MW$ in $\Delta r$, and $\Delta r^{(1)}$ is the one-loop approximation for $\Delta r$.
To obtain $\MW$ with NLO accuracy, the LO value $\MW^{(0)}$ for $\MW$ can be used in the evaluation of $\Delta r^{(1)}$.
Since $\MW^{(0)}$ and $x_\PW$ are the same in the SM and SESM, the BSM contribution
we are after is completely contained in $\Delta r$.
Both in the SM and SESM, the NLO prediction for $\Delta r$ can be written
as~\cite{Denner:1991kt,Denner:2019vbn}
\begin{equation}
\Delta r^{(1)} ={}
2\delta Z_e
-\frac{2\delta \sw}{\sw}
+\frac{\Sigma^{WW}_{\rT}(0)-\delta\MW^2}{\MW^{2}}
+\frac{2\Sigma^{AZ}_{\rT}(0)}{\sw \cw \MZ^{2}}
+\frac{\alpha_{\mathrm{em}} }{4\pi \sw^{2}}
\left(6 +  \frac{7-4\sw^{2}}{2\sw^{2}}\ln \cw^{2}\right),
\end{equation}
where we have made the dependence on the bosonic renormalization constants explicit.%
\footnote{The explicit expressions of the
wave-function corrections of the external fermions, which are identical
for the SM and the SESM in the valid approximation of neglecting external fermion masses,
have been explicitly inserted and are
contained in the last term.}
Note that differences between SM and SESM only occur in the quantities
$\delta \sw$, $\Sigma^{WW}_{\rT}(0)$, and $\delta\MW^2$, i.e.\
\begin{align}
\Delta r^{(1,\BSM)} ={}&
\Delta r^{(1,\SESM)} - \Delta r^{(1,\SM)}
=
-\frac{2\delta^{\BSM} \sw}{\sw}
+\frac{\Sigma^{WW,\BSM}_{\rT}(0)-\delta^{\BSM}\MW^2}{\MW^{2}}
\,,
\end{align}
where we have used that the Higgs sector of the SESM does not
contribute to $\delta Z_e$ and the $AZ$ self-energy at NLO.
Here, ``BSM'' indicates the difference between SESM and SM contributions
in each of the terms on the r.h.s.
In the EFT, each of the BSM contributions decomposes into
hard and soft parts,
which can be easily identified from the explicit results given in \app{CTEFT}.

For later reference, we shall just write out the hard part:
\begin{equation}
	\Delta r^{(1,\BSM,\EFT)}_h=\frac{v_2^2}{2\sw^2}
	(\cw^2C_{\Phi D}
+4\sw\cw C_{\Phi WB}
+8\sw^2C^\nonSMEFT_{2}) \,.
\label{eq:dr_EFTh}
\end{equation}
From $\Delta r^{(1,\BSM)}$ the BSM contribution at NLO to the
predicted W-boson mass can be obtained as
\begin{equation}
\Delta\MW^{\BSM} = \MW^{(0)}\,x_\PW \, \Delta r^{(1,\BSM)} \,,
\end{equation}
which is the correction that has to be added to the SM prediction at NLO for $\MW$
to obtain the NLO prediction of $\MW$ in the SESM.

The ``best'' prediction for $\MW$ in the SESM would
be obtained by adding the (small) BSM contribution $\Delta\MW^{\BSM}$
to the state-of-the-art SM prediction for $\MW$ which contains the
complete set of two-loop corrections
(see \rcite{Awramik:2003rn} and references therein).
Experimentally $\MW$ is determined to the level of
$\sim10\MeV$ by measurements at LEP2, the Tevatron, and the LHC
(see, e.g., \rcite{ParticleDataGroup:2024cfk}), delivering an important
constraint on the fit of the SM and its extensions to experimental data.

\begin{figure}
	\centering{\includegraphics[width=.9\textwidth]{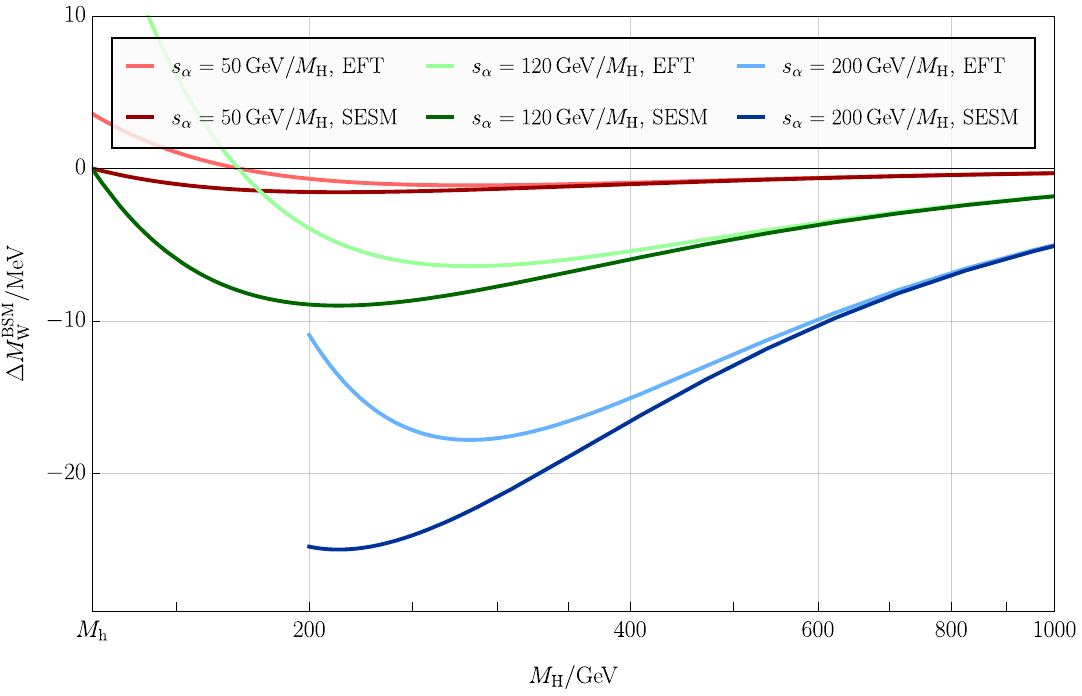}}
	\caption{BSM correction $\Delta\MW^{\BSM}$ to the W-boson mass,
shown for the full SESM and the EFT approximation.
The product $s_\alpha\MH$ is kept fixed in each curve.}
\label{fig:MW}
\end{figure}

In \reffi{fig:MW}, we show $\Delta\MW^{\BSM}$ for the full SESM as well as
the corresponding EFT approximation as a function of the mass $\MH$ of the
heavy Higgs boson~H.
To account for the $\ord(\zeta^{-1})$ scaling of the mixing angle in
our heavy-mass limit, we fix the product $s_\alpha\MH$ to some constant
while varying $\MH$.

For $\MH\to\infty,$ we can see the results of the EFT first converging
to those of the SESM before both converge to the SM result, which
is represented by the zero line in the $\Delta\MW^{\BSM}$ plot,
in line with the expected scaling behaviour of the EFT.
The precise scaling behaviour has been checked by considering
the large-$\MH$ behaviour of $\MH^2 \times \Delta\MW^{\BSM}$,
which shows that the corresponding full SESM and EFT predictions
asymptotically still approach each other and both show
the same asymptotic behaviour
$a+b\ln\MH$, with $a,b$ some constants.

Figure~\ref{fig:MW} also demonstrates
that there is still some viable parameter space of the SESM
leading to deviations from the SM prediction at the order of the current experimental
precision of $\sim10\MeV$ on $\MW$, underlining the importance of $\MW$ in constraining the SESM
phenomenologically.

\subsection{Effective weak mixing angle}
\label{subsec:sintheff}

Next, we consider the (leptonic) effective weak mixing angle $\theta_{\mathrm{eff}}^f$,
which is one of the classic pseudo-observables on the Z~resonance.
The quantity $\sin^2\theta_{\mathrm{eff}}^f$ is derived from the ratio of effective
vector and axial-vector $Zf\bar f$~couplings of some fermion~$f$
(see Refs.~\cite{Altarelli:1989hv,Bardin:1997xq,Bardin:1999gt} for details) and coincides
with the OS quantity $\sw^2$ in lowest order, but (in contrast to $\sw^2$) receives process-specific
EW radiative corrections from the $Zf\bar f$ vertex.
These corrections are sensitive to the Higgs-boson mass in the SM
and to parameters
of the Higgs sector of any SM extension.
Experimentally $\sin^2\theta_{\mathrm{eff}}^\ell$ for charged leptons~$f=\ell$
is determined to the level of
$\sim10^{-4}$ by measurements at LEP1, SLC, the Tevatron, and the LHC (see, e.g., \rcite{ParticleDataGroup:2024cfk}),
rendering this pseudo-observable a key variable in constraining scalar sectors in the
SM and its extensions.

\begin{sloppypar}
The NLO prediction for $\sin^2\theta_{\mathrm{eff}}^\ell$ can be written as
\begin{align}
\sin^2\theta_{\mathrm{eff}}^\ell = \sw^2 + \Delta\sin^2\theta_{\mathrm{eff}}^\ell
= \sw^2 + \Delta_{\text{1-loop}}\sin^2\theta_{\mathrm{eff}}^\ell + \delta\sw^2 \,,
\end{align}
where the NLO correction $\Delta\sin^2\theta_{\mathrm{eff}}^\ell$ is decomposed
into a genuine one-loop contribution
$\Delta_{\text{1-loop}}\sin^2\theta_{\mathrm{eff}}^\ell$ and the renormalization
constant $\delta\sw^2$.
In the valid approximation of taking the lepton~$\ell$ massless, there is no
contribution from Higgs bosons to the explicit one-loop correction
$\Delta_{\text{1-loop}}\sin^2\theta_{\mathrm{eff}}^\ell$ in the SM and SESM, so that
the BSM contribution $\Delta^{\BSM}\sin^2\theta_{\mathrm{eff}}^\ell$
to $\sin^2\theta_{\mathrm{eff}}^\ell$ is completely
contained in the renormalization constant $\delta\sw^2$,
\end{sloppypar}
\begin{align}
\Delta^{\BSM}\sin^2\theta_{\mathrm{eff}}^\ell = \delta^{\BSM}\sw^2
= 2\sw \delta^{\BSM}\sw \,,
\end{align}
where
\begin{align}
\delta^{\BSM}\sw = -\frac{\delta^{\BSM}\cw^2}{2\sw}
= - \frac{\cw^2}{2\sw}
\left( \frac{\delta^{\BSM}\MW^2}{\MW^2} - \frac{\delta^{\BSM}\MZ^2}{\MZ^2} \right).
\label{eq:delswBSM}
\end{align}
The hard and soft contributions to the EFT approximations for the BSM
contributions $\delta^{\BSM}M_V^2$ ($V=\PW,\PZ$) can again be read from
\app{CTEFT}.
In particular, we find for the hard part
\begin{equation}
\Delta^{\BSM,\EFT}_h\sin^2\theta_{\mathrm{eff}}^\ell=
v_2^2 \left(2\cw^2\sw^2 C_{\Phi W}-2\cw^2\sw^2 C_{\Phi B}-\frac{\cw^2}{2}C_{\Phi D}
-2\cw^3\sw C_{\Phi WB}
-2\sw^2 C^\nonSMEFT_{2}\right).
\label{eq:dsw_EFTh}
\end{equation}

\begin{figure}
\centering{\includegraphics[width=.9\textwidth]{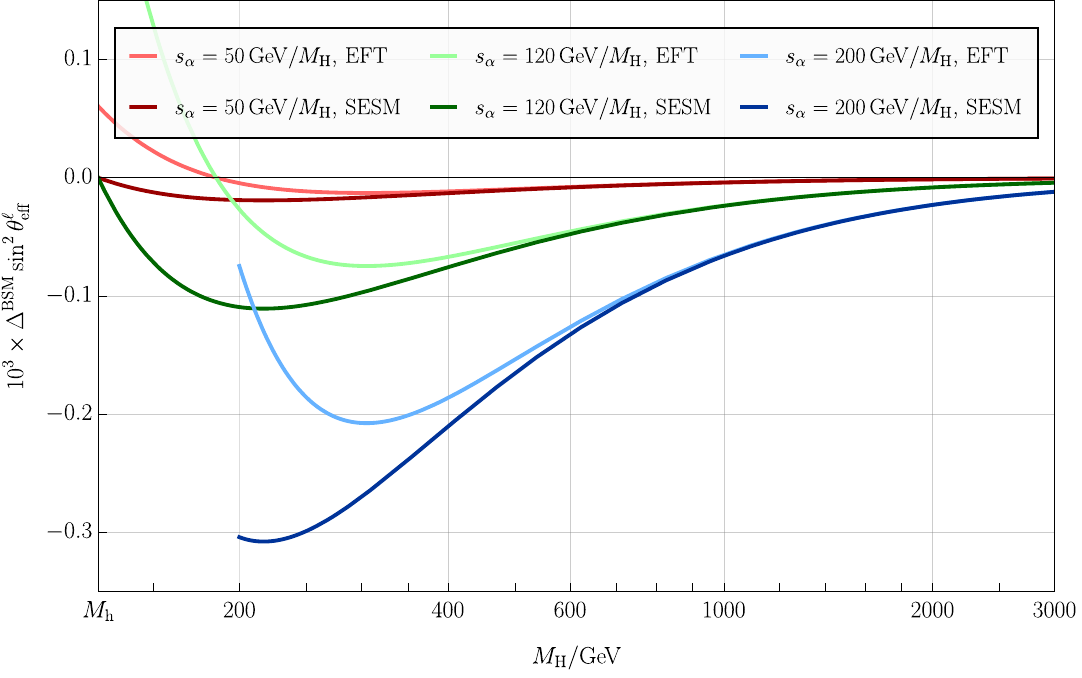}}
\caption{BSM correction to the effective weak mixing angle
$\sin^2\theta_{\mathrm{eff}}^\ell$ in the full SESM vs.\ its
EFT approximation.
The product $s_\alpha\MH$ is kept fixed in each curve.}
\label{fig:SWEff}
\end{figure}

Figure~\ref{fig:SWEff} shows the correction $\Delta^{\BSM}\sin^2\theta_{\mathrm{eff}}^\ell$
for a charged lepton~$\ell$, both for the full SESM prediction and
our EFT approximation.
The agreement of the two predictions in the asymptotic large-$\MH$ limit (with $\sa\MH$ fixed)
shows the same patterns as for $\MW$ discussed in the previous subsection, nicely
illustrating the consistency of the EFT.
The figure also demonstrates the importance of the pseudo-observable
$\sin^2\theta_{\mathrm{eff}}^\ell$ in constraining the SESM by confronting
it with experimental data, as $\Delta^{\BSM}\sin^2\theta_{\mathrm{eff}}^\ell$
is easily of the size of the experimental precision of $\sim10^{-4}$.
As in the case of the $\MW$ prediction discussed in the previous section,
the NLO prediction for $\sin^2\theta_{\mathrm{eff}}^\ell$ in the SESM
can be improved to catch the NNLO effects of the SM upon
adding $\Delta^{\BSM}\sin^2\theta_{\mathrm{eff}}^\ell$ to the
best available SM prediction
(see \rcite{Dubovyk:2019szj} and references therein).

\subsection{Decays of the Z~boson}
\label{subsec:zdecay}

Considering massless fermions~$f$,
the Higgs sector of the SM or SESM impacts the NLO
EW corrections to the vector and axial-vector
$Zf\bar f$ couplings in a universal
(i.e.\ flavour-independent) way.
While the effective weak mixing angle probes the ratio of effective
vector to axial-vector
couplings for on-shell Z~bosons, the Z-decay widths are sensitive
to the normalization of these effective couplings.
We can, thus, pick just one representative $\PZ\to f\bar f$ decay channel
in the following,
in order to check whether our EFT correctly describes $\PZ\to f\bar f$ decays
in the large-$\MH$ limit.
To this end, we choose the simplest case of the invisible decay
$\PZ\to \nu_\ell\bar\nu_\ell$, where $\nu_\ell$ is the (massless) neutrino
partner of the charged lepton~$\ell$.

The total as well as several leptonic and hadronic partial Z-decay widths
represent another cornerstone in model fits to experimental data
both in the SM and its extensions, with some of these decay widths
even known to the $0.1\%$ level~\cite{ParticleDataGroup:2024cfk}.
Here, however, we merely
investigate $\Gamma_{\PZ\to \nu_\ell\bar\nu_\ell}$ to validate our EFT.

\begin{figure}
\centering{\includegraphics[width=.9\textwidth]{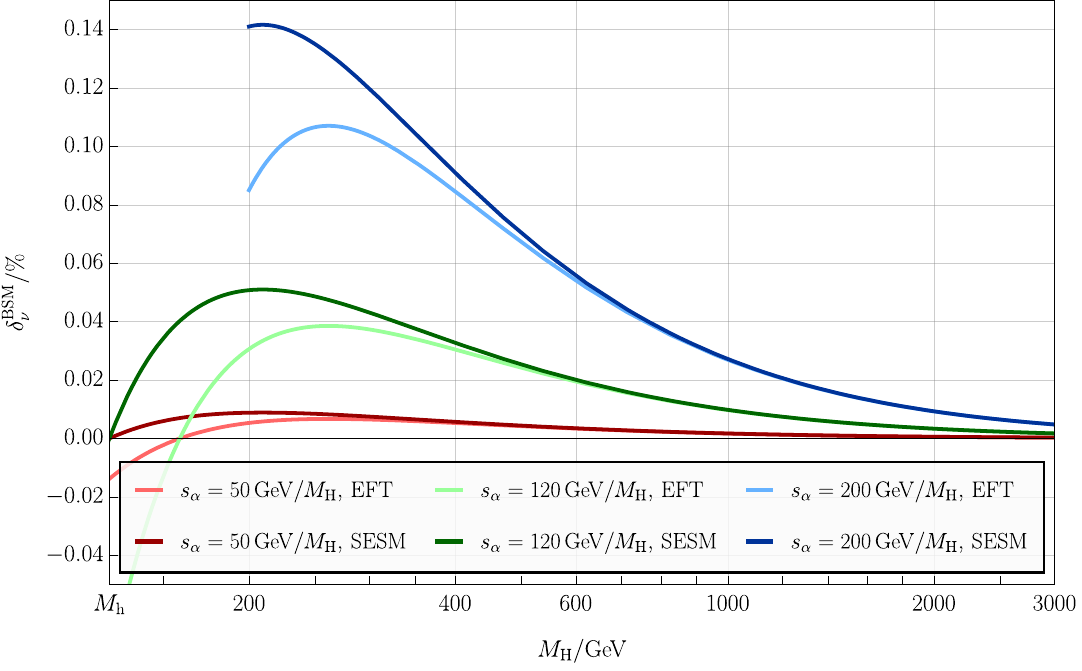}}
\caption{BSM contribution to the relative NLO correction to the
partial decay width $\Gamma_{\PZ\to \nu\bar\nu}$
in the full SESM vs.\ its EFT approximation.
The product $s_\alpha\MH$ is kept fixed in each curve.}
\label{fig:Zdecay}
\end{figure}

The NLO prediction to the partial decay width
$\Gamma_{\PZ\to \nu\bar\nu} \equiv
\Gamma_{\PZ\to \nu_\ell\bar\nu_\ell}$, which does not depend on the generation
of $\ell$, can be written as
\begin{align}
\Gamma_{\PZ\to \nu\bar\nu} =
\Gamma^{\LO}_{\PZ\to \nu\bar\nu} \left(1+\delta_{\nu\bar\nu}\right)
= \Gamma^{\LO}_{\PZ\to \nu\bar\nu}
\left[1+2\Re(\delta_{\nu\bar\nu,\text{1-loop}}) +2\delta_{\nu\bar\nu,\ct}\right],
\end{align}
with
\begin{align}
\Gamma^{\LO}_{\PZ\to \nu\bar\nu} = \frac{\alpha_{\mathrm{em}}\MZ}{24\cw^2\sw^2}\,,
\end{align}
where $\delta_{\nu\bar\nu}$ is the relative NLO correction to the partial
decay width, which in turn receives contributions from the
correction factors $\delta_{\nu\bar\nu,\text{1-loop}}$ and $\delta_{\nu\bar\nu,\ct}$
from the one-loop amplitude and the corresponding counterterm, respectively.
The correction factors appear as $\Re(...)$ and with factors of 2,
because these one-loop contributions interfere with the lowest-order diagram.
Since the decay $\PZ\to \nu\bar\nu$ does not involve external charged
particles, there are no bremsstrahlung corrections at NLO;
and if there were, they would be identical in the SESM and SM.
There are no Higgs exchange vertex diagrams at the one-loop level,
so that the difference between SESM and SM is again confined to the
counterterm contribution, and the BSM part of the SESM
prediction for $\Gamma_{\PZ\to \nu\bar\nu}$ reads
\begin{align}
\Gamma_{\PZ\to \nu\bar\nu}^{\BSM}
= \Gamma^{\LO}_{\PZ\to \nu\bar\nu}\, \delta_{\nu}^{\BSM}, \qquad
\delta_{\nu}^{\BSM} = 2 \delta_{\nu,\ct}^{\BSM}
=
\delta^{\BSM} Z_{ZZ}
+ \frac{2(\sw^2-\cw^2)}{\cw^2}\,\frac{\delta^{\BSM}\sw}{\sw} \,.
\end{align}
Here we have already exploited the fact that the neutrino
wave-function renormalization constant of the SESM coincides with the one
of the SM at NLO
and thus drops out in the BSM part of the correction.
Explicit results for the EFT contributions to the renormalization
constants can again be found in
\app{CTEFT}.
From these we compute the hard and soft
contributions originating from the one-loop
effective Lagrangian \eqref{eq:SMEFTHEFT}, with the following
result for the contribution of the hard modes,
\begin{equation}
\delta_{\nu,h}^{\BSM,\EFT} =
v_2^2 \left[
\frac{2\cw}{\sw} C_{\Phi WB}
+ \frac{1-2\sw^2}{2\sw^2} C_{\Phi D}+4C^\nonSMEFT_{2} \right].
\label{eq:dn_EFTh}
\end{equation}
Figure~\ref{fig:Zdecay} compares the BSM contribution $\delta_{\nu\bar\nu}^{\BSM}$
to the relative correction $\delta_{\nu\bar\nu}$ with its EFT approximation
$\delta_{\nu\bar\nu}^{\BSM,\EFT}$, demonstrating the validity of the EFT
in the large-$\MH$ limit.

\subsection{Decays of the W~boson}
\label{subsec:wdecay}

The partial decay widths of the W~boson can be used to probe yet another combination of EFT contributions.
As representative decay channel we consider the leptonic decay
$\PW\to \ell\bar\nu_\ell$, where again all external fermions are
taken massless.
The impact of actual measurements of W-boson decay widths on model
fits to data is less significant than
for the Z~boson, as the current
experimental precision on the total W-boson width and the leptonic
branching ratio are (only) of the order of $\sim2\%$ and $\sim0.5\%$,
respectively~\cite{ParticleDataGroup:2024cfk}.

\begin{figure}
\centering{\includegraphics[width=.9\textwidth]{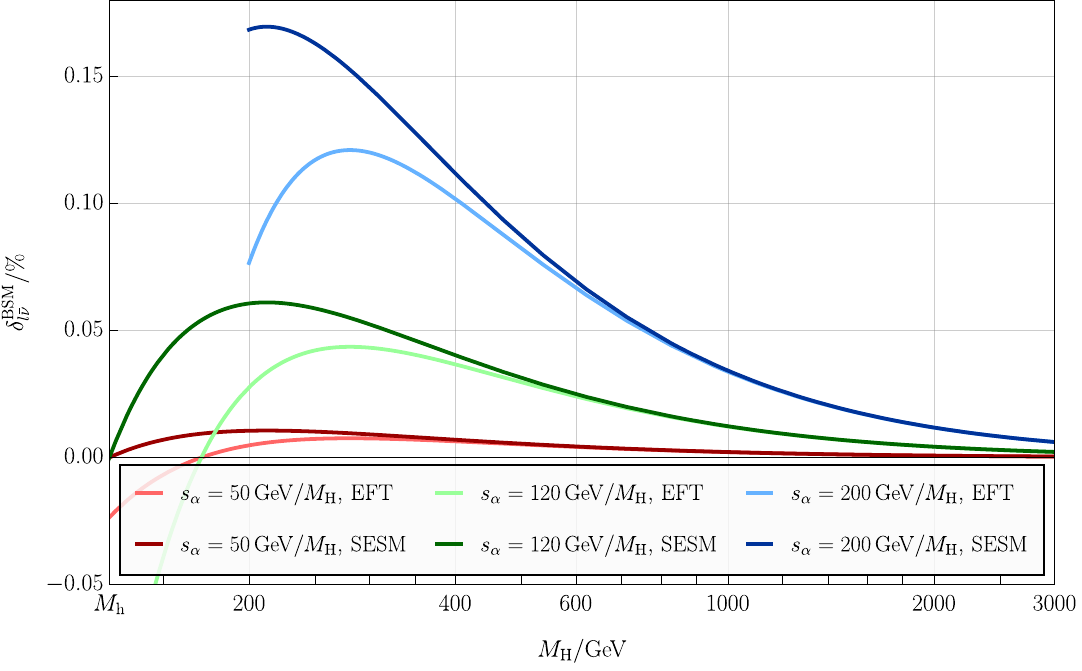}}
\caption{BSM contribution to the relative NLO correction to the
partial decay width $\Gamma_{\PW\to \ell\bar\nu}$
in the full SESM vs.\ its EFT approximation.
The product $s_\alpha\MH$ is kept fixed in each curve.}
\label{fig:Wdecay}
\end{figure}

The NLO prediction to the partial decay width
$\Gamma_{\PW\to \ell\bar\nu} \equiv
\Gamma_{\PW\to \ell\bar\nu_\ell}$, which does not depend on the generation
of the massless lepton~$\ell$, can be written as
\begin{align}
\Gamma_{\PW\to \ell\bar\nu} \equiv
\Gamma^{\LO}_{\PW\to \ell\bar\nu} \left(1+\delta_{\ell\bar\nu}\right)
= \Gamma^{\LO}_{\PW\to \ell\bar\nu}
\left[1+2\Re(\delta_{\ell\bar\nu,\text{1-loop}}) + 2\delta_{\ell\bar\nu,\ct}
+\delta_{\ell\bar\nu,\real} \right]
\end{align}
with
\begin{align}
\Gamma^{\LO}_{\PW\to \ell\bar\nu} = \frac{\alpha_{\mathrm{em}}\MW}{12\sw^2} \,,
\end{align}
where the NLO correction $\delta_{\ell\bar\nu}$ is split into the genuine
one-loop part $2\Re(\delta_{\ell\bar\nu,\text{1-loop}})$, the counterterm
contribution $2\delta_{\ell\bar\nu,\ct}$, and the contribution
$\delta_{\ell\bar\nu,\real}$ from real-photon radiation.
Again, at NLO only the counterterm contribution is sensitive to the Higgs
sector in the SM and the SESM for massless external fermions, so that
the BSM part of the NLO contribution is given by
\begin{align}
\Gamma_{\PW\to \ell\bar\nu}^{\BSM}
= \Gamma^{\LO}_{\PW\to \ell\bar\nu} \, \delta_{\ell\bar\nu}^{\BSM},
\qquad
\delta_{\ell\bar\nu}^{\BSM} = 2 \delta_{\ell\bar\nu,\ct}^{\BSM}
=
\delta^{\BSM} Z_{W}
- \frac{2\delta^{\BSM}\sw}{\sw} \,.
\label{eq:delta_en}
\end{align}
The hard part arising from the one-loop effective Lagrangian \eqref{eq:SMEFTHEFT} reads
\begin{equation}
\delta_{\ell\bar\nu,h}^{\BSM,\EFT} =
v_2^2 \left(
\frac{2\cw}{\sw} C_{\Phi WB}
+\frac{\cw^2}{2\sw^2}C_{\Phi D}+4C^\nonSMEFT_{2} \right),
\label{eq:delta_ln_EFTh}
\end{equation}
cf.~\app{CTEFT}.
Figure~\ref{fig:Wdecay} compares the BSM contribution $\delta_{\ell\bar\nu}^{\BSM}$
to the relative correction $\delta_{\ell\bar\nu}$ with its EFT counterpart
$\delta_{\ell\bar\nu}^{\BSM,\EFT}$, again confirming the validity of the EFT
in the large-$\MH$ limit.

\subsection[The four-body Higgs decay $\mathrm{h\to WW\to\nu_ee^+\mu^-\bar\nu_\mu}$]%
{\boldmath{The four-body Higgs decay $\mathrm{h\to WW\to\nu_ee^+\mu^-\bar\nu_\mu}$}}
\label{subsec:hdecay}

\begin{figure}
\setlength{\unitlength}{1pt}
\centerline{
\includegraphics{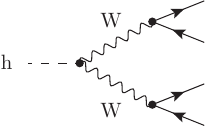}
}
\caption{Tree-level diagram for $\Ph\to \PW\PW\to\nu_{\Pe}\Pe^+\mu^-\bar\nu_\mu$.}
\label{fig:hWW_tree}
\end{figure}

In order to probe the EFT couplings that influence the kinetic Lagrangian of the light
Higgs boson~h, we have to consider Higgs production or decay processes.
Since we do not yet support massive fermions in the SESM EFT, we have to choose a
process where the Higgs boson~h couples to W or Z bosons.
We choose the decay process $\Ph\rightarrow \PW\PW\rightarrow\nu_\Pe \Pe^+\mu^-\bar{\nu}_\mu$
with massless leptons in the final state, in which at least one of the two W~bosons is off
its mass shell, because $\Mh<2\MW$.
The tree-level diagram for this process is shown in \fig{hWW_tree}.
Other four-body decays via W- or Z-boson pairs do not
provide further independent probes of the EFT Lagrangian; the difference between intermediate
W and Z~bosons is already resolved by the previously considered processes
in the SESM.
NLO predictions for $\Ph\to\PW\PW/\PZ\PZ\to4\,$fermions are provided by the
\textsc{Prophecy4f} Monte Carlo program~\cite{Denner:2018opp}
for the SM~\cite{Bredenstein:2006rh,Bredenstein:2006ha,Denner:2019fcr},
the Two-Higgs-Doublet Model~\cite{Altenkamp:2017kxk,Altenkamp:2017ldc,Denner:2018opp},
and the SESM~\cite{Altenkamp:2018bcs,Denner:2018opp}.
We have extended \textsc{Prophecy4f} by adding
an SESM EFT prediction for the
decay channel $\Ph\rightarrow \PW\PW\rightarrow\nu_\Pe \Pe^+\mu^-\bar{\nu}_\mu$,
called $\Ph\to\Pe\mu2\nu$ in short in the following.
\textsc{Prophecy4f} consistently employs the complex-mass
scheme~\cite{Denner:2005fg} (see also \rcite{Denner:2019vbn})
to treat the W~resonances and all off-shell regions at NLO accuracy, which
proceeds via an analytical continuation of the W/Z~masses and $\sw$ in all couplings
to complex quantities.
We have proceeded within the EFT exactly in the same way, without
encountering any additional complications.

Predictions for $\Ph\to\Pe\mu2\nu$ entail the new feature that the LO SM and SESM
predictions do not coincide, because the $hWW$ coupling of the SESM involves
the additional factor~$\ca$ relative to its SM counterpart.
The respective LO partial decay widths are, thus, related by
\begin{align}
\Gamma^{\SESM,\LO}_{\Ph\to\Pe\mu2\nu} = \ca^2\, \Gamma^{\SM,\LO}_{\Ph\to\Pe\mu2\nu} \,.
\label{eq:h2l2nLO}
\end{align}
The NLO predictions receive virtual one-loop corrections and real corrections from
real-photon emission,
\begin{align}
\Gamma^{\mathrm{M},\NLO}_{\Ph\to\Pe\mu2\nu}
= \Gamma^{\mathrm{M},\LO}_{\Ph\to\Pe\mu2\nu} \left( 1 + \delta^{\mathrm{M}}_{\Pe\mu2\nu} \right)
= \Gamma^{\mathrm{M},\LO}_{\Ph\to\Pe\mu2\nu}
\left( 1 + \delta^{\mathrm{M},\virt}_{\Pe\mu2\nu}+ \delta^{\mathrm{M},\real}_{\Pe\mu2\nu} \right),
\qquad \mathrm{M}=\SM,\SESM.
\end{align}
Since the real-photonic contributions to $\Gamma^{\mathrm{M},\NLO}_{\Ph\to\Pe\mu2\nu}$
of the SM and SESM are related by the same factor $\ca^2$ as the LO contributions,
made explicit in \eq{h2l2nLO}, the relative real-photonic corrections coincide,
$\delta^{\SESM,\real}_{\Pe\mu2\nu} = \delta^{\SM,\real}_{\Pe\mu2\nu}$,
and the BSM contribution to the relative correction is entirely given by the difference
SESM${}-{}$SM in the virtual corrections,
\begin{align}
\delta^{\BSM}_{\Pe\mu2\nu} = \delta^{\SESM}_{\Pe\mu2\nu} - \delta^{\SM}_{\Pe\mu2\nu}
= \delta^{\SESM,\virt}_{\Pe\mu2\nu} - \delta^{\SM,\virt}_{\Pe\mu2\nu}.
\label{eq:deltaBSM-hem2n}
\end{align}
In this BSM difference of the virtual corrections further systematic cancellations
between SESM and SM contributions occur.
In detail, the cancellation takes place between all contributions in which the
corresponding SESM and SM graphs are related by the same factor~$\ca$ as the LO
amplitudes. Since this factor occurs in all couplings of~$h$ to any other SM field,
the required factor~$\ca$ relates all corresponding SESM and SM graphs in which
in addition to the external~$h$ line no further internal $h$~lines occur.
This, in particular, includes all one-loop diagrams with closed fermion loops
as well as the corresponding one-loop contributions to counterterms.
This is particularly convenient, since contributions of graphs with Higgs bosons coupling to
massive internal fermions systematically cancel in $\delta^{\BSM}_{\Pe\mu2\nu}$;
otherwise our EFT with massless fermions would not be able to reproduce
$\delta^{\BSM}_{\Pe\mu2\nu}$ in the large-$\MH$ limit.

In the course of working out the EFT approximation for $\delta^{\SESM,\virt}_{\Pe\mu2\nu}$,
we first have repeated the calculation of this virtual correction,
which is described in some detail for the linear Higgs realization in
\rcites{Altenkamp:2018bcs,Denner:2018opp},
within the non-linear Higgs realization, and there within the framework of the BFM.
We fully confirmed the original calculation of
\rcites{Altenkamp:2018bcs,Denner:2018opp},
but the transition to the non-linear realization revealed many intriguing
rearrangements within the virtual corrections between self-energy, vertex, and box contributions.
Fortunately, there are also classes of Feynman diagrams that to not change in this transition,
such as all corrections due to closed fermion loops and all pentagon graphs, which do not
contribute to the BSM correction $\delta^{\BSM}_{\Pe\mu2\nu}$ either.

\begin{figure}
\centerline{
\includegraphics{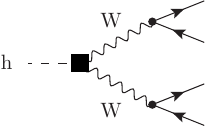}
\qquad\qquad
\includegraphics{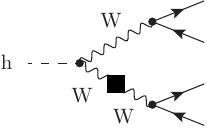}
\qquad\qquad
\includegraphics{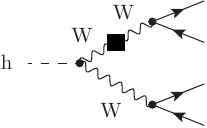}
}
\caption{\revision{Bosonic EFT operator insertions} into tree-level diagrams
for $\Ph\to \PW\PW\to\nu_{\Pe}\Pe^+\mu^-\bar\nu_\mu$
in unitary gauge. These diagrams yield
hard(-momentum) contributions to the virtual NLO corrections in the SESM.
Diagrams with counterterm insertions that contribute additional corrections from hard Wilson coefficients are not shown.}
\label{fig:hWW_hard}
\end{figure}

In order to go into more detail in the EFT approximation, we first define the
BSM contribution $\delta^{\BSM}\M$ of the one-loop amplitude $\delta^{\SESM}\M$
of the SESM by subtracting $\ca$ times the one-loop amplitude $\delta^{\SM}\M$
of the SM,
\begin{align}
\delta^{\BSM}\M = \delta^{\SESM}\M - \ca\,\delta^{\SM}\M \,.
\end{align}
This definition applies both to the full SESM and to the EFT prediction.
The quantity $\delta^{\BSM}\M$ is designed such that the LO contribution to the decay drops out; the following numerical results, thus, show genuine NLO effects in the comparison between full SESM and its EFT approximation.
From $\delta^{\BSM}\M$ we can reconstruct the BSM contribution to the
squared amplitude $|\M|^2$ by interfering it with the tree-level amplitude
$\M^{\SESM}_0$ of the SESM,
\begin{align}
\delta^{\BSM}|\M|^2 ={}& 2\Re\left\{ \bigl(\delta^{\SESM}\M - \ca\,\delta^{\SM}\M \bigr) \bigl(\M^{\SESM}_0\bigr)^{\!*}\right\}
\nn\\
={}& 2\Re\left\{ \delta^{\SESM}\M \bigl(\M^{\SESM}_0\bigr)^{\!*}\right\}
- 2\ca^2 \Re\left\{  \delta^{\SM}\M \bigl(\M^{\SM}_0\bigr)^{\!*}\right\}
\nn\\
={}& \delta^{\SESM}|\M|^2 - \ca^2 \,\delta^{\SM}|\M|^2,
\end{align}
where $\delta^{\mathrm{M}}|\M|^2$ denotes the one-loop contributions to the squared
matrix elements in model $\mathrm{M}=\SESM,\SM$ and spin summation is understood
in all interference diagrams.
Integrating $\delta^{\BSM}|\M|^2$ over the four-particle phase space $\Phi_4$
and applying the usual flux factor makes contact
to the relative correction $\delta^{\BSM}_{\Pe\mu2\nu}$ introduced above,
\begin{align}
\delta^{\BSM}_{\Pe\mu2\nu}\cdot\Gamma^{\SESM,\LO}_{\Ph\to\Pe\mu2\nu}
= \frac{1}{2\Mh}\int\rd\Phi_4\,\delta^{\BSM}|\M|^2.
\end{align}
Next, we decompose $\delta^{\BSM}\M$
into self-energy and vertex contributions
(box and pentagon contributions drop out),
\begin{align}
\delta^{\BSM}\M = \delta^{\BSM}\M_{WW}
+ \delta^{\BSM}\M_{W\ell\nu}
+ \delta^{\BSM}\M_{hWW}.
\end{align}%
The corresponding EFT operator insertions into tree-level graphs
relevant for the hard contributions are shown in \reffi{fig:hWW_hard}.
The self-energy correction $\delta^{\BSM}\M_{WW}$ receives contributions from
the W~self-energy inserted into each of the two intermediate W-boson lines of the tree graph
of the decay process
$\Ph\rightarrow \PW\PW\rightarrow\nu_\Pe \Pe^+\mu^-\bar{\nu}_\mu$,
\begin{align}
\delta^{\BSM}\M_{WW} = -\biggl[
 \frac{\Sigma^{WW,\BSM}_{\rT}\!(k_+^2)-\delta^{\BSM}\MW^2}{k_+^2-\MW^2}
+\frac{\Sigma^{WW,\BSM}_{\rT}\!(k_-^2)-\delta^{\BSM}\MW^2}{k_-^2-\MW^2}
+ 2\delta^{\BSM}Z_W \biggr]\! \M^{\SESM}_0,
\label{eq:hWW_Wself}
\end{align}
where $k_+,k_-$ are the four-momenta of the two intermediate W~bosons.
Inserting the explicit results for the hard EFT contributions to $\Sigma^{WW,\BSM}_{\rT}(k^2)$
and the renormalization constants of \app{CTEFT},
it turns out that the hard EFT part of
$\delta^{\BSM}\M_{WW}$ vanishes,
\begin{align}
\delta^{\BSM,\EFT}_h \M_{WW} = 0 \,.
\end{align}
The BSM parts of the vertex corrections to the (off-shell) $W^+\to\nu_\Pe\Pe^+$ and
$W^-\to\mu^-\bar\nu_\mu$ decays are, for the same reason as in the previous
section, entirely contained in the (lepton-generation-independent)
counterterm contributions,
\begin{align}
\delta^{\BSM}\M_{W\ell\nu}
= 2 \delta_{\ell\bar\nu,\ct}^{\BSM} \, \M^{\SESM}_0,
\end{align}
with $\delta_{\ell\bar\nu,\ct}^{\BSM}$ as given in \eq{delta_en}, so that the hard EFT part is
\revision{\begin{align}
\delta^{\BSM,\EFT}_h \M_{W\ell\nu}
= \delta_{\ell\bar\nu,h}^{\BSM,\EFT} \, \M^{\SM}_0,
\end{align}}%
with $\delta_{\ell\bar\nu,h}^{\BSM,\EFT} $ given in \eq{delta_ln_EFTh}.
The $hWW$ vertex correction $\delta^{\BSM}\M_{hWW}$ is the most involved ingredient.
The core part of this contribution to the $\Ph\to\Pe\mu2\nu$ amplitude is the
one-particle-irreducible (1PI) $hWW$ vertex function $\Gamma^{hW^-W^+}_{\mu\nu}(k_h,-k_+,-k_-)$,
which decomposes into covariant structures as follows ($V_1=W^-,V_2=W^+,k_1=-k_+,k_2=-k_-$),%
\footnote{In the considered approximation of the SESM (all fermions massless except for
the top-quark) CP symmetry is unbroken, ruling out $\epsilon$-tensor contributions in the
covariant decomposition of $\Gamma^{hV_1V_2}_{\mu\nu}$.}
\begin{align}
\label{eq:hV1V2vert}
\Gamma^{hV_1V_2}_{\mu\nu}(k_h,k_1,k_2) =
g_{\mu\nu}F_1^{hV_1V_2}
+k_{2\mu}k_{1\nu}F_2^{hV_1V_2}
+k_{1\mu}k_{1\nu}F_3^{hV_1V_2}
+k_{2\mu}k_{2\nu}F_4^{hV_1V_2}
+k_{1\mu}k_{2\nu}F_5^{hV_1V_2},
\end{align}
where (following the conventions of \rcite{Denner:2019vbn})
the fields $h,V_1,V_2$ and momenta $k_h,k_1,k_2$ are incoming.
Among the formfactors $F_i^{hV_1V_2}\equiv F_i^{hV_1V_2}(k_h^2,k_1^2,k_2^2)$,
only $F_1$ receives a tree-level and a counterterm contribution,
\begin{align}
F_{1,0}^{hW^-W^+,\SESM} ={}& \ca\,\frac{e\MW}{\sw} = \frac{\ca g_2^2 v_2}{2} \,, \qquad
F_{1,0}^{hW^-W^+,\SM} = \frac{e\MW}{\sw} = \frac{g_2^2 v_2}{2} \,,
\\
F_{1,\ct}^{hW^-W^+,\SESM} ={}& \frac{\ca g_2^2 v_2}{2}
\left( \delta Z_e
+ \frac{1}{2}\delta Z_{hh}
+ \frac{\sa}{2\ca}\delta Z_{Hh}
+ \delta Z_W
- \frac{\sa\delta\sa}{\ca^2}
+ \frac{\delta\MW}{\MW} - \frac{\delta\sw}{\sw} \right)^{\!\SESM},
\nn\\
F_{1,\ct}^{hW^-W^+,\SM} ={}& \frac{g_2^2 v_2}{2}
\left( \delta Z_e
+ \frac{1}{2}\delta Z_{hh}
+ \delta Z_W
+ \frac{\delta\MW}{\MW} - \frac{\delta\sw}{\sw} \right)^{\!\SM},
\end{align}
but all $F_i^{hW^-W^+}$ receive one-loop contributions
$F_{i,\text{1-loop}}^{hW^-W^+}$.
Since the external fermions are taken massless, the covariants in \eq{hV1V2vert}
involving $k_{1\mu}$ or $k_{2\nu}$ do not contribute by virtue of current conservation,%
\footnote{In line with the convention that fields and momenta are incoming in
$\Gamma^{hW^-W^+}_{\mu\nu}(k_h,-k_+,-k_-)$, the decaying $\PW^\pm$~bosons have momenta $k_\pm$.
This $hWW$ vertex function is contracted with the vertex functions of the decays
$g^{\mu\rho}\Gamma^{W^+\bar\nu_\Pe\Pe^-}_\rho(k_+,-p_1,-p_2)$ and
$g^{\nu\sigma}\Gamma^{W^-\mu^+\bar\nu_\mu}_\sigma(k_-,-p_3,-p_4)$.
Current conservation for the decays, which holds for massless decay leptons, implies
$k_+^\rho\Gamma^{W^+\bar\nu_\Pe\Pe^-}_\rho(k_+,-p_1,-p_2) =
k_-^\sigma\Gamma^{W^-\mu^+\bar\nu_\mu}_\sigma(k_-,-p_3,-p_4)=0$.}
so that only the formfactors $F_1^{hW^-W^+}$ and $F_2^{hW^-W^+}$ are relevant.
The hard EFT contributions to the relevant renormalized formfactors, expressed in terms of
Wilson coefficients, are given by
\begin{align}
F^{hW^-W^+}_{1,h}={}& 2v_2  \biggl[
	(C_{\Phi W}+C^\nonSMEFT_{2})(k_1^2+k_2^2)
	-(C_{\Phi W}+C^\nonSMEFT_{2}+2C^\nonSMEFT_{6})k_h^2\nn\\&
	+\MW^2
	\left(C_{\Phi\Box}  +\frac{1-2\sw^2 }{4\sw^2} C_{\Phi D}
	+\frac{\cw}{\sw} C_{\Phi WB}
+2C^\nonSMEFT_{2}\right)
	\biggr],
	\nn\\
F^{hW^-W^+}_{2,h}={}&
	4v_2 (C_{\Phi W}-C^\nonSMEFT_{3}-C^\nonSMEFT_{5}) \,.
\label{eq:FhWW_hard}
\end{align}

Again, for later reference, we explicitly give
all the hard EFT parts to $\delta^{\BSM}\M$.
To this end, we collect the results
for the W~self-energies, the
W-decay vertex corrections, and the $hW^-W^+$ vertex corrections \revision{from} Eqs.~\eqref{eq:hWW_Wself}--\eqref{eq:FhWW_hard} and obtain
\begin{align}
\delta^{\BSM,\EFT}_h\M ={}&
2 \delta_{\ell\bar\nu,\ct,h}^{\BSM} \M^{\SESM}_0
+ \frac{e^2\left( F^{hW^-W^+}_{1,h} {\cal A}_1
+ F^{hW^-W^+}_{2,h} {\cal A}_2 \right)}{2\sw^2(k_+^2-\MW^2)(k_-^2-\MW^2)} \,,
\end{align}
where we have introduced the spinor chains
\begin{align}
{\cal A}_1 = [\bar u_{\nu_\Pe}\gamma_\alpha\omega_- v_{\Pep}]\,
[\bar u_{\mu^-}\gamma^\alpha\omega_- v_{\bar\nu_\mu}] \,,
\qquad
{\cal A}_2 = [\bar u_{\nu_\Pe}\dsl{k}_-\omega_- v_{\Pep}]\,
[\bar u_{\mu^-}\dsl{k}_+\omega_- v_{\bar\nu_\mu}] \,,
\end{align}
with the chirality projector $\omega_-=(1-\gamma_5)/2$ and the usual
standard notation for the Dirac spinors $\bar u_{\nu_\Pe}$, etc.
Using
\begin{align}
\M^{\SESM}_0 = \frac{e^2 F^{hW^-W^+,\SESM}_{1,0} {\cal A}_1}{2\sw^2(k_+^2-\MW^2)(k_-^2-\MW^2)} \,,
\end{align}
in terms of Wilson coefficients we find
\begin{align}
\delta^{\BSM,\EFT}_h\M ={}&
\revision{
\frac{4\sw^2}{e^2} \biggl[
\MW^2
\biggl(
\frac{2\cw}{\sw} C_{\Phi WB}
+\frac{\cw^2}{2\sw^2}C_{\Phi D}+4C^\nonSMEFT_{2} \biggr)
}
\nn\\
& {}
	+ \Bigl(C_{\Phi W}+C^\nonSMEFT_{2} \,\Bigr)(k_+^2+k_-^2)
	- \Bigl(C_{\Phi W}+C^\nonSMEFT_{2}+2C^\nonSMEFT_{6} \,\Bigr)\Mh^2
\nn\\
& {}
	+ \MW^2
	\biggl(C_{\Phi\Box} +\frac{1-2\sw^2 }{4\sw^2} C_{\Phi D}
	+\frac{\cw}{\sw} C_{\Phi WB}
+2C^\nonSMEFT_{2}\biggr)
\biggr]
\revision{\M^{\SM}_0}
\nn\\
& {}
+  2v_2 \Bigl(C_{\Phi W}-C^\nonSMEFT_{3}-C^\nonSMEFT_{5} \,\Bigr)\,
\frac{e^2{\cal A}_2}{\sw^2(k_+^2-\MW^2)(k_-^2-\MW^2)} \,.
\label{eq:dMh4f_EFTh}
\end{align}

\begin{figure}
	\centering{\includegraphics[width=.9\textwidth]{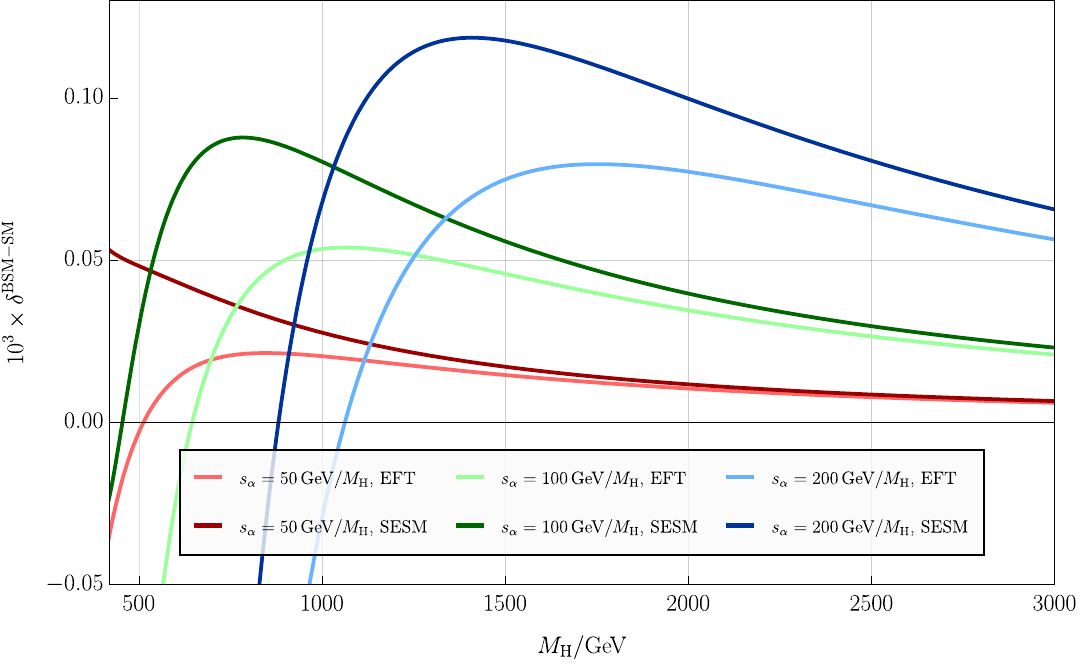}}
	\caption{BSM corrections to the decay $ h\rightarrow \PW\PW\rightarrow\nu_\Pe\bar{\nu}_\mu \Pe^+\mu^- $ in the $\GF$-scheme.
  The product $s_\alpha\MH$ is kept fixed in each curve.}
\label{fig:Hdecay}
\end{figure}

Figure~\ref{fig:Hdecay} shows the BSM contribution $\delta^{\BSM}_{\Pe\mu2\nu}$
to the relative correction as defined in \eq{deltaBSM-hem2n}
in the so-called $\GF$-scheme, where $\MW$, $\MZ$, and
$\alpha_{\mathrm{em}} \equiv \alpha_{\GF} = \sqrt{2}\GF\MW^2(1-\MW^2/\MZ^2)/\pi$
are taken as input in the gauge-boson sector
(see, e.g., \rcite{Denner:2019vbn}).
This change in the input for $\alpha_{\mathrm{em}}$ from $\alpha_{\mathrm{em}}(0)$ to $\alpha_{\GF}$
has to be compensated by the substitution
$\delta Z_e \to \delta Z_e - \Delta r^{(1)}/2$
in the charge renormalization constant, which absorbs some universal corrections
into the lowest-order prediction.
In fact the extreme smallness of $\delta^{\BSM}_{\Pe\mu2\nu}$ in \reffi{fig:Hdecay}
demonstrates that the major part of the BSM corrections is absorbed into the lowest-order
prediction in the $\GF$-scheme.
The resulting small correction $\delta^{\BSM}_{\Pe\mu2\nu}$ again shows the expected
asymptotic behaviour in the large-$\MH$ limit as already discussed in the previous
sections and proves the validity of the EFT.

\subsection{Failure of matching to bosonic SMEFT}
\label{subsec:SMEFTbreakdown}

In \subsecsm{MW}{hdecay} we have verified that the EFT with the effective Lagrangian
\eqref{eq:FinalResult}, including non-SMEFT operators,
correctly reproduces the SESM predictions
for physical observables with massless fermions to $\ord(\zeta^{-2})$.
\revision{Although our ansatz for the field transformation made in
\subsec{FieldRedefLO} was quite general, it does not exhaust all possibilities,
since the ansatz was based on polynomial field redefinitions.
It is, thus, not proven at this point that the effective Lagrangian necessarily
involves non-SMEFT operators.
Some non-polynomial field transformation may
exists that transforms the effective Lagrangian into SMEFT form, with correspondingly
adjusted Wilson coefficients.
If this is the case, the EFT predictions for
observables, obtained with our non-SMEFT-type effective Lagrangian,
will be reproducible in terms of appropriate SMEFT Wilson coefficients alone.
In the following, we show that this is impossible
if the SMEFT operator basis is restricted to operators without fermions or gluons.}
While this ``bosonic SMEFT'' may at first sight seem to be a reasonable ansatz for a
(dimension-six) low-energy EFT of the SESM with massless fermions, we stress that in order to
\revision{ultimately conclude on the equivalence or inequivalence of our EFT and SMEFT
 requires to include
also fermionic and gluonic SMEFT operators.
The latter might in principle be generated by field redefinitions}
(this time necessarily also of gauge-boson fields) transforming our non-SMEFT operators into
the SMEFT basis, although our functional derivation a priori does not yield any fermionic
or gluonic operators at all.
\revision{In \subsec{SMEFTmatch} we will come back to the question whether a general SMEFT
matching is possible and will find that this is indeed the case.
In the rest of this section we give a proof
that bosonic SMEFT fails to reproduce the SESM in our large-mass limit,
which instructively demonstrates a line of arguments for revealing inconsistencies
if an ansatz for the effective Lagrangian in the bottom-up approach is insufficient.}

Carrying out the proof on the basis of observables, i.e.\ at the level
of amplitudes, can be viewed as showing that any attempt of diagrammatically matching SMEFT to the SESM in the large-mass limit \eqref{eq:EFTlimit} necessarily has to fail,
once a sufficiently large set of different amplitudes is considered.
We emphasize that an overcomplete set of amplitudes has to be considered in the
diagrammatic matching to see this, i.e.\ the number of independent observables has to
exceed the number of SMEFT Wilson coefficients entering the corresponding amplitudes
in the diagrammatic matching. Conversely, being able to match SMEFT to any
(necessarily incomplete) set of amplitudes can never be taken as proof that SMEFT
fully reproduces an underlying theory in a given EFT limit.
Matching to additional amplitudes
could always reveal a possible incompleteness of the SMEFT operator basis.
In this sense, diagrammatic (bottom-up)
matching can never decide on the completeness of
an assumed SMEFT form of an EFT, in contrast to functional matching,
where the general form of the effective Lagrangian is part of the result.
Moreover, we emphasize that the following proof of the failure of
bottom-up matching to bosonic SMEFT is completely independent of the method by
which we have obtained our EFT. In principle, we do not even need the knowledge
of the correct EFT, we just need the large-$\MH$ expansion of the considered
amplitudes, which we have derived both in the EFT and the full SESM for the
considered processes.

Before starting the attempt of diagrammatic matching, we realize that it
is enough to focus only on the hard EFT contributions to the considered
amplitudes, since these are the only sources of corrections
$\propto\alpha_{\mathrm{em}}\ln\MH$ in the large-$\MH$ limit. In any EFT for this
limit, these terms originate from loop-induced contributions to Wilson coefficients
that enter tree-level EFT diagrams at NLO.
This means that diagrammatic SMEFT matching tries to
absorb all hard non-SMEFT contributions quantified by the Wilson coefficients
$C^\nonSMEFT_i$ of our EFT into redefined SMEFT Wilson coefficients $\hat C_i$.

We start the matching attempt
by identifying the Wilson coefficients $\hat C_i$ of SMEFT operators without fermion
or gluon fields in the Warsaw basis,
which \revision{span the space of a potential EFT.
Further,} we can drop all CP-violating operators, as the SESM
in the considered approximation does not give rise to
CP violation.
In summary, only the set of purely bosonic
\revision{CP-conserving operators
are allowed to receive non-zero Wilson coefficients, namely
\begin{equation}
  \{\hat{C}_{\Phi},
  \hat{C}_{\Phi\Box},
  \hat{C}_{\Phi D},
  \hat{C}_{W},
  \hat{C}_{\Phi W},
  \hat{C}_{\Phi B},
  \hat{C}_{\Phi WB}
  \} \,,
\end{equation}
thereby defining precisely what we mean by ``bosonic SMEFT''.}
This set of operators includes the SMEFT part of our effective
Lagrangian \eqref{eq:LeffSMEFT}.
Note that the operators $\mathcal{O}_{\Phi},$ $\mathcal{O}_{W}$ only affect Higgs
and vector-boson self-interactions, respectively.
Therefore, their coefficients $\hat{C}_{\Phi},$ $\hat{C}_{W}$ can be fixed
(e.g.\ by the amplitudes for processes for $ \Ph\Ph\rightarrow \Ph\Ph $
and $ \PZ\PZ\rightarrow \PW\PW $, respectively)
independently from the set
\begin{equation}\label{eq:FreeWC}
	\{\hat{C}_{\Phi\Box},
	\hat{C}_{\Phi D},
	\hat{C}_{\Phi W},
	\hat{C}_{\Phi B},
	\hat{C}_{\Phi WB}
	\} \,,
\end{equation}
which governs the observables discussed in \subsecsm{MW}{hdecay}
to validate our EFT at NLO.
The matching conditions imposed by these observables on the Wilson coefficients in~\eqref{eq:FreeWC} read:
\begin{itemize}
\item
for $\Delta r^{(1,\BSM,\EFT)}_h$ from \eq{dr_EFTh}:
\begin{align}
\cw^2C_{\Phi D}
+4\sw\cw C_{\Phi WB}
+8\sw^2C^\nonSMEFT_{2}
\;\overset{!}{=}\; {} &
\cw^2\hat C_{\Phi D}
+4\sw\cw \hat C_{\Phi WB} \,;
\label{eq:SMEFTmatch1}
\end{align}
\item
for $\Delta^{\BSM}\sin^2\theta_{\mathrm{eff},h}^\ell$ from \eq{dsw_EFTh}:
\begin{align}
& 2\cw^2\sw^2 C_{\Phi W}-2\cw^2\sw^2 C_{\Phi B}-\frac{\cw^2}{2}C_{\Phi D}
-2\cw^3\sw C_{\Phi WB}
-2\sw^2 C^\nonSMEFT_{2}
\nn\\
& {} \;\overset{!}{=}\; {}
2\cw^2\sw^2 \hat C_{\Phi W}-2\cw^2\sw^2 \hat C_{\Phi B}-\frac{\cw^2}{2}\hat C_{\Phi D}
-2\cw^3\sw \hat C_{\Phi WB} \,;
\label{eq:SMEFTmatch2}
\end{align}
\item
for $\delta_{\nu,h}^{\BSM}$ from \eq{dn_EFTh}:
\begin{align}
\frac{2\cw}{\sw} C_{\Phi WB}
+ \frac{1-2\sw^2}{2\sw^2} C_{\Phi D}+4C^\nonSMEFT_{2}
\;\overset{!}{=}\; {} &
\frac{2\cw}{\sw} \hat C_{\Phi WB}
+ \frac{1-2\sw^2}{2\sw^2} \hat C_{\Phi D} \,;
\label{eq:SMEFTmatch3}
\end{align}
\item
for $\delta_{\ell\bar\nu,h}^{\BSM}$ from \eq{delta_ln_EFTh}:
\begin{align}
\frac{2\cw}{\sw} C_{\Phi WB}
+\frac{\cw^2}{2\sw^2}C_{\Phi D}+4C^\nonSMEFT_{2}
\;\overset{!}{=}\; {} &
\frac{2\cw}{\sw} \hat C_{\Phi WB}
+\frac{\cw^2}{2\sw^2}\hat C_{\Phi D} \,;
\label{eq:SMEFTmatch4}
\end{align}
\item
for $\delta^{\BSM}_h\M$ from \eq{dMh4f_EFTh}:
\begin{align}
& C_{\Phi W}+C^\nonSMEFT_{2} \;\overset{!}{=}\; {} \hat C_{\Phi W},
\label{eq:SMEFTmatch5}
\\
& \MW^2 \left(
C_{\Phi\Box}
+\frac{5-6\sw^2}{4\sw^2} C_{\Phi D}
+\frac{5\cw}{\sw} C_{\Phi WB}
+10C^\nonSMEFT_{2}\right)
\nn\\
& \quad {} - \Mh^2(C_{\Phi W}+C^\nonSMEFT_{2}+2C^\nonSMEFT_{6})
\nn\\
& {} \qquad \;\overset{!}{=}\; {}
\MW^2 \left(
\hat C_{\Phi\Box}
+\frac{5-6\sw^2}{4\sw^2} \hat C_{\Phi D}
+\frac{5\cw}{\sw} \hat C_{\Phi WB}
\right)
- \Mh^2\hat C_{\Phi W} \,,
\label{eq:SMEFTmatch6}
\\
& C_{\Phi W}-C^\nonSMEFT_{3}-C^\nonSMEFT_{5} \;\overset{!}{=}\; {} \hat C_{\Phi W}\,.
\label{eq:SMEFTmatch7}
\end{align}
Note that three matching conditions result from $\delta^{\BSM}_h\M$ as
given in \eq{dMh4f_EFTh}, because the parts proportional to
$\M^{\SESM}_0$ (and thus ${\cal A}_1$) and
${\cal A}_2$ influence differential cross sections in a different way.
Similarly, the constant parts and the ones proportional to $k_\pm^2$
in the contribution proportional to $\M^{\SESM}_0$ affect observables differently.
\end{itemize}

The system of linear equations Eqs.~\eqref{eq:SMEFTmatch1}--\eqref{eq:SMEFTmatch7}
overconstrains the set of SMEFT Wilson coefficients \eqref{eq:FreeWC},
appearing on the r.h.s., as we have six independent equations
(Eqs.~\eqref{eq:SMEFTmatch1} and~\eqref{eq:SMEFTmatch4} are linearly dependent)
and only five free parameters.
It has the solution
\begin{align}
\hat C_{\Phi WB} ={} & C_{\Phi WB} + \frac{2\sw}{\cw} C^\nonSMEFT_{2},
\nn\\
\hat C_{\Phi B} ={} & C_{\Phi B} + \frac{\sw^2}{\cw^2} C^\nonSMEFT_{2},
\nn\\
\hat C_{\Phi D} ={} & C_{\Phi D} \,,
\nn\\
\hat C_{\Phi W}={} & C_{\Phi W} + C^\nonSMEFT_{2},
\nn\\
\hat C_{\Phi\Box} ={} & C_{\Phi\Box} -\frac{ 2\Mh^2}{\MW^2} C^\nonSMEFT_{6},
\label{eq:SMEFTsol}
\end{align}
but only if the following constraint is fulfilled,
\begin{align}
C^\nonSMEFT_{2}+C^\nonSMEFT_{3}+C^\nonSMEFT_{5} \;\overset{!}{=}\; 0\,.
\label{eq:SMEFTcond}
\end{align}
However, inserting our explicit results
\eqref{eq:HEFTWC} for $C^\nonSMEFT_{i}$ into the l.h.s.\ of this relation, we find
\begin{align}
  \label{eq:C235v2}
C^\nonSMEFT_{2}+C^\nonSMEFT_{3}+C^\nonSMEFT_{5}
&=\frac{(D-6)e^2s_{\alpha}^2I_{20}}{16\pi^2D\left(D^2-4\right)\sw^2v_2^2}\nn\\
&=-\frac{e^2s_{\alpha}^2}{384\pi^2\sw^2v_2^2}
\bigg[\Delta+\ln\left(\frac{\mu^2}{\MH^2}\right)+\frac{17}{6}\bigg]
+\mathcal{O}(\epsilon)
\neq0 \,,
\end{align}
proving that bosonic SMEFT cannot be matched to our large-$\MH$ limit of the SESM.
\revision{We recall that this proof is tied to the fact that we assume a
suppression of $\sa\sim\zeta^{-1}$. Any stronger suppression would invalidate the argument;
in fact the whole non-SMEFT part of the effective Lagrangian \refeq{eq:SMEFTHEFT}
would then  vanish to $\ord(\zeta^{-2})$.}

\revision{As a first step to further investigate the question whether the SESM in the considered
large-mass limit is equivalent to an
EFT of general SMEFT form (including bosonic and fermionic EFT operators), we have
dropped the fermion fields completely (i.e.\ also in the SM part). If the attempted matching of bosonic
SMEFT in this case had failed, the SESM with or without fermions certainly could not
be matched by SMEFT at all. However, inspite of including a fairly large set of
purely bosonic processes in the list of matched amplitudes, we could not provoke
a contradiction in the matching. This observation in combination with a formal hint
by Gerhard Buchalla based on his findings in \citeres{Buchalla:2013rka,Buchalla:2016bse,Buchalla:2020kdh}
triggered the successful matching of the full SMEFT to the SESM
described in the next section.}

\section{\revision{Effective Lagrangian in SMEFT form}}
\label{sec:LeffSMEFT}

\subsection{\revision{Full SMEFT matching}}
\label{subsec:SMEFTmatch}

\revision{Having proven that \textit{bosonic} SMEFT is not enough to reproduce the large-$\MH$
limit of the SESM with massless fermions, we now turn to the question whether the constructed
bosonic EFT is equivalent to the full SMEFT with bosonic and fermionic operators,
with suitably matched coefficients.
In fact, in the first preprint version of this paper this rather
non-trivial question was left open,
but in the meantime could be answered in a positive way.%
\footnote{\revision{We greatfully acknowledge Gerhard Buchalla giving a an important hint
towards solving this problem.}}
The reason why our polynomial ansatz for possible field redefinitions in
\subsec{field-redef-NLO} failed to bring $\delta\L_\eff^{\oneloop,\BSM}$ in SMEFT form
is that the necessary field transformations (of the vector fields) turn out to be
non-polynomial w.r.t.\ the Higgs field $h$.
}

\revision{
The simplest way to derive the SMEFT Lagrangian that is equivalent to our non-SMEFT Lagrangian
given in \sec{final} is to use the SM EOMs
(in the $1/\MH^2$-suppressed part of the EFT Lagrangian),
which in fact amounts to applying dedicated
field redefinitions~\cite{Manohar:2018aog,Criado:2018sdb}.
In detail, the strategy is to translate the non-SMEFT operators via IBP relations
into bosonic operators that reduce to fermionic operators or even vanish by
virtue of the EOMs at $\ord(\zeta^0)$.%
\footnote{\revision{Once this transformation is found, it is straightforward
to identify the corresponding field transformations.}}
To this end, we define the following auxiliary operators in unitary gauge,
\begin{align}
  \mathcal{O}^{\EOM}_1 ={}& (v_2+h)^3 \biggl[ \Box h
  - \frac{g_2^2}{4} (v_2+h) C_\mu^a C^{a,\mu}
  + \frac{\Mh^2}{2v_2^2}\, h(v_2+h)(2v_2+h) \biggr] ,
  \label{eq:OpEOM1}
  \\
  \mathcal{O}^{\EOM}_2 ={}&  (v_2+h)^2 C^{3,\nu}
  \biggl[ \partial^\mu B_{\mu\nu}
  + \frac{g_1 g_2}{4} (v_2+h)^2 C^3_\nu \biggr] ,
  \\
  \mathcal{O}^{\EOM}_3 ={}& (v_2+h)^2 \,C^{a,\nu}
  \biggl[ ( D_W^\mu W_{\mu\nu} )^a
  +\frac{g_2^2}{4}\,(v_2+h)^2 C^a_\nu \biggr] ,
  \\
  \mathcal{O}^{\EOM}_4 ={}& (v_2+h) \, (D_W^\nu C_\nu )^a
  \Bigl[ 2(\partial^\mu h) C^a_\mu
  + (v_2+h) (D_W^\mu C_\mu )^a \Bigr] ,
  \\
  \mathcal{O}^{\EOM}_5 ={}& (\partial^\nu h) \, C_\nu^a
  \Bigl[ 2(\partial^\mu h) C^a_\mu
  + (v_2+h) (D_W^\mu C_\mu )^a \Bigr] ,
  \label{eq:OpEOM5}
\end{align}
where according to \eq{covDW},
\begin{equation}
  (D_W^\mu X_\nu)^a = \partial^\mu X^a_\nu + g_2\, \eps^{abc} \, W^{\mu, b} X^c_\nu \,.
\end{equation}
Note also that, using \eq{Bderiv}, $(D_W^\mu C_\mu )^a = (D_B^\mu C_\mu )^a$.
The non-SMEFT operators $\mathcal{O}^{\nonSMEFT}_i$
can be reduced to these operators plus SMEFT and SM
operators by the relations
\begin{align}
  0 ={}&
  \frac{8\Mh^2}{v_2^2} \mathcal{O}_\Phi
  + \frac{2(3-2\sw^2)}{\cw^2} \mathcal{O}_{\Phi\Box}
  + \frac{8\sw^2}{\cw^2} \mathcal{O}_{\Phi D}
  - 3 \mathcal{O}_{\Phi W}
  - \frac{3\sw^2}{\cw^2} \mathcal{O}_{\Phi B}
  - \frac{6\sw}{\cw} \mathcal{O}_{\Phi WB}
  \nn\\ & {}
  + \mathcal{O}^{\nonSMEFT}_1 + \mathcal{O}^{\nonSMEFT}_2 - 2\mathcal{O}^{\nonSMEFT}_3
  - 3g_1\mathcal{O}^{\nonSMEFT}_4 - \mathcal{O}^{\nonSMEFT}_5 + \mathcal{O}^{\nonSMEFT}_6
  - \frac{g_2^2}{2}\mathcal{O}^{\nonSMEFT}_8
  \nn\\ & {}
  -  \Mh^2 (v_2+ h)^4
  - 2 \mathcal{O}^{\EOM}_1
  - \frac{2\sw}{\cw} \mathcal{O}^{\EOM}_2
  - 2\mathcal{O}^{\EOM}_3 \,,
  \label{eq:OnonSMEFTid1}
  \\
  0 ={}&
  \mathcal{O}^{\nonSMEFT}_1 + \mathcal{O}^{\nonSMEFT}_3
  + g_1\mathcal{O}^{\nonSMEFT}_4 - \mathcal{O}^{\nonSMEFT}_5
  - 2 \mathcal{O}^{\EOM}_4 \,,
  \label{eq:OnonSMEFTid2}
  \\
  0 ={}&
  \mathcal{O}^{\nonSMEFT}_1 - \mathcal{O}^{\nonSMEFT}_3
  - g_1\mathcal{O}^{\nonSMEFT}_4 + \mathcal{O}^{\nonSMEFT}_5
  - 8\mathcal{O}^{\nonSMEFT}_7
  + 4\mathcal{O}^{\EOM}_5 \,,
  \label{eq:OnonSMEFTid3}
\end{align}
which follow (exclusively) from IBP.
Specifically, these relations can be used
to eliminate three non-SMEFT operators (e.g.\ $\mathcal{O}^{\nonSMEFT}_5$,
$\mathcal{O}^{\nonSMEFT}_7$, $\mathcal{O}^{\nonSMEFT}_8$)
from $\delta\L_\eff^{\oneloop,\BSM}$ of \sec{final} at the cost of introducing
the operators $\mathcal{O}^{\EOM}_i$.
By conspiracy among the non-SMEFT Wilson coefficients $C^{\nonSMEFT}_i$ given in
\eq{HEFTWC}, not only the three chosen non-SMEFT operators are removed, but in fact
the coefficients of the remaining non-SMEFT operators, which are now linear
combinations of the $C_i^\nonSMEFT$ in \eq{CiHEFTeps}, happen to vanish exactly.
The $(v_2+h)^4$ term introduced by exploiting \eq{OnonSMEFTid1} can be absorbed in the SM part of the effective Lagrangian, up to an unphysical constant that is dropped, by adjusting the pre-renormalization constants as shown in \app{CThard}.
}

\revision{
We now have a Lagrangian of the form
\begin{align}
  \delta\L_\eff^{\oneloop,\BSM} \,=\, \hat{\delta} \L^\oneloop_\mathrm{SMEFT,bos}
  \,+\, \sum_{n=1}^{5}C^{\EOM}_n \, \mathcal{O}^{\EOM}_n \,,
\end{align}
where $\hat{\delta} \L^\oneloop_\mathrm{SMEFT,bos}$ consists of the same SMEFT operators as \eq{LeffSMEFT}, albeit with different Wilson coefficients (see below).
The subscript ``bos'' emphasizes that no fermionic operators are included.
The operators in \eqsm{OpEOM1}{OpEOM5} have been defined such that they either vanish or reduce
to fermionic SMEFT operators by employing the SM EOMs, which in unitary gauge read
\begin{align}
  0 ={}& \Box h + \frac{\Mh^2}{2v_2^2}\, h (v_2+h)(2v_2+h)
  -\frac{g_2^2}{4}\,(v_2+h) C^a_\mu C^{a,\mu} \,,
  \label{eq:EOM_h}
  \\
  0 ={}& \partial^\mu B_{\mu\nu}
  +\frac{g_1 g_2}{4}\,(v_2+h)^2 C^3_\nu
  -\frac{g_1}{2}\left( Y_{\rw,F} \overline{F}\gamma_\nu F
  +Y_{\rw,f} \overline{f}\gamma_\nu f \right),
  \label{eq:EOM_B}
  \\
  0 ={}& D_W^\mu W^a_{\mu\nu}
  +\frac{g_2^2}{4}\,(v_2+h)^2 C^a_\nu
  + \frac{g_2}{2}\,\overline{F}\tau^a \gamma_\nu F \,,
  \label{eq:EOM_W}
  \\
  0 ={}& \dsl{D} F
  = \biggl( \dsl{\partial} -\ri g_2 \dsl{W} + \frac{\ri g_1}{2}\,Y_{\rw,F}\dsl{B} \biggr)F \,,
  \label{eq:EOM_F}
  \\
  0 ={}& \dsl{D} f
  = \bigg( \dsl{\partial} + \frac{\ri g_1}{2}\,Y_{\rw,f}\dsl{B} \biggr)f \,.
  \label{eq:EOM_f}
\end{align}
Following again the conventions of \rcite{Denner:2019vbn},
the (massless) fermion fields are denoted by $F=L,Q$ for the left-handed
SU(2) doublets and $f=l,u,d$ for the right-handed singlets.
Summation over the fermion species $F$ and $f$ is understood in all terms
proportional to $\overline{F}\cdots F$ and $\overline{f}\cdots f$, respectively,
and $Y_{\rw,F}$ and $Y_{\rw,f}$ denote the respective hypercharges of $F$ and $f$
(e.g.\ $Y_{\rw,L} = -1$ and $Y_{\rw,l} = -2$).
Exploiting the EOMs in \eqsm{EOM_h}{EOM_f} together with the (covariant)
four-divergence of \eq{EOM_W},
\begin{align}
0 ={}& D_W^\mu D_W^\nu W^a_{\mu\nu}
= -\frac{g_2^2}{4}\, D_W^\mu
\left[ (v_2+h)^2 C^a_\mu \right]
-\frac{g_2}{2}\, D_W^\mu \left( \overline{F}\tau^a\gamma_\mu F \right)
\nn\\
={} &  -\frac{g_2^2}{4}\,(v_2+h)
\left[ 2(\partial^\mu h) C^a_\mu
+ (v_2+h) (D^\mu_W C^a_\mu)  \right]
-\frac{g_2}{2}\left[ \overline{(\dsl{D}F)}\tau^a F + \overline{F}\tau^a(\dsl{D}F) \right] ,
\label{eq:EOM_W2}
\end{align}
turns \eqsm{OnonSMEFTid1}{OnonSMEFTid3} into the physically equivalent form
\begin{align}
  \mathcal{O}^{\EOM}_1 \underset{\mbox{\scriptsize\refeq{eq:EOM_h}}}{=}{}& 0 \,,
  \\
  \mathcal{O}^{\EOM}_2 \underset{\mbox{\scriptsize\refeq{eq:EOM_B}}}{=}{}&
  \frac{g_1}{2} \, (v_2+h)^2 C^{3,\nu}
  \left( Y_{\rw,F} \overline{F}\gamma_\nu F
  +Y_{\rw,f} \overline{f}\gamma_\nu f \right),
  \\
  \mathcal{O}^{\EOM}_3 \underset{\mbox{\scriptsize\refeq{eq:EOM_W}}}{=}{}&
  - \frac{g_2}{2}\, (v_2+h)^2 \,C^{a,\nu} \,\overline{F}\tau^a \gamma_\nu F \,,
  \\
  \mathcal{O}^{\EOM}_4 \underset{\mbox{\scriptsize\refeq{eq:EOM_W2}}}{=}{}&
  -	\frac{2}{g_2}\, (D_W^\nu C^a_\nu)
    \left[ \overline{(\dsl{D}F)}\tau^a F + \overline{F}\tau^a(\dsl{D}F) \right]
  \underset{\mbox{\scriptsize\refeq{eq:EOM_F},\refeq{eq:EOM_f}}}{=} {}  0 \,,
  \\
  \mathcal{O}^{\EOM}_5 \underset{\mbox{\scriptsize\refeq{eq:EOM_W2}}}{=}{}&
  -\frac{2}{g_2}\,(\partial^\nu h) \, C_\nu^a \, (v_2+h)^{-1}
  \left[ \overline{(\dsl{D}F)}\tau^a F + \overline{F}\tau^a(\dsl{D}F) \right]
  \underset{\mbox{\scriptsize\refeq{eq:EOM_F}}}{=} 0 \,.
\end{align}
Note, in particular, the appearance of the non-polynomial factor $(v_2+h)^{-1}$ in the
transformation of $\mathcal{O}^{\EOM}_5$, which is the reason that the EOM operators
cannot be removed by field transformations based on an ansatz that is only polynomial
in the fields, as the one in \subsec{field-redef-NLO}.
Finally, the two non-vanishing operators $\mathcal{O}^{\EOM}_{2,3}$ can be brought
into canonical SMEFT form by using the identities
\begin{align}
  \phi_{\lin}^\dagger \ri \overset\leftrightarrow{D}_\mu \phi_{\lin} = {} &
  \ri \Bigl[\phi_{\lin}^\dagger D_\mu  \phi_{\lin} - (D_\mu \phi_{\lin} )^\dagger \phi_{\lin}    \Bigr] =
  -\frac{g_2}{2}\, (v_2+h)^2\, C^3_\mu \,,
  \nn\\
  \phi_{\lin}^\dagger \ri \overset\leftrightarrow{D}{}_\mu^a \phi_{\lin} = {} &
  \ri \Bigl[\phi_{\lin}^\dagger \tau^a D_\mu  \phi_{\lin} - (D_\mu \phi_{\lin} )^\dagger \tau^a \phi_{\lin}    \Bigr]=
  \frac{g_2}{2}\, (v_2+h)^2\, C^a_\mu \,,
\end{align}
in unitary gauge with the result
\begin{align}
  \mathcal{O}^{\EOM}_2 = {}&
  -\frac{g_1}{g_2}
  \Bigl(\phi_{\lin}^\dagger \ri \overset\leftrightarrow{D}_\mu \phi_{\lin}\Bigr)
  \left( Y_{\rw,F} \overline{F}\gamma^\mu F
  +Y_{\rw,f} \overline{f}\gamma^\mu f \right)
  \nn\\
  = {}& -\frac{g_1}{g_2} \left(
  Y_{\rw,L} \mathcal{O}^{(1)}_{\Phi L}
  + Y_{\rw,Q} \mathcal{O}^{(1)}_{\Phi Q}
  + Y_{\rw,l} \mathcal{O}_{\Phi l}
  + Y_{\rw,u} \mathcal{O}_{\Phi u}
  + Y_{\rw,d} \mathcal{O}_{\Phi d}
  \right),
  \\
  \mathcal{O}^{\EOM}_3 ={}&
  -\Bigl(\phi_{\lin}^\dagger \ri \overset\leftrightarrow{D}{}_\mu^a \phi_{\lin}\Bigr)
  (\overline{F}\tau^a \gamma^\mu F)
  = -  \mathcal{O}^{(3)}_{\Phi L} - \mathcal{O}^{(3)}_{\Phi Q} \,.
\end{align}
The explicit expressions for the fermionic SMEFT operators
$\mathcal{O}^{(1)}_{\Phi F}$, $\mathcal{O}^{(3)}_{\Phi F}$, and $\mathcal{O}_{\Phi f}$
in our field conventions are given in Tab.~1 of \rcite{Denner:2019vbn}.
}

\revision{
By means of these manipulations,
we have eventually managed to transform our effective Lagrangian of \sec{final} to SMEFT form,
\begin{align}
    \delta\L_\eff^{\oneloop,\BSM} ={}& \hat{\delta} \L^\oneloop_\mathrm{SMEFT,bos} +  \hat{\delta}\L^\oneloop_\mathrm{SMEFT,ferm}
    \nn\\
    ={}&
    \hat{C}_\Phi\mathcal{O}_\Phi
    +\hat{C}_{\Phi\Box}\mathcal{O}_{\Phi\Box}
    +\hat{C}_{\Phi D}\mathcal{O}_{\Phi D}
    +\hat{C}_{\Phi W}\mathcal{O}_{\Phi W}
    +\hat{C}_{\Phi B}\mathcal{O}_{\Phi B}
    +\hat{C}_{\Phi WB}\mathcal{O}_{\Phi WB}
     \nn\\
    &+ \hat{C}_{W}\mathcal{O}_{W}
    +\hat{C}^{(1)}_{\Phi F} \mathcal{O}^{(1)}_{\Phi F}
    +\hat{C}^{(3)}_{\Phi F} \mathcal{O}^{(3)}_{\Phi F}
    +\hat{C}_{\Phi f} \mathcal{O}_{\Phi f} \,,
\label{eq:SMEFT}
\end{align}
with the Wilson coefficients
\begin{align}
  \hat C_\Phi ={}&
  - \frac{(D^2+4) e^2 \Mh^2 \sa^2 I_{20}}{4\pi^2 D(D^2-4) \sw^2 v_2^4}
  + \frac{(D^2+6D+20) \Mh^4 \sa^2 I_{20}}{2\pi^2 D(D^2-4) v_2^6}
  \nn\\ & {}
  + \frac{(D-4)(D+2) \Mh^2 \sa^2 (\MH^2 \sa^2 - 2 \lambda_{12} v_2^2)
    I_{20}}{4\pi^2 D(D-2) v_2^6}
  + \frac{(D-4) (\MH^2 \sa^2 - 2 \lambda_{12} v_2^2)^3 I_{20}}{48\pi^2  \MH^2 v_2^6}
  \nn\\ ={}&
  -\frac{5e^2\Mh^2s_{\alpha}^2}{48\pi^2\sw^2v_2^4}\left(L_{\epsilon}+\frac{31}{30}\right)
  +\frac{5\Mh^4s_{\alpha}^2}{8\pi^2v_2^6}\left(L_{\epsilon}+\frac{41}{30}\right)
  +\frac{3s_{\alpha}^2\Mh^2\left(2\lambda_{12}v_2^2-\MH^2s_{\alpha}^2\right)}{8\pi^2v_2^6}
  \nn\\
  &{}
  -\frac{\left(\MH^2s_{\alpha}^2-2\lambda_{12}v_2^2\right)^3}{24\pi^2\MH^2v_2^6}
  +\mathcal{O(\epsilon)}\, ,
\label{eq:CPhi_SMEFT}
  \\
  \hat C_{\Phi\Box} ={}& -\frac{s_\alpha\delta s_{\alpha}}{v_2^2}
  - \frac{(D^2+4) e^2 \sa^2(3-2\sw^2) I_{20}}{16\pi^2 D(D^2-4) \sw^2 \cw^2 v_2^2}
  + \frac{(\MH^2 \sa^2 - 2 \lambda_{12} v_2^2) I_{20}}{192\pi^2 D(D-2)\MH^2 v_2^4}
  \nn\\ & \quad {}
  \times \left[(D^3-18D^2+128D-96) \MH^2 \sa^2
  - 2D(D^2-42D+44) \lambda_{12} v_2^2\right]
  \nn\\ ={}&
  -\frac{\delta s_{\alpha}s_{\alpha}}{v_2^2}
  +\frac{5e^2s_{\alpha}^2\left(2\sw^2-3\right)}{192\pi^2\sw^2\cw^2v_2^2} \left(L_{\epsilon}+\frac{31}{30}\right)
  +\frac{9\lambda_{12}\left(\MH^2s_{\alpha}^2-2\lambda_{12}v_2^2\right)}{16\pi^2\MH^2v_2^2} \left(L_{\epsilon}+\frac{10}{27}\right)
  \nn\\
  &
  +\frac{s_{\alpha}^2\left(\MH^2s_{\alpha}^2-2\lambda_{12}v_2^2\right)}{8\pi^2v_2^4} \left(L_{\epsilon}+\frac{7}{6}\right)
  +\mathcal{O(\epsilon)}
  \,,
\label{eq:CPhiBox_SMEFT}
  \\
  \hat C_{\Phi D} ={}&
  -\frac{(D^2+4) e^2 \sa^2 I_{20}}{4\pi^2 D(D^2-4) \cw^2 v_2^2}
  =
  -\frac{5e^2s_{\alpha}^2}{48\pi^2\cw^2v_2^2} \left(L_{\epsilon}+\frac{31}{30}\right)
  +\mathcal{O(\epsilon)} \,,
  \\
  \hat C_{\Phi W} ={}& -\frac{(D-4) e^2 \sa^2 I_{20}}{8\pi^2 D (D^2-4) \sw^2 v_2^2}
  =
  \frac{e^2s_{\alpha}^2}{192\pi^2\sw^2v_2^2}
  +\mathcal{O(\epsilon)}\,,
  \\
  \hat C_{\Phi B} ={}& -\frac{(D-4) e^2 \sa^2 I_{20}}{8\pi^2 D (D^2-4) \cw^2 v_2^2}
  =
  \frac{e^2s_{\alpha}^2}{192\pi^2\cw^2v_2^2}
  +\mathcal{O(\epsilon)}\,,
  \\
  \hat C_{\Phi WB} ={}& -\frac{(D-4)e^2 \sa^2 I_{20}}{4\pi^2 D(D^2-4) \sw \cw v_2^2}
  =
  \frac{e^2s_{\alpha}^2}{96\pi^2\sw\cw v_2^2}
  +\mathcal{O(\epsilon)}\,,
  \\
  \hat C_{W} ={}& 0 \,,
  \\
  \hat C^{(1)}_{\Phi F} ={}& \frac{(D-6)e^2\sa^2 I_{20}}{16\pi^2D(D^2-4)\cw^2 v_2^2} \, Y_{\rw,F}
  =
  -\frac{e^2\sa^2}{384\pi^2\cw^2 v_2^2} \, Y_{\rw,F}  \left(L_{\epsilon}+\frac{17}{6}\right)
  +\mathcal{O(\epsilon)}\,,
  \\
  \hat C^{(3)}_{\Phi F} ={}& \frac{(D-6)e^2\sa^2 I_{20}}{16\pi^2D(D^2-4)\sw^2 v_2^2}
  =
  -\frac{e^2\sa^2}{384\pi^2\sw^2 v_2^2} 
  \left(L_{\epsilon}+\frac{17}{6}\right)
  +\mathcal{O(\epsilon)}\,,
  \\
  \hat C_{\Phi f} ={}& \frac{(D-6)e^2\sa^2 I_{20}}{16\pi^2D(D^2-4)\cw^2 v_2^2} \, Y_{\rw,f}
  =
  -\frac{e^2\sa^2}{384\pi^2\cw^2 v_2^2} \, Y_{\rw,f}
  \left(L_{\epsilon}+\frac{17}{6}\right)
  +\mathcal{O(\epsilon)} \,.
\label{eq:CPhif_SMEFT}
\end{align}
}

\begin{figure}
\centerline{
	\includegraphics{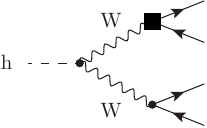}
	\qquad\qquad
	\includegraphics{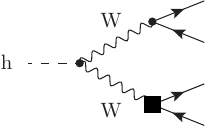}
	\qquad\qquad
	\includegraphics{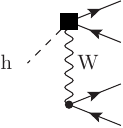}
	\qquad\qquad
	\includegraphics{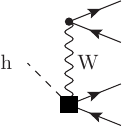}
}
\caption{\revision{Fermionic EFT operator insertions into tree-level diagrams
for $\Ph\to \PW\PW\to\nu_{\Pe}\Pe^+\mu^-\bar\nu_\mu$
in unitary gauge, producing hard(-momentum) contributions to the virtual NLO corrections
in addition to the bosonic operator insertions shown in \fig{hWW_hard}.}}
\label{fig:hWW_hardF}
\end{figure}

\revision{
Finally, we translate the hard EFT contributions to the EW precision observables
in Eqs.~\eqref{eq:dr_EFTh}, \eqref{eq:dsw_EFTh}, \eqref{eq:dn_EFTh}, \eqref{eq:delta_ln_EFTh}, and \eqref{eq:dMh4f_EFTh}, which depend on non-SMEFT coefficients,
to the new SMEFT form:
\begin{align}
  \Delta r^{(1,\BSM,\EFT)}_h ={} &
  \frac{v_2^2}{2\sw^2} \left(\cw^2\hat C_{\Phi D}+4\sw\cw \hat C_{\Phi WB}
  +4\sw^2\hat C^{(3)}_{\Phi L}\right),
  \label{eq:dr_EFThNEW}
  \\
  \Delta^{\BSM,\EFT}_h\sin^2\theta_{\mathrm{eff}}^\ell ={} &
  \frac{v_2^2}{2} \Bigl[
  -\cw^2 \hat C_{\Phi D}
  +4\cw^2\sw^2  \hat C_{\Phi W}-4\cw^2\sw^2  \hat C_{\Phi B}
  -4\cw^3\sw  \hat C_{\Phi WB}
  \nn\\ & \qquad {}
  -2\sw^2 \left( \hat C^{(1)}_{\Phi L} + \hat C^{(3)}_{\Phi L} \right)
  - (1-2\sw^2) \hat C_{\Phi l}
  \Bigr] \,,
  \label{eq:dsw_EFThNEW}
  \\
  \delta_{\nu,h}^{\BSM,\EFT} ={} &
  \frac{v_2^2}{2\sw^2} \left[
  (1-2\sw^2) \hat C_{\Phi D}
  + 4\cw\sw \hat C_{\Phi WB}
  + 4\sw^2\hat C^{(3)}_{\Phi L}
  - 4\sw^2\hat C^{(1)}_{\Phi L}
  \right] ,
  \label{eq:dn_EFThNEW}
  \\
  \delta_{\ell\bar\nu,h}^{\BSM,\EFT} ={} &
  \frac{v_2^2}{2\sw^2} \left(\cw^2\hat C_{\Phi D}+4\sw\cw \hat C_{\Phi WB}
  +4\sw^2\hat C^{(3)}_{\Phi L}\right),
  \label{eq:delta_ln_EFThNEW}
  \\
  \delta^{\BSM,\EFT}_h\M ={}&
  \frac{4\sw^2}{e^2} \biggl[
  \left(\hat C_{\Phi W}+\hat C^{(3)}_{\Phi L}\right)(k_+^2+k_-^2)
  - \hat C_{\Phi W}\Mh^2
  \nn\\
  & \qquad {}
  + \MW^2
  \biggl(\hat C_{\Phi\Box} +\frac{3-4\sw^2 }{4\sw^2} \hat C_{\Phi D}
  +\frac{3\cw}{\sw} \hat C_{\Phi WB} \biggr)
  \biggr] \M^{\SESM}_0
  \nn\\
  & {}
  +  2v_2 \hat C_{\Phi W}\, \frac{e^2{\cal A}_2}{\sw^2(k_+^2-\MW^2)(k_-^2-\MW^2)} \,.
  \label{eq:dMh4f_EFThNEW}
\end{align}
Note that in \eq{dMh4f_EFThNEW} also the diagrams shown in \fig{hWW_hardF}, which arise from the new fermionic operators contribute.}

\revision{%
We have checked by explicitly inserting the new Wilson coefficients
$\hat C_i$ given in \eqsm{CPhi_SMEFT}{CPhif_SMEFT}
that the predictions for the observables are still in agreement with the results given in
\refse{sec:pheno}.
Since the manipulations described in this section only concern hard-momentum contributions,
the (SMEFT) form of $\delta\L_{\eff}^{\tree}$ in \eq{LeffTreeSMEFT} relevant for the soft EFT contributions is not
affected by these manipulations.}

\subsection{\revision{Discussion}}
\label{subsec:discussion}

\revision{Before we compare our findings to related results in the literature,
we briefly discuss the most important aspects of the successful formulation of our EFT in SMEFT form.}

\begin{itemize}
\item
\revision{In hindsight, it might not be surprising that a theory with a heavy BSM sector
that does not directly couple to fermions, such as the SESM with massless fermions,
cannot be matched by bosonic SMEFT, but requires fermionic SMEFT operators as well.
The reason for this fact is the extensive use of EOMs to reduce the emerging EFT operators
to the standard Warsaw form, and the fact that the EOMs mix bosonic and fermionic operators.
In this sense, the occurrence of fermionic EFT operators can be viewed as an artefact of
the basis choice of SMEFT.
On the other hand, imagine a fit of SMEFT to future experimental data produces non-vanishing Wilson coefficients
for bosonic and fermionic operators, one might be tempted to assume that the BSM
sector of the underlying theory includes direct couplings to fermions. The considered
large-$\MH$ limit of the SESM with massless fermions shows that this conclusion
could be wrong.}

\item
\revision{While the non-SMEFT effective Lagrangian in \sec{final} is of HEFT type, and is thus naturally formulated in the non-linear Higgs realization, the SMEFT Lagrangian in  \subsec{SMEFTmatch} may also be translated to the linear Higgs realization, where all Higgs and Goldstone degrees of freedom are accommodated in an SU(2) doublet.
This transition, however, requires some care
owing to soft contributions which are, depending on the tadpole and renormalization schemes, encoded in Wilson coefficients that incorporate renormalization constants of BSM parameters.}

\revision{The EFT Lagrangian \refeq{eq:SMEFT}, which has been derived in the non-linear Higgs
realization, can be directly used in the linear Higgs realization as well, as long as
the renormalization scheme of the EFT loop calculation is in line with
the scheme used to define the SESM renormalization constant $\delta\sa$
appearing in $\hat C_{\Phi\Box}$, as given in \eq{CPhiBox_SMEFT}.
We recall that we have derived the EFT using the PRTS in the non-linear Higgs realization,
i.e.\ the (soft and hard contributions of the) tadpole constants have been evaluated using
the ``on-shell-like'' tadpole renormalization condition $t_\PH=t_\Ph=0$ on the
renormalized tadpole parameters. This fixes the renormalization constants
$\de t_{\PH/\Ph}$ by the values of the loop contributions to the Higgs one-point functions
according to $\de t_{\PH/\Ph}=-T^{\hat H/\hat h}_{\mathrm{nl}}$, where we have added
the suffix ``nl'' here to emphasize the use of the non-linear Higgs realization.
By design, the GIVS tadpole treatment~\cite{Dittmaier:2022maf,Dittmaier:2022ivi}
makes use of exactly this tadpole renormalization
condition in any Higgs realization, i.e.\ (as already explained in
\subsec{tadpoleren} in detail) PRTS and GIVS are identical in the non-linear Higgs realization.
In the linear Higgs realization, the part that potentially influences physical amplitudes
in the GIVS is exactly the contribution $-T^{\hat H/\hat h}_{\mathrm{nl}}$, as given as
$\de t_{\Ph/\PH,1}^\GIVS$ in \eq{dtGIVS1}, while the second part $\de t_{\Ph/\PH,2}^\GIVS$
introduced via some field shift cannot leave any effect in physical predictions.
The EFT Lagrangian \refeq{eq:SMEFT} with Wilson coefficients
\eqsm{CPhi_SMEFT}{CPhif_SMEFT} can, thus, be directly used with $\de\sa$ as given in
\eq{dsa} if the tadpole contributions in the EFT renormalization are evaluated
in the GIVS,
independently of the Higgs realization.}

\revision{Any other tadpole scheme may be used in the EFT renormalization as well
as long as $\de\sa$ is adjusted accordingly. Since we have imposed an OS condition on the
renormalized parameter $\sa$, a change in the tadpole scheme cannot lead to a
change in NLO predictions for observables as function of $\sa$; since the physical meaning
of $\sa$ is the same, not even a conversion of the input for the
OS-renormalized $\sa$ is needed.
To illustrate the change of tadpole treatment in the EFT renormalization, we give
the shift in $\de\sa$ that corresponds to the transition from the GIVS to the PRTS
in the linear Higgs realization,
  \begin{align}
    \delta s_{\alpha,\mathrm{lin,PRTS}}^{\OS} = \delta s_{\alpha}^{\OS}
    + \frac{\sa}{16\pi^2 (D-2)v_2^2}
    \left[2\MW^2 B_0(0,\MW^2,\MW^2) + \MZ^2 B_0(0,\MZ^2,\MZ^2)\right],
    \label{eq:dsalinPRTS}
  \end{align}
where $\delta \sa^{\OS}$ is given in \eq{dsa} and
``lin,PRTS'' indicates the use of the PRTS in the linear Higgs realization.%
\footnote{\revision{This change in $\de\sa$ exactly corresponds to the large-$\MH$ limit of the
shift given in Sec.~3.1.2 of \citere{Dittmaier:2022ivi} for the contribution of the
$h$-tadpole.}}
Since there is no $H$~field in the EFT anymore, the corresponding tadpole scheme for $H$
cannot be changed in the EFT retroactively.
This fact does not pose any problem, because the tadpole treatment cannot lead to any
measurable effect as long as $\sa$ is renormalized in the OS scheme.
We have successfully checked that our EFT predictions for
the Higgs decay $\Ph\to \PW\PW\to\nu_{\Pe}\Pe^+\mu^-\bar\nu_\mu$, as given in \subsec{hdecay},
do not change if we switch to the linear Higgs realization using the GIVS or PRTS
with $\de\sa$ treated as explained above.
}

\item
\revision{
We have just seen that the soft ``pollution'' via $\de\sa$ in the Wilson coefficient $\hat C_{\Phi\Box}$ in general requires adjustments of the latter, when switching the tadpole scheme or the Higgs realization in the EFT.
Analogously, the Wilson coefficient $C_{\Phi\Box}$ of the non-SMEFT effective Lagrangian \eq{SMEFTHEFT} depends on the tadpole scheme used for corresponding EFT calculations.
In order to avoid such dependencies one has to choose a renormalization and tadpole scheme for the BSM parameters that allows for a strict separation (``factorization'') of hard and soft scales in the EFT. This is also required for a consistent resummation of large logarithms $\sim \ln(\MH/\Mh)$, which in SMEFT may be carried out at the leading logarithmic level using the known one-loop anomalous dimension matrix of the dimension-six Wilson coefficients~\cite{Grojean:2013kd,Elias-Miro:2013gya,Elias-Miro:2013mua,Jenkins:2013zja,Jenkins:2013wua,Alonso:2013hga,Alonso:2014zka}.
For the case of our SESM EFT, this means that we need a scheme, where $\de \sa$ exclusively receives contributions from hard momentum modes.
Note that the $\MSbar$ scheme for $\sa$ in combination with PRTS tadpole renormalization fails in this respect, because the light and heavy Higgs tadpoles have divergent soft parts that contribute to $\de s_{\alpha, \PRTS}^\MSbar$ (in both Higgs realizations).
Another naive candidate would be the combination of the FJTS with the $\MSbar$ scheme for $\sa$.
As demonstrated in \rcite{Dittmaier:2021fls} this choice, however, leads to a breaking of the EFT power counting, caused by $\de s_{\alpha, \FJTS}^\MSbar / \sa \sim \zeta$, and thus to an invalidation of the EFT.
A feasible solution would be a hybrid scheme, where the tadpoles with a heavy Higgs loop, which is purely hard, are absorbed by renormalization constants in the PRTS and all other tadpoles, i.e.\ the ones with soft loops, are treated in the FJTS.
In that case we have $\de s_{\alpha,\mathrm{hybrid}}^\MSbar \sim \sa$ as required by a consistent power counting and at the same time all contributions to
$\de s_{\alpha, \mathrm{hybrid}}^\MSbar$, and thus to the Wilson coefficients, are purely hard.
A concrete implementation of such a hybrid scheme as well as the resummation of large logarithms is beyond the scope of this paper and left to future work.
}

\item
\revision{As discussed in more detail in the next subsection,
the possibility of matching SMEFT to the SESM in the considered large-$\MH$ limit
was already pointed out in the literature before~\cite{Haisch:2020ahr,Jiang:2018pbd,Cohen:2020xca}.
However, these analyses integrated out
the field ($h_1$) responsible for the heavy degree of freedom effectively in the unbroken phase
of the theory, i.e.\ the field that is integrated out is not aligned with the field
corresponding to the mass eigenstate of the heavy particle. By contrast, we have integrated out the field corresponding to the mass eigenstate in the broken phase of the theory.}

\revision{At tree level the equivalence of these two approaches, i.e.\ working in the broken or unbroken phase, is evident. Instead of solving and applying the EOM of the soft heavy Higgs field $\tilde{H}_s \sim \zeta^{-1}$, as we have done in \subsec{Lefftree}, one can just as well proceed analogously with $\tilde{h}_{1s} \sim \zeta^{-1}$.
The physical content of the resulting Lagrangian $\L_\eff^\tree$ is the same. In the latter case, however, the SU(2) doublet structure of the SESM Lagrangian in \eq{SESM} remains manifestly intact.
Given the decoupling property, $\L_\eff^\tree$ must therefore be of SMEFT type, which we have confirmed (upon redefining $h$ and $\Mh$) in our approach
in \subsec{nonSMEFTLeff}.}

\revision{At the one-loop level one may construct a similar line of arguments. The situation is,
however, more intricate than at tree level, since the hard scalar modes now \textit{propagate} in the loops.
Working in the $\{h_{1h}, h_{2h}\}$ basis as opposed to our mass-eigenstate basis, this means that an infinite number of loops with two-point vertex insertions ($\propto \lambda_{12}$) turning $h_1$ into $h_2$ and vice versa contribute in the matching, unless the two-point mixing vertex is parametrically suppressed.
Performing the matching in the unbroken phase may thus seem impractical.
For our scenario, however, a perturbative treatment of the mixing (within diagrammatic or functional matching) can be justified by the following power-counting argument:}

\revision{
In the hard momentum region the partial derivatives in \eq{SESM} scale like $\partial_\mu \sim \zeta$.
Consistently counting $\rd x^4 \sim \zeta^{-4}$ and demanding that the action scales like $\int \rd x^4 \L^\SESM_\mathrm{Higgs} \sim \zeta^0$ we have $h_{1h} \sim h_{2h} \sim \zeta$.
Assuming the parameter scaling of \eq{EFTlimit2}, we find that the term $\propto \lambda_{12} v_1 v_2 h_1 h_2$ encoded in \eq{SESM}, which causes the Higgs mixing, is  indeed suppressed by $\zeta^{-1}$ compared to the leading kinetic terms.
It may therefore be dropped in the leading hard one-loop contributions to the Wilson coefficients, which for our scenario first appear at $\ord(\zeta^{-2})$.
Beyond that order in the EFT expansion (a finite number of) insertions of the mixing vertex in hard loop diagrams generally have to be incorporated.
Since integrating out $h_{1h}$ once again preserves the doublet $\hat{\Phi}$, we may conclude that the emerging EFT should be of SMEFT type, now with a Higgs field $h_{2(s)}$.
Note that,
depending on whether the matching is performed in the broken or unbroken phase the role of the interpolating field that generates the physical Higgs state in the LSZ formalism is played by $h$ or $h_2$, respectively.
Nevertheless, the corresponding effective Lagrangians  should be physically equivalent and therefore be related by field redefinitions.
In this way, one may argue that the EFT emerging in our decoupling limit of the SESM must be of SMEFT type, which we have eventually verified in \subsec{SMEFTmatch}.
}

\revision{
To the best of our knowledge, this line of reasoning has so far not been spelled out explicitly in the literature. We are also not aware of any other (more) rigorous proof of the equivalence of the two approaches.
In fact, we only came up with this explanation after realizing that our result for the effective Lagrangian derived in the broken phase is equivalent to results in the literature derived in the unbroken phase
(up to renormalization-scheme-dependent terms, see below).
Our derivation therefore serves as a confirmation (or at least a strong evidence) that the two approaches indeed yield equivalent results and the arguments outlined above hold.
In the next subsection we explicitly show that our SMEFT-type effective Lagrangian  indeed agrees with some results in the literature obtained by integrating out (the equivalent of) $h_{1h}$ rather than $H_{h}$ (up to renormalization/tadpole scheme differences).}

\end{itemize}

\subsection{Comparison to the literature}
\label{subsec:Comparison}

\begin{itemize}
\item
Comparison with
Boggia, Gomez-Ambrosio, and Passarino~\cite{Boggia:2016asg}:
\\
\rcite{Boggia:2016asg} starts from the same full theory as we do:
the SESM with $\mathbb{Z}_2$ symmetry concerning sign change of the new
real scalar singlet field.
In contrast to us, they, however, work in the linear realization
of the Higgs sector.
The Higgs mass eigenstates are thus linear combinations of the new singlet and part of the original SU(2) doublet.
The heavy Higgs field is integrated out using a functional approach.
Their
explicit parametrization of the BSM sector
(using $\MH,\lambda_{12},\lambda_1$ as free parameters)
is slightly different from ours, but the large-$\MH$ limit considered there is equivalent to our limit defined in \eq{EFTlimit}, since all parameters of the Lagrangian have the
same scaling as given in \eq{EFTlimit2}.
As we do in this paper, the authors of \citere{Boggia:2016asg} emphasize that
it is important to specify the scaling behaviour of all free parameters of the model in connection
with the large-mass limit.
They also state that the functional integral should in general be
solved for the fields corresponding to the heavy mass eigenstates (rather than for gauge eigenstates)
in order to reproduce the correct EFT in the presence of Higgs mixing in the
underlying full theory.

Although the EFT presented in \rcite{Boggia:2016asg} should be compatible with our EFT,
the explicit results for the effective Lagrangians look rather different owing to the difference in the Higgs parametrization.
As also discussed in \citere{Boggia:2016asg}, the EFT Lagrangian given there is not
$\mathrm{SU(2)\times U(1)}$ invariant, since the light Higgs field~$h$ of the EFT
Lagrangian does not combine with the Goldstone fields to a proper SU(2) doublet.
The set consisting of the field~$h$ and the Goldstone fields does not even close under gauge transformations.
\revision{By contrast, our EFT Lagrangians \eqref{eq:SMEFTHEFT} and
\eqref{eq:SMEFT} are fully gauge invariant.
In Lagrangian \eqref{eq:SMEFTHEFT}}
the Goldstone fields transform within a well-defined (though non-linear) realization of SU(2), and
$h$ is a singlet field.
\revision{In Lagrangian \eqref{eq:SMEFT}, which is of SMEFT form, the Higgs and Goldstone fields
together form the usual SU(2) doublet.}
Connected to its gauge non-invariance, the EFT Lagrangian given in the appendix of
\citere{Boggia:2016asg} is rather lengthy and
depends on the gauge-fixing procedure, which also leads to the appearance of
Faddeev--Popov ghost fields in the EFT Lagrangian.
Owing to the gauge invariance of our EFT
\revision{Lagrangians,}
no dependence on the
gauge-fixing procedure remains there. Notably, the gauge-invariance properties
of the BFM allowed us to return to the conventional formalism in the final results for the EFT Lagrangians.

The EFT Lagrangian of \citere{Boggia:2016asg} is obviously not of SMEFT form,
but the question whether it could be brought into SMEFT form with appropriate field
redefinitions is not addressed there.
Since the large-mass limits of \citere{Boggia:2016asg} and of this paper are equivalent,
\revision{we give two answers to this question:
firstly, the non-existence of a purely bosonic EFT Lagrangian of SMEFT form,
as proven in \subsec{SMEFTbreakdown} without reference to any parametrization of the Higgs sector,
and secondly, as shown in \subsec{SMEFTmatch},
the possibility to transform the EFT Lagrangian into SMEFT form,
where both bosonic and fermionic operators appear.}
In practice, the application of the
\revision{very different}
(but physically equivalent)
Lagrangians of \citere{Boggia:2016asg} and of this work differs
in many subcontributions to physical predictions, involving even different types of Feynman graphs.
For instance, the EFT Lagrangian of \citere{Boggia:2016asg} leads to graphs
with EFT couplings and Faddeev--Popov ghosts that are not present in our approach;
on the other hand, the use of the non-linear Higgs realization in our EFT
\revision{Lagrangian \eqref{eq:SMEFTHEFT} of \sec{final}}
leads to non-polynomial structures with Goldstone fields and thus to
vertices with arbitrarily many Goldstone legs.%
\footnote{For a given process, of course, only finitely many graphs contribute.}
Even graphs that look the same in the two approaches in general lead to
different contributions owing to differences in the Feynman rules.

\item
Comparison with UOLEA and related results~\cite{Ellis:2017jns,Jiang:2018pbd,Haisch:2020ahr}:\\
In these papers, the more general version of the SESM with a real singlet Higgs field, which does not feature a $\mathbb{Z}_2$ symmetry, is investigated.
\revision{
Nevertheless, the results of \rcite{Haisch:2020ahr}, which employs the same notation as \rcites{Ellis:2017jns,Jiang:2018pbd}, can be compared to ours, by identifying the fundamental parameters of the respective Higgs-sector Lagrangians
\begin{align}
	\left\{ \mu_h^2, \lambda, M^2, A, \mu, \lambda_\phi, \kappa \right\}_{\text{\cite{Haisch:2020ahr}}}
	\;\longleftrightarrow\;
	\left\{ \mu_2^2, \lambda_2, \mu_1^2, \lambda_1, \lambda_{12} \right\}_{[\text{this paper}]}.
\end{align}
Using \eqs{originalparams}{EFTlimit} the explicit relations between the parameters read
\begin{align}\label{eq:HaischTransl}
	\mu_h^2&=\mu_2^2-\la_{12}v_1^2
	=\frac{\Mh^2+\sa^2\MH^2}{2}+\ord(\zeta^{-2}) \,,
	\nn\\
	\lambda&=\frac{\la_2}{2}+\frac{2\la_{12}v_1^2}{\mu_1^2-6\la_1v_1^2}
	=\frac{\Mh^2}{v_2^2}+\ord(\zeta^{-2}) \,,
	\nn\\
	M^2&=12\la_1v_1^2-2\mu_1^2
	=\MH^2+\ord(\zeta^0) \,,
	\nn\\
	A&=2\la_{12}v_1
	=\frac{\sa\MH^2}{v_2}+\ord(\zeta^{-1}) \,,
	\nn\\
	\mu&=14\la_1v_1
	=\frac{6\la_{12}v_2}{\sa}+\ord(\zeta^{-1}) \,,
	\nn\\
	\la_\phi&=14\la_1
	=\frac{12\la_{12}^2v_2^2}{\sa^2\MH^2}+\ord(\zeta^{-2}) \,,
	\nn\\
	\kappa&=2\la_{12} \,.
\end{align}
The field $\phi$ of \rcites{Ellis:2017jns,Jiang:2018pbd,Haisch:2020ahr}
plays the role of our $h_1$,
with an implicit tadpole term $t_\phi=2v_1(2\la_1v_1^2-\mu_1^2)$.
Note that \eq{HaischTransl} not only identifies parameters,
but also provides two additional constraints
that, together with the tadpole condition $t_\phi=0$, restore the $\mathbb{Z}_2$ symmetry.
From the rightmost part of \eq{HaischTransl}
we can conclude that our limit $\zeta\rightarrow\infty$
with $\MH\sim\zeta$ and $\sa\sim\zeta^{-1}$
corresponds to $M\sim A\sim\mu\rightarrow\infty$
which
is in fact the limit investigated in \rcites{Ellis:2017jns,Jiang:2018pbd,Haisch:2020ahr},
although not stated explicitly there.
However, these papers work in the unbroken phase
of the SU(2) symmetry (i.e.\ the SU(2) doublet is left intact),
thus not directly taking into account Higgs mixing effects
caused by a non-diagonal mass matrix.
Note that \rcite{Haisch:2020ahr} claims agreement of their effective Lagrangian with
\rcite{Ellis:2017jns,Jiang:2018pbd} after correcting some (omitted) terms in the latter
(\rcites{Anisha:2021hgc,Banerjee:2023qbg} confirm this result).
We therefore restrict our comparison to \rcite{Haisch:2020ahr}.}

\revision{Applying the translation~\eqref{eq:HaischTransl} to the final result of \rcite{Haisch:2020ahr},
we find agreement with our SMEFT result presented in \eq{CPhif_SMEFT},
with the exception of the one-loop Wilson coefficient $\hat C_{\Phi\Box}$.
This discrepancy may be due to differences in the input parameter
and renormalization scheme of the BSM sector and the tadpole treatment
in the full theory:
In \rcite{Haisch:2020ahr} the $\MSbar$ renormalization scheme is used for all parameters and tadpoles are treated in the FJTS,
while we use the OS scheme for the mixing angle
in combination with the GIVS for the tadpoles.%
\footnote{\revision{The renormalization of other BSM parameters does not contribute at this
order in the EFT expansion, and our effective Lagrangian does not depend on the renormalization of the SM-like parameters.}}
A more detailed comparison that accounts for this difference is intricate
due to the subtleties discussed in \subsec{discussion},
and beyond the scope of this paper.
The reason that we agree with \rcite{Haisch:2020ahr}, except for scheme differences, despite working in a different fields basis
and a different treatment of mixing effects
is laid out in the last bullet point of \subsec{discussion}.
}

\item
Comparison with Cohen, Craig, Lu, and Sutherland \cite{Cohen:2020xca}:\\
\rcite{Cohen:2020xca} specifies two cases in which SMEFT cannot be matched onto
an SM extension, so that HEFT is required:
(i) if a BSM particle gets all of its mass from the breaking of the
EW SU(2) symmetry, or
(ii) if the BSM model provides new sources of EW SU(2) breaking.
We comment on these two cases in the spirit of the present paper, where we have integrated out a field corresponding to a mass eigenstate.
Under the assumption of case~(i), the BSM particle in question is not heavy and should, thus, not be integrated out at the EW energy scale.
This case does therefore not apply to our setup.
\revision{Concerning case~(ii), it seems that the authors of \citere{Cohen:2020xca}
consider the criterion not fulfilled,
because the vev $v_2$, which generates the W-boson mass, does not receive a BSM contribution
in our EFT limit (in contrast to the light Higgs field $h$ for $\alpha\ne0$).
Otherwise criterion~(ii) would imply that our EFT should be incompatible with SMEFT, which is not the case.}

In \rcite{Cohen:2020xca} the same variant of the SESM as
in this paper (real singlet, $\mathbb{Z}_2$ symmetry) is considered as an example.
\revision{
Their parameters can be identified with ours
\begin{align}
	\left\{ \mu_H^2, \lambda_H, m^2, \lambda_S, \kappa \right\}_{\text{\cite{Cohen:2020xca}}}
	\;\longleftrightarrow\;
	\left\{ \mu_2^2, \lambda_2, \mu_1^2, \lambda_1, \lambda_{12} \right\}_{[\text{this paper}]},
\end{align}
via the relations
\begin{align}\label{eq:CohenTransl}
	\mu_H^2&=\mu_2^2
	=\frac{\sa^2\MH^4}{4\la_{12}v_2^2}+\ord(\zeta^0) \,,
	\nn\\
	\la_H&=\frac{\la_2}{4}
	=\frac{\Mh^2+\sa^2\MH^2}{2v_2^2}+\ord(\zeta^{-2}) \,,
	\nn\\
	m^2&=-2\mu_1^2
	=-\frac{\MH^2}{2}+\ord(\zeta^0) \,,
	\nn\\
	\la_S&=4\la_1
	=\frac{2\la_{12}^2v_2^2}{\sa^2\MH^2}+\ord(\zeta^{-2}) \,,
	\nn\\
	\kappa&=2\la_{12} \,.
	%\nn\\
	%v_s&=v_1=\frac{\sa\MH^2}{2\la_{12}v_2}+\ord(\zeta^{-1}).
\end{align}
As in \rcites{Ellis:2017jns,Jiang:2018pbd,Haisch:2020ahr}, the
authors of \rcite{Cohen:2020xca} do not switch to a Higgs field basis
corresponding to mass eigenstates, but keep the full SU(2) doublet structure
of the SM-like Higgs field untouched while integrating out the
original scalar singlet field.
However, the exact limit in which the BSM field is integrated out
is not stated in \rcite{Cohen:2020xca},
since the authors aim to compare different scenarios
according to whether they produce SMEFT or HEFT as a low-energy EFT.
}

\revision{
The limit we consider clearly has $m\neq0$,
which is found in \rcite{Cohen:2020xca} to yield a low-energy EFT of SMEFT form,
an expectation which we confirm.
The tree-level result given for the effective Lagrangian in Eq.~(6.17)
of \rcite{Cohen:2020xca} agrees with ours upon redefinition of the SM-like parameters
\begin{align}
	\mu_H^2\rightarrow(\mu'_H)^2=\mu_H^2+\frac{\kappa m^2}{2\la_S}
	=\frac{\Mh^2}{2}+\ord(\zeta^{-2}),\qquad
	\la_H\rightarrow\la'_H=\la_H-\frac{\kappa^2}{4\la_S}
	=\frac{\Mh^2}{2v_2^2}+\ord(\zeta^{-2}).
\end{align}
Note, in particular, that now $(\mu'_H)^2 \sim \zeta^0$ as opposed to $\mu_H = \mu_2 \sim\zeta^2$ in \eq{CohenTransl}.
A comparison of the one-loop results does not seem meaningful,
since renormalization is not considered at all in \rcite{Cohen:2020xca}.
}

\item
Comparison with Buchalla, Cata, Celis, and Krause~\cite{Buchalla:2016bse}:\\
\rcite{Buchalla:2016bse} investigates a strong-coupling and a weak-coupling scenario
of the same SESM variant as considered in this paper as well as the transition between the two
scenarios.
To this end, non-linear and linear Higgs realizations are employed, and their interplay is
investigated.
The five original parameters parametrizing the Higgs sector in \rcite{Buchalla:2016bse} are related to our
Higgs parameters as follows,
\begin{align}
\left\{ \mu_1^2, \mu_2^2, \lambda_1, \lambda_2, \lambda_3 \right\}_{\text{\cite{Buchalla:2016bse}}}
\;\longleftrightarrow\;
\left\{ 2\mu_2^2, 4\mu_1^2, \lambda_2, 16\lambda_1, 4\lambda_{12} \right\}_{[\text{this paper}]},
\end{align}
which implies the following correspondence between the derived parameters of the Higgs sector,
\begin{align}
\left\{ v, v_s, m, M, \chi \right\}_{\text{\cite{Buchalla:2016bse}}}
\;\longleftrightarrow\;
\left\{ v_2, v_1, \Mh, \MH, \alpha \right\}_{[\text{this paper}]}.
\end{align}
The authors of \rcite{Buchalla:2016bse} control the strong- and weak-coupling scenarios by specifying
hierarchies among the auxiliary parameters
\begin{align}
r = \frac{m^2}{M^2} \,, \quad \xi = \frac{v^2}{v^2+v_s^2}\,, \quad \omega = \sin^2\chi \,.
\end{align}

The strong-coupling scenario is characterized by large $\lambda_i$ (but not too large
in order not to spoil perturbativity) and $m,v,v_s\ll M$ and leads to non-decoupling of the heavy Higgs field~$H$.
The emerging EFT Lagrangian is a variant of the ``Electroweak chiral Lagrangian''
investigated by the authors in earlier work in detail~\cite{Buchalla:2013rka}.
It parametrizes the SU(2) Higgs doublet non-linearly, includes arbitrarily high
powers in the light Higgs field~$h$, and is, thus, of HEFT type.
In \rcite{Buchalla:2016bse}, the full tree-level strong-coupling EFT Lagrangian is presented explicitly,
but does not have any direct relation to our work.

The weak-coupling scenario is characterized by
$\lambda_i=\ord(1)$ and $\Mh,v\ll\MH$ with $v_s \sim \MH$, which exactly
corresponds to our large-$\MH$ limit with $\lambda_i=\ord(1)$ and
$\alpha \sim \zeta^{-1}$.
The emerging EFT Lagrangian is worked out at tree level even up to $\ord(\MH^{-4})$,
confirming the SM Lagrangian at $\ord(\MH^0)$ (decoupling of~H)
and finding operators up to dimension six at $\ord(\MH^{-2})$
in line with their expectation that their weak-coupling limit leads to an EFT of
SMEFT type%
\footnote{The authors of \rcite{Dawson:2023oce} investigate the very same model and limit. They claim to confirm this result although they nominally match HEFT to the SESM.}.
\revision{Specifically, they keep the SU(2) doublet intact and integrate out
the singlet field~$S$, corresponding to $\sigma$ in our notation,
with the argument that mixing effects are subleading and appear as corrections
if the Higgs mass matrix is not diagonalized.
At tree level, the EFT Lagrangian of \rcite{Buchalla:2013rka}
and our EFT Lagrangian obviously agree with each other.
As far as loop corrections to the matching are concerned, the\ authors of \rcite{Buchalla:2016bse} seem to expect that they
do not affect the EFT type (SMEFT versus HEFT) identified at tree level.
We confirm this expectation by explicit calculation without making assumptions.
In their App.~B of \rcite{Buchalla:2016bse}, the authors
only consider pure Higgs self-interactions, ignoring the
sector of gauge and Goldstone bosons completely, and sketch the derivation of the
corresponding part of the one-loop effective Lagrangian.
In line with our results, the one-loop-induced pure Higgs self-interactions
are of polynomial form in the light Higgs field~$h$ with operators up to dimension six,
but this does not yet imply that the full EFT is of SMEFT type.
A definite conclusion requires the inclusion of the interactions of the Higgs
field with the gauge-boson and Goldstone fields.}

In their final result for the weak-coupling scenario,
the authors of \rcite{Buchalla:2013rka} specialize to the case of an
``approximate SO(5) symmetry'' in the Higgs sector, where
$\omega/\xi-1 \ll 1$  and which corresponds to a large-$\MH$ limit
that features a further suppression of the Higgs mixing angle~$\chi\equiv\alpha$,
i.e.\ $\alpha \sim \zeta^{-2}$ or even higher.
\revision{In this special case, according to our results of \sec{final}
and in line with the statements made in \rcite{Buchalla:2013rka},
the emerging EFT is of bosonic SMEFT type,
because the non-SMEFT part
of our bosonic EFT Lagrangian~\eqref{eq:SMEFTHEFT}
as well as the fermionic part of the SMEFT Lagrangian~\eqref{eq:SMEFT}
are proportional to $s_\alpha^2 \sim \alpha^2$, i.e.\
they vanish to $\ord(\MH^{-2})$.}

\end{itemize}

\section{Conclusions}
\label{sec:conclusions}

Effective field theories offer an ideal framework to tackle the problem
of identifying experimentally viable SM extensions with heavy new particles
and, thus, to guide possible steps towards comprehensive theories solving questions left open by the SM despite its tremendous phenomenological success.
Based on the SM gauge symmetry and particle content, two different types
of EFTs may emerge in large-mass limits of SM extensions:
SMEFT and HEFT.
The two EFT types differ in the role played by the Higgs boson.
In SMEFT it is part of a complex SU(2) doublet just like in the SM,
in HEFT it appears as a gauge singlet, and the gauge symmetry is realized in a non-linear fashion.
The question whether the appropriate EFT framework for certain classes of popular BSM theories
is SMEFT or the more general HEFT has,
despite some effort and recent progress
made in the literature, not been conclusively answered to full generality so far.

In this paper we investigate this issue in detail for a simple
example of a BSM theory with extended Higgs sector which is prototypical in the
sense that it shares some salient features with more comprehensive models.
Most notably, this concerns the feature of Higgs mixing, i.e., the existence of at least one
more CP-even, heavy Higgs state, whose quantum field is a linear combination of the Higgs component of the SM-type SU(2) doublet and some new Higgs field.
Concretely, we have derived the low-energy EFT of
the Singlet Extension of the SM (SESM)
defined in \subsec{SESM}, featuring
a new real Higgs singlet field respecting a $\mathbb{Z}_2$ symmetry.
Fermions are included in the massless approximation.
The field corresponding to the heavy Higgs boson~H with mass $\MH$
emerges upon rotation of the interaction eigenbasis of the Higgs field  by the mixing angle $\alpha$.
The couplings of the light, observed Higgs boson~h with other SM particles are generically
reduced by a factor of $\ca\equiv\cos\alpha$.
To be in line with existing Higgs couplings measurements, we assume
$\MH \gg \Mh$ and $\sa\equiv\sin\alpha \sim \Mh/\MH$,
where $\Mh$ ($\approx 125$\,GeV) is the mass of the SM-like Higgs boson~h.

For our calculation of the effective Lagrangian to $\ord(1/\MH^2)$ we have integrated out the heavy Higgs field at the one-loop level directly in the path integral employing the functional ``matching'' method detailed in \rcite{Dittmaier:2021fls}.
We have included all contributions from one-loop renormalization constants of the full SESM,
\revision{employing a renormalization scheme that is well suited for phenomenological analyses.}
In contrast to diagrammatic, or more generally bottom-up matching, the functional approach
does not require an ansatz for the operator basis to be matched on, i.e.,
functional matching is agnostic with respect to the type of the emerging EFT.
This is particularly useful in situations when it is not clear whether the
effective Lagrangian will be of SMEFT form or may only be accommodated in the
much more general HEFT framework.
\revision{We emphasize that we have integrated out the field corresponding
to the heavy mass eigenstate, without making any assumption whether mixing effects
can be included perturbatively in the calculation of the EFT Lagrangian.}
At $\ord(1/\MH^0)$ the BSM effects
\revision{obviously decouple} in the considered large-mass limit of the SESM, i.e.,
the effective Lagrangian coincides with the SM one, in line with experimental observations.
\revision{The EFT we have derived in this work}
includes a complete set of operators to $\ord(1/\MH^2)$
as well as their matching (Wilson) coefficients to $\ord(e^2)$.
\revision{We have given two equivalent versions of the effective Lagrangian, one in SMEFT form
and another one that is (apart from the leading SM part) purely bosonic and
necessarily contains non-SMEFT operators.
Both Lagrangians are}
ready for phenomenological applications at current collider energies;
we have discussed some interesting examples and used them to validate our results.

We emphasize that for the consistent derivation of the EFT it is crucial to precisely specify the low-energy or, equivalently, large-mass limit, where the EFT is supposed to be valid.
In particular, we have made explicit that it depends on the assumed scaling of the mixing
angle~$\alpha$ (or equivalently $\sa$)
in the large-$\MH$ limit whether the effective Lagrangian, to a given order in $\MH^{-1}$, is of (bosonic) SMEFT form or not.
By choosing the limit $\MH \to \infty$ with $\MH \sa$ fixed we have ensured decoupling, as opposed to keeping $\sa$ fixed,
and put otherwise minimal constraints on the BSM parameters of the SESM.
We have shown that in this limit the physics of the SESM with vanishing fermion masses cannot be fully reproduced to $\ord(1/\MH^2)$ by SMEFT operators without fermion or gluon fields, independent of the choice of operator basis.
Further parametric suppression of the mixing angle by assuming
$\sa \sim (\Mh/\MH)^{a}$
with integer $a>1$ leads to an $\ord(1/\MH^2)$ EFT Lagrangian of SMEFT form
\revision{with bosonic EFT operators only,}
while non-SMEFT and/or fermionic operators will occur at higher orders in the large-$\MH$ expansion for any non-zero value of $\sa$.
\revision{As an alternative to the completely bosonic EFT Lagrangian, we have transformed
it into the usual SMEFT form, which required us to allow for fermionic EFT
operators as well, although the BSM part of the SESM with massless fermions does not
couple to fermions directly.
Technically, the transformation to SMEFT form involves field transformations that
are non-polynomial in the light Higgs field, rendering it difficult to find this
transformation algorithmically
by making a suitable ansatz.}

\revision{We have validated the two effective Lagrangians}
in \sec{pheno}
by computing several physical observables at NLO both in the EFT and the full SESM and comparing the results.
In all cases we observe the expected convergence behaviour of the EFT towards the full prediction for increasing $\MH$ with fixed $\MH \sa$.
To the best of our knowledge this is the most comprehensive phenomenological NLO analysis of such kind for a SMEFT/HEFT-type EFT.
For the discussed set of observables and massless fermions we have finally proven in
\subsec{SMEFTbreakdown} that it is impossible to reproduce the correct asymptotic behaviour
to $\ord(1/\MH^2)$ in the large-mass expansion by an effective Lagrangian exclusively consisting of SMEFT operators without fermions or gluons.
\revision{Such a procedure is helpful in a situation where it is not yet clear whether
the EFT can be transformed into SMEFT form or not.
We expect that qualitatively much of our study within the SESM carries over to other}
EFTs of BSM theories with extended Higgs sectors, which naturally feature heavy
Higgs states arising from mixing of
BSM Higgs fields and the SM-type Higgs field (with not too small mixing angle).
Even if SMEFT finally
may prove too restrictive, given the expected experimental uncertainties, we anticipate that SMEFT will still be a useful tool in the classification of future new physics signals in collider data.
One should, however, be aware of its limitations when it comes to precision fits and conceptual conclusions.

In summary, the conceptual and technical key features of the presented work are:
\begin{itemize}
  \item The heavy Higgs field corresponding to the heavy Higgs mass eigenstate of the SESM is integrated out, in contrast to many studies in the literature that are based on integrating out the original BSM Higgs singlet field before mixing.
  \revision{We nevertheless agree with some of their results, in particular those of  \rcite{Haisch:2020ahr}, up to differences that are apparently due to the chosen renormalization schemes. This agreement justifies the perturbative treatment of the Higgs mixing,
  which is an essential requirement implicitly assumed in their approach.}

  \item The large-mass (EFT) limit is defined such that the mixing angle~$\alpha$,
  and therefore $\sa$, scales like $\sa \sim \Mh/\MH$.
  This is the weakest scaling condition on $\alpha$ that still
  maintains decoupling of the heavy Higgs boson and, thus, represents
  the phenomenologically most relevant large-mass limit.

  \item Gauge invariance of the effective Lagrangian is ensured by employing a
non-linear realization of the SU(2) Higgs doublet of the SESM.
Using the BFM in the procedure of functional matching, does not only make
the separation between tree and loop effects of heavy particles most transparent.
The invariance of the Lagrangian under gauge transformations of the background fields also
offers the nice possibility to choose any convenient background gauge in intermediate
steps.
It is straightforward to restore background gauge invariance after deriving the effective Lagrangian, and
return to the conventional quantization formalism
at the end of the calculation.

  \item The employed functional matching approach is agnostic with respect to
the operator basis of the emerging EFT.

  \item The EFT is matched to the fully renormalized SESM at the one-loop level. The contributions from the SESM counterterms are crucial for phenomenological applications of the EFT.

  \item Our result for the effective Lagrangian is validated by comparing NLO predictions for several physical observables in the EFT and the full SESM and checking their asymptotic convergence
to the required level of $\ord(1/\MH^2)$.

  \item A proof is given that any EFT reproducing the SESM for
  $\MH \gg \Mh$, $\sa\sim \Mh/\MH$, and vanishing fermion masses necessarily
  contains $\ord(1/\MH^2)$ operators that cannot be accommodated in a SMEFT basis with EW bosonic fields only.
  Dimension-six SMEFT operators involving fermions or
\revision{bosonic} non-SMEFT operators (like in HEFT) must be included at $\ord(1/\MH^2)$.
  In this context, we stress that the functional approach to one-loop matching does not directly generate any fermionic
  %or gluonic
  BSM operators at $\ord(1/\MH^2)$,
  as long as the fermions are massless.

\end{itemize}

A straightforward extension of our calculation is to include the fermionic sector of the
SESM with non-zero fermion masses in the derivation of the effective Lagrangian in our large-$\MH$ limit.
It would also be interesting to apply our calculational setup to more comprehensive
BSM theories with extended Higgs sectors, such as Two-Higgs-Doublet Models,
models with even higher Higgs multiplets, or models with fermion-number violation.

\section*{Acknowledgments}
\revision{We are grateful to Gerhard Buchalla for giving us a formal hint on the
possibility to match SMEFT to our EFT, which eventually lead to the results
presented in \subsec{SMEFTmatch} (which was not yet part of the first preprint version of this work).
Moreover,} S.D. thanks Giampiero Passarino for fruitful discussions.
We acknowledge support by the German Research Foundation (DFG)
via Research Training Group RTG~2044 and DFG grant DI~784/8-1.

\bigskip
\appendix

\section*{Appendix}
\addcontentsline{toc}{section}{Appendix}

\section{Equation of motion of the heavy Higgs field}
\label{app:EOM}
As explained in \subsec{Lefftree}, the EOM of the
heavy Higgs field can be solved for its soft mode $\tilde{H}_s$
order by order in $\zeta^{-1}$.
In \eq{EOMtree} the result has been given to $\ord(\zeta^{-3})$.
In order to obtain $\delta \L_\eff^\oneloop$
from the intermediate expression~\eqref{eq:Leff1loopBeforeEOM},
the corresponding EOM for the background field $\hat{H}_s$ is needed even to $\ord(\zeta^{-5})$, while the quantum part would only contribute beyond the one-loop level. We find
\begin{align}
  \label{eq:HEOMFull}
	\hat{H}_s={}&
	-\frac{s_{\alpha}}{2v_2}\hat{h}^2
	+\frac{s_{\alpha}(-2\lambda_{12}v_2^2-2\Mh^2+\MH^2s_{\alpha}^2)}{2\MH^2v_2}\hat{h}^2
	+\frac{s_{\alpha}(-2\lambda_{12}v_2^2-\Mh^2+\MH^2s_{\alpha}^2)}{2\MH^2v_2^2}\hat{h}^3
	\nn\\& {}
	-\frac{\lambda_{12}s_{\alpha}}{4\MH^2v_2}\hat{h}^4
	+\frac{s_{\alpha}}{\MH^2v_2}\left[
	\hat{h}\Box\hat{h}
	+(\partial^{\mu}\hat{h})^2\right]
	+\frac{e^2s_{\alpha}}{4\MH^2\sw^2}(v_2+\hat{h})(\hat{C}_{\mu}^a)^2
	\nn\\& {}
	+\frac{\Mh^2s_{\alpha}(\MH^2s_{\alpha}^2-3\lambda_{12}v_2^2)}{\MH^4v_2}\hat{h}^2
	\nn\\&
	+\frac{s_{\alpha}(-8\lambda_{12}^2v_2^4
    +15\Mh^2 \MH^2s_{\alpha}^2-28\lambda_{12} \Mh^2 v_2^2
    +10\lambda_{12}\MH^2s_{\alpha}^2v_2^2
    -3\MH^4s_{\alpha}^4)}{4\MH^4v_2^2}\hat{h}^3
	\nn\\& {}
	+\frac{s_{\alpha}(-24\lambda_{12}^2v_2^4
    +14\Mh^2\MH^2s_{\alpha}^2-38\lambda_{12}\Mh^2 v_2^2
    +22\lambda_{12}\MH^2s_{\alpha}^2v_2^2
    -5\MH^4s_{\alpha}^4)}{8\MH^4v_2^3}\hat{h}^4
	\nn\\& {}
	-\frac{\lambda_{12}s_{\alpha}(6\lambda_{12}v_2^2+4\Mh^2-3\MH^2s_{\alpha}^2)}{4\MH^4v_2^2}\hat{h}^5
	-\frac{\lambda_{12}^2s_{\alpha}}{4\MH^4v_2}\hat{h}^6
	\nn\\& {}
	+\frac{s_{\alpha}(2\lambda_{12}v_2^2+2\Mh^2-\MH^2s_{\alpha}^2)}{\MH^4v_2}\hat{h}\Box\hat{h}
	+\frac{s_{\alpha}(14\lambda_{12}v_2^2+3\Mh^2-7\MH^2s_{\alpha}^2)}{2\MH^4v_2^2}\hat{h}^2\Box\hat{h}
	+\frac{3\lambda_{12}s_{\alpha}}{\MH^4v_2}\hat{h}^3\Box\hat{h}
	\nn\\& {}
	+\frac{s_{\alpha}(10\lambda_{12}v_2^2+3\Mh^2-5\MH^2s_{\alpha}^2)}{\MH^4v_2^2}
	\hat{h}(\partial^{\mu}\hat{h})^2
	+\frac{5\lambda_{12}s_{\alpha}}{\MH^4v_2}\hat{h}^2(\partial^{\mu}\hat{h})^2
	\nn\\& {}
	+\frac{s_{\alpha}(2\lambda_{12}v_2^2+2\Mh^2-\MH^2s_{\alpha}^2)}{\MH^4v_2}(\partial^{\mu}\hat{h})^2
	\nn\\& {}
	+\frac{e^2s_{\alpha}}{8\MH^4\sw^2v_2}
	\big[4\lambda_{12}v_2(2v_2+\hat{h})-5\MH^2s_{\alpha}^2\big]\hat{h}(v_2+\hat{h})(\hat{C}_{\mu}^a)^2
	-\frac{e^2s_{\alpha}}{4\MH^4\sw^2}(\hat{C}_{\mu}^a)^2\Box\hat{h}
	\nn\\& {}
	-\frac{e^2s_{\alpha}}{\MH^4\sw^2}(\partial_{\mu}\hat{h})\hat{C}_{\nu}^a(\partial^{\mu}\hat{C}^{a,\nu})
	-\frac{e^2s_{\alpha}}{2\MH^4\sw^2}(v_2+\hat{h})(\partial^{\mu}\hat{C}_{\nu}^a)^2
	-\frac{e^2s_{\alpha}}{2\MH^4\sw^2}(v_2+\hat{h})\hat{C}_{\nu}^a\Box\hat{C}_{\nu}^a
	\nn\\& {}
	-\frac{2s_{\alpha}}{\MH^4v_2}(\partial^{\mu}\partial^{\nu}\hat{h})^2
	-\frac{4s_{\alpha}}{\MH^4v_2}(\partial_{\mu}\Box\hat{h})(\partial^{\mu}\hat{h})
	-\frac{s_{\alpha}}{\MH^4v_2}(\Box\hat{h})^2
	-\frac{s_{\alpha}}{\MH^4v_2}\hat{h}\Box\Box\hat{h}
	+\mathcal{O}(\zeta^{-7}),
\end{align}
where we dropped the $s$~labels of all (SM-like background) fields
used to indicate soft modes on the right-hand side for better readability.

\section{Operator building blocks}
\label{app:XD}

The following $\mathcal{D}^{-1}$  and $\mathcal{X}$ (differential) operators are needed for the evaluation of \eq{Chains}.
The explicit expressions are given (at least) up to  the order in the EFT power counting parameter $\zeta^{-1}$ required in this work. They read

\begin{align}
	(\mathcal{D}^{-1})^{\mu\nu}_{\barW_a\barW_b} ={}&
   \frac{\delta^{ab}}{p^2}\left[g^{\mu\nu}+(\xi-1)\frac{p^\mu p^\nu}{p^2}\right]
	+\mathcal{O}\bigl(\zeta^{-1}\bigr) \,,
	\\
  %%%%%%%%%%%%%%%%%%%%
  (\mathcal{D}^{-1})_{\varphi_a\varphi_b}={}&
  -\frac{\delta^{ab}}{p^2}\bigg(1+\frac{\hat{h}}{v_2}\bigg)^{\!\!-2}
  +\frac{2 \ri}{v_2 p^4}
  \bigg(1+\frac{\hat{h}}{v_2}\bigg)^{\!\!-3}
  \delta^{ab} p^\mu (\partial_\mu h)
  -\frac{2 \ri}{p^4}
  \bigg(1+\frac{\hat{h}}{v_2}\bigg)^{\!\!-2}
  \delta^{ab} p^\mu \partial_\mu
  \nn\\& {}
  -\frac{2 \ri g_1}{p^4} \bigg(1+\frac{\hat{h}}{v_2}\bigg)^{\!\!-2} \eps^{ab3}p^\mu \hat{B}_\mu
  +\frac{\ri g_2}{p^4} \bigg(1+\frac{\hat{h}}{v_2}\bigg)^{\!\!-2} \eps^{abc} p^\mu \hat{C}^c_\mu
  \nn\\& {}
  -\frac{\xi v_2^2}{4p^4}
  \bigg(1+\frac{\hat{h}}{v_2}\bigg)^{\!\!-4}
  (g_2^2\delta^{ab} + g_1^2 \delta^{a3}\delta^{b3})
  +\frac{\delta^{ab} \sa }{v_2 p^2}
  \bigg(1+\frac{\hat{h}}{v_2}\bigg)^{\!\!-3}
  (2 \hat{H}- \sa \hat{h})
  \nn\\& {}
  +\frac{g_1g_2}{2 p^4}
  \bigg(1+\frac{\hat{h}}{v_2}\bigg)^{\!\!-2}
  \hat{B}^\mu(\delta^{a3}\hat{W}^b_\mu + \delta^{b3}\hat{W}^a_\mu)
  \nn\\& {}
  + \frac{4g_1^2}{p^6}\bigg(1+\frac{\hat{h}}{v_2}\bigg)^{\!\!-2}
  \delta^{a3} \delta^{b3} \bigl(p_\mu \hat{B}^\mu \bigr)^2
  +\frac{g_2^2}{p^6} \bigg(1+\frac{\hat{h}}{v_2}\bigg)^{\!\!-2}
  p^\mu p^\nu \hat{C}^a_\mu \hat{C}^b_\nu
  \nn\\& {}
  -\frac{2g_1 g_2}{p^6}
  \bigg(1+\frac{\hat{h}}{v_2}\bigg)^{\!\!-2}
  \bigl(\delta^{a3} \hat{C}^b_\mu + \delta^{b3} \hat{C}^a_\mu \bigr)
  p^\mu p^\nu \hat{B}_\nu
  \nn\\& {}
  -\frac{2g_1}{p^6}
  \bigg(1+\frac{\hat{h}}{v_2}\bigg)^{\!\!-2}
  \eps^{ab3}  \hat{B}^\mu
  \bigl(p^2 \partial_\mu- 4 p_\mu p^\nu \partial_\nu \bigr)
  \nn\\& {}
  + \frac{g_2}{p^6}\bigg(1+\frac{\hat{h}}{v_2}\bigg)^{\!\!-2}
  \eps^{abc}  \hat{C}^c_\mu
  \bigl(p^2 \partial^\mu - 4 p^\mu p^\nu \partial_\nu  \bigr)
  \nn\\& {}
  +\frac{\delta^{ab}}{v_2^2 p^6}
  \bigg(1+\frac{\hat{h}}{v_2}\bigg)^{\!\!-4}
  \Bigl[ 8 (p^\mu\partial_\mu \hat{h})^2 - 2 p^2 (\partial^\mu \hat{h})(\partial_\mu \hat{h})
  \Bigr]
  \nn\\& {}
  +\frac{2 \delta^{ab}}{v_2 p^6}
  \bigg(1+\frac{\hat{h}}{v_2}\bigg)^{\!\!-3}
  \Bigl[ p^2 (\Box \hat{h})
  - 2 p^\mu p^\nu (\partial_\mu \partial_\nu \hat{h})
  +p^2 (\partial^\mu \hat{h})\partial_\mu
  -4 p^\mu p^\nu (\partial_\mu \hat{h}) \partial_\nu
  \Bigr]
  \nn\\& {}
  -\frac{\delta^{ab}}{p^6}
  \bigg(1+\frac{\hat{h}}{v_2}\bigg)^{\!\!-2}
  \Bigl[
  g_2^2 \bigl( p^\mu \hat{C}^c_\mu \bigr)^2
  -4 g_1 g_2 p^\mu p^\nu \hat{B}_\mu \hat{W}^3_\nu
  + g_1 g_2 p^2 \hat{B}^\mu \hat{W}^3_\mu
  \nn\\& {}\quad
  + p^2 \Box
  -4 (p^\mu \partial_\mu)^2
  \Bigr]
  +\mathcal{O} \bigl(\zeta^{-5}\bigr) \,,
\label{eq:InvDelphi}	\\
  %%%%%%%%%%%%%%%%%%%%
	(\mathcal{D}^{-1})_{hh}={}&
    -\frac{1}{p^2}
	-\frac{1}{p^4}\bigg[
	2\ri p_\mu\partial^\mu
    +\Box+Y
	-\frac{9\Mh^2s_{\alpha}^2}{2v_2}\hat{h}
	+\frac{s_{\alpha}(2\lambda_{12}v_2^2+2\Mh^2-\MH^2s_{\alpha}^2)}{v_2}\hat{H}
	\nn\\& {}
	+\frac{3s_{\alpha}^2(4\lambda_{12}v_2^2-3\Mh^2-2\MH^2s_{\alpha}^2)}{2v_2^2}\hat{h}^2
	+\frac{3s_{\alpha}(-2\lambda_{12}v_2^2+\Mh^2+\MH^2s_{\alpha}^2)}{v_2^2}\hat{H}\hat{h}
    +\lambda_{12}\hat{H}^2
	\nn\\& {}
	+\frac{g_2^2s_{\alpha}^2}{4}(\hat{C}_{\mu}^a)^2
	\bigg]
	-\frac{1}{p^6}\bigg[
	-4(p_\mu\partial^\mu)^2
	+2\ri p_\mu\bigr(\partial^\mu Y \bigl)
	+4\ri (\Box + Y) p_\mu\partial^\mu
	+ (\Box + Y)^2
	\bigg]
	\nn\\& {}
	-\frac{1}{p^8}\bigg[
	-8\ri(p_\mu\partial^\mu)^3
	-4p_\mu p_\nu \bigr(\partial^\mu\partial^\nu Y \bigl)
    -12 p_\mu p_\nu \bigr(\partial^\mu Y \bigl)\partial^\nu
	-12p_\mu p_\nu (\Box + Y) \partial^\mu\partial^\nu
	\bigg]
	\nn\\& {}
	-\frac{16}{p^{10}} (p_\mu\partial^\mu)^4
	+\mathcal{O} \bigl(\zeta^{-7}\bigr) \,,
\end{align}
with
\begin{equation}
  	Y = \Mh^2
  +\frac{3\Mh^2}{v_2}\hat{h}
  +\frac{3(\Mh^2+\MH^2s_{\alpha}^2)}{2v_2^2}\hat{h}^2
  +\frac{\MH^2s_{\alpha}}{v_2}\hat{H}
  -\frac{g_2^2}{4}(\hat{C}_{\mu}^a)^2 \,,
\end{equation}
and
\begin{align}
  %%%%%%%%%%%%%%%%%%%%
	\mathcal{X}_{Hh} =  \mathcal{X}_{hH} ={}&
	-\frac{\MH^2s_{\alpha}}{v_2} \hat{h}
	+\biggl(-2\lambda_{12}s_{\alpha}
	-\frac{2\Mh^2s_{\alpha}}{v_2^2}
	+\frac{\MH^2s_{\alpha}^3}{v_2^2}\biggr)v_2\hat{h}
	+ 2\biggl(2\lambda_{12}
	-\frac{\MH^2s_{\alpha}^2}{v_2^2}\biggr)v_2\hat{H}
	\nn\\& {}
	+3\biggl(\lambda_{12}s_{\alpha}
	-\frac{\Mh^2s_{\alpha}}{2v_2^2}
	-\frac{\MH^2s_{\alpha}^3}{2v_2^2}\biggr)\hat{h}^2
	-2\lambda_{12}\hat{h}\hat{H}
	+\frac{g_2^2 \sa}{4} (\hat{C}_{\mu}^a)^2
	\nn\\& {}
	+\frac{2s_{\alpha}\Mh^2}{v_2} \bigg(s_{\alpha}^2
	-\frac{3\lambda_{12}v_2^2}{\MH^2}\bigg)\hat{h}
	+\biggl(\frac{6\lambda_{12}\Mh^2}{\MH^2}-2\lambda_{12}s_{\alpha}^2
	-\frac{\Mh^2s_{\alpha}^2}{v_2^2}
	+\frac{\MH^2s_{\alpha}^4}{v_2^2}\biggr)v_2\hat{H}
	\nn\\& {}
	+3s_{\alpha}\bigg(\frac{2\lambda_{12}^2v_2^2}{\MH^2}
	-\frac{5\lambda_{12}s_{\alpha}^2}{2}
	+\frac{5\Mh^2s_{\alpha}^2}{4v_2^2}
	+\frac{3\MH^2s_{\alpha}^4}{4v_2^2}\bigg)\hat{h}^2
	\nn\\& {}
	+3\biggl(-\frac{4\lambda_{12}^2v_2^2}{\MH^2}
	+4\lambda_{12}s_{\alpha}^2
	-\frac{\Mh^2s_{\alpha}^2}{v_2^2}
	-\frac{\MH^2s_{\alpha}^4}{v_2^2}\biggr)\hat{h}\hat{H}
	+3\lambda_{12} \biggl(\frac{2\lambda_{12}v_2^2}{\MH^2s_{\alpha}}
	-s_{\alpha}\biggr)\hat{H}^2
	\nn\\&
	-\frac{g_2^2 \sa^3}{8} (\hat{C}_{\mu}^a)^2
	+\mathcal{O} \bigl(\zeta^{-5}\bigr)\,,
	\\
  %%%%%%%%%%%%%%%%%%%%
	\mathcal{X}^\mu_{H\barW_a} ={}&
	\mathcal{X}^\mu_{\barW_a H} +\mathcal{O}\bigl(\zeta^{-3}\bigr)=
    \frac{g_2^2}{2}v_2s_\alpha\biggl(1+\frac{\hat{h}}{v_2}\biggr)\!
    \biggl(\hat{C}^{a,\mu}+\frac{1-\cw}{\cw}\delta^{a3}\hat{C}^{3,\mu}\biggr)
    +\mathcal{O}\bigl(\zeta^{-3}\bigr) \,,
  \\
  %%%%%%%%%%%%%%%%%%%%
	\mathcal{X}^\mu_{h\barW_a} ={}&
    \mathcal{X}^\mu_{\barW_a h} +\mathcal{O}\bigl(\zeta^{-2}\bigr) =
	\frac{g_2^2}{2} v_2\biggl(1+\frac{\hat{h}}{v_2}\biggr)\!
    \biggl(\hat{C}^{a,\mu}+\frac{1-\cw}{\cw}\delta^{a3}\hat{C}^{3,\mu}\biggr)
    +\mathcal{O}\bigl(\zeta^{-2}\bigr) \,,
	\\
  %%%%%%%%%%%%%%%%%%%%
	\mathcal{X}_{H\varphi_a} ={}&
    g_2 s_\alpha\biggl(1+\frac{\hat{h}}{v_2}\biggr)
    \biggl[g_1\hat{B}^\mu \hat{W}^b_\mu\epsilon^{ab3}-\hat{C}^a_\mu\left(\partial^{\mu}+\ri p^\mu\right)\biggr]
\nn\\ & {}
    -\frac{g_2 \sa^3}{2v_2}\biggl(\frac{2\hat{H}}{\sa}-\hat{h}\biggr) \ri \hat{C}^a_\mu p^\mu
    +\mathcal{O}\bigl(\zeta^{-3}\bigr) \,,
	\\
  %%%%%%%%%%%%%%%%%%%%
	\mathcal{X}_{\varphi_aH} ={}&
    g_2 \sa\biggl(1+\frac{\hat{h}}{v_2}\biggr)
    \Bigl(g_1\hat{B}^\mu \hat{W}^b_\mu\epsilon^{ab3}
    + (\partial^\mu \hat{C}^a_\mu)
    + \hat{C}^a_\mu \partial^\mu
    \Bigr)
    + \frac{g_2 \sa}{v_2} \hat{C}^a_\mu (\partial^\mu \hat{h})
\nn\\
    &+g_2 \sa \biggl[1+\frac{\hat{h}}{v_2}+\frac{\sa^2}{2v_2}\bigg(\frac{2\hat{H}}{\sa}-\hat{h}\bigg)\biggr]\ri \hat{C}^a_\mu p^\mu
    +\mathcal{O}\bigl(\zeta^{-3}\bigr)  \,,
  \\
  %%%%%%%%%%%%%%%%%%%%
	\mathcal{X}_{h\varphi_a} ={}&
    g_2 \bigg(1+\frac{\hat{h}}{v_2} \biggr)
    \biggl[g_1\hat{B}^\mu \hat{W}^b_\mu\epsilon^{ab3}-\hat{C}^a_\mu\left(\partial^{\mu}+\ri p^\mu\right)\biggr]
\nn\\ & {}
    +\frac{g_2\sa^2}{2v_2}\biggl(v_2+2\hat{h}-\frac{2\hat{H}}{\sa}\biggr) \ri \hat{C}^a_\mu p^\mu
	+\mathcal{O}\bigl(\zeta^{-2}\bigr) \,,
	\\
  %%%%%%%%%%%%%%%%%%%%
	\mathcal{X}_{\varphi_a h} ={}&
    g_2 \biggl(1+\frac{\hat{h}}{v_2}\biggr)
    \Bigl(g_1\hat{B}^\mu \hat{W}^b_\mu\epsilon^{ab3}
    + (\partial^\mu \hat{C}^a_\mu)
    + \hat{C}^a_\mu \partial^\mu
    \Bigr)
    + \frac{g_2}{v_2}  \hat{C}^a_\mu (\partial^\mu \hat{h})
\nn\\
    &+g_2  \biggl[1+\frac{\hat{h}}{v_2}-\frac{\sa^2}{2v_2}\biggl(v_2+2\hat{h}-\frac{2\hat{H}}{\sa}\biggr) \!\biggr]\ri \hat{C}^a_\mu p^\mu
    +\mathcal{O}\bigl(\zeta^{-2}\bigr)  \,,
	\\
  %%%%%%%%%%%%%%%%%%%%
	\mathcal{X}_{\barW_a\varphi_b}^\mu ={}&
	-\frac{\ri g_2}{2}\hat{h}\biggl(2+\frac{\hat{h}}{v_2}\biggr)
  \biggl(\delta^{ab}+\frac{1-\cw}{\cw}\delta^{a3}\delta^{b3}\biggr)
    p^\mu +\mathcal{O}\bigl(\zeta^{0}\bigr)\,,
	\\
  %%%%%%%%%%%%%%%%%%%%
	\mathcal{X}_{\varphi_a\barW_b}^\mu ={}&+
	\frac{\ri g_2}{2}\hat{h}\biggl(2+\frac{\hat{h}}{v_2}\biggr)
  \biggl(\delta^{ab}+\frac{1-\cw}{\cw}\delta^{a3}\delta^{b3}\biggr)
    p^\mu + \mathcal{O}\bigl(\zeta^{0}\bigr)\,.
\end{align}
We note that
adding to $(\mathcal{D}^{-1})_{\varphi_a\varphi_b}$ an arbitrary function (as opposed to a differential operator) that is anti-symmetric under $a \leftrightarrow b$ and at the same time proportional to an
even power of $p^\mu$ (and thus scales like an even power of $\zeta$) leaves  \eq{Chains} unchanged.
In this sense, $(\mathcal{D}^{-1})_{\varphi_a\varphi_b}$ is not uniquely determined, and the expression we give in \eq{InvDelphi} contains
no such term.

\section{Expressions for the one-loop effective Lagrangian}
\label{app:Leff1loopbeforefieldredef}

Here we give the one-loop effective Lagrangian as it results directly from \eq{LeffPi}
after loop momentum integration.
This is the form \textit{before} applying the EOM of the heavy (background) Higgs field
$\hat{H} \equiv \hat{H}_s$:
\begin{align}
  \delta \L_\eff^\oneloop={}& \frac{I_{20}}{16\pi^2(D-2)} \Bigg\{
  \frac{\MH^2}{v_2^2}\bigg[\frac{2\MH^2s_{\alpha}^2}{v_2^2}-4\lambda_{12}\bigg]
  v_2^3\hat{h}
  +6\lambda_{12}\frac{\MH^2}{v_2^2}
  \frac{v_2^3\hat{H}}{s_{\alpha}}
  +\frac{\MH^2}{v_2^2}\bigg[\lambda_{12}
  +\frac{\MH^2s_{\alpha}^2}{v_2^2}\bigg]
  v_2^2\hat{h}^2
  \nn\\&
  %O(\zeta^0)
  +\bigg[-\frac{6\lambda_{12}\Mh^2}{v_2^2}
  +\frac{2\lambda_{12}\MH^2s_{\alpha}^2}{v_2^2}
  +\frac{\Mh^2\MH^2s_{\alpha}^2}{v_2^4}
  -\frac{\MH^4s_{\alpha}^4}{v_2^4}\bigg]
  v_2^3\hat{h}
  \nn\\&
  +\bigg[\frac{6\lambda_{12}\Mh^2}{v_2^2}
  -\frac{6\lambda_{12}\MH^2s_{\alpha}^2}{v_2^2}
  +\frac{3\MH^4s_{\alpha}^4}{v_2^4}\bigg]
  \frac{v_2^3\hat{H}}{s_{\alpha}}
  \nn\\&
  +\bigg[(4D-2)\lambda_{12}^2
  -\frac{2(2D-3)\lambda_{12}\MH^2s_{\alpha}^2}{v_2^2}
  +\frac{(2D-5)\MH^4s_{\alpha}^4}{2v_2^4}
  +\frac{13\Mh^2\MH^2s_{\alpha}^2}{2v_2^4}\bigg]
  v_2^2\hat{h}^2
  \nn\\&
  +\bigg[-12(D-1)\lambda_{12}^2
  +\frac{2(3D-7)\lambda_{12}\MH^2s_{\alpha}^2}{v_2^2}
  +\frac{4\MH^4s_{\alpha}^4}{v_2^4}\bigg]
  \frac{v_2^2\hat{h}\hat{H}}{s_{\alpha}}
  +3(3D-4)\lambda_{12}^2
  \frac{v_2^2\hat{H}^2}{s_{\alpha}^2}
  \nn\\&
  +\bigg[-(2D-4)\lambda_{12}^2
  -\frac{D\lambda_{12}\MH^2s_{\alpha}^2}{v_2^2}
  +\frac{(D-1)\MH^4s_{\alpha}^4}{v_2^4}
  +\frac{6\Mh^2\MH^2s_{\alpha}^2}{v_2^4}\bigg]
  v_2\hat{h}^3
  \nn\\&
  +\bigg[3(D-2)\lambda_{12}^2
  +\frac{(3D-8)\lambda_{12}\MH^2s_{\alpha}^2}{v_2^2}
  +\frac{\MH^4s_{\alpha}^4}{v_2^4}\bigg]
  \frac{v_2\hat{h}^2\hat{H}}{s_{\alpha}}
  \nn\\&
  +\bigg[\frac{1}{4}(D-2)\lambda_{12}^2
  +\frac{(D-4)\lambda_{12}\MH^2s_{\alpha}^2}{2v_2^2}
  +\frac{D\MH^4s_{\alpha}^4}{4v_2^4}
  +\frac{3\Mh^2\MH^2s_{\alpha}^2}{2v_2^4}\bigg]
  \hat{h}^4
  \nn\\&
  +\frac{(D-4)\MH^2s_{\alpha}^2}{Dv_2^2}
  \hat{h}\Box\hat{h}
  -\frac{(D-4)e^2\MH^2s_{\alpha}^2}{4D\sw^2v_2^2}
  (v_2 + \hat{h})^2(\hat{C}_{\mu}^a)^2
  \nn\\&
  %O(\zeta^2)
  +\bigg[-\frac{6\lambda_{12}\Mh^4}{\MH^2v_2^2}
  +\frac{3\lambda_{12}\Mh^2s_{\alpha}^2}{v_2^2}
  -\frac{\Mh^2\MH^2s_{\alpha}^4}{2v_2^4}
  +\frac{\lambda_{12}\MH^2s_{\alpha}^4}{2v_2^2}
  -\frac{\MH^4s_{\alpha}^6}{4v_2^4}\bigg]
  v_2^3\hat{h}
  \nn\\&
  +\bigg[\frac{6\lambda_{12}\Mh^4}{\MH^2v_2^2}
  -\frac{6\lambda_{12}\Mh^2s_{\alpha}^2}{v_2^2}\bigg]
  \frac{v_2^3\hat{H}}{s_{\alpha}}
  \nn\\&
  +\bigg[\frac{12(D-1)\lambda_{12}^2\Mh^2}{\MH^2}
  -(4D-6)\lambda_{12}^2s_{\alpha}^2
  +\frac{(4D-6)\lambda_{12}\MH^2s_{\alpha}^4}{v_2^2}
  -\frac{8(D-5)\lambda_{12}\Mh^2s_{\alpha}^2}{v_2^2}
  \nn\\&\quad
  +\frac{(D-15)\Mh^2\MH^2s_{\alpha}^4}{v_2^4}
  -\frac{(2D-3)\MH^4s_{\alpha}^6}{2v_2^4}
  +\frac{9\Mh^4s_{\alpha}^2}{v_2^4}\bigg]
  v_2^2\hat{h}^2
  \nn\\&
  +\bigg[-\frac{6(5D-6) \lambda_{12}^2 \Mh^2}{\MH^2}
  +\lambda_{12}^2(18D-34)s_{\alpha}^2
  +\frac{9(D-6)\lambda_{12}\Mh^2s_{\alpha}^2}{v_2^2}
  \nn\\&\quad
  -\frac{5(3D-7)\lambda_{12}\MH^2s_{\alpha}^4}{v_2^2}
  +\frac{3(D-3)\MH^4s_{\alpha}^6}{v_2^4}
  +\frac{17\Mh^2\MH^2s_{\alpha}^4}{v_2^4}\bigg]
  \frac{v_2^2\hat{h}\hat{H}}{s_{\alpha}}
  \nn\\&
  +\bigg[\frac{6(3D-4)\lambda_{12}^2\Mh^2}{\MH^2}
  -2(9D-20)\lambda_{12}^2s_{\alpha}^2
  +\frac{(9D-28)\lambda_{12}\MH^2s_{\alpha}^4}{v_2^2}
  +\frac{4\MH^4s_{\alpha}^6}{v_2^4}\bigg]
  \frac{v_2^2\hat{H}^2}{s_{\alpha}^2}
  \nn\\&
  +\bigg[-\frac{4(D-2)(2D+1)\lambda_{12}^3v_2^2}{3\MH^2}
  -\frac{(3D-6)\lambda_{12}^2\Mh^2}{\MH^2}
  -\frac{(31D-122)\lambda_{12}\Mh^2s_{\alpha}^2}{2v_2^2}
  \nn\\&\quad
  +\frac{(7D-39)\Mh^2\MH^2s_{\alpha}^4}{v_2^4}+(4D^2-13D+2)\lambda_{12}^2s_{\alpha}^2
  -\frac{(4D^2-23D+12)\lambda_{12}\MH^2s_{\alpha}^4}{2v_2^2}
  \nn\\&\quad
  +\frac{(2D^2-18D+13)\MH^4s_{\alpha}^6}{6v_2^4}
  +\frac{27\Mh^4s_{\alpha}^2}{v_2^4}\bigg]
  v_2\hat{h}^3
  \nn\\&
  +\bigg[-(12D^2-55D+62)\lambda_{12}^2s_{\alpha}^2
  +\frac{(3D^2-31D+54)\lambda_{12}\MH^2s_{\alpha}^4}{v_2^2}
  \nn\\&\quad
  +\frac{6(D-2)(2D-1)\lambda_{12}^3v_2^2}{\MH^2}
  +\frac{3(D-2)\lambda_{12}^2\Mh^2}{\MH^2}
  +\frac{3(15D-74)\lambda_{12}\Mh^2s_{\alpha}^2}{2v_2^2}
  \nn\\&\quad
  +\frac{32\Mh^2\MH^2s_{\alpha}^4}{v_2^4}
  +\frac{(11D-26)\MH^4s_{\alpha}^6}{2v_2^4}\bigg]
  \frac{v_2\hat{h}^2\hat{H}}{s_{\alpha}}
  \nn\\&
  +\bigg[-\frac{6(3D^2-10D+8)\lambda_{12}^3v_2^2}{\MH^2}
  +(9D^2-54D+92)\lambda_{12}^2s_{\alpha}^2
  +\frac{6(2D-7)\lambda_{12}\MH^2s_{\alpha}^4}{v_2^2}
  \nn\\&\quad
  +\frac{4\MH^4s_{\alpha}^6}{v_2^4}\bigg]
  \frac{v_2\hat{h}\hat{H}^2}{s_{\alpha}^2}
  +\frac{9(D-2)^2\lambda_{12}^3v_2^2}{\MH^2}
  \frac{v_2\hat{H}^3}{s_\alpha^3}
  \nn\\&
  +\bigg[\frac{(2D^2-9D+10)\lambda_{12}^3v_2^2}{\MH^2}
  -\frac{(6D^2-15D-38)\lambda_{12}\MH^2s_{\alpha}^4}{4v_2^2}
  -\frac{(35D-106)\lambda_{12}\Mh^2s_{\alpha}^2}{4v_2^2}
    \nn\\&\quad
  +\frac{(2D^2\!-\!9D\!-\!3)\MH^4s_{\alpha}^6}{4v_2^4}+(6D\!-\!21)\lambda_{12}^2s_{\alpha}^2
  +\frac{(37D \!-\! 136)\Mh^2\MH^2s_{\alpha}^4}{4v_2^4}
  +\frac{117\Mh^4s_{\alpha}^2}{4v_2^4}\bigg]
  \hat{h}^4
  \nn\\&
  +\bigg[-\frac{6(D^2-5D+6)\lambda_{12}^3v_2^2}{\MH^2}
  -3(D^2-3D-2)\lambda_{12}^2s_{\alpha}^2
  +\frac{3(D^2-6D+6)\lambda_{12}\MH^2s_{\alpha}^4}{v_2^2}
  \nn\\&\quad
  +\frac{6(3D-14)\lambda_{12}\Mh^2s_{\alpha}^2}{v_2^2}
  +\frac{3(D-2)\MH^4s_{\alpha}^6}{v_2^4}
  +\frac{18\Mh^2\MH^2s_{\alpha}^4}{v_2^4}\bigg]
  \frac{\hat{h}^3\hat{H}}{s_{\alpha}}
  \nn\\&
  +\bigg[\frac{3(3D^2-16D+20)\lambda_{12}^3v_2^2}{2\MH^2}
  +\frac{9D^2-60D+104}{2}\lambda_{12}^2s_{\alpha}^2
  \nn\\&\quad
  +\frac{(3D-13)\lambda_{12}\MH^2s_{\alpha}^4}{v_2^2}
  +\frac{\MH^4s_{\alpha}^6}{v_2^4}\bigg]
  \frac{\hat{h}^2\hat{H}^2}{s_{\alpha}^2}
  \nn\\&
  +\bigg[-\frac{(D^2-6D+8)\lambda_{12}^3v_2^2}{2\MH^2}
  -\frac{3D^2-22D+40}{4}\lambda_{12}^2s_{\alpha}^2
  +\frac{(D^2-2D-6)\MH^4s_{\alpha}^6}{4v_2^4}
  \nn\\&\quad
  +\frac{3(3D-8)\Mh^2\MH^2s_{\alpha}^4}{2v_2^4}
  -\frac{6\lambda_{12}\Mh^2s_{\alpha}^2}{v_2^2}
  -\frac{(5D-18)\lambda_{12}\MH^2s_{\alpha}^4}{2v_2^2}
  +\frac{27\Mh^4s_{\alpha}^2}{2v_2^4}\bigg]
  \frac{\hat{h}^5}{v_2}
  \nn\\&
    +\bigg[
    \frac{3 (D^2-6 D+8)\lambda_{12}^3 v_2^2}{4 M_H^2}
    +\frac{ (3 D^2-26 D+56)\lambda_{12}^2 \sa^2}{2}
    +\frac{3 (3 D-14)\lambda_{12}  M_h^2 \sa^2}{2 v_2^2}
  \nn\\&\quad
    +\frac{D (3 D-14)\lambda_{12} M_H^2 \sa^4}{4 v_2^2}
    +\frac{(D-2) M_H^4 \sa^6}{2  v_2^4}
    +\frac{3 M_h^2 M_H^2 \sa^4}{v_2^4}
    \bigg]
    \frac{\hat{h}^4\hat{H}}{v_2\, \sa}
  \nn\\&
  +\bigg[\frac{(D^2-6D+8)\lambda_{12}^3v_2^2}{24\MH^2}
  +\frac{D^2-10D+24}{8}\lambda_{12}^2s_{\alpha}^2
  +\frac{(D^2-10)\MH^4s_{\alpha}^6}{24v_2^4}
  \nn\\&\quad
  +\frac{3(D-6)\lambda_{12}\Mh^2s_{\alpha}^2}{4v_2^2}
  +\frac{(D-6)(D-2)\lambda_{12}\MH^2s_{\alpha}^4}{8v_2^2}
  \nn\\&\quad
  +\frac{3(D-2)\Mh^2\MH^2s_{\alpha}^4}{4v_2^4}
  +\frac{9\Mh^4s_{\alpha}^2}{4v_2^4}\bigg]
  \frac{\hat{h}^6}{v_2^2}
  \nn\\&
  %Derivative Terms
  +\bigg[-\frac{D^3-6D^2-4D+48}{3D}\lambda_{12}s_{\alpha}^2
  +\frac{(D-4)(D-2)\lambda_{12}^2v_2^2}{3\MH^2}
  \nn\\&\quad
  +\frac{(6D-28)\Mh^2s_{\alpha}^2}{Dv_2^2}
  +\frac{(D-6)(D-4)(D+4)\MH^2s_{\alpha}^4}{12Dv_2^2}\bigg]
  \hat{h}\Box\hat{h}
  \nn\\&
  +\bigg[\frac{D^3-6D^2-8D+64}{2D}\lambda_{12}s_{\alpha}^2
  -\frac{(D-4)(D-2)\lambda_{12}^2v_2^2}{\MH^2}
  +\frac{4(D-4)\MH^2s_{\alpha}^4}{Dv_2^2}\bigg]
  \frac{\hat{h}\Box\hat{H}}{s_{\alpha}}
  \nn\\&
  +\frac{3(D-4)(D-2)\lambda_{12}^2v_2^2}{4\MH^2}
  \frac{\hat{H}\Box\hat{H}}{s_{\alpha}^2}
  \nn\\&
  +\bigg[\frac{(D^3-40D+96)\MH^2s_{\alpha}^4}{12Dv_2^2}
  -\frac{(D-4)(D-2)\lambda_{12}^2v_2^2}{6\MH^2}
  \nn\\&\quad
  -\frac{(D-4)(D^2+10D-48)\lambda_{12}s_{\alpha}^2}{12D}
  +\frac{3(3D-14)\Mh^2s_{\alpha}^2}{Dv_2^2}\bigg]
  \frac{\hat{h}^2\Box\hat{h}}{v_2}
  \nn\\&
  +\bigg[\frac{3D^2-26D+56}{D}\lambda_{12}s_{\alpha}^2
  +\frac{2(D-6)\MH^2s_{\alpha}^4}{Dv_2^2}\bigg]
  \frac{\hat{h}\hat{H}\Box\hat{h}}{s_{\alpha}v_2}
  \nn\\&
  +\bigg[\frac{(D-4)(D-2)\lambda_{12}^2v_2^2}{4\MH^2}
  +\frac{(D-6)(D-4)(D-2)\lambda_{12}s_{\alpha}^2}{4D}
  +\frac{2\MH^2s_{\alpha}^4}{Dv_2^2}\bigg]
  \frac{\hat{h}^2\Box\hat{H}}{s_{\alpha}v_2}
  \nn\\&
  +\bigg[\frac{(D^3-16D-24)\MH^2s_{\alpha}^4}{36Dv_2^2}
  +\frac{(D-4)(D-2)\lambda_{12}^2v_2^2}{36\MH^2}
  \nn\\&\quad
  +\frac{(D-6)(D-4)(D+7)\lambda_{12}s_{\alpha}^2}{18D}
  +\frac{(3D-14)\Mh^2s_{\alpha}^2}{Dv_2^2}\bigg]
  \frac{\hat{h}^3\Box\hat{h}}{v_2^2}
  \nn\\&
  +\frac{(D-6)(D-4)s_{\alpha}^2}{D(D+2)}
  \frac{\hat{h}\Box\Box\hat{h}}{v_2^2}
  \nn\\&
  %O(e^2)Terms
  -	\bigg[\frac{(3D^2-26D+56)\lambda_{12}s_{\alpha}^2}{4D\sw^2}(v_2 + \hat{h})^2
  +\frac{(D-4)\MH^2s_{\alpha}^4}{2D\sw^2v_2^2}(\hat{h}^2+3\hat{h}v_2+2v_2^2)\bigg]
  \frac{e^2\hat{H}(\hat{C}_{\mu}^a)^2}{\sa v_2}
  \nn\\&
  - \frac{(D-4)\sa^2 }{8 D \sw^2}	 \bigg[\lambda_{12}
  (v_2 + \hat{h})^2\hat{h}\big[(D-6)\hat{h}-4(D-4)v_2\big]
  +\frac{6\Mh^2 }{v_2^2}
  (v_2 + \hat{h})^2\hat{h}(2v_2 + \hat{h})
 	\nn\\&\quad
  +\frac{\MH^2s_{\alpha}^2}{v_2^2}
  (v_2 + \hat{h})\hat{h}\big[(D-2)\hat{h}^2
  +(3D-10)\hat{h}v_2+2(D-5)v_2^2\big]\bigg]
  \frac{e^2(\hat{C}_{\mu}^a)^2}{v_2^2}
  \nn\\&
  +\frac{s_{\alpha}^2}{D(D+2)\sw^2}
  \Big[-(D-6)\hat{h}^2+2(D-2)\hat{h}v_2+(D-2)v_2^2\Big]
  \frac{e^2\hat{C}_{\mu}^a\Box\hat{C}^{a,\mu}}{v_2^2}
  \nn\\&
  +\frac{s_{\alpha}^2}{D(D+2)\sw^2}
  \Big[-(D^2-2D+4)\hat{h}^2-2(D^2-2D+4)\hat{h}v_2
  \nn\\&\quad
  +(D-2)Dv_2^2\Big]
  \frac{e^2(\partial^{\mu}\hat{C}_{\mu}^a)(\partial^{\nu}\hat{C}_{\nu}^a)}{v_2^2}
  \nn\\&
  +\frac{2s_{\alpha}^2(D^2-2D+2)}{D(D+2)\sw^2}
  \hat{h}(2v_2 + \hat{h})
  \frac{e^2(\partial^{\nu}\hat{C}_{\mu}^a)(\partial^{\mu}\hat{C}_{\nu}^a)}{v_2^2}
  -\frac{2s_{\alpha}^2(D-4)}{D(D+2)\sw^2}
  \hat{h}^2
  \frac{e^2(\partial_{\mu}\hat{C}_{\nu}^a)^2}{v_2^2}
  \nn\\&
  -\frac{s_{\alpha}^2}{2D(D+2)\sw^2}
  \big[(D^2-8D+28)\hat{h}+(D^2-4D+12)v_2\big] \,(\Box\hat{h})\,
  \frac{e^2(\hat{C}_{\mu}^a)^2}{v_2^2}
  \nn\\&
  +\frac{4(D^2-2D+4)s_{\alpha}^2}{D(D+2)\sw^2}
  (\partial^{\mu}\hat{h})(\partial^{\nu}\hat{h})
  \frac{e^2 \hat{C}_{\mu}^a\hat{C}_{\nu}^a}{v_2^2}
  \nn\\&
  +\frac{4(D^2-3D+6)s_{\alpha}^2}{D(D+2)\sw^2}
  (v_2 + \hat{h})(\partial^{\nu}\hat{h})
  \frac{e^2\hat{C}_{\mu}^a(\partial^{\mu}\hat{C}_{\nu}^a)}{v_2^2}
  \nn\\&
  %O(e^3Terms)
  +\frac{2s_{\alpha}^2}{ D(D+2)\cw\sw^2}
  \big[(D^2-2D-4)v_2^2 - (D-2)\hat{h}^2 - 2(D-2)\hat{h}v_2\big]
  \frac{e^3\eps^{3ab}\hat{B}^{\nu}(\partial^{\mu}\hat{C}_{\mu}^a)\hat{C}_{\nu}^b}{v_2^2}
  \nn\\&
  -\frac{2(D-2)s_{\alpha}^2}{ D(D+2)\cw\sw^2}
  (v_2 + \hat{h})^2
  \frac{e^3\eps^{3ab}\hat{B}_{\mu}(\partial^{\mu}\hat{C}_{\nu}^a)\hat{C}^{b,\nu}}{v_2^2}
  \nn\\&
  +\frac{2s_{\alpha}^2}{D (D+2) \cw \sw^2}
  \big[(D^2-D-2)\hat{h}(2 v_2+\hat{h}) +4v_2^2\big]
  \frac{e^3\eps^{3ab}\hat{B}^{\nu}\hat{C}_{\mu}^b(\partial^{\mu}\hat{C}_{\nu}^a)}{v_2^2}
  \nn\\&
  +\frac{2(D-3)s_{\alpha}^2}{ D\,\cw\sw^2}
  \hat{h}(2v_2 + \hat{h})
  \frac{e^3\eps^{3ab}\hat{C}_{\nu}^a\hat{C}_{\mu}^b(\partial^{\mu}\hat{B}^{\nu})}{v_2^2}
  +\frac{s_{\alpha}^2}{D\sw^3}
  (v_2 + \hat{h})^2
  \frac{e^3\eps^{abc}\hat{C}_{\mu}^c\hat{C}_{\nu}^b(\partial^{\mu}\hat{C}_{\nu}^a)}{v_2^2}
  \nn\\&
  %O(e^4)
  +\frac{(D-2)s_{\alpha}^2}{D(D+2)\cw^2\sw^2}
  (v_2 + \hat{h})^2
  \frac{e^4(\hat{B}_{\nu})^2\big[(\hat{C}_{\mu}^3)^2-(\hat{C}_{\mu}^a)^2\big]}{v_2^2}
  \nn\\&
  +\frac{(D-2)s_{\alpha}^2}{(D+2)\cw^2\sw^2}
  (v_2 + \hat{h})^2
  \frac{e^4\hat{B}^{\mu}\hat{B}^{\nu}\big[\hat{C}_{\mu}^a\hat{C}_{\nu}^a
    -\hat{C}_{\mu}^3\hat{C}_{\nu}^3\big]}{v_2^2}
  \nn\\&
  +\frac{s_{\alpha}^2}{D\,\cw\sw^3}
  (v_2 + \hat{h})^2
  \frac{e^4\hat{B}^{\nu}\big[\hat{C}_{\nu}^3(\hat{C}_{\mu}^a)^2
    -\hat{C}^{3,\mu}\hat{C}_{\mu}^a\hat{C}_{\nu}^a\big]}{v_2^2}
  \nn\\&
  -\frac{(D-1)s_{\alpha}^2}{4D\,\cw^2\sw^4}
  (v_2 + \hat{h})^4
  \frac{e^4\big[\sw^2(\hat{C}_{\mu}^3)^2+\cw^2(\hat{C}_{\mu}^a)^2\big]}{v_2^2}
  +\frac{(D-6)s_{\alpha}^2}{16(D+2)\sw^4}
  (v_2 + \hat{h})^2
  \frac{e^4(\hat{C}_{\mu}^b)^2(\hat{C}_{\nu}^a)^2}{v_2^2}
  \nn\\&
  +\frac{2s_{\alpha}^2}{D(D+2)\sw^4}
  (v_2 + \hat{h})^2
  \frac{e^4\hat{C}_{\mu}^a\hat{C}^{b,\mu}\hat{C}_{\nu}^a\hat{C}^{b,\nu}}{v_2^2}
  \Bigg\}
  \,.
  \label{eq:Leff1loopBeforeEOM}
\end{align}

In particular, this includes all contributions resulting from the hard momentum regions
of the full theory, which we later reuse in \app{CThard} to read off
  the hard parts of (pre-)renormalization constants.

The one-loop effective Lagrangian \textit{after} applying the EOM of the heavy
(background) Higgs field $\hat{H}$ using \eq{HEOMFull} and
subsequently expanding to $\ord(\zeta^{-2})$ reads:
\begin{align}
  \delta\L^\oneloop_\eff={}&  \frac{I_{20}}{16\pi^2(D-2)} \Bigg\{
  \frac{\MH^2}{v_2^2}\bigg[\frac{2\MH^2s_{\alpha}^2}{v_2^2}- 4\lambda_{12}\bigg]
  v_2^3\hat{h}
  +\frac{\MH^2}{v_2^2}\bigg[\frac{\MH^2s_{\alpha}^2}{v_2^2}-2\lambda_{12}\bigg]
  v_2^2\hat{h}^2
  \nn\\&
  %O(zeta^0)
  +\biggl[-\frac{6\lambda_{12}\Mh^2}{v_2^2}
  +\frac{2\lambda_{12}\MH^2s_{\alpha}^2}{v_2^2}
  +\frac{\Mh^2\MH^2s_{\alpha}^2}{v_2^4}
  -\frac{\MH^4s_{\alpha}^4}{v_2^4}\biggr]
  v_2^3\hat{h}
  \nn\\&
  +\bigg[4(D-2)\lambda_{12}^2
  -\frac{4(D-3)\lambda_{12}\MH^2s_{\alpha}^2}{v_2^2}
  -\frac{9\lambda_{12}\Mh^2}{v_2^2}
  \nn\\&\quad
  +\frac{(D-4)\MH^4s_{\alpha}^4}{v_2^4}
  +\frac{13\Mh^2\MH^2s_{\alpha}^2}{2v_2^4}\bigg]
  v_2^2\hat{h}^2
  \nn\\&
  +\bigg[4(D-2)\lambda_{12}^2
  -\frac{(4D-10)\lambda_{12}\MH^2s_{\alpha}^2}{v_2^2}
  +\frac{(D-3)\MH^4s_{\alpha}^4}{v_2^4}
  \nn\\&\quad
  +\frac{6\Mh^2\MH^2s_{\alpha}^2}{v_2^4}
  -\frac{3\Mh^2\lambda_{12}}{v_2^2}\bigg]
  v_2\hat{h}^3
  \nn\\&
  +\bigg[(D-2)\lambda_{12}^2
  -\frac{(D-2)\lambda_{12}\MH^2s_{\alpha}^2}{v_2^2}
  +\frac{(D-2)\MH^4s_{\alpha}^4}{4v_2^4}
  +\frac{3\Mh^2\MH^2s_{\alpha}^2}{2v_2^4}\bigg]
  \hat{h}^4
  \nn\\&
  %Derivative Terms
  +\frac{(D-4)\MH^2s_{\alpha}^2}{Dv_2^2}
  \hat{h}\Box\hat{h}
  +\bigg[\frac{3\lambda_{12}}{2\sw^2}v_2(v_2 + \hat{h})
  -\frac{(D-4)\MH^2s_{\alpha}^2}{4D\sw^2v_2^2}(v_2 + \hat{h})^2\bigg]
  e^2(\hat{C}_{\mu}^a)^2
  \nn\\&
  %O(zeta^2)
  +\biggl[-\frac{6\lambda_{12}\Mh^4}{\MH^2v_2^2}
  +\frac{3\lambda_{12}\Mh^2s_{\alpha}^2}{v_2^2}
  -\frac{\Mh^2\MH^2s_{\alpha}^4}{2v_2^4}
  +\frac{\lambda_{12}\MH^2s_{\alpha}^4}{2v_2^2}
  -\frac{\MH^4s_{\alpha}^6}{4v_2^4}\biggr]
  v_2^3\hat{h}
  \nn\\&
  +\bigg[\frac{12(D-3)\lambda_{12}^2\Mh^2}{\MH^2}
  -4(D-3)\lambda_{12}^2s_{\alpha}^2
  -\frac{(8D-58)\lambda_{12}\Mh^2s_{\alpha}^2}{v_2^2}
  +\frac{4(D-3)\lambda_{12}\MH^2s_{\alpha}^4}{v_2^2}
  \nn\\&\quad
  -\frac{9\lambda_{12}\Mh^4}{\MH^2v_2^2}
  +\frac{(D-18)\Mh^2\MH^2s_{\alpha}^4}{v_2^4}
  -\frac{(D-3)\MH^4s_{\alpha}^6}{v_2^4}
  +\frac{9\Mh^4s_{\alpha}^2}{v_2^4}\bigg]
  v_2^2\hat{h}^2
  \nn\\&
  +\biggl[-\frac{8(D-4)(D-2)\lambda_{12}^3v_2^2}{3\MH^2}
  +\frac{24(D-3)\lambda_{12}^2\Mh^2}{\MH^2}
  -\frac{(52D-261)\lambda_{12}\Mh^2s_{\alpha}^2}{2v_2^2}
  \nn\\&\quad
  +\frac{(7D-53)\Mh^2\MH^2s_{\alpha}^4}{v_2^4}
  +2(D-6)(2D-5)\lambda_{12}^2s_{\alpha}^2
  -\frac{(2D^2-22D+45)\lambda_{12}\MH^2s_{\alpha}^4}{v_2^2}
  \nn\\&\quad
  +\frac{(2D^2-27D+61)\MH^4s_{\alpha}^6}{6v_2^4}
  +\frac{27\Mh^4s_{\alpha}^2}{v_2^4}
 -\frac{3\lambda_{12} \Mh^4}{\MH^2v_2^2}\biggr]
  v_2\hat{h}^3
  \nn\\&
  +\bigg[(6D^2-44D+69)\lambda_{12}^2s_{\alpha}^2
  +\frac{15(D-3)\lambda_{12}^2\Mh^2}{\MH^2}
  -\frac{4(D-4)(D-2)\lambda_{12}^3v_2^2}{\MH^2}
  \nn\\&\quad
  -\frac{(52D-215)\lambda_{12}\Mh^2s_{\alpha}^2}{2v_2^2}
  -\frac{(3D^2-26D+45)\lambda_{12}\MH^2s_{\alpha}^4}{v_2^2}
  +\frac{(37D-212)\Mh^2\MH^2s_{\alpha}^4}{4v_2^4}
  \nn\\&\quad
  +\frac{(2 D^2-20D+37)\MH^4s_{\alpha}^6}{4v_2^4}
  +\frac{117\Mh^4s_{\alpha}^2}{4v_2^4}\bigg]
  \hat{h}^4
  \nn\\&
  +\biggl[-\frac{2(D-4)(D-2)\lambda_{12}^3v_2^2}{\MH^2}
  +\frac{3(D-3)\lambda_{12}^2\Mh^2}{\MH^2}
  -\frac{(21D-80)\lambda_{12}\Mh^2s_{\alpha}^2}{2v_2^2}
  \nn\\&\quad
  +\frac{(9D-43)\Mh^2\MH^2s_{\alpha}^4}{2v_2^4}
  +(3D^2-20D+29)\lambda_{12}^2s_{\alpha}^2
  -\frac{(3D^2-22D+33)\lambda_{12}\MH^2s_{\alpha}^4}{2v_2^2}
  \nn\\&\quad
  +\frac{(D-6)(D-2)\MH^4s_{\alpha}^6}{4v_2^4}
  +\frac{27\Mh^4s_{\alpha}^2}{2v_2^4}\biggr]
  \frac{\hat{h}^5}{v_2}
  \nn\\&
  +\biggl[-\frac{(D-4)(D-2)\lambda_{12}^3v_2^2}{3\MH^2}
  +\frac{(D-4)(D-2)}{2}\lambda_{12}^2s_{\alpha}^2
  -\frac{3(D-4)\lambda_{12}\Mh^2s_{\alpha}^2}{2v_2^2}
  \nn\\&\quad
  -\frac{(D-4)(D-2)\lambda_{12}\MH^2s_{\alpha}^4}{4v_2^2}
  +\frac{3(D-4)\Mh^2\MH^2s_{\alpha}^4}{4v_2^4}
  +\frac{(D-4)(D-2)\MH^4s_{\alpha}^6}{24v_2^4}
  \nn\\&\quad
  +\frac{9\Mh^4s_{\alpha}^2}{4v_2^4}\biggr]
  \frac{\hat{h}^6}{v_2^2}
  \nn\\&
  +\biggl[
  -\frac{D^3-6D^2-4D+48}{3D}\lambda_{12}s_{\alpha}^2
  +\frac{(D-4)(D-2)\lambda_{12}^2v_2^2}{3\MH^2}
  \nn\\&\quad
  +\frac{(6D-28)\Mh^2s_{\alpha}^2}{Dv_2^2}
  +\frac{(D-6)(D-4)(D+4)\MH^2s_{\alpha}^4}{12Dv_2^2}\biggr]
  \hat{h}\Box\hat{h}
  \nn\\&
  +\bigg[\frac{(D^3-40D+192)\MH^2s_{\alpha}^4}{12Dv_2^2}
  -\frac{D^3-12D^2+11D+96}{3D}\lambda_{12}s_{\alpha}^2
  \nn\\&\quad
  +\frac{(D^2 -24D+62)\lambda_{12}^2v_2^2}{3\MH^2}
  +\frac{(9D-42)\Mh^2s_{\alpha}^2}{Dv_2^2}\bigg]
  \frac{\hat{h}^2\Box\hat{h}}{v_2}
  \nn\\&
  +\bigg[\frac{(D^3-28D+144)\MH^2s_{\alpha}^4}{36Dv_2^2}
  +\frac{(D^2-42D+116)\lambda_{12}^2v_2^2}{9\MH^2}
  \nn\\&\quad
  -\frac{(D^3-21 D^2+20 D+96)\lambda_{12}s_{\alpha}^2}{9D}
  +\frac{(3D-14)\Mh^2s_{\alpha}^2}{Dv_2^2}\bigg]
  \frac{\hat{h}^3\Box\hat{h}}{v_2^2}
  \nn\\&
  +\frac{(D^2-10D+24)\sa^2}{D(D+2)}\frac{\hat{h}\Box\Box\hat{h}}{v_2^2}
  \nn\\&
  %O(e^2)
  +\biggl[-\frac{\MH^2s_{\alpha}^4}{8D\sw^2v_2^2}
  (v_2 + \hat{h})^2\big[(D-4)^2\hat{h}^2+2(D^2-10D+20)\hat{h}v_2-6Dv_2^2\big]
  \nn\\&\quad
  +\frac{\lambda_{12}s_{\alpha}^2}{4D\sw^2}
  (v_2 + \hat{h})\Big[(D-4)^2\hat{h}^3+(6D^2-32D+48)\hat{h}^2v_2
  +(8D^2-45D+32)\hat{h}v_2^2
  \nn\\&\quad\quad
  -6Dv_2^3\Big]
  +\frac{3\lambda_{12}\Mh^2}{2\MH^2\sw^2}v_2^3(v_2 + \hat{h})
  -\frac{3(D-3)\lambda_{12}^2v_2^2}{2\MH^2\sw^2}
  v_2(v_2 + \hat{h})\hat{h}(2v_2 + \hat{h})
  \nn\\&\quad
  -\frac{3(D-4)\Mh^2s_{\alpha}^2}{4D\sw^2v_2^2}
  (v_2 + \hat{h})^2\hat{h}(2v_2 + \hat{h})\biggr]
  \frac{e^2(\hat{C}_{\mu}^a)^2}{v_2^2}
  \nn\\&
  -\frac{s_{\alpha}^2}{D(D+2)\sw^2}
  \Big[(D-6)\hat{h}^2 - 2(D-2)\hat{h}v_2 - (D-2)v_2^2\Big]
  \frac{e^2\hat{C}_{\mu}^a\Box\hat{C}^{a,\mu}}{v_2^2}
  \nn\\&
  -\frac{s_{\alpha}^2}{D(D+2)\sw^2}
  \Big[(D^2-2D+4)\hat{h}^2 + 2(D^2-2D+4)\hat{h}v_2
  \nn\\&\quad
  - (D-2)Dv_2^2\Big]
  \frac{e^2(\partial^{\mu}\hat{C}_{\mu}^a)(\partial^{\nu}\hat{C}_{\nu}^a)}{v_2^2}
  -\frac{2(D-4)s_{\alpha}^2}{D(D+2)\sw^2}
  \hat{h}^2
  \frac{e^2(\partial_{\mu}\hat{C}_{\nu}^a)^2}{v_2^2}
  \nn\\&
  -\frac{s_{\alpha}^2}{2D(D+2)\sw^2}
  \big[(D^2-8D+28)\hat{h}+(D^2-4D+12)v_2\big](\Box\hat{h})
  \frac{e^2(\hat{C}_{\mu}^a)^2}{v_2^2}
  \nn\\&
  +\frac{4(D^2-3D+6)s_{\alpha}^2}{D(D+2)\sw^2}
  (v_2 + \hat{h})(\partial^{\nu}\hat{h})
  \frac{e^2\hat{C}_{\mu}^a\partial^{\mu}\hat{C}_{\nu}^a}{v_2^2}
  \nn\\&
  +\frac{2(D^2-2D+2)s_{\alpha}^2}{D(D+2)\sw^2}
  \hat{h}(2v_2 + \hat{h})
  \frac{e^2(\partial^{\nu}\hat{C}_{\mu}^a)\partial^{\mu}\hat{C}_{\nu}^a}{v_2^2}
  \nn\\&
  +\frac{4(D^2-2D+4)s_{\alpha}^2}{D(D+2)\sw^2}
  (\partial^{\mu}\hat{h})(\partial^{\nu}\hat{h})
  \frac{e^2\hat{C}_{\mu}^a\hat{C}_{\nu}^a}{v_2^2}
  \nn\\&
  %O(e^3)
  -\frac{2s_{\alpha}^2}{ D(D+2)\cw\sw^2}
  \Big[(D-2)\hat{h}^2+2(D-2)\hat{h}v_2-   (D^2-2D-4)v_2^2   \Big]
  \frac{e^3\eps^{3ab}\hat{B}^{\nu}\hat{C}_{\nu}^b\partial^{\mu}\hat{C}_{\mu}^a}{v_2^2}
  \nn\\&
  +\frac{2s_{\alpha}^2}{ D(D+2)\cw\sw^2}
  \Big[(D-2)(D+1)\hat{h}^2+2(D-2)(D+1)\hat{h}v_2+4v_2^2\Big]
  \frac{e^3\eps^{3ab}\hat{B}^{\nu}\hat{C}_{\mu}^b\partial^{\mu}\hat{C}_{\nu}^a}{v_2^2}
  \nn\\&
  -\frac{2(D-2)s_{\alpha}^2}{ D(D+2)\cw\sw^2}
  (v_2 + \hat{h})^2
  \frac{e^3\eps^{3ab}\hat{B}_{\mu}\hat{C}_{\nu}^b(\partial^{\mu}\hat{C}_{\nu}^a)}{v_2^2}
  \nn\\&
  +\frac{2(D-3)s_{\alpha}^2}{ D\,\cw \sw^2}
  (2v_2 + \hat{h})\hat{h}
  \frac{e^3\eps^{3ab}\hat{C}_{\mu}^b\hat{C}_{\nu}^a\partial^{\mu}\hat{B}^{\nu}}{v_2^2}
  +\frac{s_{\alpha}^2}{D\sw^3}
  (v_2 + \hat{h})^2
  \frac{e^3\eps^{abc}\hat{C}_{\mu}^c\hat{C}_{\nu}^b\partial^{\mu}\hat{C}^{a,\nu}}{v_2^2}
  \nn\\&
  %O(e^4)
  +\frac{(D-2)s_{\alpha}^2}{D(D+2)\cw^2\sw^2}
  (v_2 + \hat{h})^2
  \frac{e^4(\hat{B}_{\nu})^2\big[(\hat{C}_{\mu}^3)^2-(\hat{C}_{\mu}^a)^2\big]}{v_2^2}
  \nn\\&
  +\frac{(D-2)s_{\alpha}^2}{(D+2)\cw^2\sw^2}
  (v_2 + \hat{h})^2
  \frac{e^4\hat{B}^{\mu}\hat{B}^{\nu}
    \big[\hat{C}_{\mu}^a\hat{C}_{\nu}^a-\hat{C}_{\mu}^3\hat{C}_{\nu}^3\big]}{v_2^2}
  \nn\\&
  +\frac{s_{\alpha}^2}{ D\,\cw\sw^3}
  (v_2 + \hat{h})^2
  \frac{e^4\hat{B}^{\nu}
    \big[\hat{C}_{\nu}^3(\hat{C}_{\mu}^a)^2-\hat{C}_{\mu}^3\hat{C}^{a,\mu}\hat{C}_{\nu}^a\big]}{v_2^2}
  \nn\\&
  -\frac{(D-1)s_{\alpha}^2}{4D\,\cw^2\sw^4}
  (v_2 + \hat{h})^4
  \frac{e^4\big[\cw^2(\hat{C}_{\mu}^a)^2+\sw^2(\hat{C}_{\mu}^3)^2\big]}{v_2^2}
  +\frac{(D-6)s_{\alpha}^2}{16(D+2)\sw^4}
  (v_2 + \hat{h})^2
  \frac{e^4(\hat{C}_{\mu}^b)^2(\hat{C}_{\nu}^a)^2}{v_2^2}
  \nn\\&
  +\frac{2s_{\alpha}^2}{D(D+2)\sw^4}
  (v_2 + \hat{h})^2
  \frac{e^4\hat{C}_{\mu}^a\hat{C}^{b,\mu}\hat{C}_{\nu}^a\hat{C}^{b,\nu}}{v_2^2}
  \Bigg\}
  \,.
  \label{eq:Leff1loopAfterEOM}
\end{align}

This part of the Lagrangian, together with the tree-level contribution $\L_\eff^\tree$ obtained in \subsec{Lefftree}, is
further processed by
the (pre-)renormalization of parameters and fields,
and field redefinitions as described in \secs{renormalization}{fieldredef}.

\section{Counterterm Lagrangian}
\label{app:CTLagrangian}

The effective counterterm Lagrangian $\delta\L_\eff^{\SESM,\ct}$ is derived from the one of the full theory $\delta \L_\SESM^\ct$ as detailed in Section~\ref{subsec:LeffRen}.
In particular, the derivation is in direct analogy to the one of $\L^\tree_\eff$, i.e.\ by expanding to the desired order in $ \zeta^{-1} $ \textit{after} applying the EOM for $\tilde{H}_s$
and before further field redefinitions.
In order to perform this calculation, however, we
have to know the scaling of the $\delta c_i$ and $\delta Z_i$
of the full SESM in $\zeta^{-1}$,
which are determined by the hard-momentum contributions encoded
in the effective Lagrangians $\delta \L_\eff^\oneloop$ given in \app{Leff1loopbeforefieldredef}.
The relevant contributions, which are mostly contained in self-energies,
are easy to identify from the OS/PRTS renormalization conditions, but eventually
the scaling of the renormalization constants should be checked again after the
(pre-)renormalization step (cf.~\sec{renormalization}, \app{CThard}).
We write $\delta\L_\eff^{\SESM,\ct}$ in terms of contributions corresponding
to the individual renormalization constants,
\begin{align}
	\delta\L_\eff^{\SESM,\ct} ={}&L_{t_\mathrm{h}}\delta t_\mathrm{h}
	+L_{t_\mathrm{H}}\delta t_\mathrm{H}
	+L_{v}\delta v_2
	+L_{\Mh}\delta \Mh^2
	+L_{e}\delta Z_e
	+L_{\sw}\delta \sw
	+L_{\sa} \delta \sa \nn \\&
	+L_{hh}\delta Z_{hh}
	+L_{hH}\delta Z_{hH}
	+L_{W}\delta Z_{W}
	+L_{B}\delta Z_{B}
  + \ord(\zeta^{-3}) \,,
  \label{eq:Lct}
\end{align}
and give the coefficients $L_{\dots}$ in unitary background gauge below.
Note that $\delta Z_{hH}$ term can be eliminated from the final result for
the effective Lagrangian by redefining  the SM-like Higgs field~$h$, see \eq{HiggsFieldRedef}.
For this reason, in principle $\delta Z_{hH}$ could be set to zero
to simplify intermediate expressions; we prefer to keep the $\delta Z_{hH}$
for completeness.
\begin{align}
	L_{t_\mathrm{h}}={}&
	\hat{h}
	-\frac{\hat{h}^3}{2v_2^2}
	-\frac{\hat{h}^4}{8v_2^3}
	+\frac{s_{\alpha}^2}{v_2^2}\hat{h}^3
	+\frac{17s_{\alpha}^2}{16v_2^3}\hat{h}^4
	+\frac{s_{\alpha}^2}{4v_2^4}\hat{h}^5
	\nn\\&
	-\frac{s_{\alpha}^4}{2v_2^2}\hat{h}^3
	+\frac{3s_{\alpha}^2}{2} \bigg(\frac{\lambda_{12}v_2^2}{\MH^2}
	+\frac{\Mh^2}{\MH^2}
	-\frac{45s_{\alpha}^2}{32}\bigg)\frac{\hat{h}^4}{v_2^3}
	+s_{\alpha}^2 \bigg(\frac{2\lambda_{12}v_2^2}{\MH^2}
	+\frac{5\Mh^2}{4\MH^2}
	-\frac{15s_{\alpha}^2}{8}\bigg)\frac{\hat{h}^5}{v_2^4}
	\nn\\&
	+\frac{s_{\alpha}^2}{4} \bigg(\frac{7\lambda_{12}v_2^2}{2\MH^2}
	+\frac{\Mh^2}{\MH^2}
	-\frac{7s_{\alpha}^2}{4}\bigg)\frac{\hat{h}^6}{v_2^5}
	+\frac{\lambda_{12}s_{\alpha}^2}{8\MH^2 v_2^4} \hat{h}^7
	-\frac{s_{\alpha}^2}{8\MH^2 v_2^4}(8v_2+3\hat{h})\hat{h}^3\Box\hat{h}
	\nn\\&
	-\frac{g_2^2s_{\alpha}^2}{8\MH^2 v_2^3}
	\hat{h}^2(v_2+\hat{h})(3v_2+\hat{h})
	(\hat{C}_{\mu}^a)^2
 + \ord\bigl(\zeta^{-5}\bigr)
	\,,\\
	L_{t_\mathrm{H}}={}&
	-\frac{s_{\alpha}}{2v_2} \hat{h}^2
	-\frac{s_{\alpha}}{2v_2^2} \hat{h}^3
	-\frac{s_{\alpha}}{8v_2^3} \hat{h}^4
	\nn\\& {}
	+\frac{\sa}{v_2} \bigg(-\frac{\lambda_{12}v_2^2}{\MH^2}
	-\frac{\Mh^2}{\MH^2}
	+\frac{s_{\alpha}^2}{2}\bigg) \hat{h}^2
	+\frac{\sa}{v_2^2} \bigg(-\frac{\lambda_{12}v_2^2}{\MH^2}
	-\frac{\Mh^2}{2\MH^2}
	+\frac{5s_{\alpha}^2}{4}\bigg) \hat{h}^3
	\nn\\& {}
	+ \frac{\sa}{v_2^3} \bigg(s_{\alpha}^2
	-\frac{\lambda_{12}v_2^2}{4\MH^2}\bigg) \hat{h}^4
	+\frac{s_{\alpha}^3}{4v_2^4} \hat{h}^5
	+\frac{g_2^2 s_{\alpha}}{4\MH^2}
	(v_2+\hat{h}) (\hat{C}_{\mu}^a)^2
	\nn\\& {}
	+ \frac{\sa\Mh^2}{\MH^2 v_2} \bigg( s_{\alpha}^2
	-\frac{3\lambda_{12}v_2^2}{\MH^2}\bigg) \hat{h}^2
	+ \frac{\sa}{v_2^2} \bigg(\frac{5\lambda_{12}s_{\alpha}^2v_2^2}{2\MH^2}
	-\frac{7\lambda_{12}\Mh^2v_2^2}{\MH^4}
	+\frac{15\Mh^2s_{\alpha}^2}{4\MH^2}
	-\frac{15s_{\alpha}^4}{16}\bigg) \hat{h}^3
	\nn\\& {}
	+ \frac{\sa}{4v_2^3} \bigg(\frac{13\Mh^2s_{\alpha}^2}{\MH^2}
	-\frac{19\lambda_{12}\Mh^2v_2^2}{\MH^4}
	+\frac{17\lambda_{12}s_{\alpha}^2v_2^2}{\MH^2}
	-9s_{\alpha}^4\bigg) \hat{h}^4
	\nn\\& {}
	+ \frac{\sa}{v_2^4} \bigg(\frac{5\Mh^2s_{\alpha}^2}{4\MH^2}
	-\frac{\lambda_{12}\Mh^2v_2^2}{\MH^4}
	+\frac{11\lambda_{12}s_{\alpha}^2v_2^2}{4\MH^2}
	-\frac{7s_{\alpha}^4}{4}\bigg) \hat{h}^5
	+ \frac{\sa^3}{v_2^5} \bigg(\frac{7\lambda_{12}v_2^2}{8\MH^2}
	+\frac{\Mh^2}{4\MH^2}
	-\frac{7s_{\alpha}^2}{16}\bigg) \hat{h}^6
	\nn\\& {}
	+\frac{\lambda_{12}s_{\alpha}^3}{8\MH^2 v_2^4} \hat{h}^7
	+\frac{\sa}{\MH^2v_2^2} \bigg(\frac{2\lambda_{12}v_2^2}{\MH^2}
	-s_{\alpha}^2 \bigg) \hat{h}^2\Box\hat{h}
	+\frac{\sa}{\MH^2 v_2^3} \bigg(\frac{4\lambda_{12}v_2^2}{3\MH^2}
	-s_{\alpha}^2 \bigg) \hat{h}^3\Box\hat{h}
	\nn\\& {}
	-\frac{3s_{\alpha}^3}{8\MH^2 v_2^4} \hat{h}^4\Box\hat{h}
	+ \frac{g_2^2\sa}{2\MH^2} \bigg[\frac{\lambda_{12}}{\MH^2} (2v_2+\hat{h})
	-\frac{s_{\alpha}^2}{4v_2^3}
	( 5v_2^2 +3\hat{h}v_2 +\hat{h}^2)\bigg]
	\hat{h}(v_2+\hat{h})
	(\hat{C}_{\mu}^a)^2
	\nn\\& {}
    + \ord\bigl(\zeta^{-6}\bigr)
  \,,	\label{eq:LtH}\\
	L_v={}&
	\frac{\Mh^2}{2v_2^2} \hat{h}^3
	+\frac{\Mh^2}{4v_2^3} \hat{h}^4
    +\frac{g_2^2}{4}(v_2+\hat{h}) (\hat{C}_{\mu}^a)^2
	\nn\\& {}
	-\frac{3\Mh^2s_{\alpha}^2}{4v_2^2} \hat{h}^3
	-\frac{7\Mh^2s_{\alpha}^2}{4v_2^3} \hat{h}^4
	-\frac{3\Mh^2s_{\alpha}^2}{4v_2^4} \hat{h}^5
	+\frac{s_{\alpha}^2}{3v_2^3} \hat{h}^3\Box\hat{h}
	-\frac{g_2^2s_{\alpha}^2}{8v_2^2}
	\hat{h}(v_2^2-\hat{h}^2) (\hat{C}_{\mu}^a)^2
    + \ord\bigl(\zeta^{-3}\bigr)
  \,,	\\
	L_{\Mh}={}&
	-\frac{1}{2} \hat{h}^2
	-\frac{1}{2v_2} \hat{h}^3
	-\frac{1}{8v_2^2} \hat{h}^4
	+\frac{3s_{\alpha}^2}{4v_2} \hat{h}^3
	+\frac{7s_{\alpha}^2}{8v_2^2} \hat{h}^4
	+\frac{s_{\alpha}^2}{4v_2^3} \hat{h}^5
	\nn\\& {}
	+ \frac{\sa^2}{v_2} \bigg(\frac{\lambda_{12}v_2^2}{\MH^2}
	-\frac{3s_{\alpha}^2}{16}\bigg) \hat{h}^3
	+ \frac{\sa^2}{v_2^2} \bigg(\frac{5\lambda_{12}v_2^2}{2\MH^2}
	+\frac{\Mh^2}{\MH^2}
	-\frac{11s_{\alpha}^2}{8}\bigg) \hat{h}^4
	\nn\\& {}
	+ \frac{\sa^2}{v_2^3} \bigg(\frac{9\lambda_{12}v_2^2}{4\MH^2}
	+\frac{\Mh^2}{\MH^2}
	-\frac{3s_{\alpha}^2}{2}\bigg) \hat{h}^5
	+ \frac{\sa^2}{4v_2^4} \bigg(\frac{7\lambda_{12}v_2^2}{2\MH^2}
	+\frac{\Mh^2}{\MH^2}
	-\frac{7s_{\alpha}^2}{4}\bigg) \hat{h}^6
	+\frac{\lambda_{12}s_{\alpha}^2v_2^2}{8\MH^2v_2^5} \hat{h}^7
	\nn\\& {}
	-\frac{2s_{\alpha}^2}{3\MH^2v_2^2} \hat{h}^3\Box\hat{h}
	-\frac{3s_{\alpha}^2}{8\MH^2v_2^3} \hat{h}^4\Box\hat{h}
	-\frac{g_2^2s_{\alpha}^2}{8\MH^2v_2^2}\hat{h}^2
	( v_2+\hat{h}) ( 2v_2+\hat{h}) (\hat{C}_{\mu}^a)^2
   + \ord\bigl(\zeta^{-5}\bigr)
  \,,	\\
	L_e={}&
	- g_2 \varepsilon^{abc} (\partial^\mu \hat{W}^{a,\nu}) \hat{W}_{\mu}^b  \hat{W}_{\nu}^c
	+ \frac{g_2^2}{2}\biggl[ \hat{W}_{\mu}^a \hat{W}^{b,\mu} \hat{W}_{\nu}^a \hat{W}^{b,\nu}
	- (\hat{W}_{\mu}^a)^2 (\hat{W}_{\nu}^b)^2 \biggr]
	\nn\\&
	+ \frac{g_2^2}{4} (v_2+\hat{h})^2 (\hat{C}_{\mu}^a)^2
   + \ord(\zeta^{-1})
  \,, \\
	L_{\sw}={}& -\frac{L_e}{\sw}
	+ \frac{g_2^2}{4\cw^3} (v_2+\hat{h})^2 \hat B^\mu \hat{C}_{\mu}^3
   + \ord(\zeta^{-1})
	\,, \\
	L_{\sa}={}&
	\frac{3\Mh^2s_{\alpha}}{2v_2} \hat{h}^3
	+\frac{7\Mh^2s_{\alpha}}{4v_2^2} \hat{h}^4
	+\frac{\Mh^2s_{\alpha}}{2v_2^3} \hat{h}^5
	-\frac{s_{\alpha}}{3v_2^2} \hat{h}^3\Box\hat{h}
	-\frac{g_2^2s_{\alpha}}{4v_2}\hat{h} (v_2 + \hat{h})^2 (\hat{C}_{\mu}^a)^2
   + \ord\bigl(\zeta^{-2}\bigr)
  \,, \label{eq:L_sa}
  \\
	L_{hh}={}&
	-\frac{\Mh^2}{2} \hat{h}^2
	-\frac{3\Mh^2}{4v_2} \hat{h}^3
	-\frac{\Mh^2}{4v_2^2}\hat{h}^4
	-\frac{1}{2}\hat{h}\Box\hat{h}
	+\frac{g_2^2}{8}\hat{h}(v_2+\hat{h}) (\hat{C}_{\mu}^a)^2
	\nn\\&
	+\frac{9\Mh^2s_{\alpha}^2}{8v_2} \hat{h}^3
	+\frac{7\Mh^2s_{\alpha}^2}{4v_2^2} \hat{h}^4
	+\frac{5\Mh^2s_{\alpha}^2}{8v_2^3} \hat{h}^5
	-\frac{s_{\alpha}^2}{3v_2^2} \hat{h}^3\Box\hat{h}
	-\frac{g_2^2 s_{\alpha}^2}{16v_2}
	\hat{h}(v_2+\hat{h})(v_2+3\hat{h}) (\hat{C}_{\mu}^a)^2
  \nn\\&
   + \ord\bigl(\zeta^{-3}\bigr)
  \,,\\
	L_{hH}={}&
	\frac{\sa\Mh^2}{4v_2} \hat{h}^3
	+\frac{3\sa\Mh^2}{8v_2^2} \hat{h}^4
	+\frac{\sa\Mh^2}{8v_2^3} \hat{h}^5
	+\frac{s_{\alpha}}{4v_2} \hat{h}^2\Box\hat{h}
	-\frac{g_2^2 s_{\alpha}}{16v_2}
	(v_2 + \hat{h})\hat{h}^2 (\hat{C}_{\mu}^a)^2
	\nn\\& {}
	+ \frac{\sa\Mh^2}{2v_2} \bigg(
	  \frac{\lambda_{12}v_2^2}{\MH^2}
	+\frac{\Mh^2}{\MH^2}
	- \frac{s_{\alpha}^2}{2} \bigg) \hat{h}^3
	+ \frac{\sa\Mh^2}{v_2^2} \bigg(\frac{5\lambda_{12}v_2^2}{4\MH^2}
	+\frac{\Mh^2}{\MH^2}
	-\frac{19s_{\alpha}^2}{16}\bigg) \hat{h}^4
	\nn\\& {}
	+\frac{\sa\Mh^2}{8v_2^3} \bigg(\frac{9\lambda_{12}v_2^2}{\MH^2}
	+\frac{5\Mh^2}{\MH^2}
	-11s_{\alpha}^2 \bigg) \hat{h}^5
	+ \frac{\sa\Mh^2}{8v_2^4} \bigg(\frac{7\lambda_{12}v_2^2}{2\MH^2}
	+\frac{\Mh^2}{\MH^2}
	-\frac{7s_{\alpha}^2}{2} \bigg) \hat{h}^6
	\nn\\& {}
	+\frac{\lambda_{12}\Mh^2s_{\alpha}}{16\MH^2v_2^3} \hat{h}^7
	+ \frac{\sa}{2v_2} \bigg(\frac{\lambda_{12}v_2^2}{\MH^2}
	+\frac{\Mh^2}{2\MH^2}
	-\frac{s_{\alpha}^2}{2}\bigg) \hat{h}^2\Box\hat{h}
	\nn\\& {}
	+ \frac{\sa}{2v_2^2} \bigg(\frac{\lambda_{12}v_2^2}{\MH^2}
	-\frac{\Mh^2}{2\MH^2}
	-\frac{s_{\alpha}^2}{2}\bigg) \hat{h}^3\Box\hat{h}
	+ \frac{\sa}{8v_2^3} \bigg(\frac{\lambda_{12}v_2^2}{\MH^2}
	-\frac{3\Mh^2}{2\MH^2}
	+\frac{3s_{\alpha}^2}{2}\bigg) \hat{h}^4\Box\hat{h}
	\nn\\& {}
	- \frac{\sa}{2\MH^2 v_2} (\Box \hat h)\left[  (\partial_\mu \hat h)^2
		+(\Box \hat h) \hat h \right]
	+\frac{g_2^2 \sa}{8v_2^2} \biggl[
    -  \frac{\Mh^2}{\MH^2} v_2^2
	 + \biggl( - \frac{\lambda_{12} v_2^2}{\MH^2} - \frac{5\Mh^2}{2\MH^2}
		+\frac{3\sa^2}{4} \biggr) v_2 \hat{h}
	\nn\\ & \quad {}
	+ \biggl(-\frac{\Mh^2}{\MH^2} + \frac{5\sa^2}{4}
		- \frac{\lambda_{12}v_2^2}{\MH^2}\biggr) \hat h^2
	- \frac{\lambda_{12}v_2}{4\MH^2} \hat h^3
	\biggr] \hat h (v_2+\hat h) (\hat{C}_{\mu}^a)^2
	\nn\\& {}
	+\frac{g_2^2 s_{\alpha}}{16\MH^2v_2}
	( -2v_2^2 -2\hat{h}v_2 +\hat{h}^2)(\Box\hat{h}) (\hat{C}_{\mu}^a)^2
	+\frac{g_2^2 s_{\alpha}v_2^2}{24\MH^2v_2^3}
	\hat{h}^2(3v_2+\hat{h})
	(\partial_{\mu}\hat{C}_{\nu}^a)^2
	\nn\\& {}
	+\frac{g_2^2 s_{\alpha}}{24\MH^2v_2}
	\hat{h}^2(3v_2+\hat{h})
	\hat{C}_{\mu}^a\Box\hat{C}^{a,\mu}
	+\frac{g_2^4s_{\alpha}}{32\MH^2}
	(v_2+\hat{h})^2
	(\hat{C}_{\mu}^b)^2(\hat{C}_{\nu}^a)^2
       + \ord\bigl(\zeta^{-4}\bigr)
  \,, \\
	L_{W}={}&
	\frac{1}{2} \hat{W}_{\mu}^a \Box \hat{W}^{a,\mu}
	+\frac{1}{2} (\partial_\mu\hat{W}_{\nu}^a) (\partial^\nu\hat{W}^{a,\mu})
	- \frac{3g_2}{2} \varepsilon^{abc} (\partial^\mu \hat{W}^{a,\nu})
		\hat{W}_{\mu}^b  \hat{W}_{\nu}^c
	\nn\\&
	+ \frac{g_2^2}{2}\biggl[ \hat{W}_{\mu}^a \hat{W}^{b,\mu} \hat{W}_{\nu}^a \hat{W}^{b,\nu}
	- (\hat{W}_{\mu}^a)^2 (\hat{W}_{\nu}^b)^2 \biggr]
	+ \frac{g_2^2}{8} (v_2+\hat{h})^2
		\left[ \frac{\sw}{\cw} \hat B^\mu \hat{W}^3_\mu + (\hat{W}_{\mu}^a)^2 \right]
		\nn\\& {}
    + \ord(\zeta^{-1})
  \,, \\
	L_{B}={}&
	-\frac{1}{4}\hat{B}_{\mu\nu}\hat{B}^{\mu\nu}
	+\frac{g_1 g_2}{8}   (v_2+\hat{h})^2 \hat{B}^{\mu}\hat{C}_{\mu}^3
   + \ord(\zeta^{-1})
  \,.
\end{align}

\section{Hard contributions to SESM counterterms}
\label{app:CThard}

In this appendix we give the explicit results for the (pre-)renormalization
constants of the SM-like parameters and fields of the SESM used
in the pre-renormalization step described in \sec{renormalization}.
Recall that these renormalization constants are just chosen to
achieve a simple form of the effective Lagrangian obtained after
the field transformation described in \sec{fieldredef},
\revision{and adjusted to obtain the SMEFT form of the EFT Lagrangian in \subsec{SMEFTmatch}}, i.e.\
they do not have any direct physical meaning.
The (actually relevant) renormalization constants of the EFT guaranteeing the
imposed renormalization conditions on the EFT parameters and fields
are given \revision{(in terms of the Wilson coefficients of \sec{final})} in \refapp{app:CTEFT} below.
Also recall that, by contrast, the genuine BSM renormalization constants
are already fixed once and for all during the derivation of the EFT,
cf.~\subsec{BSMRenCond}.

Technically, the following renormalization constants are determined
in the course of fixing the field redefinition described in
\subsec{field-redef-NLO}, to bring the one-loop EFT Lagrangian \revision{in the form of \eq{SMEFTHEFT}}.
Except for the contributions from $\delta\sa$ and $\delta_s t_{\PH}$,
the pre-renormalization constants of the SM-like quantities contain
only hard-momentum contributions,
\begin{align}
\delta t_{\Ph}={}&
	-\frac{\MH^2(\MH^2\sa^2-2\lambda_{12}v_2^2) I_{20}}{8(D-2)\pi^2v_2}
	+ \frac{[\MH^2\sa^2(\MH^2\sa^2-2\lambda_{12}v_2^2)
		-\Mh^2(\MH^2\sa^2-6\lambda_{12} v_2^2)] I_{20}}{16(D-2)\pi^2v_2}
\nn\\ &{}
	+\left(\MH^2\sa^2-2\lambda_{12}v_2^2\right) \left[
		\frac{(D-3)^2 \MH^2\sa^4 }{4(D-2)}
		+ \frac{(D-4) \lambda_{12}v_2^2 (-\MH^2\sa^2+\lambda_{12}v_2^2)
			}{\MH^2} \right] \frac{I_{20}}{16\pi^2 v_2}
\nn\\ &{}
	+ \left[ (6D^2-11D-48)\MH^2\sa^2 - 6(2D^2-3D-16)\lambda_{12} v_2^2 \right]
		\frac{\Mh^2\sa^2 I_{20}}{32D(D-2)\pi^2 v_2}
\nn\\ &{}
	+ \left[ \frac{(D^2+6D+20)\MH^2\sa^2}{D(D+2)} + \lambda_{12} v_2^2 \right]
		\frac{3\Mh^4 I_{20}}{8(D-2)\pi^2\MH^2 v_2}
	-\frac{3(D-1)e^2\Mh^2\sa^2 v_2 I_{20}}{16D(D-2)\pi^2\sw^2} \,,
\\
\delta M_{\Ph}^2={}&
	- \frac{\sa}{v_2} \left[
		1 + \frac{2\Mh^2-\MH^2\sa^2+2\lambda_{12}v_2^2}{\MH^2}
		\right] \delta_s t_{\PH}
	-2\Mh^2\sa\delta\sa
\nn\\ &{}
  +\frac{\MH^2(\MH^2\sa^2+\lambda_{12}v_2^2) I_{20}}{8(D-2)\pi^2v_2^2}
	+ \frac{(11D+8)\Mh^2\MH^2\sa^2 I_{20}}{16D(D-2)\pi^2v_2^2}
\nn\\ &{}
	+ \left(\MH^2\sa^2 - 2\lambda_{12} v_2^2\right)
		\left[ (2D-5)\MH^2\sa^2 - 2(2D-1)\lambda_{12} v_2^2 \right]
		\frac{I_{20}}{16(D-2)\pi^2v_2^2}
\nn\\ &{}
	- \left( \MH^2\sa^2 - 2\lambda_{12}v_2^2 \right)^2
		\left[ \frac{(5D^2-22D+28)\MH^2\sa^2}{D-2} - 10(D-4)\lambda_{12} v_2^2 \right]
	\frac{I_{20}}{64\pi^2\MH^2 v_2^2}
\nn\\ &{}
	+  \sa^2\Mh^2\left[(-15D^2+24D+88)\MH^2\sa^2 + 2(15D^2-20D-88)\lambda_{12} v_2^2 \right]
		\frac{I_{20}}{16D(D-2)\pi^2v_2^2}
\nn\\ &{}
	- \frac{11(D^2+6D+20)\Mh^4\sa^2	I_{20}}{8D(D^2-4)\pi^2v_2^2}
	+\frac{9(D-1)e^2s_\alpha^2\Mh^2 I_{20}}{16D(D-2)\pi^2\sw^2} \,,
\label{eq:dMh2_preren}
\\
\delta Z_{\hat h\hat h}={}&
	2\sa \delta\sa
	+ \frac{(D-4)\MH^2\sa^2 I_{20}}{8D(D-2)\pi^2 v_2^2}
\nn\\ &{}
	+ \left(\MH^2\sa^2-2\lambda_{12}v_2^2\right) \left[
		(D^2-12D+16)\MH^2\sa^2 - 6D(D-1) \lambda_{12} v_2^2 \right]
		\frac{I_{20}}{8D(D-2)\pi^2 \MH^2 v_2^2}
\nn\\ &{}
	+ \frac{(5D^2-6D-80)\Mh^2\sa^2 I_{20}}{8D(D^2-4)\pi^2 v_2^2}
	+ \frac{3(D-1)e^2\sa^2 I_{20}}{8D(D-2)\pi^2\sw^2} \,,
\\
\frac{\delta v_2}{v_2}={}&
	- \frac{\sa}{\MH^2 v_2} \delta_s t_{\PH}
	+\frac{(D-4)s_\alpha^2\MH^2I_{20}}{16D(D-2)\pi^2v_2^2}
	+\frac{(D-1)e^2s_\alpha^2I_{20}}{16D(D-2)\pi^2\sw^2} \,,
\label{eq:dv2_preren}
\\
\delta Z_{e}={}& \delta \sw = \delta Z_{\hat W} = \delta Z_{\hat B} = 0 \,.
\end{align}

\revision{
In order to bring the EFT Lagrangian in SMEFT form, as described in \subsec{SMEFTmatch}, we need the following slightly modified pre-renormalization constants (now indicated by $\hat{\delta}$):
\begin{align}
  \hat\delta t_{\Ph}={}&
  \delta t_{\Ph} + \frac{(D-6)e^2v_2 \Mh^2\sa^2 I_{20}}{8D(D^2-4)\pi^2\sw^2} \,,
  \\
  \hat\delta M_{\Ph}^2={}&
  \delta M_{\Ph}^2 - \frac{3(D-6)e^2 \Mh^2\sa^2 I_{20}}{8D(D^2-4)\pi^2\sw^2} \,,
  \label{eq:dMh2_preren_new}
  \\
  \hat\delta Z_{\hat h\hat h}={}&
  \delta Z_{\hat h\hat h} \,,
  \\
  \frac{\hat\delta v_2}{v_2}={}&
  \frac{\delta v_2}{v_2} \,,
  \label{eq:dv2_preren_new}
  \\
  \hat\delta Z_{e}={}& \hat\delta \sw = \hat\delta Z_{\hat W} = \hat\delta Z_{\hat B} = 0 \,.
\end{align}
}

\section{EFT renormalization constants}
\label{app:CTEFT}

For our phenomenological applications in \sec{pheno} we need EFT renormalization constants for the EFT
defined by the
\revision{tree-level and one-loop parts} of the effective Lagrangian
given in \eqs{LeffTreeFinal}{SMEFTHEFT}, respectively.
These can be calculated fully within the EFT.
As detailed in \sec{pheno} we employ the OS renormalization scheme
and the PRTS.
\revision{The EFT described in \sec{final} still necessarily uses the
non-linear parametrization of the Higgs sector, but is}
fully gauge independent, meaning a gauge can be chosen freely.
The following renormalization constants have been calculated in the
't~Hooft--Feynman gauge \revision{($\xi=1$)}.
We recall that it is most convenient to use the form~\eq{LeffTreeFinal}
for the tree-level effective Lagrangian $\delta\L_{\eff}^{\tree}$,
where the SM-like Higgs field is canonically normalized, rather than the
more compact form~\eq{Ctree}, where
$\delta\L_{\SMEFT}^{\tree}=C^\tree_{\Phi\Box}\mathcal{O}_{\Phi\Box}$
with $C^\tree_{\Phi\Box}=-\sa^2/(2v_2^2)$.

\begin{figure}
	\centerline{
	\includegraphics{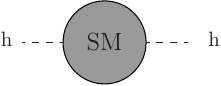}
	\hspace{2em}
	\includegraphics{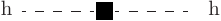}
	}
	\vspace{1em}
	\centerline{
	\includegraphics{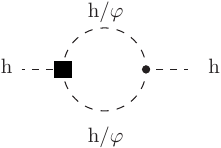}
	\hspace{2em}
	\includegraphics{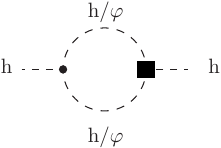}
	\hspace{2em}
	\includegraphics{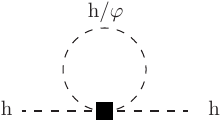}
	}
	\vspace{1em}
	\centerline{
	\includegraphics{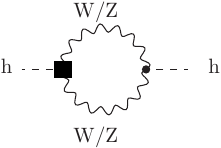}
	\hspace{2em}
	\includegraphics{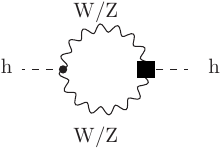}
	\hspace{2em}
	\includegraphics{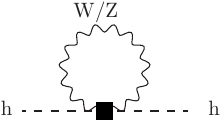}
}
\caption{Diagrams contributing to the EFT mass renormalization constant $ \deft\Mh^2 $ for the SM-like
Higgs field in the OS scheme. Black squares denote an insertion of a SMEFT or a non-SMEFT operator.
The second graph corresponds to the contribution from hard modes, and the
remaining explicit loop diagrams correspond to the soft EFT modes with insertions
of the tree-level EFT couplings contained in $\delta\L_{\eff}^{\tree}$
as given in~\eq{LeffTreeFinal}.
}
\label{fig:EFTRenconst}
\end{figure}

We exemplary sketch
the diagrams contributing to the mass renormalization constant
of the SM-like Higgs boson in \reffi{fig:EFTRenconst}.
The first diagram symbolizes all one-loop SM contributions to the Higgs self-energy
and, thus, all diagrams contributing to
the SM renormalization constant $\delta^\SM\Mh^2$.
At NLO, the difference in the results obtained within the BFM or the conventional
quantization formalism is confined in this subset of diagrams.
The second graph of \fig{EFTRenconst}
represents all diagrams with one insertion of the SMEFT and non-SMEFT operators from $\delta\L_\eff^{\oneloop,\BSM}$, denoted by the black square.
The remaining diagrams comprise all possible one-loop topologies with one insertion of the
tree-level operators in $\delta\L_{\eff}^{\tree}$, again marked by a black square.
For $\deft\Mh^2$, all diagrams of \reffi{fig:EFTRenconst}
need to be evaluated at $ p^2=\Mh^2 $.
The calculations for the $W$- and $Z$-boson self-energies follow the same line.
In total we find
\begin{align}
\nn
	\deft\Mh^2={}&
	\delta^\SM\Mh^2
	+ \frac{C_{\Phi\Box}^{\tree}}{16\pi^2}
		\biggl[
	27 \Mh^4 B_0(\Mh^2, \Mh^2, \Mh^2)
\nn\\ & \qquad {}
	+ 2 \big(4 (D-1) \MW^4 - 4 \MW^2 \Mh^2 + \Mh^4\big) B_0(\Mh^2, \MW^2, \MW^2)
\nn\\ & \qquad {}
	+ \big(4 (D-1) \MZ^4 - 4 \MZ^2 \Mh^2 + \Mh^4\big) B_0(\Mh^2, \MZ^2, \MZ^2)
\nn\\ & \qquad {}
	+ 25 \Mh^4 B_0(0, 0, \Mh^2)
	+ 4 \MW^2 \big(2(D-1) \MW^2 - \Mh^2\big) B_0(0, 0, \MW^2)
\nn\\ & \qquad {}
	+ 2 \MZ^2 \big(2(D-1) \MZ^2 - \Mh^2\big) B_0(0, 0, \MZ^2)
	\biggr]
\nn\\ & {}
	+\frac{v_2^2}{4}\bigl[15v_2^2C_{\Phi}+2\Mh^2(C_{\Phi D}-4C_{\Phi\Box})\bigr] \,,
\\\nn
	\deft\MW^2={}&
	\delta^\SM\MW^2
	+ \frac{C_{\Phi\Box}^{\tree}}{8\pi^2(D-1)}
		\biggl[ \big(\Mh^4 - 4\Mh^2\MW^2 + 4(D-1)\MW^4\big) B_0(\MW^2, \Mh^2, \MW^2)
\nn\\ & \qquad {}
			-\Mh^2(\Mh^2 - 2(D-1) \MW^2) B_0(0,0,\Mh^2)
			+\MW^2 (\Mh^2 - 2 \MW^2) B_0(0,0,\MW^2) \biggr]
\nn\\ & {}
	-2\MW^2v_2^2(C_{\Phi W}+C^\nonSMEFT_{2}) \,,
	\\\nn
	\deft\MZ^2={}&
	\delta^\SM\MZ^2
	+ \frac{C_{\Phi\Box}^{\tree}}{8\pi^2(D-1)}
		\biggl[ \big(\Mh^4 - 4\Mh^2\MZ^2 + 4(D-1)\MZ^4\big) B_0(\MZ^2, \Mh^2, \MZ^2)
\nn\\ & \qquad {}
			-\Mh^2(\Mh^2 - 2(D-1) \MZ^2) B_0(0,0,\Mh^2)
			+\MZ^2 (\Mh^2 - 2 \MZ^2) B_0(0,0,\MZ^2) \biggr]
\nn\\ & {}
	-\frac{\MZ^2 v_2^2}{2\cw^2} \big[ 4\cw^4C_{\Phi W} +4\cw^2\sw^2C_{\Phi B}
	+4\cw^3\sw C_{\Phi WB} +\cw^2C_{\Phi D} +4C^\nonSMEFT_{2} \big] \,,
\\
	\deft Z_{hh}={}&\delta^\SM Z_{hh}
	+ \frac{C_{\Phi\Box}^{\tree}}{16\pi^2}
		\biggl[
		- 4 (\Mh^2 - 2 \MW^2) B_0(\Mh^2, \MW^2, \MW^2)
	\nn\\& \qquad {}
		- 2 (\Mh^2 - 2 \MZ^2) B_0(\Mh^2, \MZ^2, \MZ^2)
		+ 4 \MW^2 B_0(0, 0, \MW^2)
	\nn\\& \qquad {}
		+ 2 \MZ^2 B_0(0, 0, \MZ^2)
		- 2 \Mh^2 B_0(0, 0, \Mh^2)
		- 27 \Mh^4 B'_0(\Mh^2, \Mh^2, \Mh^2)
	\nn\\& \qquad {}
		- 2 \big(\Mh^4 - 4 \Mh^2 \MW^2 + 4 (D-1) \MW^4\big) B'_0(\Mh^2, \MW^2, \MW^2)
	\nn\\& \qquad {}
		- \big(\Mh^4 - 4 \Mh^2 \MZ^2 + 4 (D-1) \MZ^4\big) B'_0(\Mh^2, \MZ^2, \MZ^2)
		\biggr]
	\nn\\&
	+\frac{v_2^2}{2}(4C_{\Phi\Box}-C_{\Phi D}) \,,
	\\\nn
	\deft Z_W={}&\delta^\SM Z_{W}
	+ \frac{C_{\Phi\Box}^{\tree}}{8\pi^2(D-1)\MW^2}
		\biggl[ \Mh^2 (\Mh^2 - 2 \MW^2) B_0(\MW^2, \Mh^2, \MW^2)
	\nn\\& \qquad {}
		+ \Mh^2 (\MW^2 - \Mh^2) B_0(0, 0, \Mh^2)
		+ \MW^2 (\Mh^2 - \MW^2) B_0(0, 0, \MW^2)
	\nn\\& \qquad {}
		- \MW^2 \big(\Mh^4 - 4 \Mh^2 \MW^2 + 4 (D-1) \MW^4\big)  B'_0(\MW^2, \Mh^2, \MW^2)
		\biggr]
	\nn\\& {}
	+2v_2^2(C_{\Phi W}+C^\nonSMEFT_{2}) \,,
\\\nn
	\deft Z_{ZZ}={}&\delta^\SM Z_{ZZ}
	+ \frac{C_{\Phi\Box}^{\tree}}{8\pi^2(D-1)\MZ^2}
		\biggl[ \Mh^2 (\Mh^2 - 2 \MZ^2) B_0(\MZ^2, \Mh^2, \MZ^2)
	\nn\\& \qquad {}
		+ \Mh^2 (\MZ^2 - \Mh^2) B_0(0, 0, \Mh^2)
		+ \MZ^2 (\Mh^2 - \MZ^2) B_0(0, 0, \MZ^2)
	\nn\\& \qquad {}
		- \MZ^2 \big(\Mh^4 - 4 \Mh^2 \MZ^2 + 4 (D-1) \MZ^4\big)  B'_0(\MZ^2, \Mh^2, \MZ^2)
		\biggr]
	\nn\\& {}
	+\frac{2v_2^2}{\cw^2}
	\big[\cw^4C_{\Phi W}+\cw^2\sw^2C_{\Phi B}
	+\cw^3\sw C_{\Phi WB}+C^\nonSMEFT_{2}\big] \,,
	\\
	\deft Z_{ZA}={}&\delta^\SM Z_{ZA} \,,
	\\
	\deft Z_{AZ}={}&\delta^\SM Z_{AZ}
	+ 2 v_2^2 \left[ -2 \cw \sw C_{\Phi W} + 2 \cw \sw C_{\Phi B}
	+ (\cw^2-\sw^2) C_{\Phi WB} \right]
	=\delta^\SM Z_{AZ} \,,
	\\
	\deft Z_{AA}={}&\delta^\SM Z_{AA}
	+ 2 v_2^2 \left[ \sw^2 C_{\Phi W} + \cw^2 C_{\Phi B}
	- \cw \sw C_{\Phi WB} \right]
	=\delta^\SM Z_{AA} \,.
\end{align}

Note that the mass renormalization constants $\deft\Mh^2$, $\deft\MW^2$, and $\deft\MZ^2$
contain contributions from diagrams that originate from explicit $H$-tadpole loops in the full
SESM despite the use of the PRTS, because the $H$~propagator of the tadpole diagram
shrinks to a point, rendering the tadpole diagram irreducible.
In the full SESM, using the PRTS these contributions are fully cancelled by the
(soft parts of the) tadpole renormalization constant $\delta t_{\PH}$.
This cancellation does not take place within the mass renormalization constants
in our formulation of the EFT, since
we have absorbed $\delta_s t_{\PH}$ into $\delta v_2$ and $\delta\Mh^2$ in the
pre-renormalization step, cf.~\eqs{dMh2_preren}{dv2_preren}.
Instead, the corresponding cancellation happens between self-energy contributions and
related mass renormalization constants in the calculation of observables.

We also provide the EFT expression for the transversal part of the $W$~self-energy,
which is needed in the calculation of
$\Delta r$ in \sec{pheno}:
\begin{align}
	\Sigma^{WW}_\rT(p^2) ={}&
	\Sigma^{WW}_{\rT,\SM}(p^2)
	+ p^2\bigg[\frac{\MW^2}{8\pi^2(D-1)}C_{\Phi\Box}^{\tree}B_0(p^2,\Mh^2,\MW^2)
	-2v_2^2C_{\Phi W}-2v_2^2C^\nonSMEFT_{2}\bigg]
	\nn\\& {}
	+\frac{\MW^2}{8\pi^2(D-1)}
	C_{\Phi\Box}^{\tree}
	\Big[2 \left((2D-3)\MW^2-\Mh^2\right)B_0(p^2,\Mh^2,\MW^2)
	\nn\\&\qquad {}
	+ (2D-3)\Mh^2B_0(0,0,\Mh^2)
	-\MW^2B_0(0,0,\MW^2)\Big]
	\nn\\&
	+\frac{\MW^2(\Mh^2-\MW^2)}{8\pi^2(D-1)p^2}
	C_{\Phi\Box}^{\tree}
	\Big[(\Mh^2-\MW^2)B_0(p^2,\Mh^2,\MW^2)
	\nn\\& \qquad {}
	-\Mh^2B_0(0,0,\Mh^2)+\MW^2B_0(0,0,\MW^2)\Big]
	+ \ord(\zeta^{-4}) \,.
\end{align}

\revision{We emphasize that the results shown in this appendix are obtained with the EFT Lagrangian of \subsec{nonSMEFTLeff} formulated in the non-linear Higgs realization.
The corresponding results derived from the SMEFT Lagrangian in \subsec{SMEFTmatch}
in the linear Higgs realization differ from those given here, both in their SM and BSM parts.
For the self-energies and the renormalization constants related to the gauge bosons,
i.e.\ $\Sigma^{WW}_\rT$, $\deft\MW^2$, $\deft\MZ^2$,
$\deft Z_W$, $\deft Z_{AA}$, $\deft Z_{AZ}$, $\deft Z_{ZA}$, and $\deft Z_{ZZ}$
as given above, the transition to SMEFT in the linear Higgs realization
(using the GIVS or PRTS) can simply
be performed by replacing the corresponding SM parts and the
substitutions $C_i\to\hat C_i$ and $C^\nonSMEFT_i\to0$ for the SMEFT and non-SMEFT
Wilson coefficients.
For the Higgs-boson-related constants $\deft\Mh^2$ and $\deft Z_{hh}$,
additionally some soft contributions change in this transition.
In particular, in the linear Higgs realization for all diagrams shown
in \fig{EFTRenconst} involving W or Z-boson fields there are additional diagrams
involving the corresponding Goldstone-boson fields.
We have explicitly checked, however, that all EFT predictions discussed in
\sec{pheno} are the same using either formulation of the EFT.
}

\addcontentsline{toc}{section}{References}
\bibliographystyle{jhep}
\bibliography{largeMH}

\providecommand{\href}[2]{#2}\begingroup\raggedright\begin{thebibliography}{100}

\bibitem{Haisch:2020ahr}
U.~Haisch, M.~Ruhdorfer, E.~Salvioni, E.~Venturini and A.~Weiler,
  \emph{{Singlet night in Feynman-ville: one-loop matching of a real scalar}},
  \href{https://doi.org/10.1007/JHEP04(2020)164}{\emph{JHEP} {\bfseries 04}
  (2020) 164} [\href{https://arxiv.org/abs/2003.05936}{{\ttfamily
  2003.05936}}].

\bibitem{Carmona:2021xtq}
A.~Carmona, A.~Lazopoulos, P.~Olgoso and J.~Santiago, \emph{{Matchmakereft:
  automated tree-level and one-loop matching}},
  \href{https://doi.org/10.21468/SciPostPhys.12.6.198}{\emph{SciPost Phys.}
  {\bfseries 12} (2022) 198}
  [\href{https://arxiv.org/abs/2112.10787}{{\ttfamily 2112.10787}}].

\bibitem{DeAngelis:2023bmd}
S.~De~Angelis and G.~Durieux, \emph{{EFT matching from analyticity and
  unitarity}},
  \href{https://doi.org/10.21468/SciPostPhys.16.3.071}{\emph{SciPost Phys.}
  {\bfseries 16} (2024) 071}
  [\href{https://arxiv.org/abs/2308.00035}{{\ttfamily 2308.00035}}].

\bibitem{Li:2023edf}
X.~Li and S.~Zhou, \emph{{One-loop Matching and Running via On-shell
  Amplitudes}},  \href{https://arxiv.org/abs/2309.10851}{{\ttfamily
  2309.10851}}.

\bibitem{Chala:2024llp}
M.~Chala, J.~L{\'o}pez~Miras, J.~Santiago and F.~Vilches, \emph{{Efficient
  on-shell matching}},
  \href{https://doi.org/10.21468/SciPostPhys.18.6.185}{\emph{SciPost Phys.}
  {\bfseries 18} (2025) 185}
  [\href{https://arxiv.org/abs/2411.12798}{{\ttfamily 2411.12798}}].

\bibitem{LopezMiras:2025gar}
J.~L{\'o}pez~Miras and F.~Vilches, \emph{{Automation of a matching on-shell
  calculator}}, \href{https://doi.org/10.1016/j.cpc.2025.109935}{\emph{Comput.
  Phys. Commun.} {\bfseries 320} (2026) 109935}
  [\href{https://arxiv.org/abs/2505.21353}{{\ttfamily 2505.21353}}].

\bibitem{Dittmaier:1995cr}
S.~Dittmaier and C.~Grosse-Knetter, \emph{{Deriving nondecoupling effects of
  heavy fields from the path integral: A Heavy Higgs field in an SU(2) gauge
  theory}}, \href{https://doi.org/10.1103/PhysRevD.52.7276}{\emph{Phys. Rev. D}
  {\bfseries 52} (1995) 7276}
  [\href{https://arxiv.org/abs/hep-ph/9501285}{{\ttfamily hep-ph/9501285}}].

\bibitem{Dittmaier:1995ee}
S.~Dittmaier and C.~Grosse-Knetter, \emph{{Integrating out the standard Higgs
  field in the path integral}},
  \href{https://doi.org/10.1016/0550-3213(95)00551-X}{\emph{Nucl. Phys. B}
  {\bfseries 459} (1996) 497}
  [\href{https://arxiv.org/abs/hep-ph/9505266}{{\ttfamily hep-ph/9505266}}].

\bibitem{Drozd:2015rsp}
A.~Drozd, J.~Ellis, J.~Quevillon and T.~You, \emph{{The Universal One-Loop
  Effective Action}},
  \href{https://doi.org/10.1007/JHEP03(2016)180}{\emph{JHEP} {\bfseries 03}
  (2016) 180} [\href{https://arxiv.org/abs/1512.03003}{{\ttfamily
  1512.03003}}].

\bibitem{Boggia:2016asg}
M.~Boggia, R.~Gomez-Ambrosio and G.~Passarino, \emph{{Low energy behaviour of
  standard model extensions}},
  \href{https://doi.org/10.1007/JHEP05(2016)162}{\emph{JHEP} {\bfseries 05}
  (2016) 162} [\href{https://arxiv.org/abs/1603.03660}{{\ttfamily
  1603.03660}}].

\bibitem{Henning:2016lyp}
B.~Henning, X.~Lu and H.~Murayama, \emph{{One-loop Matching and Running with
  Covariant Derivative Expansion}},
  \href{https://doi.org/10.1007/JHEP01(2018)123}{\emph{JHEP} {\bfseries 01}
  (2018) 123} [\href{https://arxiv.org/abs/1604.01019}{{\ttfamily
  1604.01019}}].

\bibitem{Fuentes-Martin:2016uol}
J.~Fuentes-Martin, J.~Portoles and P.~Ruiz-Femenia, \emph{{Integrating out
  heavy particles with functional methods: a simplified framework}},
  \href{https://doi.org/10.1007/JHEP09(2016)156}{\emph{JHEP} {\bfseries 09}
  (2016) 156} [\href{https://arxiv.org/abs/1607.02142}{{\ttfamily
  1607.02142}}].

\bibitem{Buchalla:2016bse}
G.~Buchalla, O.~Cata, A.~Celis and C.~Krause, \emph{{Standard Model Extended by
  a Heavy Singlet: Linear vs. Nonlinear EFT}},
  \href{https://doi.org/10.1016/j.nuclphysb.2017.02.006}{\emph{Nucl. Phys. B}
  {\bfseries 917} (2017) 209}
  [\href{https://arxiv.org/abs/1608.03564}{{\ttfamily 1608.03564}}].

\bibitem{Zhang:2016pja}
Z.~Zhang, \emph{{Covariant diagrams for one-loop matching}},
  \href{https://doi.org/10.1007/JHEP05(2017)152}{\emph{JHEP} {\bfseries 05}
  (2017) 152} [\href{https://arxiv.org/abs/1610.00710}{{\ttfamily
  1610.00710}}].

\bibitem{Ellis:2017jns}
S.A.R.~Ellis, J.~Quevillon, T.~You and Z.~Zhang, \emph{{Extending the Universal
  One-Loop Effective Action: Heavy-Light Coefficients}},
  \href{https://doi.org/10.1007/JHEP08(2017)054}{\emph{JHEP} {\bfseries 08}
  (2017) 054} [\href{https://arxiv.org/abs/1706.07765}{{\ttfamily
  1706.07765}}].

\bibitem{DasBakshi:2018vni}
S.~Das~Bakshi, J.~Chakrabortty and S.K.~Patra, \emph{{CoDEx: Wilson coefficient
  calculator connecting SMEFT to UV theory}},
  \href{https://doi.org/10.1140/epjc/s10052-018-6444-2}{\emph{Eur. Phys. J. C}
  {\bfseries 79} (2019) 21} [\href{https://arxiv.org/abs/1808.04403}{{\ttfamily
  1808.04403}}].

\bibitem{Jiang:2018pbd}
M.~Jiang, N.~Craig, Y.-Y.~Li and D.~Sutherland, \emph{{Complete one-loop
  matching for a singlet scalar in the Standard Model EFT}},
  \href{https://doi.org/10.1007/JHEP02(2019)031}{\emph{JHEP} {\bfseries 02}
  (2019) 031} [\href{https://arxiv.org/abs/1811.08878}{{\ttfamily
  1811.08878}}].

\bibitem{Cohen:2020fcu}
T.~Cohen, X.~Lu and Z.~Zhang, \emph{{Functional Prescription for EFT
  Matching}}, \href{https://doi.org/10.1007/JHEP02(2021)228}{\emph{JHEP}
  {\bfseries 02} (2021) 228}
  [\href{https://arxiv.org/abs/2011.02484}{{\ttfamily 2011.02484}}].

\bibitem{Dittmaier:2021fls}
S.~Dittmaier, S.~Schuhmacher and M.~Stahlhofen, \emph{{Integrating out heavy
  fields in the path integral using the background-field method: general
  formalism}},
  \href{https://doi.org/10.1140/epjc/s10052-021-09587-7}{\emph{Eur. Phys. J. C}
  {\bfseries 81} (2021) 826}
  [\href{https://arxiv.org/abs/2102.12020}{{\ttfamily 2102.12020}}].

\bibitem{Fuentes-Martin:2022jrf}
J.~Fuentes-Mart{\'\i}n, M.~K{\"o}nig, J.~Pag{\`e}s, A.E.~Thomsen and F.~Wilsch,
  \emph{{A proof of concept for matchete: an automated tool for matching
  effective theories}},
  \href{https://doi.org/10.1140/epjc/s10052-023-11726-1}{\emph{Eur. Phys. J. C}
  {\bfseries 83} (2023) 662}
  [\href{https://arxiv.org/abs/2212.04510}{{\ttfamily 2212.04510}}].

\bibitem{Fuentes-Martin:2024agf}
J.~Fuentes-Mart{\'\i}n, A.~Moreno-S{\'a}nchez, A.~Palavri{\'c} and
  A.E.~Thomsen, \emph{{A guide to functional methods beyond one-loop order}},
  \href{https://doi.org/10.1007/JHEP08(2025)099}{\emph{JHEP} {\bfseries 08}
  (2025) 099} [\href{https://arxiv.org/abs/2412.12270}{{\ttfamily
  2412.12270}}].

\bibitem{Cohen:2020xca}
T.~Cohen, N.~Craig, X.~Lu and D.~Sutherland, \emph{{Is SMEFT Enough?}},
  \href{https://doi.org/10.1007/JHEP03(2021)237}{\emph{JHEP} {\bfseries 03}
  (2021) 237} [\href{https://arxiv.org/abs/2008.08597}{{\ttfamily
  2008.08597}}].

\bibitem{Dawson:2023oce}
S.~Dawson, D.~Fontes, C.~Quezada-Calonge and J.J.~Sanz-Cillero, \emph{{Is the
  HEFT matching unique?}},  \href{https://arxiv.org/abs/2311.16897}{{\ttfamily
  2311.16897}}.

\bibitem{Ge:2026qfa}
Z.~Ge, H.~Song and X.~Wan, \emph{{Establishing the Primary HEFT as a Precision
  Benchmark for UV-HEFT Matching}},
  \href{https://arxiv.org/abs/2602.14418}{{\ttfamily 2602.14418}}.

\bibitem{Asiain:2026sio}
{\'I}.~Asi{\'a}in, R.~Gr{\"o}ber and L.~Tiberi, \emph{{Is the Standard Model
  Effective Field Theory Enough for Higgs Pair Production?}},
  \href{https://arxiv.org/abs/2602.16288}{{\ttfamily 2602.16288}}.

\bibitem{Lee:1972yfa}
B.~Lee and J.~Zinn-Justin, \emph{{Spontaneously Broken Gauge Symmetries Part 3:
  Equivalence}}, \href{https://doi.org/10.1103/PhysRevD.5.3155}{\emph{Phys.
  Rev. D} {\bfseries 5} (1972) 3155}.

\bibitem{Grosse-Knetter:1992tbp}
C.~Grosse-Knetter and R.~{K\"ogerler}, \emph{{Unitary gauge, Stuckelberg
  formalism and gauge invariant models for effective lagrangians}},
  \href{https://doi.org/10.1103/PhysRevD.48.2865}{\emph{Phys. Rev. D}
  {\bfseries 48} (1993) 2865}
  [\href{https://arxiv.org/abs/hep-ph/9212268}{{\ttfamily hep-ph/9212268}}].

\bibitem{Dittmaier:2022maf}
S.~Dittmaier and H.~Rzehak, \emph{{Electroweak renormalization based on
  gauge-invariant vacuum expectation values of non-linear Higgs
  representations. Part I. Standard Model}},
  \href{https://doi.org/10.1007/JHEP05(2022)125}{\emph{JHEP} {\bfseries 05}
  (2022) 125} [\href{https://arxiv.org/abs/2203.07236}{{\ttfamily
  2203.07236}}].

\bibitem{Buchmuller:1985jz}
W.~{Buchm\"uller} and D.~Wyler, \emph{{Effective Lagrangian Analysis of New
  Interactions and Flavor Conservation}},
  \href{https://doi.org/10.1016/0550-3213(86)90262-2}{\emph{Nucl. Phys. B}
  {\bfseries 268} (1986) 621}.

\bibitem{Grzadkowski:2010es}
B.~Grzadkowski, M.~Iskrzynski, M.~Misiak and J.~Rosiek, \emph{{Dimension-Six
  Terms in the Standard Model Lagrangian}},
  \href{https://doi.org/10.1007/JHEP10(2010)085}{\emph{JHEP} {\bfseries 10}
  (2010) 085} [\href{https://arxiv.org/abs/1008.4884}{{\ttfamily 1008.4884}}].

\bibitem{LHCHiggsCrossSectionWorkingGroup:2013rie}
{\scshape LHC Higgs Cross Section Working Group} collaboration, \emph{{Handbook
  of LHC Higgs Cross Sections: 3. Higgs Properties}},
  \href{https://arxiv.org/abs/1307.1347}{{\ttfamily 1307.1347}}.

\bibitem{LHCHiggsCrossSectionWorkingGroup:2016ypw}
{\scshape LHC Higgs Cross Section Working Group} collaboration, \emph{{Handbook
  of LHC Higgs Cross Sections: 4. Deciphering the Nature of the Higgs Sector}},
  \href{https://doi.org/10.23731/CYRM-2017-002}{\emph{CERN Yellow Rep. Monogr.}
  {\bfseries 2} (2017) 1} [\href{https://arxiv.org/abs/1610.07922}{{\ttfamily
  1610.07922}}].

\bibitem{Brivio:2017vri}
I.~Brivio and M.~Trott, \emph{{The Standard Model as an Effective Field
  Theory}}, \href{https://doi.org/10.1016/j.physrep.2018.11.002}{\emph{Phys.
  Rept.} {\bfseries 793} (2019) 1}
  [\href{https://arxiv.org/abs/1706.08945}{{\ttfamily 1706.08945}}].

\bibitem{Grinstein:2007iv}
B.~Grinstein and M.~Trott, \emph{{A Higgs-Higgs bound state due to new physics
  at a TeV}}, \href{https://doi.org/10.1103/PhysRevD.76.073002}{\emph{Phys.
  Rev. D} {\bfseries 76} (2007) 073002}
  [\href{https://arxiv.org/abs/0704.1505}{{\ttfamily 0704.1505}}].

\bibitem{Alonso:2012px}
R.~Alonso, M.B.~Gavela, L.~Merlo, S.~Rigolin and J.~Yepes, \emph{{The Effective
  Chiral Lagrangian for a Light Dynamical ''Higgs Particle''}},
  \href{https://doi.org/10.1016/j.physletb.2013.04.037}{\emph{Phys. Lett. B}
  {\bfseries 722} (2013) 330}
  [\href{https://arxiv.org/abs/1212.3305}{{\ttfamily 1212.3305}}].

\bibitem{Contino:2013kra}
R.~Contino, M.~Ghezzi, C.~Grojean, M.~Muhlleitner and M.~Spira,
  \emph{{Effective Lagrangian for a light Higgs-like scalar}},
  \href{https://doi.org/10.1007/JHEP07(2013)035}{\emph{JHEP} {\bfseries 07}
  (2013) 035} [\href{https://arxiv.org/abs/1303.3876}{{\ttfamily 1303.3876}}].

\bibitem{Buchalla:2013rka}
G.~Buchalla, O.~Cat{\`a} and C.~Krause, \emph{{Complete Electroweak Chiral
  Lagrangian with a Light Higgs at NLO}},
  \href{https://doi.org/10.1016/j.nuclphysb.2014.01.018}{\emph{Nucl. Phys. B}
  {\bfseries 880} (2014) 552}
  [\href{https://arxiv.org/abs/1307.5017}{{\ttfamily 1307.5017}}].

\bibitem{Brivio:2013pma}
I.~Brivio, T.~Corbett, O.J.P.~{\'E}boli, M.B.~Gavela, J.~Gonzalez-Fraile,
  M.C.~Gonzalez-Garcia et~al., \emph{{Disentangling a dynamical Higgs}},
  \href{https://doi.org/10.1007/JHEP03(2014)024}{\emph{JHEP} {\bfseries 03}
  (2014) 024} [\href{https://arxiv.org/abs/1311.1823}{{\ttfamily 1311.1823}}].

\bibitem{Ghezzi:2015vva}
M.~Ghezzi, R.~Gomez-Ambrosio, G.~Passarino and S.~Uccirati, \emph{{NLO Higgs
  effective field theory and {\ensuremath{\kappa}}-framework}},
  \href{https://doi.org/10.1007/JHEP07(2015)175}{\emph{JHEP} {\bfseries 07}
  (2015) 175} [\href{https://arxiv.org/abs/1505.03706}{{\ttfamily
  1505.03706}}].

\bibitem{Brivio:2016fzo}
I.~Brivio, J.~Gonzalez-Fraile, M.C.~Gonzalez-Garcia and L.~Merlo, \emph{{The
  complete HEFT Lagrangian after the LHC Run I}},
  \href{https://doi.org/10.1140/epjc/s10052-016-4211-9}{\emph{Eur. Phys. J. C}
  {\bfseries 76} (2016) 416}
  [\href{https://arxiv.org/abs/1604.06801}{{\ttfamily 1604.06801}}].

\bibitem{Alonso:2015fsp}
R.~Alonso, E.E.~Jenkins and A.V.~Manohar, \emph{{A Geometric Formulation of
  Higgs Effective Field Theory: Measuring the Curvature of Scalar Field
  Space}}, \href{https://doi.org/10.1016/j.physletb.2016.01.041}{\emph{Phys.
  Lett. B} {\bfseries 754} (2016) 335}
  [\href{https://arxiv.org/abs/1511.00724}{{\ttfamily 1511.00724}}].

\bibitem{Alonso:2016oah}
R.~Alonso, E.E.~Jenkins and A.V.~Manohar, \emph{{Geometry of the Scalar
  Sector}}, \href{https://doi.org/10.1007/JHEP08(2016)101}{\emph{JHEP}
  {\bfseries 08} (2016) 101}
  [\href{https://arxiv.org/abs/1605.03602}{{\ttfamily 1605.03602}}].

\bibitem{Helset:2020yio}
A.~Helset, A.~Martin and M.~Trott, \emph{{The Geometric Standard Model
  Effective Field Theory}},
  \href{https://doi.org/10.1007/JHEP03(2020)163}{\emph{JHEP} {\bfseries 03}
  (2020) 163} [\href{https://arxiv.org/abs/2001.01453}{{\ttfamily
  2001.01453}}].

\bibitem{Schabinger:2005ei}
R.M.~Schabinger and J.D.~Wells, \emph{{A Minimal spontaneously broken hidden
  sector and its impact on Higgs boson physics at the large hadron collider}},
  \href{https://doi.org/10.1103/PhysRevD.72.093007}{\emph{Phys. Rev. D}
  {\bfseries 72} (2005) 093007}
  [\href{https://arxiv.org/abs/hep-ph/0509209}{{\ttfamily hep-ph/0509209}}].

\bibitem{Patt:2006fw}
B.~Patt and F.~Wilczek, \emph{{Higgs-field portal into hidden sectors}},
  \href{https://arxiv.org/abs/hep-ph/0605188}{{\ttfamily hep-ph/0605188}}.

\bibitem{Bowen:2007ia}
M.~Bowen, Y.~Cui and J.D.~Wells, \emph{{Narrow trans-TeV Higgs bosons and H
  $\to$ hh decays: Two LHC search paths for a hidden sector Higgs boson}},
  \href{https://doi.org/10.1088/1126-6708/2007/03/036}{\emph{JHEP} {\bfseries
  03} (2007) 036} [\href{https://arxiv.org/abs/hep-ph/0701035}{{\ttfamily
  hep-ph/0701035}}].

\bibitem{Pruna:2013bma}
G.M.~Pruna and T.~Robens, \emph{{Higgs singlet extension parameter space in the
  light of the LHC discovery}},
  \href{https://doi.org/10.1103/PhysRevD.88.115012}{\emph{Phys. Rev. D}
  {\bfseries 88} (2013) 115012}
  [\href{https://arxiv.org/abs/1303.1150}{{\ttfamily 1303.1150}}].

\bibitem{Kanemura:2015fra}
S.~Kanemura, M.~Kikuchi and K.~Yagyu, \emph{{Radiative corrections to the Higgs
  boson couplings in the model with an additional real singlet scalar field}},
  \href{https://doi.org/10.1016/j.nuclphysb.2016.04.005}{\emph{Nucl. Phys. B}
  {\bfseries 907} (2016) 286}
  [\href{https://arxiv.org/abs/1511.06211}{{\ttfamily 1511.06211}}].

\bibitem{Bojarski:2015kra}
F.~Bojarski, G.~Chalons, D.~Lopez-Val and T.~Robens, \emph{{Heavy to light
  Higgs boson decays at NLO in the Singlet Extension of the Standard Model}},
  \href{https://doi.org/10.1007/JHEP02(2016)147}{\emph{JHEP} {\bfseries 02}
  (2016) 147} [\href{https://arxiv.org/abs/1511.08120}{{\ttfamily
  1511.08120}}].

\bibitem{Altenkamp:2018bcs}
L.~Altenkamp, M.~Boggia and S.~Dittmaier, \emph{{Precision calculations for $h
  \to WW/ZZ \to 4$ fermions in a Singlet Extension of the Standard Model with
  Prophecy4f}}, \href{https://doi.org/10.1007/JHEP04(2018)062}{\emph{JHEP}
  {\bfseries 04} (2018) 062}
  [\href{https://arxiv.org/abs/1801.07291}{{\ttfamily 1801.07291}}].

\bibitem{Denner:2018opp}
A.~Denner, S.~Dittmaier and J.-N.~Lang, \emph{{Renormalization of mixing
  angles}}, \href{https://doi.org/10.1007/JHEP11(2018)104}{\emph{JHEP}
  {\bfseries 11} (2018) 104}
  [\href{https://arxiv.org/abs/1808.03466}{{\ttfamily 1808.03466}}].

\bibitem{DeWitt:1967ub}
B.S.~DeWitt, \emph{{Quantum Theory of Gravity. 2. The Manifestly Covariant
  Theory}}, \href{https://doi.org/10.1103/PhysRev.162.1195}{\emph{Phys. Rev.}
  {\bfseries 162} (1967) 1195}.

\bibitem{DeWitt:1980jv}
B.S.~DeWitt, \emph{{A gauge invariant effective action}},  in \emph{{Oxford
  Conference on Quantum Gravity}}, pp.~449--487, 7, 1980.

\bibitem{Abbott:1980hw}
L.~Abbott, \emph{{The Background Field Method Beyond One Loop}},
  \href{https://doi.org/10.1016/0550-3213(81)90371-0}{\emph{Nucl. Phys. B}
  {\bfseries 185} (1981) 189}.

\bibitem{Denner:1994xt}
A.~Denner, G.~Weiglein and S.~Dittmaier, \emph{{Application of the background
  field method to the electroweak standard model}},
  \href{https://doi.org/10.1016/0550-3213(95)00037-S}{\emph{Nucl. Phys. B}
  {\bfseries 440} (1995) 95}
  [\href{https://arxiv.org/abs/hep-ph/9410338}{{\ttfamily hep-ph/9410338}}].

\bibitem{Beneke:1997zp}
M.~Beneke and V.A.~Smirnov, \emph{{Asymptotic expansion of Feynman integrals
  near threshold}},
  \href{https://doi.org/10.1016/S0550-3213(98)00138-2}{\emph{Nucl. Phys. B}
  {\bfseries 522} (1998) 321}
  [\href{https://arxiv.org/abs/hep-ph/9711391}{{\ttfamily hep-ph/9711391}}].

\bibitem{Smirnov:2002pj}
V.A.~Smirnov, \emph{{Applied asymptotic expansions in momenta and masses}},
  {\emph{Springer Tracts Mod. Phys.} {\bfseries 177} (2002) 1}.

\bibitem{Dittmaier:2022ivi}
S.~Dittmaier and H.~Rzehak, \emph{{Electroweak renormalization based on
  gauge-invariant vacuum expectation values of non-linear Higgs
  representations. Part II. Extended Higgs sectors}},
  \href{https://doi.org/10.1007/JHEP08(2022)245}{\emph{JHEP} {\bfseries 08}
  (2022) 245} [\href{https://arxiv.org/abs/2206.01479}{{\ttfamily
  2206.01479}}].

\bibitem{Bohm:1986rj}
M.~B{\"o}hm, H.~Spiesberger and W.~Hollik, \emph{{On the 1-Loop Renormalization
  of the Electroweak Standard Model and its Application to Leptonic
  Processes}}, \href{https://doi.org/10.1002/prop.19860341102}{\emph{Fortsch.
  Phys.} {\bfseries 34} (1986) 687}.

\bibitem{Denner:1991kt}
A.~Denner, \emph{{Techniques for calculation of electroweak radiative
  corrections at the one loop level and results for W physics at LEP-200}},
  \href{https://doi.org/10.1002/prop.2190410402}{\emph{Fortsch. Phys.}
  {\bfseries 41} (1993) 307} [\href{https://arxiv.org/abs/0709.1075}{{\ttfamily
  0709.1075}}].

\bibitem{Denner:2019vbn}
A.~Denner and S.~Dittmaier, \emph{{Electroweak Radiative Corrections for
  Collider Physics}},
  \href{https://doi.org/10.1016/j.physrep.2020.04.001}{\emph{Phys. Rept.}
  {\bfseries 864} (2020) 1} [\href{https://arxiv.org/abs/1912.06823}{{\ttfamily
  1912.06823}}].

\bibitem{Appelquist:1974tg}
T.~Appelquist and J.~Carazzone, \emph{{Infrared Singularities and Massive
  Fields}}, \href{https://doi.org/10.1103/PhysRevD.11.2856}{\emph{Phys. Rev. D}
  {\bfseries 11} (1975) 2856}.

\bibitem{FERBER2024104105}
T.~Ferber, A.~Grohsjean and F.~Kahlhoefer, \emph{Dark higgs bosons at
  colliders},
  \href{https://doi.org/https://doi.org/10.1016/j.ppnp.2024.104105}{\emph{Progress
  in Particle and Nuclear Physics} {\bfseries 136} (2024) 104105}.

\bibitem{cms_collaboration_2026_rn1g0-73405}
C.~Collaboration, \emph{Higgs singlet model (tan$\beta$ = 0.1)},  Mar., 2026.
\newblock 10.17181/rn1g0-73405.

\bibitem{cms_collaboration_2026_90xht-tpc60}
C.~Collaboration, \emph{Higgs singlet model (tan$\beta$ = 4.0)},  Mar., 2026.
\newblock 10.17181/90xht-tpc60.

\bibitem{Stueckelberg:1938zz}
E.~Stueckelberg, \emph{{Interaction forces in electrodynamics and in the field
  theory of nuclear forces}}, {\emph{Helv. Phys. Acta} {\bfseries 11} (1938)
  299}.

\bibitem{Stueckelberg:1957zz}
E.~Stueckelberg, \emph{{Theory of the radiation of photons of small arbitrary
  mass}}, {\emph{Helv. Phys. Acta} {\bfseries 30} (1957) 209}.

\bibitem{Kunimasa:1967zza}
T.~Kunimasa and T.~Goto, \emph{{Generalization of the Stueckelberg Formalism to
  the Massive Yang-Mills Field}},
  \href{https://doi.org/10.1143/PTP.37.452}{\emph{Prog. Theor. Phys.}
  {\bfseries 37} (1967) 452}.

\bibitem{Denner:2017vms}
A.~Denner, J.-N.~Lang and S.~Uccirati, \emph{{NLO electroweak corrections in
  extended Higgs Sectors with RECOLA2}},
  \href{https://doi.org/10.1007/JHEP07(2017)087}{\emph{JHEP} {\bfseries 07}
  (2017) 087} [\href{https://arxiv.org/abs/1705.06053}{{\ttfamily
  1705.06053}}].

\bibitem{Denner:2016etu}
A.~Denner, L.~Jenniches, J.-N.~Lang and C.~Sturm, \emph{{Gauge-independent
  $\overline{MS}$ renormalization in the 2HDM}},
  \href{https://doi.org/10.1007/JHEP09(2016)115}{\emph{JHEP} {\bfseries 09}
  (2016) 115} [\href{https://arxiv.org/abs/1607.07352}{{\ttfamily
  1607.07352}}].

\bibitem{Sirlin:1985ux}
A.~Sirlin and R.~Zucchini, \emph{{Dependence of the Quartic Coupling H(m) on
  M($H$) and the Possible Onset of New Physics in the Higgs Sector of the
  Standard Model}},
  \href{https://doi.org/10.1016/0550-3213(86)90096-9}{\emph{Nucl. Phys. B}
  {\bfseries 266} (1986) 389}.

\bibitem{Fleischer:1980ub}
J.~Fleischer and F.~Jegerlehner, \emph{{Radiative Corrections to Higgs Decays
  in the Extended Weinberg-Salam Model}},
  \href{https://doi.org/10.1103/PhysRevD.23.2001}{\emph{Phys. Rev. D}
  {\bfseries 23} (1981) 2001}.

\bibitem{Chisholm:1961tha}
J.S.R.~Chisholm, \emph{{Change of variables in quantum field theories}},
  \href{https://doi.org/10.1016/0029-5582(61)90106-7}{\emph{Nucl. Phys.}
  {\bfseries 26} (1961) 469}.

\bibitem{Kamefuchi:1961sb}
S.~Kamefuchi, L.~O'Raifeartaigh and A.~Salam, \emph{{Change of variables and
  equivalence theorems in quantum field theories}},
  \href{https://doi.org/10.1016/0029-5582(61)90056-6}{\emph{Nucl. Phys.}
  {\bfseries 28} (1961) 529}.

\bibitem{Passarino:2016saj}
G.~Passarino, \emph{{Field reparametrization in effective field theories}},
  \href{https://doi.org/10.1140/epjp/i2017-11291-5}{\emph{Eur. Phys. J. Plus}
  {\bfseries 132} (2017) 16}
  [\href{https://arxiv.org/abs/1610.09618}{{\ttfamily 1610.09618}}].

\bibitem{Cohen:2023ekv}
T.~Cohen, X.~Lu and D.~Sutherland, \emph{{On amplitudes and field
  redefinitions}}, \href{https://doi.org/10.1007/JHEP06(2024)149}{\emph{JHEP}
  {\bfseries 06} (2024) 149}
  [\href{https://arxiv.org/abs/2312.06748}{{\ttfamily 2312.06748}}].

\bibitem{Manohar:2018aog}
A.V.~Manohar, \emph{{Introduction to Effective Field Theories}},
  \href{https://arxiv.org/abs/1804.05863}{{\ttfamily 1804.05863}}.

\bibitem{Criado:2018sdb}
J.C.~Criado and M.~P{\'e}rez-Victoria, \emph{{Field redefinitions in effective
  theories at higher orders}},
  \href{https://doi.org/10.1007/JHEP03(2019)038}{\emph{JHEP} {\bfseries 03}
  (2019) 038} [\href{https://arxiv.org/abs/1811.09413}{{\ttfamily
  1811.09413}}].

\bibitem{LONGHITANO1981118}
A.C.~Longhitano, \emph{Low-energy impact of a heavy higgs boson sector},
  \href{https://doi.org/https://doi.org/10.1016/0550-3213(81)90109-7}{\emph{Nuclear
  Physics B} {\bfseries 188} (1981) 118}.

\bibitem{Grojean:2013kd}
C.~Grojean, E.E.~Jenkins, A.V.~Manohar and M.~Trott, \emph{{Renormalization
  Group Scaling of Higgs Operators and $\Gamma(h \to\gamma\gamma)$}},
  \href{https://doi.org/10.1007/JHEP04(2013)016}{\emph{JHEP} {\bfseries 04}
  (2013) 016} [\href{https://arxiv.org/abs/1301.2588}{{\ttfamily 1301.2588}}].

\bibitem{Elias-Miro:2013gya}
J.~Elias-Mir\'o, J.R.~Espinosa, E.~Masso and A.~Pomarol, \emph{{Renormalization
  of dimension-six operators relevant for the Higgs decays $h\rightarrow
  \gamma\gamma,\gamma Z$}},
  \href{https://doi.org/10.1007/JHEP08(2013)033}{\emph{JHEP} {\bfseries 08}
  (2013) 033} [\href{https://arxiv.org/abs/1302.5661}{{\ttfamily 1302.5661}}].

\bibitem{Elias-Miro:2013mua}
J.~Elias-Miro, J.R.~Espinosa, E.~Masso and A.~Pomarol, \emph{{Higgs windows to
  new physics through d=6 operators: constraints and one-loop anomalous
  dimensions}}, \href{https://doi.org/10.1007/JHEP11(2013)066}{\emph{JHEP}
  {\bfseries 11} (2013) 066} [\href{https://arxiv.org/abs/1308.1879}{{\ttfamily
  1308.1879}}].

\bibitem{Jenkins:2013zja}
E.E.~Jenkins, A.V.~Manohar and M.~Trott, \emph{{Renormalization Group Evolution
  of the Standard Model Dimension Six Operators I: Formalism and lambda
  Dependence}}, \href{https://doi.org/10.1007/JHEP10(2013)087}{\emph{JHEP}
  {\bfseries 10} (2013) 087} [\href{https://arxiv.org/abs/1308.2627}{{\ttfamily
  1308.2627}}].

\bibitem{Jenkins:2013wua}
E.E.~Jenkins, A.V.~Manohar and M.~Trott, \emph{{Renormalization Group Evolution
  of the Standard Model Dimension Six Operators II: Yukawa Dependence}},
  \href{https://doi.org/10.1007/JHEP01(2014)035}{\emph{JHEP} {\bfseries 01}
  (2014) 035} [\href{https://arxiv.org/abs/1310.4838}{{\ttfamily 1310.4838}}].

\bibitem{Alonso:2013hga}
R.~Alonso, E.E.~Jenkins, A.V.~Manohar and M.~Trott, \emph{{Renormalization
  Group Evolution of the Standard Model Dimension Six Operators III: Gauge
  Coupling Dependence and Phenomenology}},
  \href{https://doi.org/10.1007/JHEP04(2014)159}{\emph{JHEP} {\bfseries 04}
  (2014) 159} [\href{https://arxiv.org/abs/1312.2014}{{\ttfamily 1312.2014}}].

\bibitem{Alonso:2014zka}
R.~Alonso, H.-M.~Chang, E.E.~Jenkins, A.V.~Manohar and B.~Shotwell,
  \emph{{Renormalization group evolution of dimension-six baryon number
  violating operators}},
  \href{https://doi.org/10.1016/j.physletb.2014.05.065}{\emph{Phys. Lett. B}
  {\bfseries 734} (2014) 302}
  [\href{https://arxiv.org/abs/1405.0486}{{\ttfamily 1405.0486}}].

\bibitem{Sirlin:1980nh}
A.~Sirlin, \emph{{Radiative Corrections in the SU(2)-L x U(1) Theory: A Simple
  Renormalization Framework}},
  \href{https://doi.org/10.1103/PhysRevD.22.971}{\emph{Phys. Rev. D} {\bfseries
  22} (1980) 971}.

\bibitem{Awramik:2003rn}
M.~Awramik, M.~Czakon, A.~Freitas and G.~Weiglein, \emph{{Precise prediction
  for the W boson mass in the standard model}},
  \href{https://doi.org/10.1103/PhysRevD.69.053006}{\emph{Phys.\ Rev.\ D}
  {\bfseries 69} (2004) 053006}
  [\href{https://arxiv.org/abs/hep-ph/0311148}{{\ttfamily hep-ph/0311148}}].

\bibitem{ParticleDataGroup:2024cfk}
{\scshape Particle Data Group} collaboration, \emph{{Review of particle
  physics}}, \href{https://doi.org/10.1103/PhysRevD.110.030001}{\emph{Phys.
  Rev. D} {\bfseries 110} (2024) 030001}.

\bibitem{Altarelli:1989hv}
G.~Altarelli, R.~Kleiss and C.~Verzegnassi, eds., \emph{{Z~physics at LEP~1.
  Vol. 1: Standard Physics}}, (Geneva), CERN, 1989.
\newblock 10.5170/CERN-1989-008-V-1.

\bibitem{Bardin:1997xq}
D.Y.~Bardin et~al., \emph{{Electroweak working group report}},  in
  \emph{{Reports of the Working Group on Precision Calculations for the
  Z~Resonance}}, {D. Bardin, W. Hollik, G. Passarino}, ed., pp.~7--162, 1995,
  \href{https://doi.org/10.5170/CERN-1995-003}{DOI}
  [\href{https://arxiv.org/abs/hep-ph/9709229}{{\ttfamily hep-ph/9709229}}].

\bibitem{Bardin:1999gt}
D.Y.~Bardin, M.~Gr{\"u}newald and G.~Passarino, ``{Precision calculation
  project report}.'' 1999.

\bibitem{Dubovyk:2019szj}
I.~Dubovyk, A.~Freitas, J.~Gluza, T.~Riemann and J.~Usovitsch,
  \emph{{Electroweak pseudo-observables and Z-boson form factors at two-loop
  accuracy}}, \href{https://doi.org/10.1007/JHEP08(2019)113}{\emph{JHEP}
  {\bfseries 08} (2019) 113}
  [\href{https://arxiv.org/abs/1906.08815}{{\ttfamily 1906.08815}}].

\bibitem{Bredenstein:2006rh}
A.~Bredenstein, A.~Denner, S.~Dittmaier and M.M.~Weber, \emph{{Precise
  predictions for the Higgs-boson decay $H \to WW/ZZ \to4\,$leptons}},
  \href{https://doi.org/10.1103/PhysRevD.74.013004}{\emph{Phys. Rev. D}
  {\bfseries 74} (2006) 013004}
  [\href{https://arxiv.org/abs/hep-ph/0604011}{{\ttfamily hep-ph/0604011}}].

\bibitem{Bredenstein:2006ha}
A.~Bredenstein, A.~Denner, S.~Dittmaier and M.M.~Weber, \emph{{Radiative
  corrections to the semileptonic and hadronic Higgs-boson decays $H \to WW /
  ZZ \to4\,$fermions}},
  \href{https://doi.org/10.1088/1126-6708/2007/02/080}{\emph{JHEP} {\bfseries
  02} (2007) 080} [\href{https://arxiv.org/abs/hep-ph/0611234}{{\ttfamily
  hep-ph/0611234}}].

\bibitem{Denner:2019fcr}
A.~Denner, S.~Dittmaier and A.~M{\"u}ck, \emph{{PROPHECY4F 3.0: A Monte Carlo
  program for Higgs-boson decays into four-fermion final states in and beyond
  the Standard Model}},
  \href{https://doi.org/10.1016/j.cpc.2020.107336}{\emph{Comput. Phys. Commun.}
  {\bfseries 254} (2020) 107336}
  [\href{https://arxiv.org/abs/1912.02010}{{\ttfamily 1912.02010}}].

\bibitem{Altenkamp:2017kxk}
L.~Altenkamp, S.~Dittmaier and H.~Rzehak, \emph{{Precision calculations for $h
  \to WW/ZZ \to 4$ fermions in the Two-Higgs-Doublet Model with Prophecy4f}},
  \href{https://doi.org/10.1007/JHEP03(2018)110}{\emph{JHEP} {\bfseries 03}
  (2018) 110} [\href{https://arxiv.org/abs/1710.07598}{{\ttfamily
  1710.07598}}].

\bibitem{Altenkamp:2017ldc}
L.~Altenkamp, S.~Dittmaier and H.~Rzehak, \emph{{Renormalization schemes for
  the Two-Higgs-Doublet Model and applications to h \textrightarrow{} WW/ZZ
  \textrightarrow{} 4 fermions}},
  \href{https://doi.org/10.1007/JHEP09(2017)134}{\emph{JHEP} {\bfseries 09}
  (2017) 134} [\href{https://arxiv.org/abs/1704.02645}{{\ttfamily
  1704.02645}}].

\bibitem{Denner:2005fg}
A.~Denner, S.~Dittmaier, M.~Roth and L.H.~Wieders, \emph{{Electroweak
  corrections to charged-current $e^+ e^- \to 4\,$fermion processes: Technical
  details and further results}},
  \href{https://doi.org/10.1016/j.nuclphysb.2011.09.001}{\emph{Nucl. Phys. B}
  {\bfseries 724} (2005) 247}
  [\href{https://arxiv.org/abs/hep-ph/0505042}{{\ttfamily hep-ph/0505042}}].

\bibitem{Buchalla:2020kdh}
G.~Buchalla, O.~Cat{\`a}, A.~Celis, M.~Knecht and C.~Krause,
  \emph{{Higgs-electroweak chiral Lagrangian: One-loop renormalization group
  equations}}, \href{https://doi.org/10.1103/PhysRevD.104.076005}{\emph{Phys.
  Rev. D} {\bfseries 104} (2021) 076005}
  [\href{https://arxiv.org/abs/2004.11348}{{\ttfamily 2004.11348}}].

\bibitem{Anisha:2021hgc}
Anisha, S.~Das~Bakshi, S.~Banerjee, A.~Biek{\"o}tter, J.~Chakrabortty,
  S.~Kumar~Patra et~al., \emph{{Effective limits on single scalar extensions in
  the light of recent LHC data}},
  \href{https://doi.org/10.1103/PhysRevD.107.055028}{\emph{Phys. Rev. D}
  {\bfseries 107} (2023) 055028}
  [\href{https://arxiv.org/abs/2111.05876}{{\ttfamily 2111.05876}}].

\bibitem{Banerjee:2023qbg}
U.~Banerjee, J.~Chakrabortty, C.~Englert, W.~Naskar, S.U.~Rahaman and
  M.~Spannowsky, \emph{{EFT, decoupling, Higgs boson mixing, and higher
  dimensional operators}},
  \href{https://doi.org/10.1103/PhysRevD.109.055035}{\emph{Phys. Rev. D}
  {\bfseries 109} (2024) 055035}
  [\href{https://arxiv.org/abs/2303.05224}{{\ttfamily 2303.05224}}].

\end{thebibliography}\endgroup

\end{document}